\PassOptionsToPackage{dvipsnames}{xcolor}
\documentclass[12pt]{book}
\usepackage[height=8.85in,width=6.45in]{geometry}

\usepackage{times}

\pdfoutput=1

\usepackage[utf8]{inputenc}
\usepackage[svgnames]{xcolor}
\usepackage{graphicx}
\usepackage{appendix}

\usepackage{ascmac}
\usepackage{amsmath}
\usepackage{amssymb}
\usepackage{mathtools}
\usepackage{bm}

\usepackage{braket}
\usepackage{cancel}
\usepackage{fancybox}
\usepackage{slashed}

\usepackage{multirow}
\usepackage{tikz}
\usepackage{tikz-cd}
\usetikzlibrary{fit}
\usetikzlibrary{shapes.misc}
\usetikzlibrary{decorations.pathmorphing}

\usepackage{array}
\usepackage{extarrows}
\usepackage{colortbl}
\usepackage{cite}
\usepackage
[
	bookmarks=true,
	colorlinks=true,
	citecolor=refcolor,
	linkcolor=refcolor
]{hyperref}
\usepackage{subfig}
\usetikzlibrary{shapes.geometric}
\usetikzlibrary{positioning}
\usepackage[sectionbib]{chapterbib}
\usepackage{floatrow}
\DeclareColorBox{gray}{\colorbox{black!7}}
\usepackage{dashbox}
\usepackage{CJKutf8}

\usepackage{booktabs}

\definecolor{Blue}{rgb}{0,0,1}
\definecolor{blue}{rgb}{0.0, 0.4, 1}
\definecolor{darkgreen}{rgb}{0.,0.6,0.}
\definecolor{lightyellow}{rgb}{1.0, 0.95, 0.7}
\definecolor{lightblue}{rgb}{0.7, 0.9, 1.0}
\definecolor{lightpink}{rgb}{1.0, 0.85, 0.95}
\definecolor{lightgreen}{rgb}{0.7, 1.0, 0.4}
\definecolor{refcolor}{rgb}{0.3,0.3,1}

\newcommand*{\magenta}[1]{\textcolor{magenta}{#1}}

\newcommand*{\green}[1]{\textcolor{darkgreen}{#1}}

\definecolor{colorF}{rgb}{0,0.72,0.92}
\definecolor{colorG}{rgb}{0,0,1}
\definecolor{colorH}{rgb}{0.72, 0, 0.92}
\definecolor{colorI}{rgb}{0.92, 0.2, 0.8}

\definecolor{colorA}{rgb}{0,0,1}
\definecolor{colorB}{rgb}{1,0.5,0}
\definecolor{colorC}{rgb}{0,0.72,0.92}
\definecolor{colorD}{rgb}{1,0.5,0}
\definecolor{colorE}{rgb}{0,0.72,0.92}

\usepackage{tikz}
\usetikzlibrary{patterns}
\usetikzlibrary{decorations.markings}
\usetikzlibrary{decorations.pathmorphing}
\tikzset{snake it/.style={decorate, decoration=snake}}
\usetikzlibrary{matrix,calc}

\usetikzlibrary{quotes,angles}

\tikzset{middlearrow/.style={
        decoration={markings,
            mark= at position 0.5 with {\arrow{#1}} ,
        },
        postaction={decorate}
    }
}

\makeatletter
\newcommand{\mylabel}[2]{#2\def\@currentlabel{#2}\label{#1}}
\makeatother

\newcommand{\RomanNumeralCaps}[1]
    {\MakeUppercase{\romannumeral #1}}

\usepackage{colortbl}
\definecolor{shadecolor}{rgb}{0.90,0.90,0.90}

\newtheorem{axiom_}{Axiom}
{\begin{axiom_}\begin{shaded}}%
{\end{shaded}\end{axiom_}}

\usepackage{blkarray}

\def\ssymb#1{\mbox{\strut\rlap{\smash{\scriptsize$#1$}}\quad}}
\usepackage{nicematrix}
\NiceMatrixOptions%
{code-for-first-row = \scriptstyle,
code-for-first-col = \scriptstyle }

\edef\restoreparindent{\parindent=\the\parindent\relax}
\usepackage{parskip}
\restoreparindent
\usepackage{ascmac}
\usepackage{mathrsfs}
\usepackage{tikz}
\usetikzlibrary{intersections, calc,positioning,decorations.pathreplacing,decorations.pathmorphing}
\tikzset{>=latex}

\def\d{{\rm d}}

\def\i{{\rm i}}

\def\CA{{\cal A}}

\def\CD{{\cal D}}

\def\CJ{{\cal J}}
\def\CR{{\cal R}}

\def\CO{{\cal O}}

\def\CS{{\cal S}}

\def\CW{{\cal W}}

\def\b0{\bm{0}_\perp}

\usepackage{cancel}

\usepackage{fancyhdr}
\usepackage[multiple]{footmisc}

\begin{document}

\begin{titlepage}

	\begin{center}
		Doctoral Dissertation\vspace{2mm}\\
		\begin{CJK}{UTF8}{ipxm}
			博士論文
		\end{CJK}
	\end{center}
	
	\bigskip
	
	\begin{center}
	{\Large {\bfseries
		Aspects of critical O$(N)$ model with boundary and defect}\\
		\vspace*{20pt}
		\begin{CJK}{UTF8}{ipxm}
			\large
			(境界や欠陥を持つ臨界O$(N)$模型の諸相)
		\end{CJK}
	}\\

	\vspace*{270pt}

	Submitted for the Degree of Doctor of Philosophy on December 2023\vspace{2mm}\\
	\begin{CJK}{UTF8}{ipxm}
		令和5年12月 博士 (理学) 申請
	\end{CJK}
	
	\vspace*{50pt}
	
	Department of Physics, Graduate School of Science,
	University of Tokyo\vspace{2mm}\\
	\begin{CJK}{UTF8}{ipxm}
		東京大学 大学院 理学系研究科 物理学専攻
	\end{CJK}
	
	\bigskip
	
	\large
	Yoshitaka Okuyama\vspace{2mm}\\
	\begin{CJK}{UTF8}{ipxm}
	奥山 義隆
	\end{CJK}
	\end{center}
	
\end{titlepage}

\newpage
\thispagestyle{empty}

\vspace*{40pt}
\begin{center}
	\textbf{Abstract}
\end{center}
In this thesis, we explore the critical phenomena in the presence of extended objects, which we call \emph{defects}, aiming for a better understanding of the properties of non-local objects ubiquitous in our world and a more practical and realistic study of criticality. 
To this end, we study the statistical O$(N)$ vector model in $(4-\epsilon)$ dimensions with three kinds of defects: a line defect constructed by smearing an O$(N)$ vector field along one direction and Dirichlet and Neumann boundaries. 
A conventional approach to critical phenomena would be to perform perturbative calculations using Feynman diagrams and doing renormalization group analysis. But we here also take a different but complementary approach based on three axioms that include conformal symmetry of the theory at the criticality. We apply this axiomatic framework to the critical O$(N)$ model with a defect and reproduce the perturbative results at the leading non-trivial order in $\epsilon$, substantiating the validity of our approach. Along the way, we develop and refine the axiomatic framework to derive anomalous dimensions of the composite operators on the defect that have not been accessible in the existing literature by focusing on the analyticity of the correlation functions. 

\setcounter{tocdepth}{2}
\tableofcontents
\thispagestyle{empty}

\addtocontents{toc}{\protect\thispagestyle{empty}}

\chapter{Introduction}\label{chap:Introduction}

Critical phenomena have been one of the central subjects in science since the first discovery by Andrews in 1869, who observed critical opalescence of the carbon dioxide at about $31^\circ$C and 73atm. Thanks to their enriched properties, they have been a pivotal milestone in extending the horizon of and bridging across different branches of physics, ranging from condensed matter to statistical physics, particle physics, and string theory. 

The critical opalescence at the second-order phase transition, which Andrews observed, is due to the fluctuations at all scales allowed by the container and the lack of the typical length scale of the system. This scale invariance of critical phenomena was made clear theoretically in going through Onsager's exact solution of two-dimensional statistical Ising model \cite{Onsager:1943jn} and Wilson's renormalization group (RG) studies of critical phenomena \cite{Wilson:1973jj,Wilson:1971dc} where all beta functions vanish on the RG fixed point and the theory becomes scale invariant.
The other notable feature of the critical phenomena is the power-law behaviors of physical observables near critical points, such as specific heat and magnetization of magnets. Their exponents, called critical exponents, are insensitive to the microscopic details of the system. Many systems, which have different microscopic descriptions, tend to fall into the same universality class at criticality, as is implied by Wilson's RG analysis. These insights have opened the way to study the critical phenomena of magnets through the $\phi^4$-model of Euclidean Quantum Field Theory (QFT), or statistical field theory.

Polyakov proposed in \cite{Polyakov:1970xd} that one can use conformal invariance, a local version of scale invariance, as a guiding principle to investigate critical phenomena.\footnote{In many models, one expects the enhancement from scale invariance to conformal invariance. There are many arguments on the differences between scale and conformal invariance. See \cite{Nakayama:2013is} and references therein for details on this point. The conformal invariance of the critical Ising model is very plausible but not shown rigorously except in two dimensions \cite{smirnov2007conformal}.}
Conformally invariant Quantum Field Theory is called Conformal Field Theory (CFT) and has led to the success of the field theoretical study of two-dimensional critical phenomena by the authors of \cite{Zamolodchikov:1986gt} making full use of the infinite-dimensional conformal symmetry in two dimensions.
There had been less progress for years in three or more dimensions where, unlike two-dimensional spacetime, the conformal symmetry is finite-dimensional. However, the authors of \cite{Rattazzi:2008pe} revived the conformal bootstrap program \cite{Mack:1975jr,Ferrara:1973yt,Polyakov:1974gs} and made possible the numerical studies of crossing equations of conformal four-point functions. Afterward, we have seen significant numerical and analytical progress, prompting more precise theoretical prediction of critical exponents than ever down to the fifth or sixth decimal place (see \cite{Poland:2018epd} and references therein).

For more detailed tests of theoretical prediction to experimental data, one cannot ignore the finite size effect due to the container, impurity, and other extended objects, which we call defects \cite{Andrei:2018die}. Defect Conformal Field Theory (DCFT) is a framework to study critical phenomena in the presence of a defect \cite{Billo:2016cpy,Gadde:2016fbj}. We may particularly call Boundary Conformal Field Theory (BCFT) in the case of a boundary. In DCFT, we first make the planer or spherical approximation of the defect. The validity of this approximation comes from the insensitivity of critical phenomena to the microscopic structure of the system. One can always utilize conformal transformation to make a spherical defect into a planer shape (up to Weyl factors). Hence, it is often the case that one only considers planer defects, and so do we in this thesis. The defect at criticality partially breaks the conformal symmetry down to its subgroup and seems to lessen the predictivity of the theory. But they make the structure of the critical systems more amusing. First of all, for each bulk universal class, there exist several defect universality classes, making the phase diagram richer. Correspondingly, critical exponents and operator spectrum on the defect are generically different from those in the bulk. Moreover, two-point functions of bulk fields in DCFT already depend on cross ratios and contain model-dependent information of bulk CFTs, unlike ordinary CFTs where dynamical dependence comes in from four- and higher-point functions. 

The other motivation to study DCFT is to grasp and classify extended objects that have played pivotal roles in physics, such as Wilson and 't Hooft operators serving as probes of confinement in gauge theories \cite{Wilson:1974sk,tHooft:1977nqb}, spin impurities that drive the Kondo effect \cite{Kondo:1964,Wilson:1974mb,Affleck:1995ge}, and D-branes in string theory \cite{Dai:1989ua,Horava:1989ga}. As DCFTs lie at the endpoints of the RG flow of QFTs with extended objects, uncovering the structure of DCFTs leads to the classification of non-local objects in general QFTs in light of defect central charges \cite{Herzog:2015ioa,Herzog:2017kkj,Herzog:2017xha,FarajiAstaneh:2021foi,Jensen:2018rxu,Chalabi:2021jud} and defect $C$-theorems \cite{Kobayashi:2018lil,Jensen:2015swa,Cuomo:2021rkm,Wang:2021mdq}. All these imply that studying DCFT would facilitate the theoretical and experimental understanding of our world.

The O$(N)$ vector model is a simple but significant model of statistical field theory.
Its bare action in $d=4-\epsilon$ dimensions is given by
\begin{align}
    I = \int_{\mathbb{R}^{d}}\,\d^d x \,\left[\frac{1}{2}\, |\partial \Phi_{1} |^2\, +\,   \frac{\lambda_0}{4!}\, |\Phi_1|^4 \right] \, ,\qquad |\Phi_1|^4 \equiv (\Phi_{1}^{\alpha}\Phi_{1}^{\alpha})^2\ , 
\end{align}
with $\Phi_{1}^{\alpha}$ ($\alpha=1,\cdots, N$) being an O$(N)$ vector field.
This O$(N)$ vector model has played a benchmarking role in exploring critical phenomena (see \cite{Henriksson:2022rnm} for a detailed review), and the same goes even in the presence of defects. 
One of the IR fixed points of this model without defects under the RG flow is called the Wilson-Fisher fixed point, which covers a wide range of systems at criticality, giving good agreements with experimental data: statistical Ising model, simple molecular fluids such as water and carbon dioxide, and binary alloy for $N=1$, XY ferromagnets and statistical XY model for $N=2$, and statistical Heisenberg model and isotropic magnets for $N=3$.\footnote{We note that the Wilson-Fisher fixed point in three dimensions is unstable for $N>N_{\text{crit.}}$ due to the presence of a relevant O$(N)$ tensorial operator. As that tensorial operator breaks the O$(N)$ down to the cubic symmetry group, this phenomenon is termed cubic symmetry breaking. Conformal boostrap reveals numerically that $N_{\text{crit.}}<3$ \cite{Chester:2020iyt}. We also remark here that the critical O$(2)$ model is expected to describe the lambda point of liquid ${}^4\text{He}$, but with a minor but meaningful discrepancy between theoretical predictions and experimental measurements \cite{lipa2003specific}. The interested readers are referred to \cite{Chester:2019ifh} for this topic.} One can analytically continue the positive integer parameter $N$ to $-2$. Then, the O$(N)$ vector model with $N=0$ ($N=-2$) can be associated with the model of self-avoiding (loop-erased) random walks.

The boundary critical phenomena of the O$(N)$ vector model have a long history and numerous theoretical studies have been done so far (see e.g., \cite{Bray:1977tk,Reeve1980,Cardy:1984bb,Diehl:1996kd,Diehl:1981zz,Diehl:1981jgg,McAvity:1995zd,McAvity:1993ue} and references therein).\footnote{See also \cite{Padayasi:2021sik,Metlitski:2020cqy,Toldin:2023fny} for recent developments of critical O$(N)$ with boundary.} Previous studies have shown that, even with a Dirichlet (free) boundary, one still finds that theoretical predictions agree well with experimental realizations for $N=1$ \cite{PhysRevLett.57.2191,PhysRevLett.71.1188}, $N=2$ \cite{Dosch1991SynchrotronXS} and $N=3$ \cite{PhysRevLett.48.51} (see table \ref{tab:on surface exponent}). We can tell from these papers that the experimental measurement of the defect (or boundary and surface) critical exponents relies highly on how we identify macroscopic observables on the defect of our interest within the experimental setup. Hence, more acute predictions of the critical exponents on defects will lead to a deeper insight into the microscopic features of the experimental systems with extended objects.
\begin{table}[t]
    \centering
   \scalebox{0.85}{\begin{tabular}{c|c|c|c}
         & $N=1$ & $N=2$ & $N=3$ \\\hline
           Theoretical prediction  \cite{Reeve1980,Diehl:1996kd}    & 0.79 & 0.81 & 0.82 \\\hline
            Experimental data & \begin{tabular}{c}
               $0.83\pm0.05$ \cite{PhysRevLett.57.2191} \\ $0.8\pm0.1$ \cite{PhysRevLett.71.1188}  \\
                  
            \end{tabular} &  $0.75\pm 0.02$ \cite{Dosch1991SynchrotronXS} & $0.825^{+0.025}_{-0.040}$ \cite{PhysRevLett.48.51}\\\hline
    \end{tabular}}
    \caption{For the surface critical exponent $\widehat{\beta}$ defined for the Dirichlet boundary as in \eqref{eq:surface magnetization definition}, we list experimental data in three dimensions for $N=1,2,3$ and the standard perturbative results in $d=4-\epsilon$ dimensional spacetime at order $\epsilon^2$ extrapolated to three dimensions by setting $\epsilon=1$. Surprisingly, the theoretical estimate agrees well with the experimental measurements, even at the second order in perturbation theory (see \cite[section 6.4]{LACES2021} for a review on this point).}
   \label{tab:on surface exponent}
\end{table}

For the past ten years, various types of defects have been introduced in the O$(N)$ model, such as monodromy \cite{Billo:2013jda,Gaiotto:2013nva}, line \cite{Allais:2014fqa,Cuomo:2021kfm}, surface \cite{Trepanier:2023tvb,Giombi:2023dqs} and Replica twist defects \cite{SoderbergRousu:2023pbe}.\footnote{The readers who are interested in recent developments of conformal bootstrap and numerical studies using fuzzy sphere regularization \cite{Zhu:2022gjc,Hu:2023xak,Han:2023yyb,Han:2023lky} are referred to \cite{Liendo:2012hy,Gimenez-Grau:2022czc} and \cite{Hu:2023ghk}, respectively.} These researches have been serving as testing grounds for further consideration of defects in theoretical physics. See \cite{Barrat:2023ivo} for models of fermionic theories, \cite{Cuomo:2022xgw,Aharony:2023amq,Aharony:2022ntz} for spin impurities in Wilson-Fisher and gauge theories, and \cite{Giombi:2018hsx,Giombi:2018qox} for integrable models of DCFT.
Moreover, the Ising model ($N=1$) with a line defect has the potential for experimental realization \cite{PhysRevB.95.014401} and quantum simulation \cite{Ebadi:2020ldi}.
This thesis aims to refine theoretical studies of the critical behavior of defects using the critical O$(N)$ vector model for a better understanding of extended objects in nature.

To this end, we focus on the axiomatic framework of the critical phenomena in this thesis and explore DCFT data of the O$(N)$ model in $(4-\epsilon)$ dimensions with line defect and Neumann/Dirichlet boundary.\footnote{We are ultimately interested in the DCFT data of the O$(N)$ model in three dimensions with line defect and Neumann/Dirichlet boundary. Hence, throughout our analysis, we will fix the dimension of the defect to one in the line defect case and the co-dimension of the defect to one for the Neumann/Dirichlet boundary. When $N=1$, the critical O$(N)$ model with a line defect in three dimensions also describes the quantum critical phenomena of the two-dimensional transverse Ising model with a point-like impurity.} The authors of \cite{Rychkov:2015naa} proposed the axiomatic framework to test the compatibility of the two approaches: conformal bootstrap and the epsilon expansion via conventional perturbative calculations based on Feynman diagrams and RG analysis. 
Their original paper postulated three mild assumptions, including conformal symmetry. And they derived anomalous dimensions of the operators in the critical O$(N)$ model in $(4-\epsilon)$ dimensions up to the leading non-trivial order in $\epsilon$. Their methodology is applied to numerous models \cite{Basu:2015gpa,Ghosh:2015opa,Raju:2015fza,Nii:2016lpa,Giombi:2017rhm}, and later generalized to the case with a monodromy defect in \cite{Yamaguchi:2016pbj,Soderberg:2017oaa} and subsequently to boundary and interface \cite{Giombi:2020rmc,Dey:2020jlc,Herzog:2022jlx}.
As this axiomatic approach is manifestly conformal invariant, it gives us a complementary look compared with the standard perturbative framework that only postulates the scale invariance of the theory. Besides, it is also a powerful analytical tool to study critical phenomena in the presence of a defect where numerical bootstrap techniques are not so effective as, unlike ordinary CFTs without defects, one fails to solve defect crossing equations \eqref{eq:defect crossing equation} numerically by applying the semidefinite programming methods due to technical reasons (see e.g., \cite[section 4.1]{Liendo:2012hy} for details on this point).

All these existing researches of axiomatic approach only deal with the conformal dimensions of the lowest-lying defect local operators and fail to talk about composite operators on the defect. In this thesis, we demonstrate that it is possible to access anomalous dimensions of defect composite operators within the axiomatic framework by relying on the analyticity of correlation functions in Euclidean QFTs.
The interacting O$(N)$ model with a line defect is not scale invariant unless the defect coupling constant takes an appropriate value \cite{Allais:2014fqa,Cuomo:2021kfm}. We have successfully singled out the critical defect coupling using our axioms. We have also calculated the conformal dimensions of various operators on the line defect and checked that they agree with the perturbative results.
No research addressed conformal dimensions of composite operators on the Neumann/Dirichlet boundary except the lowest-lying one in the Neumann case. In this thesis, we have investigated a family of boundary composite operators in the conventional perturbative framework and the axiomatic approach and confirmed their agreements.

\begin{table}[ht]
\centering
\renewcommand{\arraystretch}{1.4}
\begin{tabular}{cccccc}
\toprule
Operators  & Dimension & SO$(d-1)$ rep. & O$(N-1)$ rep. & Free limit & Notes \\ \midrule
$\widehat{W}_1^{\,1}$  & \eqref{conf dim: defect local scalar}  & scalar & singlet &$\widehat{\Phi}^{\,1}_{1} $ &-----\\
$\widehat{W}_1^{\hat{\alpha}}$  & \eqref{conf dim: defect local scalar}  & scalar & vector&$\widehat{\Phi}_{1}^{\hat{\alpha}}$ & only for $N\geq 2$\\
 $\widehat{U}_{i_1\cdots i_s}^{\,1}$  & \eqref{eq:conf dim:transverse spin}  & tensor & singlet &$\widehat{\Phi}_{s+1,i_1\cdots i_s}^{\,1}$ &-----\\
$\widehat{U}_{i_1\cdots i_s}^{\,\hat{\alpha}}$  & \eqref{eq:conf dim:transverse spin}  & tensor & vector &$\widehat{\Phi}_{s+1,i_1\cdots i_s}^{\,\hat{\alpha}}$& only for $N\geq 2$\\
$\widehat{S}_\pm$  & \eqref{conf dim: scalars}  & scalar & singlet &$\{ |\widehat{\Phi}_{1}^{\, 1}|^2, |\widehat{\Phi}_{1}^{\, \hat{\alpha}}|^2 \}$ &only for $N\geq 2$\\
$\widehat{V}^{\hat{\alpha}}$  &\eqref{conf dim: vector} & scalar & vector &$\widehat{\Phi}_{1}^{\,1}\widehat{\Phi}_{1}^{\,\hat{\alpha}}$ & only for $N\geq 2$ \\
$\widehat{T}^{\hat{\alpha}\hat{\beta}}$ & \eqref{conf dim: tensor} & scalar & tensor & $\displaystyle\widehat{\Phi}_{1}^{\,(\hat{\alpha}}\widehat{\Phi}_{1}^{\,\hat{\beta})}$ & only for $N\geq 3$\\
$\widehat{W}_p$ & \eqref{eq:Ising DCFT composite} & scalar & ----- & $\displaystyle|\Phi_{1}|^p$ & only for $N=1$\\
\bottomrule
\end{tabular}
\caption{List of defect local operators treated in this thesis with the classification based on the representation theory of the residual symmetry group on the line defect \eqref{eq:residual symmetry group on the defect}. The hatted operators in the free theory side are defect local ones defined by bringing the corresponding bulk fields closer to the line defect such as $\widehat{\Phi}^{\,1}_{1}=\lim_{|x_\perp|\to0}\,\Phi^{\,1}_{1}$ with $|x_\perp|$ being the perpendicular distance from the defect. See section \ref{chap:line defect} for details of their definitions and treatments.
The free limits of $\widehat{S}_+$ and $\widehat{S}_-$ are the linear combinations of $|\widehat{\Phi}_{1}^{\, 1}|^2$ and $|\widehat{\Phi}_{1}^{\, \hat{\alpha}}|^2$.}
\label{tab:list of anomalous dimensions line defect}
\end{table}

\paragraph{Summary of the result.}
We here summarize our results:
\begin{itemize}
    \item We utilize axiomatic framework to study the critical O$(N)$ model in $(4-\epsilon)$ dimensions. To be more specific, we did the following analysis at the leading order in $\epsilon$:
    \begin{itemize}
     \item[-] We invent a methodology to study the composite operator spectrum on the defect by requiring the removal of the unphysical singularities of bulk-defect-defect three-point correlators.
        \item[-]  For a line defect, we have derived the critical defect coupling constant \eqref{eq:aphi in O(N) model WFFP} and the conformal dimensions of various defect local operators listed in table \ref{tab:list of anomalous dimensions line defect} and reproduced the perturbative results \cite{Cuomo:2021kfm}.
           \item[-]  In the case of the Dirichlet/Neumann boundary, we have computed the conformal dimensions of the boundary local operators as summarized in table \ref{TableEps2}. We have also performed perturbative calculations to confirm the validity of our machinery.
    \end{itemize}
\end{itemize}

\begin{table}[t]
\renewcommand{\arraystretch}{2}
\centering
\begin{tabular}{>{\centering}m{2.7cm}>{\centering}m{2.5cm}>{\centering}m{6cm}>{\centering\arraybackslash}m{2.5cm}}
\toprule
    Boundary condition & Boundary operators & 
    Conformal dimension & Free limit 
    \\
  \midrule 
    &$\widehat{W}_{2p}$  & $\displaystyle 2p +\frac{6p\, (2p-3)}{N+8}\,\epsilon$ &$\widehat{\Phi}_{2p}$  \\
    \multirow{-2}{*}{Neumann} &$\widehat{W}_{2p+1}^{\,\alpha}$  & $\displaystyle  2p+1-\frac{N+6p\,(1-2p)+5}{N+8}\,\epsilon$ &$\widehat{\Phi}_{2p+1}^{\,\alpha}$  \\
    \hline
    &$\widehat{\CW}_{4p}$  & $\displaystyle 4p-\frac{p\, (N-6p+14)}{N+8}\,\epsilon$ &$\widehat{\Psi}_{4p}$  \\
    \multirow{-2}{*}{Dirichlet}&$\widehat{\CW}_{4p+2}^{\,\alpha}$  & $\displaystyle  4p+2-\frac{N-6p^2 +p\,(N+8)+5}{N+8}\,\epsilon$ &$\widehat{\Psi}_{4p+2}^{\,\alpha}$  \\
    \bottomrule
\end{tabular}
\caption{Listed are the conformal dimensions of boundary local operators subject to the Neumann and Dirichlet boundary conditions. See \eqref{eq:Neumann lowest-lying defect local def} and \eqref{eq:Neumann composite defect local def} for the definitions of boundary local operators in free theory for the Neumann case. Under Dirichlet boundary conditions, we define the free theory operators as in \eqref{eq:Dirichlet lowest-lying defect local def} and \eqref{eq:Dirichlet composite defect local def}.
}
\label{TableEps2}
\end{table}

\subsection*{Organization of the thesis}
The rest of this thesis consists of seven chapters and three appendices. The first three chapters are preliminaries for our analysis. In chapter \ref{chap:Foundations of conformal symmetry without defects} and \ref{chap:Elements of Defect Conformal Field Theory}, we introduce standard technologies of CFT and DCFT. In chapter \ref{chap:Review of Rychkov-Tan}, we describe the axiomatic framework with a quick review of the original paper \cite{Rychkov:2015naa} and clarify how to modify it in the presence of a defect. We then proceed to our original work and investigate DCFT data of the critical O$(N)$ model with line defect (chapter \ref{chap:line defect}), Neumann boundary (chapter \ref{chap:Neumann boundary}), and Dirichlet boundary (chapter \ref{chap:Dirichlet}). At the beginning of each chapter (section \ref{sec:line defect known} and the preface of chapter \ref{chap:Neumann boundary} and \ref{chap:Dirichlet}), we attach the review of previous studies of the model while setting notations for the operators of our interest. Chapter \ref{chap:line defect} is based on \cite{Nishioka:2022qmj}, whereas chapter \ref{chap:Neumann boundary} and \ref{chap:Dirichlet} are on \cite{Nishioka:2022odm}.

Appendix \ref{app:Useful identities} enumerates several identities used in this thesis. We record a detailed review of the unitarity bound in appendix \ref{app:unitatiry bound}. Appendix \ref{app:Conformal block expansion of bulk-defect-defect three-point function} is for derivations of the conformal block expansions of bulk-defect-defect three-point functions. 

\newpage

\subsection*{Special notes}
This thesis is based on the following papers
\begin{center}
	\begin{tabular}{ll}
 \cite{Nishioka:2022qmj}&
T.\,Nishioka, Y.\,Okuyama, S.\,Shimamori,\\
&``The epsilon expansion of the O$(N)$ model with line defect from conformal field theory,''\\
&		 \href{https://arxiv.org/abs/2212.04076}{
			\magenta{arXiv:2212.04076 [hep-th]}},
\href{https://link.springer.com/article/10.1007/JHEP03(2023)203}{
			\green{JHEP \textbf{03} (2023), 203}}\\
\cite{Nishioka:2022odm}&
T.\,Nishioka, Y.\,Okuyama, S.\,Shimamori,\\
&``Comments on epsilon expansion of the O$(N)$ model with boundary,''\\
&		 \href{https://arxiv.org/abs/2212.04078}{
			\magenta{arXiv:2212.04078 [hep-th]}},
\href{https://link.springer.com/article/10.1007/JHEP03(2023)051}{
			\green{JHEP \textbf{03} (2023), 051}}
	\end{tabular}
\end{center}

\chapter{Foundations of conformal symmetry without defects}\label{chap:Foundations of conformal symmetry without defects}
This chapter is devoted to a brief review of conformal constraints on Euclidean QFT in three and more dimensions $\mathbb{R}^{d\geq3}$, referencing several review articles \cite{Simmons-Duffin:2016gjk,Rychkov:2016iqz,Polchinski:1998rq}. We first introduce conformal symmetry in flat Euclidean spacetime (section \ref{sec:Conformal transformation and conformal algebra}) and deliver a conventional way to describe local excitations in CFT by primary and descendant operators (section \ref{sec:Local operators in conformal field theory}).
We then describe the notion of path integral in QFT on $\mathbb{R}^d$ (section \ref{sec:Path integral in QFT}) and expand on how it is utilized in CFT, leading to principal concepts such as radial quantization (section \ref{sec:Path integral in CFT}), state/operator correspondence (section \ref{sec:State/operator correspondence CFT}) and Operator Product Expansion (section \ref{sec:Operator Product Expansion}). Finally, we illustrate how correlation functions of local operators are constrained from conformal symmetry using the celebrated embedding space formalism (section \ref{sec:Correlation functions in CFT}).

\section{Conformal transformation and conformal algebra}\label{sec:Conformal transformation and conformal algebra}
Conformal transformations are coordinate transformations $x^\mu\mapsto x'^\mu$ that keep angles between any pair of curves. In $d\geq3$ dimensional flat Euclidean spacetime $\mathbb{R}^{d}$ with the metric $\delta_{\mu\nu}=\mathrm{diag}(1,\cdots,1)$ ($\mu,\nu=1,\cdots,d$), they consist of translations, rotations, dilatation and special conformal transformations (SCTs), which are enumerated with their corresponding generators in bold font as follows (see e.g., \cite{DiFrancesco:1997nk,Simmons-Duffin:2016gjk,Rychkov:2016iqz}):
\begin{align}
    \text{translation : }& \mathbf{P}_\mu & x^\mu\mapsto x'^\mu=&x^\mu+ a^\mu\ ,\\
       \text{rotation : }& \mathbf{M}_{\mu\nu} & x^\mu\mapsto x'^\mu=&w^{\mu\nu}\,x_\nu\quad \text{with}\quad w^{\mu\nu}=-w^{\nu\mu}\ ,\\
           \text{dilatation : }& \mathbf{D} & x^\mu\mapsto x'^\mu=&\lambda\,x^\mu\ ,\\
               \text{SCT : }& \mathbf{K}_\mu & x^\mu\mapsto x'^\mu=&\frac{x^\mu+x^2\,b^\mu}{1+2b\cdot x+b^2x^2}\ .
\end{align}
Notice that SCT can move a point on $\mathbb{R}^{d}$ to infinity. Hence, for finite conformal transformations to be transitive, we are led to add a point at infinity $\{\infty\}$ to $\mathbb{R}^d$ and consider the one-point compactification of Euclidean spacetime $\widehat{\mathbb{R}}^d=\mathbb{R}^d\cup\{\infty\}\cong\mathbb{S}^d$.\footnote{This isomorphism follows from the standard stereographic projection.}

The generators of the conformal group are subject to the following commutation relations:\footnote{We work in the same notations for conformal generators as in \cite{Simmons-Duffin:2016gjk}.}
\begin{align}\label{eq:com rel conf}
\begin{aligned}
        [\mathbf{M}_{\mu\nu}, \mathbf{P}_{\rho}]&=\delta_{\nu\rho}\,\mathbf{P}_{\mu}-\delta_{\mu\rho}\,\mathbf{P}_{\nu}\ ,\\
    [\mathbf{M}_{\mu\nu}, \mathbf{K}_{\rho}]&=\delta_{\nu\rho}\,\mathbf{P}_{\mu}-\delta_{\mu\rho}\,\mathbf{K}_{\nu}\ ,\\
      [\mathbf{M}_{\mu\nu},\mathbf{M}_{\rho\sigma}]&=\delta_{\nu\rho}\,\mathbf{M}_{\mu\sigma}-\delta_{\mu\rho}\,\mathbf{M}_{\nu\sigma}+\delta_{\mu\sigma}\,\mathbf{M}_{\nu\rho}-\delta_{\nu\sigma}\,\mathbf{M}_{\mu\rho}\ ,\\
         [\mathbf{D}, \mathbf{P}_{\mu}]&=\mathbf{P}_{\mu}\ , \\
                  [\mathbf{D}, \mathbf{K}_{\mu}]&=-\mathbf{K}_{\mu}\ ,\\
          [\mathbf{P}_{\mu}, \mathbf{K}_{\nu}]&=2\,\delta_{\mu\nu}\,\mathbf{D}-2\,\mathbf{M}_{\mu\nu}\ ,\\
            [\mathbf{D}, \mathbf{M}_{\mu\nu}]&=  [\mathbf{P}_{\mu}, \mathbf{P}_{\nu}]=  [\mathbf{K}_{\mu}, \mathbf{K}_{\nu}]=0\ .
\end{aligned}
\end{align}
One can organize them into the following $(d+2)\times(d+2)$ antisymmetric matrix $\mathbf{J}_{MN}$:
\begin{align}
  \mathbf{J}_{MN}= \begin{pNiceMatrix}[first-row,first-col]
\ssymb{\substack{\ \\ \ \\ M\backslash\, \\\downarrow}} \ssymb{\substack{ N\to\\ \, }} & -1 & 0 & \nu \\
-1 & 0 & \mathbf{D} & \frac{1}{2}(\mathbf{P}_\nu-\mathbf{K}_\nu) \\
0 & -\mathbf{D}& 0 & \frac{1}{2}(\mathbf{P}_\nu+\mathbf{K}_\nu) \\
\mu &-\frac{1}{2}(\mathbf{P}_\mu-\mathbf{K}_\mu)& - \frac{1}{2}(\mathbf{P}_\mu+\mathbf{K}_\mu) & \mathbf{M}_{\mu\nu}
\end{pNiceMatrix}\ ,
\end{align}
satisfying the commutation relations of $\text{SO}(1,d+1)$:
\begin{align}
    [\mathbf{J}_{KL},\mathbf{J}_{MN}]=\eta_{LM}\,\mathbf{J}_{KN}-\eta_{KM}\,\mathbf{J}_{LN}+\eta_{KN}\,\mathbf{J}_{LM}-\eta_{LN}\,\mathbf{J}_{KM}\ ,
\end{align}
with the metric $\eta_{MN}=\text{diag}(-1,1,1,\cdots ,1)$.
Hence, the conformal group in $d$-dimensional Eulidean spacetime $\mathbb{R}^{d}$ is isomorphic to the Lorentz group in $(d+2)$-dimensional Minkowski spacetime $\mathbb{R}^{1,d+1}$.

\section{Local operators in conformal field theory}\label{sec:Local operators in conformal field theory}
Next, let us consider how the generators of the conformal group act on local operators $\CO(x)$. Because all conformal transformations other than translations leave the origin of spacetime invariant, we first specify how $\mathbf{D},\mathbf{M}_{\mu\nu}$ and $\mathbf{K}_\mu$ act on the operators at the origin and then recover their actions at general position using the standard relation:
\begin{align}\label{eq:translation law}
    \CO(x)=e^{x\cdot \mathbf{P}}\,\CO(0)\,e^{-x\cdot \mathbf{P}}\ ,\qquad  [\mathbf{P}_{\mu} , \CO(x)]=\partial_{\mu} \,\CO(x)\ .
\end{align}
Because $[\mathbf{D},\mathbf{M}_{\mu\nu}]=0$, we can simultaneously diagonalize the action of $\mathbf{D}$ and $\mathbf{M}_{\mu\nu}$ on a local operator:
\begin{align}
      [\mathbf{D} , \CO(0)]&=\Delta \,\CO(0)\ , \label{eq:dilatation acting on O} \\
            [\mathbf{M}_{\mu\nu} , \CO(0)]&=\mathcal{S}_{\mu\nu} \,\CO(0)\ .\label{eq:rotation acting on O}
\end{align}
Here, $\Delta$ is referred to as \textbf{conformal dimension} of $\CO$ and $\mathcal{S}_{\mu\nu}$ is a spin matrix of $\mathrm{SO}(d)$ irreducible representation associated with $\CO$.
It follows from the commutation relations \eqref{eq:com rel conf} and Jacobi identity $[A,[B,C]]+[B,[C,A]]+[C,[A,B]]=0$ that the action of $\mathbf{P}_{\mu}$ ($\mathbf{K}_{\mu}$) on an operator increments (decrements) its conformal dimensions by one:
\begin{align}
    \begin{aligned}
      [\mathbf{D},  [\mathbf{P}_{\mu} , \CO(0)] ]&=[\mathbf{P}_{\mu},  [\mathbf{D} , \CO(0)] ]+[[\mathbf{D} ,\mathbf{P}_{\mu}] ,\CO(0) ]=(\Delta+1)\,[\mathbf{P}_{\mu} , \CO(0)]\ ,\\
            [\mathbf{D},  [\mathbf{K}_{\mu} , \CO(0)] ]&=[\mathbf{K}_{\mu},  [\mathbf{D} , \CO(0)] ]+[[\mathbf{D} ,\mathbf{K}_{\mu}] ,\CO(0) ]=(\Delta-1)\,[\mathbf{K}_{\mu} , \CO(0)]\ .
    \end{aligned}
\end{align}

Unitarity requires conformal dimensions to be bounded below. First of all, for any operators, we have $\Delta\geq 0$. And the operator with $\Delta=0$ corresponds to the identity operator $\bm{1}$. Furthermore, when $d\geq2$, the conformal dimension of an $\mathrm{SO}(d)$ symmetric traceless tensor of rank-$J$ must satisfy the inequality (\textbf{unitarity bound}) (see \cite{Dobrev:1977qv,Mack:1975je,Minwalla:1997ka} and appendix \ref{app:unitatiry bound} for a review):
\begin{align}\label{eq:unitarity bound}
\Delta\geq \begin{dcases}
    d/2-1 & \text{for } J=0\\
       d+J-2 & \text{for } J=1,2,\cdots 
\end{dcases}\ ,
\end{align}
where the equalities are satisfied for free Klein-Golden fields $\CO_{\Delta}$ ($J=0$) and conserved currents $\CO_{\Delta,\mu_1\cdots\mu_{J}}$ ($J=1,2,\cdots$) such as the stress tensor $T_{\mu\nu}$ ($J=2$):
\begin{align}
    \partial_\mu\partial^\mu\,\CO_{\Delta}(x)&=0\quad   \text{ for}\quad\Delta=d/2-1\ , J=0\ ,\\
       \partial_{\mu_1}\CO_{\Delta}^{\mu_1\mu_2\cdots\mu_J}(x)&=0\quad  \text{ for}\quad \Delta=d+J-2\ , J=1,2,\cdots\ .
\end{align}
In other words, the conformal dimensions of the Klein-Golden field and conserved currents are \emph{protected} and do not change from the canonical (engineering) dimensions in free field theories, even with interactions.

The unitarity bound implies that, for any operators, there must be some positive integer $n$ such that
\begin{align}
    [\mathbf{K}_{\mu_1},\cdots,[ \mathbf{K}_{\mu_n},\CO(0) ]\cdots]=0\ .
\end{align}
If not, there must be operators whose conformal dimensions are lower than the bound required by unitarity, leading to a contradiction. 

Consider a class of operators that are annihilated by the SCT generator $\mathbf{K}_{\mu}$ and call them \textbf{primary operators}, or simply \textbf{primaries}:
\begin{align}\label{eq:SCT acting on O}
    [\mathbf{K}_{\mu},\CO(0) ]=0\ \qquad \text{for a primary operator } \CO\ .
\end{align}
The operators made out from a primary $\CO$ by acting $\mathbf{P}_{\mu}$'s are called \textbf{descendants} of $\CO$. A family of operators consisting of a primary $\CO$ and its descendants is termed \textbf{conformal multiplet} of $\CO$. In CFT, one can express arbitrary local operators as a linear combination of primaries and descendants. Because primaries are more fundamental than descendants, we focus on primary operators later in this thesis.

From \eqref{eq:translation law} and \eqref{eq:com rel conf}, we find that the infinitesimal transformations act on a primary $\CO(x)$ at generic position in the following manner \cite{Mack:1969rr}:
\begin{align}\label{eq:infinitesimal phys}
\begin{aligned}
       [\mathbf{P}_\mu , \CO(x)]&=\partial_{\mu} \,\CO(x)\ ,\\
       [\mathbf{M}_{\mu\nu} , \CO(x)]&=\left(x_\nu\, \partial_{\mu}-x_\mu\, \partial_{\nu}+\mathcal{S}_{\mu\nu}\right) \,\CO(x)\ ,\\
      [\mathbf{D} , \CO(x)]&=\left(x\cdot \partial+\Delta\right) \,\CO(x)\ ,\\
               [\mathbf{K}_\mu , \CO(x)]&=[2x_\mu\,(x\cdot \partial+\Delta)-x^2\,\partial_{\mu}^2+x^\nu\,\mathcal{S}_{\mu\nu}]\,\CO(x)\ .
\end{aligned}
\end{align}
The above expressions are complicated and clumsy, particularly due to the quadratic terms in $x$. However, as will be explained in section \ref{sec:Correlation functions in CFT}, the embedding space formalism provides us a unified look at the conformal transformation laws of primary operators and helps to calculate their correlation functions.

\section{Path integral in quantum field theory}\label{sec:Path integral in QFT}
We here review the basics of the path integral approach and the equal-time quantization in Euclidean QFT based on \cite{Peskin:1995ev,Polchinski:1998rq} to set the stage for the next section. 

A Hilbert space of quantum states is associated with each co-dimension one equal-time surface $\Sigma$. Throughout this subsection, we decompose spacetime coordinates into Euclidean time direction $\tau$ and the other: 
\begin{align}
    x^\mu=(\vec{x},\tau)\ ,\qquad \vec{x}\in\mathbb{R}^{d-1}\ ,\qquad  \tau\in\mathbb{R}\ .
\end{align}
And we often drop the dependence of the spatial coordinates $\vec{x}$ to shorten equations.

Consider a quantum field $\bm{\varphi}(\vec{x})$ at $\tau=0$.\footnote{Here, we only consider scalar field for simplicity.} We denote its ket and bra eigenstates with the eigenvalue $\varphi$ by $|\varphi(\vec{x})\rangle$ and $\langle\varphi(\vec{x})|$ respectively:
\begin{align}
    \bm{\varphi}(\vec{x})\,|\varphi(\vec{x})\rangle= \varphi(\vec{x})\,|\varphi(\vec{x})\rangle\ ,\qquad \langle\varphi(\vec{x})|\,\bm{\varphi}(\vec{x})=\varphi(\vec{x})\,\langle\varphi(\vec{x})|\ .
\end{align}
The quantum field in the Heisenberg picture $\bm{\varphi}(\vec{x},\tau)$ is related to $ \bm{\varphi}(\vec{x})$ by:
\begin{align}\label{eq:Heisenberg picture quantum field}
    \bm{\varphi}(\vec{x},\tau)=e^{\tau\mathbf{H}}\,\bm{\varphi}(\vec{x})\,e^{-\tau\mathbf{H}}\ ,
\end{align}
with $\mathbf{H}$ being the Hamiltonian of the theory.
We then introduce the instant eigenstates of the quantum field $\bm{\varphi}(\vec{x},\tau)$ by the relations:
\begin{align}
|\varphi(\vec{x},\tau)\rangle= e^{\tau\mathbf{H}}\,|\varphi(\vec{x})\rangle\ ,\qquad \langle\varphi(\vec{x},\tau)|= \langle\varphi(\vec{x})|\,e^{-\tau\mathbf{H}}\ .
\end{align}

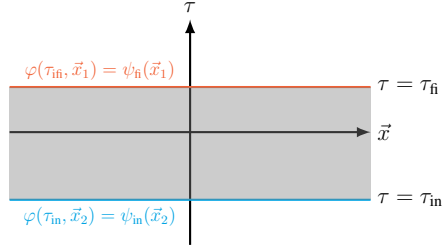
\begin{figure}[ht!]
	\centering
\begin{tikzpicture}[transform shape,scale=0.6]

\draw[black!100, thick,->] (-4,1.5) -- (4,1.5) node [right, font=\large] {$\vec{x}$};

\draw[black!100, thick,->] (0,-1) -- (0,4) node [above, font=\large] {$\tau$};

      \draw[RedOrange, thick] (-4,2.5) -- (4,2.5) node [black!100,right, font=\large] {$\tau=\tau_{\text{fi}}$};
\node[RedOrange,above,thick] at (-2,2.5) {$\varphi(\tau_{\text{ifi}},\vec{x}_1)=\psi_{\text{fi}}(\vec{x}_1)$};
      
      \draw[Cerulean, thick] (-4,0) -- (4,0) node [black!100,right, font=\large] {$\tau=\tau_{\text{in}}$};
\node[Cerulean,below,thick] at (-2,0) {$\varphi(\tau_{\text{in}},\vec{x}_2)=\psi_{\text{in}}(\vec{x}_2)$};

\fill[gray,opacity=0.4] (-4,0) -- (4,0) -- (4,2.5) -- (-4,2.5);
\end{tikzpicture}
	\caption{In the equation \eqref{eq:transtion amp QFT}, the classical fields are integrated over the shaded region in this figure.}
	\label{fig:path int QFT 1}
\end{figure}

Let $\psi_{\text{fi}}$ and $\psi_{\text{in}}$ be some boundary state configurations at $\tau=\tau_{\text{fi}}$ and $\tau=\tau_{\text{in}}$ with $\tau_{\text{in}}<0<\tau_{\text{fi}}$, respectively. One can calculate the transition amplitude between the two through the path integral:
\begin{align}\label{eq:transtion amp QFT}
      \langle\psi_{\text{fi}}(\tau_{\text{fi}})|\psi_{\text{in}}(\tau_{\text{in}})\rangle=\langle\psi_{\text{fi}}|\,e^{-(\tau_{\text{fi}}-\tau_{\text{in}})\mathbf{H}}\,|\psi_{\text{in}}\rangle=\int_{\varphi(\tau_{\text{fi}})=\psi_{\text{fi}}}^{\varphi(\tau_{\text{in}})=\psi_{\text{in}}}\,\CD\varphi_{\text{inside}}\, e^{-S_{\mathrm{E}}[\varphi_{\text{inside}}]} \ .
\end{align}
Here, the subscript of $\varphi_{\text{inside}}$ means that the path-integral and the Euclidean action are on the classical fields $\varphi$ living in the shaded domain in figure \ref{fig:path int QFT 1} with the particular boundary conditions. Stated differently, the path integral over the particular domain induces the time evolution between the two boundaries.
This implies that the initial state $|\psi_{\text{in}}(\tau_{\text{in}})\rangle$ is created by the path integral with the boundary configuration at $\tau=0$ unspecified:
\begin{equation}\label{eq:QFT ket path int}    
|\psi_{\text{in}}(\tau_{\text{in}})\rangle=\int_{\varphi(\tau_{\text{in}})=\psi}^{\varphi(0)=?}\,\CD\varphi_{\text{inside}}\, e^{-S_{\mathrm{E}}[\varphi_{\text{inside}}]}=
    \begin{tikzpicture}[transform shape,scale=0.5,baseline]
\draw[black!100, thick,->] (0,-0.5) -- (0,2) node [above, font=\large] {$\tau$};

      \draw[thick,dotted] (-2.5,1.5) -- (2.5,1.5) node [black!100,right, font=\large] {$\tau=0$};

      \draw[Cerulean, thick] (-2.5,0) -- (2.5,0) node [black!100,right, font=\large] {$\tau=\tau_{\text{in}}$};
\node[Cerulean,below,thick,font=\Large] at (-1,0) {$\psi$};

\fill[gray,opacity=0.4] (-2.5,0) -- (2.5,0) -- (2.5,1.5) -- (-2.5,1.5);
\end{tikzpicture}\ .
\end{equation}
Similarly, for the bra state $\langle\psi(\tau')|$, we have the schematic expression:
\begin{equation}\label{eq:QFT bra path int}     \langle\psi_{\text{fi}}(\tau_{\text{fi}})|=\int_{\varphi(0)=?}^{\varphi(\tau_{\text{fi}})=\psi_{\text{fi}}}\,\CD\varphi_{\text{inside}}\, e^{-S_{\mathrm{E}}[\varphi_{\text{inside}}]}=
    \begin{tikzpicture}[transform shape,scale=0.5,baseline]
\draw[black!100, thick,->] (0,-0.5) -- (0,2) node [above, font=\large] {$\tau$};

      \draw[RedOrange, thick] (-2.5,1.5) -- (2.5,1.5) node [black!100,right, font=\large] {$\tau=\tau_{\text{fi}}$};
\node[RedOrange,above,thick,font=\Large] at (-1,1.5) {$\psi_{\text{fi}}$};
      
      \draw[thick,dotted] (-2.5,0) -- (2.5,0) node [black!100,right, font=\large] {$\tau=0$};

\fill[gray,opacity=0.4] (-2.5,0) -- (2.5,0) -- (2.5,1.5) -- (-2.5,1.5);
\end{tikzpicture}\ .
\end{equation}
One can glue these path integrals together to obtain the transition amplitude \eqref{eq:transtion amp QFT} by summing over all possible boundary configurations along the equal-time slice at $\tau=0$:
\begin{equation}    
    \begin{tikzpicture}[transform shape,scale=0.5,baseline]
    \begin{scope}
\draw[black!100, thick,->] (0,-1.25) -- (0,2.25) node [above, font=\large] {$\tau$};

      \draw[RedOrange, thick] (-2.5,1.5) -- (2.5,1.5) node [black!100,right, font=\large] {$\tau=\tau_{\text{fi}}$};
            \draw[Cerulean, thick] (-2.5,-1) -- (2.5,-1) node [black!100,right, font=\large] {$\tau=\tau_{\text{in}}$};
\node[RedOrange,above,thick,font=\Large] at (-1,1.5) {$\psi_{\text{fi}}$};
\node[Cerulean,below,thick,font=\Large] at (-1,-1) {$\psi_{\text{in}}$};

\fill[gray,opacity=0.4] (-2.5,-1) -- (2.5,-1) -- (2.5,1.5) -- (-2.5,1.5);

    \end{scope}
\end{tikzpicture}
=\int\CD\psi\,
      \begin{tikzpicture}[transform shape,scale=0.5,baseline]
    \begin{scope}
\draw[black!100, thick,->] (0,-1.25) -- (0,2.25) node [above, font=\large] {$\tau$};

      \draw[RedOrange, thick] (-2.5,1.5) -- (2.5,1.5) node [black!100,right, font=\large] {$\tau=\tau_{\text{fi}}$};
            \draw[Cerulean, thick] (-2.5,-1) -- (2.5,-1) node [black!100,right, font=\large] {$\tau=\tau_{\text{in}}$};
\node[RedOrange,above,thick,font=\Large] at (-1,1.5) {$\psi_{\text{fi}}$};
\node[Cerulean,below,thick,font=\Large] at (-1,-1) {$\psi_{\text{in}}$};    
\node[black!100,above,thick,font=\Large] at (-1,0) {$\psi$};    
      \draw[thick,dotted] (-2.5,0) -- (2.5,0) node [black!100,right, font=\large] {$\tau=0$};

\fill[gray,opacity=0.4] (-2.5,-1) -- (2.5,-1) -- (2.5,-0.03) -- (-2.5,-0.03);

\fill[gray,opacity=0.4] (-2.5,0.03) -- (2.5,0.03) -- (2.5,1.5) -- (-2.5,1.5);
    \end{scope}
\end{tikzpicture}\ .
\end{equation}
In other words, we have:
\begin{align}\label{eq:all over boundary states}
  \langle\psi_{\text{fi}}(\tau_{\text{fi}})|\psi_{\text{in}}(\tau_{\text{in}})\rangle=\int\CD\psi\,  \langle\psi_{\text{fi}}(\tau_{\text{fi}})|\psi\rangle\langle\psi|\psi_{\text{in}}(\tau_{\text{in}})\rangle\  .
\end{align} 

If we perform path integral against classical fields $\CO_{\alpha}^{\text{cl}}(\tau_{\alpha})$ ($\alpha=1,\cdots,n$) on some specified domain, we obtain the transition amplitude with operator insertions, such as:\footnote{In general, quantum fields inside transition amplitudes are Euclidean time-ordered by construction. See \cite[Appendix A]{Polchinski:1987dy} and \cite[Section 9.2]{Peskin:1995ev} for details.}
\begin{align}
\begin{aligned}
          \langle\psi_{\text{fi}}(\tau_{\text{fi}})|\,\mathrm{T}_{\text{E}}&\,\{\CO(\tau_1)\cdots\CO(\tau_n)\}\,|\psi_{\text{in}}(\tau_{\text{in}})\rangle\\
          &=\int_{\varphi(\tau_{\text{in}})=\psi_{\text{in}}}^{\varphi(\tau_{\text{fi}})=\psi_{\text{fi}}}\,\CD\varphi_{\text{inside}}\, e^{-S_{\mathrm{E}}[\varphi_{\text{inside}}]}\,\CO_{1}^{\text{cl}}(\tau_{1})\cdots \CO_{n}^{\text{cl}}(\tau_{n})\ ,
\end{aligned}
\end{align}
where $\tau_{\text{in}}<\tau_{\alpha} <\tau_{\text{fi}}$ ($\alpha=1,\cdots,n$) and the symbol $\mathrm{T}_{\text{E}}$ indicates that the quantum fields are arranged in the ascending order in Euclidean time. For two operators, we have:
\begin{align}\label{eq:def of time ordering}
    \mathrm{T}_{\text{E}}\,\{\CO_1(\vec{x}_1,\tau_1)\,\CO_2(\vec{x}_2,\tau_2)\}=\begin{dcases}
       \CO_1(\vec{x}_1,\tau_1)\,\CO_2(\vec{x}_2,\tau_2) & \text{for}\quad\tau_1>\tau_2\\
                \CO_2(\vec{x}_2,\tau_2) \, \CO_1(\vec{x}_1,\tau_1) & \text{for}\quad\tau_2>\tau_1
    \end{dcases}\ .
\end{align}

\paragraph{The ground state and the vacuum correlation functions.}
Let us investigate the following limit of the ket state:
\begin{align}
|\psi(-\infty)\rangle=\lim_{\tau'\to-\infty}\,e^{\tau'\mathbf{H}}\,|\psi\rangle  \ .
\end{align}
Expanding this expression in the energy eigenbasis by use of the completeness relation $\bm{1}=\sum_n\,|E_n\rangle\langle E_n|$, we find that the dominant contribution comes from the lowest energy state, i.e., the vacuum $|E_0\rangle\equiv |\Omega\rangle$:
\begin{align}
\lim_{\tau'\to-\infty}\,|\psi(\tau')\rangle=\lim_{\tau'\to-\infty}\,\sum_n\,\langle E_n|\psi\rangle\,e^{\tau' E_n}\, |E_n\rangle\simeq \lim_{\tau'\to-\infty}\,\langle \Omega|\psi\rangle\,e^{\tau' E_0}\, |\Omega\rangle \ .
\end{align}
Alternatively, the vacuum state $|\Omega\rangle$ can be expressed by the limiting form:
\begin{align}
    |\Omega\rangle=\lim_{\tau'\to-\infty}\,(\langle\Omega|\psi\rangle)^{-1}\,e^{-\tau' E_0}\,|\psi(\tau')\rangle\ .
\end{align}
After putting extra factors into the integral measure, we conclude that the path integral over the lower half-space produces the in-vacuum:
\begin{equation}
     |\Omega\rangle=\int^{\varphi(0)=?}\,\CD\varphi_{\text{inside}}\, e^{-S_{\mathrm{E}}[\varphi_{\text{inside}}]}= 
     \begin{tikzpicture}[transform shape,scale=0.5,baseline]
     
      \draw[thick,dotted] (-2.5,0) -- (2.5,0) node [black!100,right, font=\large] {$\tau=0$};
      
      \draw[Cerulean, thick] (-2.5,-1.5) -- (2.5,-1.5) node [black!100,right, font=\large] {$\tau=-\infty$};

\fill[gray,opacity=0.4] (-2.5,0) -- (2.5,0) -- (2.5,-1.5) -- (-2.5,-1.5);
\end{tikzpicture}\ .
\end{equation}
A similar representation for the out-vacuum is given by:
\begin{equation}
     \langle\Omega|=\int_{\varphi(0)=?}\,\CD\varphi_{\text{inside}}\, e^{-S_{\mathrm{E}}[\varphi_{\text{inside}}]}= 
     \begin{tikzpicture}[transform shape,scale=0.5,baseline]

      \draw[thick,dotted] (-2.5,0) -- (2.5,0) node [black!100,right, font=\large] {$\tau=0$};
     
      \draw[Cerulean, thick] (-2.5,1.5) -- (2.5,1.5) node [black!100,right, font=\large] {$\tau=\infty$};

\fill[gray,opacity=0.4] (-2.5,0) -- (2.5,0) -- (2.5,1.5) -- (-2.5,1.5);
\end{tikzpicture}\ .
\end{equation}
The unit-normalized inner product of the in- and out-vacuum states is associated with the integral of the classical field over the whole spacetime:
\begin{equation}
     \langle\Omega|\Omega\rangle=\int\,\CD\varphi\, e^{-S_{\mathrm{E}}[\varphi]}= 
     \begin{tikzpicture}[transform shape,scale=0.5,baseline]

      \draw[Cerulean, thick] (-2.5,-1.5) -- (2.5,-1.5) node [black!100,right, font=\large] {$\tau=-\infty$};

      \draw[Cerulean, thick] (-2.5,1.5) -- (2.5,1.5) node [black!100,right, font=\large] {$\tau=\infty$};

\fill[gray,opacity=0.4] (-2.5,-1.5) -- (2.5,-1.5) -- (2.5,1.5) -- (-2.5,1.5);
\end{tikzpicture}=1\ .
\end{equation}
We denote the vacuum correlation function of quantum fields by:
\begin{align}
    \langle\,\CO(x_1)\cdots\CO(x_n)\rangle\equiv\langle\Omega|\,\mathrm{T}_{\text{E}}\,\{\CO(x_1)\cdots\CO(x_n)\}\,|\Omega\rangle \ .
\end{align}
Clearly, this corresponds to the following path integral with operator insertions:
\begin{align}\label{eq:path integral correlation QFT}
    \langle\,\CO(x_1)\cdots\CO(x_n)\rangle=\int\,\CD\varphi\, e^{-S_{\mathrm{E}}[\varphi]}\,\CO_{1}^{\text{cl}}(x_{1})\cdots \CO_{n}^{\text{cl}}(x_{n})\ .
\end{align}

\section{Path integral in conformal field theory}\label{sec:Path integral in CFT}
Leveraging the lessons from the last section, let us see the path integral in CFT. A key difference from QFT is that we choose the dilatation operator $\mathbf{D}$ as the Hamiltonian of CFT rather than one component of the translation generator $\mathbf{P}^{\mu}$. The quantization via dilatation operator is called \textbf{radial quantization} and is of great use in CFT \cite{Polchinski:1998rq,Luscher:1974ez}.

We perform Weyl transformation from the (one-point compactification of) flat $d$-dimensional spacetime $\widehat{\mathbb{R}}^d=\mathbb{R}^d\cup\{\infty\}$ to the Euclidean cylinder $\mathbb{R}\times\mathbb{S}^{d-1}$ (see figure \ref{fig:cylinder map}):
\begin{align}
\begin{aligned}
        \d s^2_{\mathbb{R}^{d}}&=\delta_{\mu\nu}\,\d x^{\mu}\d x^{\nu}\\
        &=\d r^2+r^2\,\d\Omega_{\mathbb{S}^{d-1}}^2\qquad (x^\mu=r\,\Omega_{\mathbb{S}^{d-1}}^\mu,r=|x|)\\
        &=e^{2\tau}(\d \tau^2+\d\Omega_{\mathbb{S}^{d-1}}^2)\qquad (r=e^\tau)\\
        \xrightarrow[\text{transf}]{\text{Weyl}}&\,\d s^2_{\mathbb{R}\times\mathbb{S}^{d-1}}=\d \tau^2+\d\Omega_{\mathbb{S}^{d-1}}^2\ .
\end{aligned}
\end{align}
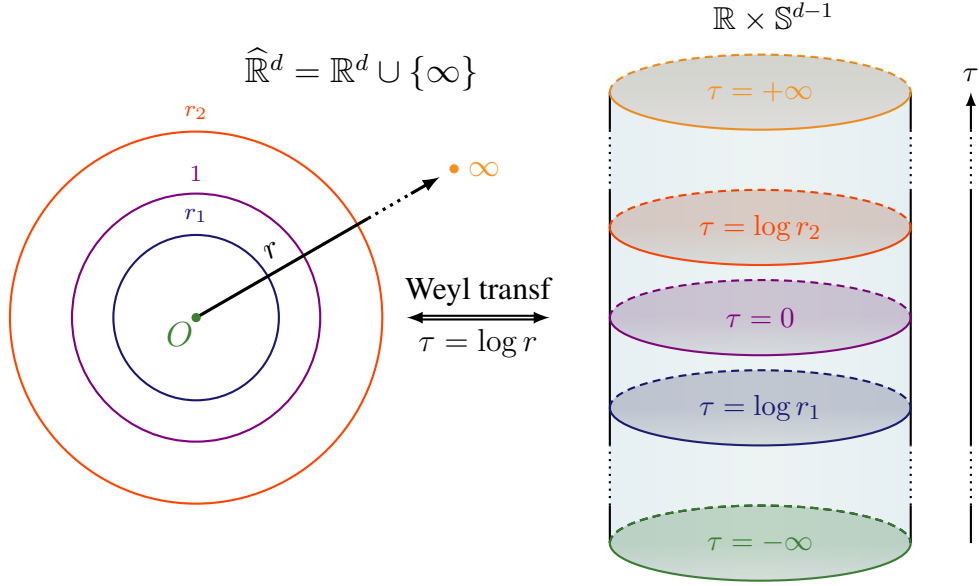
\begin{figure}[ht!]
	\centering
\begin{tikzpicture}[transform shape]

\begin{scope}[xshift=-3.5cm,transform shape,scale=1.1]
\node[black!100] at (2,3) {$\widehat{\mathbb{R}}^d=\mathbb{R}^d\cup\{\infty\}$};

\draw[thick,MidnightBlue!100] (0,0) circle (1cm);
\draw[thick,Purple!100] (0,0) circle (1.5cm);
\draw[thick,OrangeRed!100] (0,0) circle (2.25cm);   

 \node[thick,MidnightBlue!100,above,font=\scriptsize] at (0,1) {$r_1$};
  \node[thick,Purple!100,above,font=\scriptsize] at (0,1.5) {$1$};
    \node[thick,OrangeRed!100,above,font=\scriptsize] at (0,2.25) {$r_2$};

    \draw[very thick,black!100] (0:0) -- (30:2.4cm) node [midway,above, font=\small,rotate=30] {$r$}; 
        \draw[very thick,black!100,dotted] (30:2.4cm) -- (30:3cm);  
   \draw[very thick,black!100,->] (30:3cm) -- (30:3.4cm) ;

  \node[OliveGreen!100,ultra thick] at (-0.2,-0.2) {$O$};
\filldraw[ultra thick,OliveGreen!100,font=\small] (0,0) circle (0.03cm);

\filldraw[ultra thick,BurntOrange!100,font=\small] (30:3.6cm) circle (0.03cm) node [font=\small,right,BurntOrange!100] {$\infty$};

\end{scope}

         \draw[<->, thick,black, double] (-0.7,0) -- (1.2,0) node [midway,above, font=\normalsize] {$\text{Weyl transf}$}; 
      \node[below] at (0.25,0) {$\tau=\log r$};

 \fill[left color=cyan!50!blue,right color=cyan!50!blue,middle color=cyan!50,shading=axis,opacity=0.05] (6,-3) -- (6,3) arc (360:180:2cm and 0.5cm) -- (2,-3) arc (180:360:2cm and 0.5cm);
 \fill[top color=cyan!90!,bottom color=cyan!2,middle color=cyan!30,shading=axis,opacity=0.05] (4,3) circle (2cm and 0.5cm);

\node[black!100] at (4,4) {$\,\,\,\,\,\mathbb{R}\times\mathbb{S}^{d-1}$};

\draw[thick,black!100] (2,-3) -- (2,-2.5);
\draw[thick,black!100,dotted] (2,-2.5) -- (2,-1.7);
\draw[thick,black!100] (2,-1.7) -- (2,1.7);
\draw[thick,black!100,dotted] (2,2.5) -- (2,1.7);
\draw[thick,black!100] (2,3) -- (2,2.5);

\draw[thick,black!100] (6,-3) -- (6,-2.5);
\draw[thick,black!100,dotted] (6,-2.5) -- (6,-1.7);
\draw[thick,black!100] (6,-1.7) -- (6,1.7);
\draw[thick,black!100,dotted] (6,2.5) -- (6,1.7);
\draw[thick,black!100] (6,3) -- (6,2.5);

\draw[densely dashed,thick,black!100] (2,-3) arc (180:0:2cm and 0.5cm);

\draw[thick,MidnightBlue!100] (2,-1.2) arc (180:360:2cm and 0.5cm);
\draw[densely dashed,thick,MidnightBlue!100] (2,-1.2) arc (180:0:2cm and 0.5cm);
 \fill[top color=MidnightBlue!50!MidnightBlue,bottom color=MidnightBlue!10,middle color=MidnightBlue,shading=axis,opacity=0.1] (4,-1.2) circle (2cm and 0.5cm);
  \node[MidnightBlue!100,ultra thick,font=\small] at (4,-1.2) {$\tau=\log r_1$};

\draw[thick,OrangeRed!100] (2,1.2) arc (180:360:2cm and 0.5cm);
\draw[densely dashed,thick,OrangeRed!100] (2,1.2) arc (180:0:2cm and 0.5cm);
 \fill[top color=OrangeRed!50!OrangeRed,bottom color=OrangeRed!10,middle color=OrangeRed,shading=axis,opacity=0.1] (4,1.2) circle (2cm and 0.5cm);
  \node[OrangeRed!100,ultra thick,font=\small] at (4,1.2) {$\tau=\log r_2$};

\draw[thick,OliveGreen!100] (2,-3) arc (180:360:2cm and 0.5cm);
\draw[densely dashed,thick,OliveGreen!100] (2,-3) arc (180:0:2cm and 0.5cm);
 \fill[top color=OliveGreen!50!OliveGreen,bottom color=OliveGreen!10,middle color=OliveGreen,shading=axis,opacity=0.1] (4,-3) circle (2cm and 0.5cm);
  \node[OliveGreen!100,ultra thick,font=\small] at (4,-3) {$\tau=-\infty$};

\draw[thick,BurntOrange!100] (2,3) arc (180:360:2cm and 0.5cm);
\draw[densely dashed,thick,BurntOrange!100] (2,3) arc (180:0:2cm and 0.5cm);
 \fill[top color=BurntOrange!50!Tan,bottom color=BurntOrange!10,middle color=Tan,shading=axis,opacity=0.1] (4,3) circle (2cm and 0.5cm);
  \node[BurntOrange!100,ultra thick,font=\small] at (4,3) {$\tau=+\infty$};

\draw[thick,Purple!100] (2,0) arc (180:360:2cm and 0.5cm);
\draw[densely dashed,thick,Purple!100] (2,0) arc (180:0:2cm and 0.5cm);
 \fill[top color=Purple!50!Purple,bottom color=Purple!10,middle color=Purple,shading=axis,opacity=0.1] (4,0) circle (2cm and 0.5cm);
  \node[Purple!100,ultra thick,font=\small] at (4,0) {$\tau=0$};

\draw[thick,black!100] (6.8,-3) -- (6.8,-2.5);
\draw[thick,black!100,dotted] (6.8,-2.5) -- (6.8,-1.7);
\draw[thick,black!100] (6.8,-1.7) -- (6.8,1.7);
\draw[thick,black!100,dotted] (6.8,2.5) -- (6.8,1.7);
\draw[thick,black!100,->] (6.8,2.5) -- (6.8,3) node [above, font=\small] {$\tau$};
        
\end{tikzpicture}
	\caption{This figure illustrates the Weyl transformation from the (one-point compactification of) Euclidean flat spacetime $\widehat{\mathbb{R}}^d=\mathbb{R}^d\cup\{\infty\}$ to the Euclidean cylinder $\mathbb{R}\times\mathbb{S}^{d-1}$. Equal-time slice at $\tau$ on the cylinder side corresponds to the $(d-1)$-sphere with radius $r=e^\tau$ on the flat spacetime. }
	\label{fig:cylinder map}
\end{figure}
Recalling the Weyl transformation law for scalar primaries:
\begin{align}
    \CO_{\Delta}'(x')=\omega^{-\Delta}(x)\,\CO_{\Delta}(x)\ ,\qquad \text{as}\qquad \d s'^2(x')=\omega^2(x)\,\d s^2(x)\ ,
\end{align}
we find that a scalar primary operator on the Euclidean cylinder is related to the flat space one as follows:
\begin{align}\label{eq:Weyl transf operator}
    \CO_{\Delta}^{\text{cyl}}(\tau,\Omega_{\mathbb{S}^{d-1}})=r^{\Delta}\cdot \CO_{\Delta}(x)\ .
\end{align}
Then, from \eqref{eq:infinitesimal phys}, the dilatation generator $\mathbf{D}$ acts on $ \CO^{\text{cyl}}(\tau,\Omega_{\mathbb{S}^{d-1}})$ as:
\begin{align}
  [\mathbf{D} ,  \CO_{\Delta}^{\text{cyl}}(\tau,\Omega_{\mathbb{S}^{d-1}})]=\partial_\tau \, \CO_{\Delta}^{\text{cyl}}(\tau,\Omega_{\mathbb{S}^{d-1}})\ .
\end{align}
Hence, we have:
\begin{align}
\CO_{\Delta}^{\text{cyl}}(\tau,\Omega_{\mathbb{S}^{d-1}})=e^{\tau\mathbf{D}}\,\CO_{\Delta}^{\text{cyl}}(0,\Omega_{\mathbb{S}^{d-1}})\,e^{\tau\mathbf{D}}\ .
\end{align}
Therefore, by comparing this expression with \eqref{eq:Heisenberg picture quantum field}, we can naturally identify the logarithm of the radial coordinate in the flat spacetime $\tau=\log r$ with the Euclidean cylinder time and the dilatation operator $\mathbf{D}$ as the Hamiltonian of the theory.

Let us perform Weyl transformation to go back to the original spacetime $\widehat{\mathbb{R}}^d$. There, ordinary equal-time quantization on Euclidean cylinder turns into radial quantization, where the spacetime is foliated with $(d-1)$-spheres centered at the origin to each of which Hilbert space quantum states are assigned. And the time evolution between them is governed by the dilatation operator $\mathbf{D}$. Be aware that the unit $(d-1)$-sphere at $\tau=-\infty$ ($\tau=+\infty$) on the Euclidean cylinder is squeezed into a single point $O$ ($\infty$) on $\widehat{\mathbb{R}}^d$, allowing us to establish the state/operator correspondence in CFT as we will see shortly.

\section{State/operator correspondence}\label{sec:State/operator correspondence CFT}
Here, we would like to expand on the \textbf{state/operator correspondence} in CFT, following \cite{Polchinski:1998rq}. After reviewing a simple proof, we give more on the essence of the correspondence for later use.

\paragraph{Proof of the state/operator correspondence.}
Let us perform path integral over a unit $d$-ball against some operator at the origin $\CA(0)$. Then, we obtain the ket state $|\CA\rangle$ at $r=1$ that depends on the operator $\CA$ we inserted:
\begin{equation}\label{eq:so corr 1}
     |\CA\rangle=\int^{\varphi(0)=?}\,\CD\varphi_{\text{inside}}\, e^{-S_{\mathrm{E}}[\varphi_{\text{inside}}]}\,\CA(0)= 
     \begin{tikzpicture}[baseline,transform shape,scale=0.6]

    \draw[thick,dotted] (0,0) circle (1cm); 
        \fill[thick,gray!100,opacity=0.4] (0,0) circle (0.99cm); 

          \node[black!100,thick,font=\small] at (-0.2,-0.2) {$O$};
                    \node[OrangeRed!100,thick,font=\small] at (-0.2,0.3) {$\CA(0)$};
\filldraw[thick,OrangeRed!100,font=\small] (0,0) circle (0.03cm);

    \draw[black!100,<->] (0:0.05) -- (0:0.95cm) node [midway,above, font=\small] {$1$}; 

\end{tikzpicture}\ .
\end{equation}

\begin{figure}[ht!]
	\centering
\begin{tikzpicture}[transform shape,scale=0.8]
\draw[thick,MidnightBlue!100] (0,0) circle (1cm);
\draw[thick,Purple!100] (0,0) circle (1.5cm);

 \node[thick,MidnightBlue!100,below,font=\scriptsize] at (0,1) {$\psi'$};
  \node[thick,Purple!100,above,font=\scriptsize] at (0,1.5) {$\psi$};

   \draw[thick,black!100,->] (30:0cm) -- (30:1cm) node[midway,above, font=\small,rotate=30] {$r'$};
      \draw[thick,black!100,->] (-30:0cm) -- (-30:1.5cm) node[midway,above, font=\small,rotate=-30] {$1$};

  \node[OliveGreen!100,very thick] at (-0.2,-0.2) {$O$};
\filldraw[very thick,OliveGreen!100,font=\small] (0,0) circle (0.03cm);

        \fill[thick,gray!100,opacity=0.4,even odd rule] (0,0) circle[radius=0.98cm] (0,0) circle[radius=1.48cm];    
\end{tikzpicture}
	\caption{Shaded domain is the spherical shell with inner radius $r'$ and outer radius $1$ on $d$-dimensional Euclidean spacetime. Boundary conditions are imposed both for inner and outer shells.}
	\label{fig:annulus}
\end{figure}
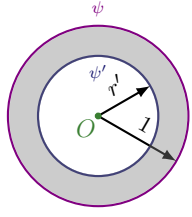

On the other way, given an arbitrary ket state $|\Psi\rangle$ at $r=1$, we perpare the inner product $\langle\psi|\Psi\rangle$ and evaluate the following quantity in two ways:
\begin{align}\label{eq:spherical shell path int}
\int\CD\psi'\,\int_{\varphi(r=r')=\psi'}^{\varphi(r=1)=\psi}\CD\varphi_{\text{inside}}   \, r'^{-\mathbf{D}} \,\langle\psi'|\Psi\rangle\ .
\end{align}
Firstly, let us focus our attention on the path integral over spherical shell $r'<r<1$ (shaded domain in figure \ref{fig:annulus}) in the middle of \eqref{eq:spherical shell path int}. This is equivalent to acting $r'^{\mathbf{D}}$ on the ket state $|\psi'\rangle$ which will cancel out by $r'^{-\mathbf{D}}$. As a result, we find that:
\begin{align}
    \int_{\varphi(r=r')=\psi'}^{\varphi(r=1)=\psi}\CD\varphi_{\text{inside}}   \, r'^{-\mathbf{D}} =\langle\psi|\psi'\rangle\ .
\end{align}
Plugging this expression into \eqref{eq:spherical shell path int}, we conclude that:
\begin{align}\label{eq:spherical shell path int 1}
\langle\psi|\Psi\rangle=\int\CD\psi'\,\int_{\varphi(r=r')=\psi'}^{\varphi(r=1)=\psi}\CD\varphi_{\text{inside}}   \, r'^{-\mathbf{D}} \,\langle\psi'|\Psi\rangle\ .
\end{align}
By taking the limit $r'\to0$, the spherical shell associated with the path integral in the right-hand side of \eqref{eq:spherical shell path int 1} (see figure \ref{fig:annulus}) becomes a unit $d$-ball. Hence, we can interpret that the quantity $\langle\psi|\Psi\rangle$ is created by acting some local operator at the origin and performing path integral over the unit $d$-ball as in \eqref{eq:so corr 1}. Because the choice of the bra state $\langle\psi|$ is arbitrary, we can associate every state with some local operator located at the radial quantization origin, substantiating a one-to-one correspondence between a state and a local operator.

\paragraph{More on state/operator correspondence.}
The in-vacuum state at the unit $(d-1)$-sphere corresponds to the identity operator by construction and is created by performing the path integral inside of the unit $d$-ball:
\begin{equation}
     |\Omega\rangle=\int^{\varphi(0)=?}\,\CD\varphi_{\text{inside}}\, e^{-S_{\mathrm{E}}[\varphi_{\text{inside}}]}= 
     \begin{tikzpicture}[baseline,transform shape,scale=0.6]

    \draw[thick,dotted] (0,0) circle (1cm); 
        \fill[thick,gray!100,opacity=0.4] (0,0) circle (0.99cm); 

          \node[black!100,thick,font=\small] at (-0.2,-0.2) {$O$};
\filldraw[thick,Black!100,font=\small] (0,0) circle (0.03cm);

    \draw[black!100,<->] (0:0.05) -- (0:0.95cm) node [midway,above, font=\small] {$1$}; 

\end{tikzpicture}\ .
\end{equation}
Also, for the out-vacuum state, we have:
\begin{equation}
     \langle\Omega|=\int_{\varphi(0)=?}\,\CD\varphi_{\text{inside}}\, e^{-S_{\mathrm{E}}[\varphi_{\text{inside}}]}= 
     \begin{tikzpicture}[baseline,transform shape,scale=0.6]

    \draw[thick,dotted] (0,0) circle (1cm); 
        \fill[thick,gray!100,opacity=0.4,even odd rule] (0,0) circle[radius=1.01cm] (-1.5,-1.5) rectangle(1.5,1.5); 

          \node[black!100,thick,font=\small] at (-0.2,-0.2) {$O$};
\filldraw[thick,Black!100,font=\small] (0,0) circle (0.03cm);

    \draw[black!100,<->] (0:0.05) -- (0:0.95cm) node [midway,above, font=\small] {$1$}; 
\end{tikzpicture}\ .
\end{equation}
These two expressions are consistent with the fact that the operator with the lowest conformal dimension is the identity operator $\bm{1}$ in unitary CFTs. 

One can compute correlation functions of local operators in CFT through the path integral as in QFT \eqref{eq:path integral correlation QFT}. However, as we have applied radial quantization, the operators are ordered in the radial direction:
\begin{align}
    \langle\,\CO(x_1)\cdots\CO(x_n)\rangle\equiv\langle\Omega|\,\mathrm{R}\,\{\CO(x_1)\cdots\CO(x_n)\}\,|\Omega\rangle \ .
\end{align}
The symbol $\mathrm{R}$ stands for radial ordering. For two arbitrary operators, we have:
\begin{align}
    \mathrm{R}\,\{\CO(x_1)\,\CO(x_2)\}=\begin{dcases}
        \CO(x_1)\,\CO(x_2) & \text{for}\quad |x_1|>|x_2|\\
                \CO(x_2)\,\CO(x_1) & \text{for}\quad |x_2|>|x_1|
    \end{dcases}\ .
\end{align}

The vacuum of CFT is conformal invariant and annihilated by any generators of the conformal group:
\begin{align}\label{eq:CFT vacuum condition}
\mathbf{J}_{MN}\,|\Omega\rangle=0\ ,\qquad  \langle\Omega|\,\mathbf{J}_{MN}=0\ .
\end{align}
A primary operator $\CO(0)$ can be identified with the primary state $|\CO\rangle\equiv \CO(0)\,|\Omega\rangle$ created by inserting a primary $\CO$ at the origin:
\begin{align}
    \CO(0)\longleftrightarrow |\CO\rangle\equiv \CO(0)\,|\Omega\rangle\ .
\end{align}
Similarly to primary operators, one can characterize the primary state by the action of conformal generators: it is an eigenstate of the dilatation $\mathbf{D}$ and rotation generators $\mathbf{M}_{\mu\nu}$ and annihilated by the SCT generators $\mathbf{K}_{\mu}$:
\begin{align}
   \mathbf{D}\, |\CO\rangle &=\mathbf{D}\, \CO(0)\,|\Omega\rangle=[\mathbf{D}, \CO(0)]\,|\Omega\rangle=\Delta \,|\CO\rangle\ ,\\
   \mathbf{M}_{\mu\nu}\, |\CO\rangle &=\mathbf{M}_{\mu\nu}\, \CO(0)\,|\Omega\rangle=[\mathbf{M}_{\mu\nu}, \CO(0)]\,|\Omega\rangle=\CS_{\mu\nu} \,|\CO\rangle\ ,\\
      \mathbf{K}_{\mu}\, |\CO\rangle &=\mathbf{K}_{\mu}\, \CO(0)\,|\Omega\rangle=[\mathbf{K}_{\mu}, \CO(0)]\,|\Omega\rangle=0\ ,
\end{align}
where we used \eqref{eq:dilatation acting on O}, \eqref{eq:rotation acting on O}, \eqref{eq:SCT acting on O} and the conformal invariance of the vacuum \eqref{eq:CFT vacuum condition}. We can make descendant states associated with $\CO$, that constitute the conformal multiplet of $\CO$, by acting translation generators on $|\CO\rangle$:
\begin{align}
    \partial_{\mu}\partial_{\nu}\cdots\CO(0)\longleftrightarrow \mathbf{P}_{\mu}\mathbf{P}_{\nu}\cdots|\CO\rangle\equiv \partial_{\mu}\partial_{\nu}\cdots\CO(0)\,|\Omega\rangle\ .
\end{align}

\section{Operator Product Expansion}\label{sec:Operator Product Expansion}
Recall that every local operator can be represented by a linear combination of primaries and descendants as demonstrated in section \ref{sec:Local operators in conformal field theory}. Combining this fact with the state/operator correspondence and the completeness of the Hilbert space, it turns out that primary and descendant states span a complete basis. Let us take $\CO(x)\,|\Omega\rangle$ ($|x|<1$) for an example. One can express this state by the infinite linear combination of $|\CO\rangle$ and its descendant states in the following manner:
\begin{align}
    \CO(x)\,|\Omega\rangle=e^{x\cdot \mathbf{P}}\,\CO(0)\,e^{x\cdot \mathbf{P}}\,|\Omega\rangle=\sum_{n=0}^{\infty}\,\frac{1}{n!}\, (x\cdot \mathbf{P})^n\,|\CO\rangle\ ,
\end{align}
or
\begin{equation}
\begin{tikzpicture}[baseline,transform shape,scale=0.7]

    \draw[thick,dotted] (0,0) circle (1cm); 
        \fill[thick,gray!100,opacity=0.4] (0,0) circle (0.99cm); 

          \node[black!100,thick,font=\small] at (-0.2,-0.2) {$O$};
\filldraw[thick,Black!100,font=\small] (0,0) circle (0.03cm);

    \draw[black!100,<->] (0:0.05) -- (0:0.95cm) node [midway,above, font=\small] {$1$}; 
   \node[OrangeRed!100,thick,font=\small] at (-0.3,0.6) {$\CO(x)$};
   \filldraw[thick,OrangeRed!100,font=\small] (-0.1,0.3) circle (0.03cm);
\end{tikzpicture}
=\sum_{n=0}^{\infty}\,\frac{1}{n!}\, (x\cdot \mathbf{P})^n\,
     \begin{tikzpicture}[baseline,transform shape,scale=0.7]

    \draw[thick,dotted] (0,0) circle (1cm); 
        \fill[thick,gray!100,opacity=0.4] (0,0) circle (0.99cm); 

          \node[black!100,thick,font=\small] at (-0.2,-0.2) {$O$};
                    \node[OrangeRed!100,thick,font=\small] at (-0.2,0.3) {$\CO(0)$};
\filldraw[thick,OrangeRed!100,font=\small] (0,0) circle (0.03cm);

    \draw[black!100,<->] (0:0.05) -- (0:0.95cm) node [midway,above, font=\small] {$1$}; 

\end{tikzpicture}\ .
\end{equation}
We are now in a position to consider the state created by acting two primary operators on the vacuum: $\CO_1(x_1)\,\CO_2(x_2)\,|\Omega\rangle$ ($|x_2|<|x_1|<1$). This state should again be some linear combination of primary and descendant states:
\begin{equation}
\begin{tikzpicture}[baseline,transform shape,scale=0.7]

    \draw[thick,dotted] (0,0) circle (1cm); 
        \fill[thick,gray!100,opacity=0.4] (0,0) circle (0.99cm); 

          \node[black!100,thick,font=\small] at (-0.2,-0.2) {$O$};
\filldraw[thick,Black!100,font=\small] (0,0) circle (0.03cm);

    \draw[black!100,<->] (0:0.05) -- (0:0.95cm) node [midway,above, font=\small] {$1$}; 
   \node[RoyalBlue!100,thick,font=\small] at (-0.1,0.6) {$\CO_2(x_2)$};
   \filldraw[thick,RoyalBlue!100,font=\small] (0.1,0.3) circle (0.03cm);
      \node[Sepia!100,thick,font=\small] at (-0.7,-1.1) {$\CO_1(x_1)$};
   \filldraw[thick,Sepia!100,font=\small] (-0.5,-0.7) circle (0.03cm);
\end{tikzpicture}
=\sum_{\CO}\,C_{\CO_1\CO_2\CO}(x_1,x_2,\mathbf{P})\,
     \begin{tikzpicture}[baseline,transform shape,scale=0.7]

    \draw[thick,dotted] (0,0) circle (1cm); 
        \fill[thick,gray!100,opacity=0.4] (0,0) circle (0.99cm); 

          \node[black!100,thick,font=\small] at (-0.2,-0.2) {$O$};
                    \node[OrangeRed!100,thick,font=\small] at (-0.2,0.3) {$\CO(0)$};
\filldraw[thick,OrangeRed!100,font=\small] (0,0) circle (0.03cm);

    \draw[black!100,<->] (0:0.05) -- (0:0.95cm) node [midway,above, font=\small] {$1$}; 

    \end{tikzpicture}\ , 
\end{equation}
that is:
\begin{align}\label{eq:OPE state}
\CO_1(x_1)\,\CO_2(x_2)\,|\Omega\rangle=\sum_{\CO}\,C_{\CO_1\CO_2\CO}(x_1,x_2,\mathbf{P})\,|\CO\rangle\ .
\end{align}
Here, the summation is for all possible primaries, and $C_{\CO_1\CO_2\CO}(x_1,x_2,\mathbf{P})$ is some function that packages the contributions from the single conformal multiplet associated with $\CO$.
We can utilize conformal transformation inside correlators to set two operators inside a unit $d$-ball $r<1$, isolating the rest. And the choice of radial quantization origin is arbitrary. Hence, one can upgrade \eqref{eq:OPE state} to an operator identity termed the \textbf{Operator Product Expansion} (\textbf{OPE}):
\begin{align}\label{eq:OPE operator}
\CO_1(x_1)\,\CO_2(x_2)=\sum_{\CO}\,C_{\CO_1\CO_2\CO}(x_1-x,x_2-x,\partial/\partial x)\,\CO(x)\ .
\end{align}
It is convenient to take the quantization origin at the same position as either of the two operators. Then, the OPE takes the form:
\begin{align}\label{eq:OPE operator simplified}
\CO_1(x_1)\,\CO_2(x_2)=\sum_{\CO}\,C_{\CO_1\CO_2\CO}(x_{12},\partial/\partial x_2)\,\CO(x_2)\ ,
\end{align}
where $x_{ij}=x_i-x_j$.
One can make use of the infinitesimal conformal transformation laws by acting $\mathbf{J}_{MN}$'s on both sides of \eqref{eq:OPE operator simplified} to determine the function $C_{\CO_1\CO_2\CO}(x_{12},\partial/\partial x_2)$. An alternative way to do this is to utilize the fact that primary operators are orthogonal to each other and three- and lower-point functions are fixed up to model-dependent coefficients (see section \ref{sec:Correlation functions in CFT}). Then, the function $C_{\CO_1\CO_2\CO}(x_{12},\partial/\partial x_2)$ are derived to satisfy the following equality:
\begin{align}\label{eq:three-point functions OPE}
\langle \,\CO_1(x_1)\,\CO_2(x_2)\,\CO(x_3)\rangle=C_{\CO_1\CO_2\CO}(x_{12},\partial/\partial x_2)\,\langle\,\CO(x_2)\,\CO(x_3)\,\rangle\ .
\end{align}
It is clear from this expression that a primary operator $\CO$ appears in the OPE of $\CO_1$ and $\CO_2$ if and only if the three-point function $\langle\,\CO_1\,\CO_2\,\CO\,\rangle$ is non-zero.

The OPE of generic primaries are complicated due to the tensor structure, but only symmetric and traceless tensors appear in the scalar OPEs, and their explicit forms have been studied in many pieces of literature (see e.g., \cite{Ferrara:1971vh,Dobrev:1975ru,Ferrara:1972uq,Ferrara:1972ay}).\footnote{Let $\CO_{\Delta,\mu_1\cdots\mu_J}$ be a generic tensorial primary operator having $J$ indices of SO$(d)$. We first make such that $\CO_{\Delta,\mu_1\cdots\mu_J}$ is traceless because one can decompose any tensorial operators into a sum of traceless tensors. For instance, we have $O^{\mu\nu\rho}=O_3^{\mu\nu\rho}+\frac{1}{d+2}\,[\delta^{\nu\rho}\,O_{1;1}^\mu+\delta^{\mu\rho}\,O_{1;2}^\nu+\delta^{\mu\nu}\,O_{1;3}^\rho]$ with $O_{1;1}^{\mu}=\delta_{\nu\rho}\CO^{\mu\nu\rho},O_{1;2}^{\nu}=\delta_{\mu\rho}O^{\mu\nu\rho},O_{1;3}^{\rho}=\delta_{\mu\nu}O^{\mu\nu\rho}$ and $O^{\mu\nu}=O_2^{\mu\nu}-\frac{1}{d}\,\delta^{\mu\nu}\,O_0$ with $O_0=\delta_{\mu\nu}O_2^{\mu\nu}$. The tensorial primary operator $\CO_{\Delta,\mu_1\cdots\mu_J}$ can appear in the scalar OPE: $\CO_{\Delta_1} \times\CO_{\Delta_2}$ if and only if $\langle\,\CO_{\Delta_1}(x) \,\CO_{\Delta_2}(-x)\,\CO_{\Delta,\mu_1\cdots\mu_J}(0)\rangle\neq 0$. (It is always possible to go to this configuration using conformal transformations.) The tensor structure of the three-point functions can be made out of $x^\mu$ and $\delta^{\mu\nu}$:
\begin{align}\label{eq:ST tensor argument}
    \langle\,\CO_{\Delta_1}(x) \,\CO_{\Delta_2}(-x)\,\CO_{\Delta,\mu_1\cdots\mu_J}(0)\,\rangle=x^{\mu_1}\cdots x^{\mu_J}\cdot  f_0(|x|)+\delta^{\mu_1\mu_2}\, x^{\mu_3}\cdots x^{\mu_J}\cdot  f_1(|x|)+\cdots \ .
\end{align}
Because the first term is symmetric concerning its indices and the others are uniquely determined from the requirement of the tracelessness, the entire tensor structure in the right-hand side of \eqref{eq:ST tensor argument} must be symmetric and traceless, so are the indices of the operator $\CO_{\Delta,\mu_1\cdots\mu_J}$.
Hence, only symmetric and traceless tensors have a non-vanishing three-point function against two scalar primaries, and we can verify that only symmetric and traceless tensors appear in the scalar OPE. If $\CO_{\Delta_1}=\CO_{\Delta_2}$, Bose symmetry requires the invariance of \eqref{eq:ST tensor argument} under the replacement $x\leftrightarrow-x$, implying that only even spin operators can appear in the OPE of identical scalars. \label{fot:symmetric and traceless argument}
} 
We here record a few leading terms of the scalar channel OPE of two scalar primaries:
\begin{align}\label{eq:CFT OPE scalars}
\begin{aligned}
     \CO_{\Delta_1}(x_1) \times&\CO_{\Delta_2}(x_2)\supset \frac{c(\CO_{\Delta_1},\CO_{\Delta_2},\CO_{\Delta})/c(\CO_{\Delta},\CO_{\Delta})}{|x_{12}|^{\Delta_1+\Delta_2-\Delta}}\\
     &\cdot \left(1+q_1\, x_{12}^{\mu} \partial_{2,\mu}+q_2\, x_{12}^{\mu}x_{12}^{\nu}\partial_{2,\mu}\partial_{2,\nu}+q_3\, x_{12}^2\,\Box+\cdots\right)\, \CO_{\Delta}(x_2)\ ,
\end{aligned}
\end{align}
where the symbol $\Box$ stands for the Laplacian on the $d$-dimensional flat Euclidean spacetime that acts on the operator located at $x$ as the differential operator $\frac{\partial}{\partial x^{\mu}}\cdot\frac{\partial}{\partial x_{\mu}}$ and the expansion coefficients such as $q_1,q_2,q_3$ are rational functions of conformal dimensions involved:
\begin{align}
    q_1&=\frac{\Delta_1-\Delta_2+\Delta}{2\Delta}\ ,\\
    q_2&= \frac{(\Delta_1-\Delta_2+\Delta)(\Delta_1-\Delta_2+\Delta+2)}{8\Delta(\Delta+1)}\ ,\\
    q_3&= \frac{(\Delta_1-\Delta_2+\Delta)(\Delta_1-\Delta_2-\Delta)}{16\Delta(\Delta+1)(\Delta+1-d/2)}\ .\label{eq:coefficient q3}
\end{align}
We further denoted scalar two- and three-point coefficients by $c(\CO_{\Delta},\CO_{\Delta})$ 
 and $c(\CO_{\Delta_1},\CO_{\Delta_2},\CO_{\Delta})$ that will be defined in \eqref{eq:conformal inv of 2pt CFT correlator phy} and \eqref{eq:conformal inv of 3pt CFT correlator phy} in respective ways.

\section{Correlation functions}\label{sec:Correlation functions in CFT}
It is virtually impossible to determine correlation functions in general interacting QFTs. However, in CFTs, the conformal symmetry imposes strong constraints and allows to fix correlators up to model-dependent coefficients. 
Because the conformal transformations are not linear as illustrated in \eqref{eq:infinitesimal phys}, it is troublesome to constrain correlation functions. To overcome this difficulty, we first introduce the \textbf{embedding space formalism} \cite{Dirac:1936fq,Costa:2011mg,Ferrara:1971vh,Weinberg:2010fx} that simplifies the situation a great deal. It will turn out that one can still use this methodology even in the presence of a conformal defect.

\paragraph{Embedding space formalism.}
We embed the $d$-dimensional Euclidean spacetime $\mathbb{R}^d$ into the projective null cone of the fictitious $(d+2)$-dimensional Minkowski spacetime $X^M\in\mathbb{R}^{1,d+1}$ with the metric $\d s^2_{\mathbb{R}^{1,d+1}}=\eta_{MN}\,\d X^M\,\d X^N=-(\d X^{-1})^2+(\d X^0)^2+(\d X^1)^2+\cdots +(\d X^d)^2$:
\begin{align}\label{eq:embedding condition}
    X^2=X^M \,\eta_{MN}\,X^N=0 \ ,\qquad X^M\sim\lambda(X)\cdot X^M\quad \text{with}\quad\lambda(X)>0\ . 
\end{align}
Let $\CO_{\Delta}(x)$ be a scalar primary $\CO_{\Delta}(x)$ with conformal dimension $\Delta$. In the embedding space formalism, it turns into an operator that depends on the embedding space coordinate $P^M$ is subject to the homogeneity condition:
\begin{align}\label{eq:homogeneity condition}
  \CO_{\Delta}(\lambda\, P)=\lambda^{-\Delta}\, \CO_{\Delta}(P)\quad \leftrightarrow \quad P^M\frac{\partial}{\partial P^M}\,\CO_{\Delta}(\lambda\, P)=-\Delta\,\CO_{\Delta}(P)\ .
\end{align}
Recall that the Euclidean conformal group in $d$-dimensions is isomorphic to the Lorentz group in $(d+2)$-dimensions.
One can identify the conformal group with the Lorentz group in the embedding space where the infinitesimal transformation law \eqref{eq:infinitesimal phys} takes a much simpler form:
\begin{align}\label{eq:generator action embed}
   [ \mathbf{J}_{MN},\CO_{\Delta}(P)]=\CJ_{MN}(P)\,\CO_{\Delta}(P)=-\left(P_M\,\frac{\partial}{\partial P_N}-P_N\,\frac{\partial}{\partial P_M}\right)\,\CO_{\Delta}(P)\ .
\end{align}
One can pull the embedding space coordinates and operators back to the physical Euclidean spacetime using the relations:
\begin{align}
    x^\mu=P^\mu/P^+ \ ,\qquad \CO_{\Delta}(x)=|P^+|^{-\Delta}\,\CO_{\Delta}(x^\mu=P^\mu/P^+)\ .
\end{align}
Here, we introduced light-cone coordinates $X^\pm=X^{-1}\pm X^0$ to simplify the equation. Because the embedding space coordinate is modded out by $\mathbb{R}_{+}$-scalings $P^M\sim \lambda\,P^M$, we can always set it back to the physical space by making the following substitution:
\begin{align}\label{eq:emb to phys 1}
    P^M=(P^+,P^-,P^\mu)=(1,x^2,x^\mu)\ .
\end{align}

\paragraph{Conformal Ward identities.}
Consider an $n$-point correlation function of scalar primaries:
\begin{align}\label{eq:n-point CFT}
\langle\, \CO_{\Delta_1}(x_1)\cdots \CO_{\Delta_n}(x_n)\,\rangle=   \langle\Omega|\,\mathrm{R}\left\{ \CO_{\Delta_1}(x_1)\cdots \CO_{\Delta_n}(x_n)\right\}\,|\Omega\rangle\ .
\end{align}
By use of the embedding space formalism, we can regard this $n$-point correlation function as a function of embedding space coordinates $P_{\alpha}^A$ ($\alpha=1,2,\cdots ,n$) with the homogeneity condition inherited from \eqref{eq:homogeneity condition}:
\begin{align}\label{eq:homogeneity CFT correlator}
     \langle\, \CO_{\Delta_1}(\lambda_1\,P_1)\cdots \CO_{\Delta_n}(\lambda_n\,P_n)\,\rangle=\lambda_1^{-\Delta_1}\cdots \lambda_n^{-\Delta_n}\,\langle\, \CO_{\Delta_1}(P_1)\cdots \CO_{\Delta_n}(P_n)\,\rangle\ .
\end{align}
Let us utilize conformal invariance to constrain the correlator. To this end, consider the following quantity 
\begin{align}\label{eq:deriv conformal Ward 0}
     \langle\Omega|\,\mathbf{J}_{MN}\,\mathrm{R}\left\{ \CO_{\Delta_1}(P_1)\cdots \CO_{\Delta_n}(P_n)\right\}\,|\Omega\rangle\ .
\end{align}
Assuming that the generator $\mathbf{J}_{MN}$ acts on the out-vacuum, due to \eqref{eq:CFT vacuum condition}, we have 
\begin{align}\label{eq:deriv conformal Ward 1}
\langle\Omega|\,\mathbf{J}_{MN}\,\mathrm{R}\left\{ \CO_{\Delta_1}(P_1)\cdots \CO_{\Delta_n}(P_n)\right\}\,|\Omega\rangle=0\ .
\end{align}
Doing the other way around, we find that:
\begin{align}\label{eq:deriv conformal Ward 2}
\begin{aligned}
 \langle\Omega|\,\mathbf{J}_{MN}\,\mathrm{R}\left\{ \CO_{\Delta_1}(P_1)\cdots \CO_{\Delta_n}(P_n)\right\}\,|\Omega\rangle
      &=\langle\Omega|\,[\mathbf{J}_{MN},\mathrm{R}\left\{ \CO_{\Delta_1}(P_1)\cdots \CO_{\Delta_n}(P_n)\right\}]\,|\Omega\rangle\\
      &=\sum_{\beta=1}^n\,\CJ_{MN}(P_\beta)\, \langle\, \CO_{\Delta_1}(P_1)\cdots \CO_{\Delta_n}(P_n)\,\rangle\ ,
\end{aligned}
\end{align}
where we used $[A,B\,C]=[A,B]\,C + A\,[B,C]$ and \eqref{eq:generator action embed}.
Therefore, we obtain the conformal Ward identities:
\begin{align}\label{eq:conformal inv of CFT correlator}
    \sum_{\beta=1}^n\,\CJ_{MN}(P_\beta)\, \langle\, \CO_{\Delta_1}(P_1)\cdots \CO_{\Delta_n}(P_n)\,\rangle=0\ , \qquad M,N=-1,0,1,\cdots d\  .
\end{align}
These identities imply that the $n$-point function $ \langle\, \CO_{\Delta_1}(P_1)\cdots \CO_{\Delta_n}(P_n)\,\rangle$ must be composed of $\mathrm{SO}(1,d+1)$ invariant products in the embedding space $P_\alpha\cdot P_\beta=P_\alpha^M \eta_{MN} P_\beta^N$ ($\alpha,\beta=1,\cdots,n$). Combining this $\mathrm{SO}(1,d+1)$ invariance with the homogeneity condition \eqref{eq:homogeneity CFT correlator}, one can constrain scalar lower-point functions as follows.  
\paragraph{One-point functions.}
Due to the null condition $P^2=0$, one fails to make $\mathrm{SO}(1,d+1)$ invariant inner products out of a single embedding space coordinate. Hence, the one-point function should be zero unless the operator is proportional to the identity operator with $\Delta=0$:
\begin{align}
    \langle\,\CO_{\Delta}(x)\,\rangle=\begin{dcases}
        0 & \text{if}\quad\Delta\neq 0\\
        \text{const.}& \text{if}\quad\Delta=0
    \end{dcases}\ .
\end{align}

\paragraph{Two-point functions.}
In this case, we have one building block $P_1\cdot P_2$. The homogeneity condition \eqref{eq:homogeneity CFT correlator} fixes the scalar two-point function as follows:
\begin{align}\label{eq:conformal inv of 2pt CFT correlator emb}
\langle\, \CO_{\Delta_1}(P_1)\,\CO_{\Delta_2}(P_2)\,\rangle
=
\begin{dcases}
  0  &\text{if}\quad \Delta_1\neq \Delta_2\\
   c(\CO_{\Delta_1},\CO_{\Delta_1})\cdot \frac{1}{(-2P_1\cdot P_2)^{\Delta_1}}  &\text{if}\quad \Delta_1=\Delta_2
\end{dcases}
\  ,
\end{align}
leading to the orthogonality of scalar primaries with different conformal dimensions. We get the physical space correlator by making the replacement $-2P_1\cdot P_2\mapsto x_{12}^2$:\footnote{From \eqref{eq:emb to phys 1}, we have $P_1\cdot P_2=-\frac{1}{2}\,(P_1^+ P_2^- + P_1^- P_2^+)+\delta_{\mu\nu} P_1^\mu P_2^\nu=-\frac{1}{2}\,(x_1-x_2)^2$.}
\begin{align}\label{eq:conformal inv of 2pt CFT correlator phy}
\langle\, \CO_{\Delta}(x_1)\,\CO_{\Delta}(x_2)\,\rangle=\frac{c(\CO_{\Delta},\CO_{\Delta})}{|x_{12}|^{2\Delta_1}}\ . 
\end{align}
This orthogonality holds even for spinning primaries, and their two-point functions are also fixed by conformal symmetry. For instance, the two-point function of rank-$J$ symmetric traceless tensors is given by:
\begin{align}\label{eq:2pt CFT spinning}
\langle\, \CO_{\Delta,\mu_1\cdots\mu_J}(x_1)\,\CO_{\Delta}^{\nu_1\cdots\nu_J}(x_2)\,\rangle=c(\CO_{\Delta,J},\CO_{\Delta,J})\cdot \frac{I_{\mu_1}^{(\nu_1}(x_{12})\cdots I_{\mu_J}^{\nu_J)}(x_{12})}{|x_{12}|^{2\Delta}}\  ,
\end{align}
where 
\begin{align}\label{eq:def of I tensor}
    I_{\mu}^{\nu}(x)=\delta_{\mu}^{\nu}-\frac{2x_{\mu} x^{\nu}}{x^2}\ .
\end{align}
Throughout this thesis, the indices enclosed in parentheses are symmetrized and their traces are subtracted. We also note that the two-point coefficients $c(\CO_{\Delta},\CO_{\Delta})$ and $c(\CO_{\Delta,J},\CO_{\Delta,J})$ are positive in unitary CFTs.\footnote{We can perform field redefinitions so that two-point functions are unit-normalized, except for operators having canonical definitions. One of the prominent examples of such operators is the stress tensor that is defined directly from the action as its response to the metric perturbation:
\begin{align}
    T^{\mu\nu}(x)=\left.\frac{2}{\sqrt{\text{det}\,g}}\cdot\frac{\delta S_{\text{E}}[\varphi,g]}{\delta g_{\mu\nu}}\right|_{g_{\mu\nu}\to\delta_{\mu\nu}}\ .
\end{align}
The coefficient that appears in the stress tensor two-point function is conventionally denoted by $C_T$ and related to the trace anomaly coefficients through Ward identities \cite{Osborn:1993cr}. In this thesis, we will not unit-normalize two-point functions of composite operators that are defined canonically as coincident limits of non-composite operators, except otherwise noted.}

\paragraph{Three-point functions.} 
Now, we can use three $\mathrm{SO}(1,d+1)$ invariants: $P_1\cdot P_2,P_2\cdot P_3,P_1\cdot P_1$. Armed with the homogeneity condition \eqref{eq:homogeneity CFT correlator}, we can fix the form of the scalar three-point function up to a model-dependent constant:
\begin{align}\label{eq:conformal inv of 3pt CFT correlator emb}
\begin{aligned}
    \langle\, \CO_{\Delta_1}&(P_1)\,\CO_{\Delta_2}(P_2)\,\CO_{\Delta_3}(P_3)\,\rangle\\
    &=\frac{c(\CO_{\Delta_1},\CO_{\Delta_2},\CO_{\Delta_3})}{(-2P_1\cdot P_2)^{\frac{\Delta_1+\Delta_2-\Delta_3}{2}}(-2P_2\cdot P_3)^{\frac{\Delta_2+\Delta_3-\Delta_1}{2}}(-2P_1\cdot P_3)^{\frac{\Delta_1+\Delta_3-\Delta_2}{2}}}\  .
\end{aligned}
\end{align}
In the physical space, we have:
\begin{align}\label{eq:conformal inv of 3pt CFT correlator phy}
    \langle\, \CO_{\Delta_1}(x_1)\,\CO_{\Delta_2}(x_2)\,\CO_{\Delta_3}(x_3)\,\rangle=\frac{c(\CO_{\Delta_1},\CO_{\Delta_2},\CO_{\Delta_3})}{|x_{12}|^{\Delta_1+\Delta_2-\Delta_3}|x_{23}|^{\Delta_2+\Delta_3-\Delta_1}|x_{13}|^{\Delta_1+\Delta_3-\Delta_2}}\  .
\end{align}
Likewise, we can determine spinning three-point correlators from conformal symmetry up to three-point coefficients. As an illustration, a scalar-scalar-spin-$J$ three-point function becomes:
\begin{align}\label{eq:scalar-scalar-spin J}
\begin{aligned}
    \langle\, &\CO_{\Delta_1}(x_1)\,\CO_{\Delta_2}(x_2)\,\CO_{\Delta}^{\mu_1\cdots\mu_J}(x_3)\,\rangle\\
    &=c(\CO_{\Delta_1},\CO_{\Delta_2},\CO_{\Delta,J})\cdot \frac{H^{(\mu_1}(x_{13},x_{12})\cdots H^{\mu_J)}(x_{13},x_{12})}{|x_{12}|^{\Delta_1+\Delta_2-\Delta+J}|x_{13}|^{\Delta_1-\Delta_2+\Delta+J}|x_{23}|^{-\Delta_1+\Delta_2+\Delta+J}} \ ,
\end{aligned}
\end{align}
with 
\begin{align}
    H^{\mu}(x,y)=\frac{x^\mu}{x^2}-\frac{y^\mu}{y^2}\ .
\end{align}
Though only a single structure appears in \eqref{eq:scalar-scalar-spin J}, several terms may be possible for more general spinning three-point correlators, to each of which structure constants are associated.

\paragraph{Higher-point functions.} 
So far, we have seen the power of conformal symmetry to fix lower-point functions. Unlike three- and less-point functions, we cannot determine higher-point functions solely from conformal symmetry, as we can make conformal invariants out of four embedding space coordinates such as:
\begin{align}
\frac{(P_i\cdot P_j)\,(P_k\cdot P_l)}{(P_i\cdot P_k)\,(P_j\cdot P_l)}\ .   
\end{align}
To rephrase it, higher-point functions are non-trivial functions of conformal invariant structures called \textbf{cross ratios}. For instance, a scalar four-point function takes the form:
\begin{align}\label{eq:four-point function CFT}
    \langle\, \CO_{\Delta_1}(x_1)\,\CO_{\Delta_2}(x_2)\,\CO_{\Delta_3}(x_3)\,\CO_{\Delta_4}(x_4) \,\rangle=T^{(\Delta_1,\Delta_2)}_{(\Delta_3,\Delta_4)}(x_1,x_2,x_3,x_4)\cdot g(\mathfrak{u},\mathfrak{v})\ .
\end{align}
The function $T^{(\Delta_1,\Delta_2)}_{(\Delta_3,\Delta_4)}(x_1,x_2,x_3,x_4)$ transforms covariantly with local primaries on the left-hand side of \eqref{eq:four-point function CFT}:
\begin{align}
    T^{(\Delta_1,\Delta_2)}_{(\Delta_3,\Delta_4)}(x_1,x_2,x_3,x_4)=\frac{1}{|x_{12}|^{\Delta^+_{12}}|x_{34}|^{\Delta^+_{34}}}\cdot \frac{|x_{24}|^{\Delta^-_{12}}}{|x_{14}|^{\Delta^-_{12}}}\cdot \frac{|x_{14}|^{\Delta^-_{34}}}{|x_{13}|^{\Delta^-_{34}}}\ ,\qquad \Delta^{\pm}_{ij}=\Delta_i\pm \Delta_j\ ,
\end{align}
whereas $g(\mathfrak{u},\mathfrak{v})$ is some function of cross ratios:
\begin{align}
   \mathfrak{u}=\frac{x_{12}^2x_{34}^2}{x_{13}^2 x_{24}^2}\ ,\qquad \mathfrak{v}=\frac{x_{14}^2x_{23}^2}{x_{13}^2 x_{24}^2}\ .
\end{align}
However, with the help of the OPE \eqref{eq:OPE operator simplified}, one can translate an $n$-point function into the sum of $(n-1)$-point functions. By doing this operation many times, one can evaluate any-point functions. Hence, all we need is the knowledge of the operator spectrum $\Delta_i$ and the three-point coefficients $c(\CO_i,\CO_j,\CO_k)$ that govern the fusion rule of primaries. In this sense, model-dependent CFT data is none other than the operator spectrum and a set of three-point coefficients.\footnote{Be aware that one can regard a two-point coefficient $c(\CO,\CO)$ as the three-point coefficient that includes the identity operator: $c(\CO,\CO)=c(\CO,\CO,\bm{1})$.} 

There may be many ways to take OPEs, but all of which should give the same result, leading to non-trivial constraints on the CFT data as described below. Let us take scalar four-point function $ \langle\, \CO_{\Delta_1}\,\CO_{\Delta_2}\,\CO_{\Delta_3}\,\CO_{\Delta_4} \,\rangle$ as the simplest example.
Taking the OPEs of $\CO_1\times\CO_2$ and $\CO_3\times\CO_4$, we obtain the following decomposition:
\begin{align}\label{eq:four-point function CFT 1}
\begin{aligned}
        \langle\, \CO_{\Delta_1}&(x_1)\,\CO_{\Delta_2}(x_2)\,\CO_{\Delta_3}(x_3)\,\CO_{\Delta_4}(x_4) \,\rangle\\
        &=\sum_{\CO}\,C_{\CO_{\Delta_1}\CO_{\Delta_2}\CO}(x_{12},\partial/\partial x_2)\,C_{\CO_{\Delta_3}\CO_{\Delta_4}\CO}(x_{34},\partial/\partial x_4)\,\langle\, \CO(x_2)\,\CO(x_4)\,\rangle\\
        &=T^{(\Delta_1,\Delta_2)}_{(\Delta_3,\Delta_4)}(x_1,x_2,x_3,x_4)\cdot \sum_{\CO}\,\frac{c(\CO_{\Delta_1},\CO_{\Delta_2},\CO)\,c(\CO_{\Delta_3},\CO_{\Delta_4},\CO)}{c(\CO,\CO)}\cdot g^{\Delta^-_{12},\Delta^-_{34}}_{\CO}(\mathfrak{u},\mathfrak{v})\ .
\end{aligned}
\end{align}
Here, we implicitly used the orthogonality of primary operators, and the sum is taken over all the possible symmetric and traceless tensors. In the last line of \eqref{eq:four-point function CFT 1}, we have defined \textbf{conformal block} $g^{\Delta^-_{12},\Delta^-_{34}}_{\CO}(\mathfrak{u},\mathfrak{v})$ that packages the contribution from a single conformal multiplet. We refer to this type of expansion in CFT, which divides higher-point functions into individual contributions of primaries, as the \textbf{conformal block expansion}.

If we swap the first and the third scalar primaries and choose to take the OPEs $\CO_{\Delta_3}\times\CO_{\Delta_2}$ and $\CO_{\Delta_1}\times\CO_{\Delta_4}$, we end up with an alternative conformal block expansion:
\begin{align}\label{eq:four-point function CFT 2}
\begin{aligned}
        \langle\, \CO_{\Delta_1}&(x_1)\,\CO_{\Delta_2}(x_2)\,\CO_{\Delta_3}(x_3)\,\CO_{\Delta_4}(x_4) \,\rangle\\
        &=\sum_{\CO}\,C_{\CO_{\Delta_3}\CO_{\Delta_2}\CO}(x_{32},\partial/\partial x_2)\,C_{\CO_{\Delta_1}\CO_{\Delta_4}\CO}(x_{14},\partial/\partial x_4)\,\langle\, \CO(x_2)\,\CO(x_4)\,\rangle\\
        &=T^{(\Delta_3,\Delta_2)}_{(\Delta_1,\Delta_4)}(x_3,x_2,x_1,x_4)\cdot \sum_{\CO}\,\frac{c(\CO_{\Delta_3},\CO_{\Delta_2},\CO)\,c(\CO_{\Delta_1},\CO_{\Delta_4},\CO)}{c(\CO,\CO)}\cdot g^{\Delta^-_{32},\Delta^-_{14}}_{\CO}(\mathfrak{v},\mathfrak{u}) \ .
\end{aligned}
\end{align}
Equating these two expressions \eqref{eq:four-point function CFT 1} and \eqref{eq:four-point function CFT 2}, we obtain the following equality:
\begin{equation}
 \langle\, \CO_{\Delta_1}(x_1)\,\CO_{\Delta_2}(x_2)\,\CO_{\Delta_3}(x_3)\,\CO_{\Delta_4}(x_4) \,\rangle=\sum_{\CO}\,
\begin{tikzpicture}[baseline,transform shape,scale=0.7]

  \draw[black!100] (-0.5,0) -- (-1,0.5) node [above, font=\small] {$\CO_{\Delta_1}$}; 
  \draw[black!100] (-0.5,0) -- (-1,-0.5) node [below, font=\small] {$\CO_{\Delta_2}$}; 
    \draw[black!100] (0.5,0) -- (1,0.5) node [above, font=\small] {$\CO_{\Delta_4}$}; 
  \draw[black!100] (0.5,0) -- (1,-0.5) node [below, font=\small] {$\CO_{\Delta_3}$}; 
    \draw[black!100,thick,decorate,decoration={snake,amplitude=0.4mm,segment length=2mm,post length=0.1mm}] (-0.5,0) -- (0.5,0) node [midway,above, font=\small] {$\CO$}; 
   
\end{tikzpicture}
=\sum_{\CO'}\,
     \begin{tikzpicture}[baseline,transform shape,scale=0.7]

  \draw[black!100] (0,0.5) -- (-0.5,1) node [above, font=\small] {$\CO_{\Delta_1}$}; 
  \draw[black!100] (0,-0.5) -- (-0.5,-1) node [below, font=\small] {$\CO_{\Delta_2}$}; 
    \draw[black!100] (0,0.5) -- (0.5,1) node [above, font=\small] {$\CO_{\Delta_4}$}; 
  \draw[black!100] (0,-0.5) -- (0.5,-1) node [below, font=\small] {$\CO_{\Delta_3}$}; 
    \draw[black!100,thick,decorate,decoration={snake,amplitude=0.4mm,segment length=2mm,post length=0.1mm}] (0,-0.5) -- (0,0.5) node [midway,right, font=\small] {$\CO'$}; 
\end{tikzpicture}\ ,
\end{equation}
or
\begin{align}\label{eq:crossing equation CFT}
\begin{aligned}
\sum_{\CO}\,&\frac{c(\CO_{\Delta_1},\CO_{\Delta_2},\CO)\,c(\CO_{\Delta_3},\CO_{\Delta_4},\CO)}{c(\CO,\CO)}\cdot \frac{g^{\Delta^-_{12},\Delta^-_{34}}_{\CO}(\mathfrak{u},\mathfrak{v})}{\mathfrak{u}^{\Delta^+_{12}}}\\
&\qquad\qquad\qquad\qquad =\sum_{\CO'}\,\frac{c(\CO_{\Delta_3},\CO_{\Delta_2},\CO')\,c(\CO_{\Delta_1},\CO_{\Delta_4},\CO')}{c(\CO',\CO')}\cdot \frac{g^{\Delta^-_{32},\Delta^-_{14}}_{\CO'}(\mathfrak{v},\mathfrak{u}) }{\mathfrak{v}^{\Delta^+_{32}}}\ ,
\end{aligned}
\end{align}
which imposes severe constraints on the CFT data. In the analogy of scattering amplitudes in QFT, this identity that comes essentially from the OPE associativity is called the \textbf{crossing equation}.

\chapter{Elements of Defect Conformal Field Theory}\label{chap:Elements of Defect Conformal Field Theory}
So far, in the previous chapter, we have seen the power of conformal symmetry in QFT. Thanks to the conformal constraints on the theory, one can fix correlation functions up to model-dependent coefficients. The dilatation operator $\mathbf{D}$ works as the Hamiltonian in CFT, resulting in the radial quantization being much more flexible than the equal-time quantization in ordinary QFTs. It simplifies the structure of the Hilbert space quite a lot and enables us to have such concepts as the state/operator correspondence and the convergent Operator Product Expansion. All of these advantages would have never been possible without conformal symmetry.

In this chapter, we give a comprehensive exposition of DCFT mainly based on two seminal papers \cite{Billo:2016cpy} and \cite{Gadde:2016fbj}. In DCFT, we put a conformal defect operator on the CFT vacuum that keeps the maximal subgroup of the conformal group. Though the conformal defect partially breaks the whole conformal symmetry and makes the theory less constraining (section \ref{sec:Defect conformal symmetry}), many CFT techniques are still available with few modifications. 
We have two types of local operators in DCFT: bulk and defect local operators (section \ref{sec:Bulk and defect local primaries}). The bulk local operators are essentially the same as in the case without defects. But defect local operators, which stands for local excitation on the defect, are quantitatively different from bulk ones and should be treated differently. In section \ref{sec:Bulk and defect local primaries}, we also exemplify two universal defect local operators that any DCFTs have in their spectrum as the stress tensor in generic CFTs. Performing radial quantization in the presence of the defect, in addition to the ordinary OPEs, one has the Defect Operator Expansions (DOEs) of arbitrary bulk operators (section \ref{sec:Radial quantization revisited}). We then explain how we can constrain DCFT correlators by defect conformal symmetry (section \ref{sec:Correlation functions in DCFT}). Lastly, we write down in section \ref{eq:DOE of bulk scalar primary} and \ref{sec:Defect Operator Expansion spectrum of free scalar field} the explicit form of the DOE of a bulk scalar primary as well as the DOE spectrum of the Klein-Gordon field, both of which will play pivotal roles in our analysis.

\section{Defect conformal symmetry}\label{sec:Defect conformal symmetry}
Without loss of generality, we can place a $p$-dimensional planer conformal defect operator $\CD^{(p)}$ at $x^{\mu}=0$ ($\mu=p+1,\cdots,d$). In what follows, we divide the flat $d$-dimensional coordinate $x^\mu$ into two parts (see figure \ref{fig:illustration of line defect} as an illustration):
\begin{align}
        x^\mu=(\hat{x}^a,x_{\perp}^i)\ , \qquad \d s^2_{\mathbb{R}^d}=\d\hat{x}^a\d \hat{x}_a+\d x_{\perp}^i\d x_{\perp,i} \ .
\end{align}
    Here, $\hat{x}^a$ ($a=1,\cdots,p$) and $x_{\perp}^i$ ($i=p+1,\cdots,d$.) stand for parallel and transverse coordinates to the defect, respectively. Moreover, we use $\hat{y}$ for a point on the defect:
\begin{align}
    \hat{y}^\mu=(\hat{y}^a,0)\ .
\end{align}
 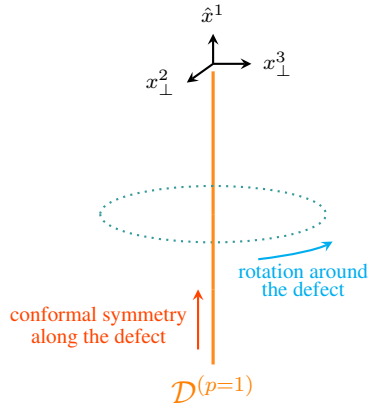
\begin{figure}[ht!]
	\centering
    \begin{tikzpicture}[transform shape,scale=1]
        \draw[thick, black!100,->,>=stealth] (0, 2)  -- (0, 2.4) node[above, black] {\scriptsize $\hat{x}^1$};
     
        \draw[very thick, orange, 
        opacity=0.9] (0,-2)  -- (0,-1);
        \draw[very thick, orange, opacity=0.9] (0,0)  -- (0,1.9);
     
        \draw[thick, black!100,->,>=stealth] (0,2)  -- (0.51,2) node[right, black] {\scriptsize $x_\perp^3$};
        \draw[thick, black!100,->,>=stealth] (0,2)  -- (-0.35, 1.75) node[left, black] {\scriptsize $x_\perp^2$};
        
        \begin{scope}[yshift=1cm]
             \draw[teal!80, thick, dotted] (-1.5,-1) arc (180:450:1.5cm and 0.375cm);
            \draw[teal!80, thick, dotted] (0,-0.625) arc (90:180:1.5cm and 0.375cm);
            \draw[cyan, thick,->, >=stealth 
            ] (0.6,-1.6) arc (-60:-20:2.25cm and 0.45cm) node[midway, below,font=\small,cyan] {${\substack{\text{rotation around}\\ \text{the defect}}}$}; 
        \end{scope} 
        
        \draw[very thick, orange, opacity=0.9] (0,-1)  -- (0,0);
    
        \begin{scope}[xshift=0.2cm, yshift=-1.3cm]
            \draw[thick,->,>=stealth,OrangeRed
            ] (-0.4,-0.5) to (-0.4,0.3);  
            \node[left ,font=\small,OrangeRed] at (-0.4,-0.2) {${\substack{\text{conformal symmetry}\\ \text{along the defect}}}$};
        \end{scope}

        \node[below, orange] at (0,-2) {\normalsize $\CD^{(p=1)}$};

    \end{tikzpicture}
    	\caption{Illustrated is a one-dimensional line defect $\CD^{(p=1)}$ in three-dimensional spacetime $(d=3)$. In this case, the spacetime symmetry of defect conformal field theory is conformal symmetry along the defect $\mathrm{SO}(1,2)\cong\mathrm{SL}(2,\mathbb{R})$ and rotation around the defect $\mathrm{SO}(2)$.}
	\label{fig:illustration of line defect}
\end{figure}
The conformal defect does not change its shape under parallel conformal transformations and transverse rotations. Hence, the conformal defect operator $\CD^{(p)}$ commutes with corresponding generators:\footnote{It is possible for $\CD^{(p)}$ to carry a transverse spin. But, in this thesis, we only consider scalar conformal defects that are singlet under transverse rotations.}
\begin{align}\label{eq:defect conformal generator on defect operator}
    [\mathbf{J}_{AB},\CD^{(p)}]&=0\ ,\qquad A,B=-1,0,1,\cdots ,p\ ,\\
     [\mathbf{J}_{IJ},\CD^{(p)}]&=0\ ,\qquad I,J=p+1,\cdots,d\ .
\end{align}
On the other hand, for other generators of conformal transformations that do not keep the shape of the defect, we have:
\begin{align}\label{eq:broken generator on defect operator}
    [\mathbf{J}_{AI},\CD^{(p)}]\neq0 \ ,\qquad  \begin{dcases}
    A=-1,0,1,\cdots,p\\
    I=p+1,\cdots, d
\end{dcases}\ .
\end{align}
Hence, in the presence of the $p$-dimensional defect, the whole conformal symmetry is broken down to the direct product of \textcolor{OrangeRed}{parallel conformal symmetry} and \textcolor{Cyan}{transverse rotational symmetry}:
\begin{align}
    \mathrm{SO}(1,d+1) \longrightarrow   {\color{OrangeRed}\mathrm{SO}(1,p+1)}\times {\color{Cyan}\mathrm{SO}(d-p)}\ .
\end{align}
In terms of conformal generators, we illustrate the symmetry-breaking pattern as follows:
\begin{align}\label{eq:defect conformal group embed}
  \mathbf{J}_{MN}= \begin{pNiceMatrix}[first-row,first-col]
  \CodeBefore
\rectanglecolor{OrangeRed!30}{1-1}{3-3}
\rectanglecolor{Cyan!30}{4-4}{4-4}
\rectanglecolor{Gray!30}{1-4}{3-4}
\rectanglecolor{Gray!30}{4-1}{4-3}
\Body
\ssymb{\substack{\ \\ \ \\ M\backslash\, \\\downarrow}} \ssymb{\substack{ N\to\\ \, }} & -1 & 0 & b & j \\
-1 & 0 & \mathbf{D} & \frac{1}{2}(\mathbf{P}_b-\mathbf{K}_b)& \cancel{\frac{1}{2}(\mathbf{P}_j-\mathbf{K}_j)} \\
0 & -\mathbf{D}& 0 & \frac{1}{2}(\mathbf{P}_b+\mathbf{K}_b) & \cancel{\frac{1}{2}(\mathbf{P}_j+\mathbf{K}_j)} \\
a &-\frac{1}{2}(\mathbf{P}_a-\mathbf{K}_a)& - \frac{1}{2}(\mathbf{P}_a+\mathbf{K}_a) & \mathbf{M}_{ab} &\cancel{\mathbf{M}_{aj}} \\
i &  \cancel{-\frac{1}{2}(\mathbf{P}_i-\mathbf{K}_i)}& \cancel{- \frac{1}{2}(\mathbf{P}_i+\mathbf{K}_i)} & \cancel{\mathbf{M}_{ib}} & \mathbf{M}_{ij}
\end{pNiceMatrix}\ .
\end{align}
In what follows, the remaining symmetry group in the presence of the defect is called the \textbf{defect conformal group}.

\section{Bulk and defect local primaries}\label{sec:Bulk and defect local primaries}
In DCFT, we start with the fixed background, in which a conformal defect sits on the CFT vacuum. We express localized excitations on and away from the defect by bulk and defect local operators in respective ways (see figure \ref{fig:local excitations on line defect}). The \textbf{bulk local operator} $\CO(x)$ is classified and denoted in the same way as ordinary CFTs according to the whole conformal group $\mathrm{SO}(1,d+1)$. So, one can express any bulk local operator by a linear combination of bulk primaries and descendants. On the other hand, the \textbf{defect local operator} $\widehat{\CO}(\hat{y})$ is organized along with the defect conformal group $\mathrm{SO}(1,p+1)\times\mathrm{SO}(d-p)$. The former part of the defect conformal group $\mathrm{SO}(1,p+1)$ acts on defect local operators as the conformal group parallel to the defect, and the latter part $\mathrm{SO}(d-p)$ works just as an \emph{internal symmetry}.
 \begin{figure}[ht!]
	\centering
    \begin{tikzpicture}[transform shape,scale=1]
        \draw[thick, black!100,->,>=stealth] (0, 2)  -- (0, 2.4) node[above, black] {\scriptsize $\hat{x}^1$};
     
        \draw[very thick, orange, 
        opacity=0.9] (0,-2)  -- (0,-1);
        \draw[very thick, orange, opacity=0.9] (0,0)  -- (0,1.9);
     
        \draw[thick, black!100,->,>=stealth] (0,2)  -- (0.51,2) node[right, black] {\scriptsize $x_\perp^3$};
        \draw[thick, black!100,->,>=stealth] (0,2)  -- (-0.35, 1.75) node[left, black] {\scriptsize $x_\perp^2$};
        
        \begin{scope}[yshift=1cm]
             \draw[teal!80, thick, dotted] (-1.5,-1) arc (180:450:1.5cm and 0.375cm);
            \draw[teal!80, thick, dotted] (0,-0.625) arc (90:180:1.5cm and 0.375cm);
            \draw[Emerald, thick,->, >=stealth 
            ] (0.6,-1.6) arc (-60:-20:2.25cm and 0.45cm) node[midway, below,font=\small,Emerald] {${\substack{\text{rotation around}\\ \text{the defect}}}$}; 
        \end{scope} 
        
        \draw[very thick, orange, opacity=0.9] (0,-1)  -- (0,0);
    
        \begin{scope}[xshift=0.2cm, yshift=-1.3cm]
            \draw[thick,->,>=stealth,Emerald
            ] (-0.4,-0.5) to (-0.4,0.3);  
            \node[left ,font=\small,Emerald] at (-0.4,-0.2) {${\substack{\text{conformal symmetry}\\ \text{along the defect}}}$};
        \end{scope}

        \node[below, orange] at (0,-2) {\normalsize $\CD^{(p=1)}$};
     
        \filldraw[black,very thick] (0,1.2) circle (0.05) node [left, font=\scriptsize,blue] {$\widehat{\CO} $};
    
        \filldraw[black,very thick] (0.5,0.82) circle (0.05);
        \node[font=\scriptsize,red] at (0.8,1.05) {$\CO $};

    \end{tikzpicture}
    	\caption{The bulk local operator ${\color{red}\CO}$ stands for the local excitations in the bulk, while the defect local operator ${\color{blue}\widehat{\CO}}$ represents local excitations on the defect.}
	\label{fig:local excitations on line defect}
\end{figure}
 
 We denote the conformal dimensions of defect primaries such as $\widehat{\Delta}$ to make a distinction from bulk conformal dimensions $\Delta$. The defect primaries can carry two kinds of spins: One comes from the parallel rotation group $\mathrm{SO}(p)$ associated with the spacetime symmetry for the defect primaries. The other is related to the transverse rotation group $\mathrm{SO}(d-p)$ that behaves just as an internal symmetry group for local operators on the defect. Throughout this dissertation, we only consider defect scalars that are $\mathrm{SO}(p)$ singlet but may carry $\mathrm{SO}(d-p)$ flavor indices. For interface and boundary ($p=d-1$), the defect local primaries fail to have transverse spin indices as the transverse rotation group becomes trivial.

The stress tensor $T_{\mu\nu}$ is often referred to as a \textbf{universal primary operator} because it is a conserved current $\partial_\mu \, T^{\mu\nu}=0$ having the protected conformal dimension $\Delta(T_{\mu\nu})=d$ and exists in any CFTs.
Here, as concrete examples of local excitations on the defect, we introduce two \textbf{universal defect local primaries}: \textbf{displacement operator} and \textbf{tilt operator}. The defect operator spectrum always contains the displacement operator. For each internal symmetry breaking on the defect, there will always be corresponding tilt operators in the theory. These two operators have protected conformal dimensions like the stress tensor. 

\paragraph{Displacement operator.}
In DCFT, the defect breaks the translational symmetry in its perpendicular directions. However, as this violation of translational symmetry is only localized on the defect, one can express it by the following anomalous conservation law of the stress tensor \cite{Gaiotto:2013nva,Billo:2016cpy}:
\begin{align}
    \partial^{\mu}\, T_{\mu a}(x)&=0 \ ,\\
    \partial^{\mu}\, T_{\mu i}(x)&=\delta^{(d-p)}(x_\perp) \cdot \widehat{D}_i(\hat{x})\ .\label{eq:anomalous conservation stress tensor}
\end{align}
We call the defect local operator appearing on the right-hand side of \eqref{eq:anomalous conservation stress tensor} the displacement operator. It represents the local deformation of the defect, and its conformal dimension is protected to be $(p+1)$ from the scaling property of both sides of \eqref{eq:anomalous conservation stress tensor}.

In the case of the co-dimension one defects $(p=d-1)$, the displacement operator no longer carries transverse spin indices, and the anomalous conservation law \eqref{eq:anomalous conservation stress tensor} can be integrated in the following manner:
\begin{align}\label{eq:displacement BCFT}
    \lim_{|x_\perp|\to0}\,T_{\perp\perp}(x)=\widehat{D}(\hat{x}) \ ,
\end{align}
where we have denoted $T_{\perp\perp}(x)$ instead of $T_{dd}(x)$.
This identity \eqref{eq:displacement BCFT} indicates that one can regard the displacement operator in BCFT as the limiting form of the stress tensor.

\paragraph{Tilt operator.}
Consider the case where the continuous global symmetry group $G$ breaks down to its subgroup $H$ on the defect. Let $J_A^{\mu}(x)$ be the broken currents, with $A=1,2,\cdots, \mathrm{dim}(G/H)$. Then, we have the following anomalous conservation law \cite{Padayasi:2021sik,Cuomo:2021cnb,AJBray_1980,Gimenez-Grau:2022czc}:
\begin{align}\label{eq:def of tilt}
  \partial_\mu\, J_A^{\mu} (x)=\delta^{(d-p)}(x_\perp)\cdot \widehat{t}_{A} (\hat{x})\ ,\qquad A=1,\cdots ,\mathrm{dim}\,(G/H)\ .
\end{align}
We call $\widehat{t}_{A} (\hat{y})$ the tilt operators. They are associated with violation of the continuous global symmetry and have protected conformal dimensions $\widehat{\Delta}(\widehat{t}_{A})=p$.\footnote{Recall that the conformal dimensions of spin-one conserved currents are $(d-1)$ due to the unitarity bound \eqref{eq:unitarity bound}.} When $p=d-1$, we have $\lim_{|x_\perp|\to0}\, J_A^{\perp} (x)=\widehat{t}_{A} (\hat{x})$ just as the displacement operator (see \eqref{eq:displacement BCFT}).

\section{Radial quantization revisited}\label{sec:Radial quantization revisited}
When a conformal defect operator $\CD^{(p)}$ is on the CFT vacuum, one has two ways of quantization: One is to perform radial quantization apart from the defect, while the other is right on the defect. The former allows us to validate the OPE for bulk local operators in the same manner as without defects. Meanwhile, the latter makes it possible to expand arbitrary bulk operators in terms of defect local ones.\footnote{Though it is not relevant in this thesis, we have an alternative operator expansion in DCFT: Let us perform conformal transformation and map the planer defect to a $p$-sphere with radius $R(<1)$ centered at the origin. Owing to the state/operator correspondence, one has the expansion of the conformal defect operator $\CD^{(p)}$ in terms of bulk local operators:
\begin{equation}
\begin{tikzpicture}[baseline,transform shape,scale=0.5]

    \draw[thick,dotted] (0,0) circle (1.5cm); 
        \fill[thick,gray!100,opacity=0.4] (0,0) circle (1.49cm); 
\filldraw[thick,black,font=\small] (0,0) circle (0.03cm);

                      \draw[thick,orange] (0,0) circle (1cm); 
         \node[above, orange] at (0,-1) {\large $\CD^{(p)}$};
\end{tikzpicture}
=\sum_{\CO}\,A_{\CD\CO}(R,\mathbf{P})\,
     \begin{tikzpicture}[baseline,transform shape,scale=0.5]

    \draw[thick,dotted] (0,0) circle (1.5cm); 
        \fill[thick,gray!100,opacity=0.4] (0,0) circle (1.49cm); 
        
                    \node[OrangeRed!100,thick,font=\small] at (-0.2,0.3) {$\CO(0)$};
\filldraw[thick,OrangeRed!100,font=\small] (0,0) circle (0.03cm);
                  
\end{tikzpicture}\ .
\end{equation}
This operator expansion makes it possible to regard the conformal defect operator $\CD^{(p)}$ as a coherent excitation of bulk local operators \cite{Gadde:2016fbj,Fukuda:2017cup}.
}

\paragraph{Radial quantization apart from the defect.}
Let us choose the radial quantization origin such that the unit $d$-ball integrated to create the in-vacuum does not intersect with the defect:
\begin{equation}
     |\Omega\rangle
     = 
     \begin{tikzpicture}[baseline,transform shape,scale=0.6]

    \draw[thick,dotted] (0,0) circle (1cm); 
        \fill[thick,gray!100,opacity=0.4] (0,0) circle (0.99cm); 

\filldraw[thick,Black!100,font=\small] (0,0) circle (0.03cm);

              \draw[very thick, orange, opacity=0.9] (1.5,-1.8)  -- (1.5,1.8);    
                  \node[above left, orange] at (1.5,-1.8) {\large $\CD^{(p)}$};

\end{tikzpicture}\ .
\end{equation}
The in-vacuum state so obtained is the same one as in ordinary CFTs and is conformally invariant:
\begin{align}
    \mathbf{J}_{MN}\,|\Omega\rangle=0 \ ,\qquad M,N=-1,0,1,\cdots, d\ .
\end{align}
Hence, on this in-vacuum, we can establish the state/operator correspondence and the OPEs for bulk primaries in the same way as in ordinary CFTs without defects:
\begin{equation}
\begin{tikzpicture}[baseline,transform shape,scale=0.7]

    \draw[thick,dotted] (0,0) circle (1cm); 
        \fill[thick,gray!100,opacity=0.4] (0,0) circle (0.99cm); 

   \node[RoyalBlue!100,thick,font=\small,above] at (0,0) {$\CO_2(x_2)$};
   \filldraw[thick,RoyalBlue!100,font=\small] (0,0) circle (0.03cm);
      \node[Sepia!100,thick,font=\small] at (-0.7,-1.1) {$\CO_1(x_1)$};
   \filldraw[thick,Sepia!100,font=\small] (-0.5,-0.7) circle (0.03cm);
                 \draw[very thick, orange, opacity=0.9] (1.5,-1.8)  -- (1.5,1.8);    
                  \node[above left, orange] at (1.5,-1.8) {\large $\CD^{(p)}$};
\end{tikzpicture}
=\sum_{\CO}\,C_{\CO_1\CO_2\CO}(x_{12},\mathbf{P})\,
     \begin{tikzpicture}[baseline,transform shape,scale=0.7]

    \draw[thick,dotted] (0,0) circle (1cm); 
        \fill[thick,gray!100,opacity=0.4] (0,0) circle (0.99cm); 

                    \node[OrangeRed!100,thick,font=\small] at (-0.2,0.3) {$\CO(x_2)$};
\filldraw[thick,OrangeRed!100,font=\small] (0,0) circle (0.03cm);

              \draw[very thick, orange, opacity=0.9] (1.5,-1.8)  -- (1.5,1.8);    
                  \node[above left, orange] at (1.5,-1.8) {\large $\CD^{(p)}$};
\end{tikzpicture}\ ,
\end{equation}
leading to the \textbf{bulk OPE}, or the OPE for bulk primaries:
\begin{align}\label{eq:OPE bulk operator DCFT}
\CO_1(x_1)\,\CO_2(x_2)=\sum_{\CO}\,C_{\CO_1\CO_2\CO}(x_{12},\partial/\partial x)\,\CO(x)\ .
\end{align}
The function $C_{\CO_1\CO_2\CO}(x_{12},\partial/\partial x)$ is again fixed up to all orders in $|x_{12}|$ by conformal symmetry and is proportional to the same three-point coefficients $c(\CO_1,\CO_2,\CO)$ that appear in the absence of defects. Thus, the fusion rule of bulk primaries in the presence of a defect is the same without defects.

Let us now create an out-vacuum by integrating over the exteriors of the unit $d$-ball. As the integrated region includes the defect operator $\CD^{(p)}$, this procedure defines the out-state decorated with the defect operator $\langle\CD^{(p)}|$:
\begin{equation}
   \langle\CD^{(p)}|
   = 
     \begin{tikzpicture}[baseline,transform shape,scale=0.6]

    \draw[thick,dotted] (0,0) circle (1cm); 
        \fill[thick,gray!100,opacity=0.4,even odd rule] (0,0) circle[radius=1.01cm] (-1.8,-1.8) rectangle(1.8,1.8); 
        
\filldraw[thick,Black!100,font=\small] (0,0) circle (0.03cm);

    \draw[very thick, orange, opacity=0.9] (1.5,-1.8)  -- (1.5,1.8);    
                  \node[above left, orange] at (1.5,-1.8) {\large $\CD^{(p)}$};
    
\end{tikzpicture}\ ,
\end{equation}
or equivalently $\langle\CD^{(p)}|=\langle\Omega|\,\CD^{(p)}$. It follows from \eqref{eq:defect conformal generator on defect operator} and \eqref{eq:broken generator on defect operator} that this decorated out-vacuum is invariant under defect conformal transformations rather than the whole conformal transformations (see \eqref{eq:defect conformal group embed} for the symmetry breaking pattern due to the defect):
\begin{align}
    \langle\CD^{(p)}|\,\mathbf{J}_{AB}&=0 \ ,\qquad A,B=-1,0,1,\cdots,p \ ,\\
   \langle\CD^{(p)}|\,\mathbf{J}_{IJ}&=0\ ,\qquad   I,J=p+1,\cdots, d\ , \\
\langle\CD^{(p)}|\,\mathbf{J}_{AI}&\neq0 \ ,\qquad  \begin{dcases}
    A=-1,0,1,\cdots,p\\
    I=p+1,\cdots, d
\end{dcases}\ .
\end{align}

\paragraph{Radial quantization on the defect and Defect Operator Expansion.}
If we take the radial quantization origin on the defect, both in- and out-vacuum states are affected by the defect as the integral domains intersect with the defect:
\begin{equation}
     |\widehat{\CD}^{(p)}\rangle=
     \begin{tikzpicture}[baseline,transform shape,scale=0.6]

    \draw[thick,dotted] (0,0) circle (1cm); 
        \fill[thick,gray!100,opacity=0.4] (0,0) circle (0.99cm);

              \draw[very thick, orange, opacity=0.9] (0,-1.8)  -- (0,1.8);    
                  \node[above left, orange] at (0,-1.8) {\large $\CD^{(p)}$};
\filldraw[thick,Black!100,font=\small] (0,0) circle (0.03cm);
\end{tikzpicture}\ ,\qquad   
\langle\widehat{\CD}^{(p)}|=
\begin{tikzpicture}[baseline,transform shape,scale=0.6]

    \draw[thick,dotted] (0,0) circle (1cm); 
        \fill[thick,gray!100,opacity=0.4,even odd rule] (0,0) circle[radius=1.01cm] (-1.8,-1.8) rectangle(1.8,1.8); 

    \draw[very thick, orange, opacity=0.9] (0,-1.8)  -- (0,1.8);    
                  \node[above left, orange] at (0,-1.8) {\large $\CD^{(p)}$};
 \filldraw[thick,Black!100,font=\small] (0,0) circle (0.03cm);   
\end{tikzpicture}\ .
\end{equation}
They are only invariant under the defect conformal transformations:
\begin{align}
    \mathbf{J}_{AB}\,  |\widehat{\CD}^{(p)}\rangle&=\langle\widehat{\CD}^{(p)}|\,\mathbf{J}_{AB}=0 \ ,\qquad A,B=-1,0,1,\cdots,p \ ,\\
   \mathbf{J}_{IJ}\,  |\widehat{\CD}^{(p)}\rangle&=\langle\widehat{\CD}^{(p)}|\,\mathbf{J}_{IJ}=0\ ,\qquad   I,J=p+1,\cdots, d\ , 
\end{align}
and
\begin{align}
     \mathbf{J}_{AI}\,  |\widehat{\CD}^{(p)}\rangle\neq0\ ,\qquad  \langle\widehat{\CD}^{(p)}|\,\mathbf{J}_{AI}\neq0\ ,\qquad  \begin{dcases}
    A=-1,0,1,\cdots,p\\
    I=p+1,\cdots, d
\end{dcases}\ .
\end{align}
Similarly to section \ref{sec:State/operator correspondence CFT}, one can establish a one-to-one correspondence between a state on the unit $d$-ball and a defect local operator inserted at the quantization origin. In particular, a defect local primary $\widehat{\CO}$ corresponds to the defect primary state in the following manner:
\begin{align}
    \widehat{\CO}(0)\longleftrightarrow | \widehat{\CO}\rangle\equiv  \widehat{\CO}(0)\,|\widehat{\CD}^{(p)}\rangle\ .
\end{align}
This correspondence makes it possible to expand arbitrary operators in terms of defect local ones. 

For a state created by acting two defect primaries on $|\widehat{\CD}^{(p)}\rangle$, we have:
\begin{equation}
\begin{tikzpicture}[baseline,transform shape,scale=0.7]

    \draw[thick,dotted] (0,0) circle (1cm); 
        \fill[thick,gray!100,opacity=0.4] (0,0) circle (0.99cm); 
                 \draw[very thick, orange, opacity=0.9] (0,-1.8)  -- (0,1.8);    
                  \node[above left, orange] at (0,-1.8) {\large $\CD^{(p)}$};

   \node[RoyalBlue!100,thick,font=\small,below right] at (0,0) {$\widehat{\CO}_2(\hat{y}_2)$};
   \filldraw[thick,RoyalBlue!100,font=\small] (0,0) circle (0.03cm);
      \node[Sepia!100,thick,font=\small,right] at (0,0.7) {$\widehat{\CO}_1(\hat{y}_1)$};
   \filldraw[thick,Sepia!100,font=\small] (0,0.7) circle (0.03cm);

\end{tikzpicture}
=\sum_{\widehat{\CO}}\,\widehat{C}_{\widehat{\CO}_1\widehat{\CO}_2\widehat{\CO}}(\hat{y}_{12},\widehat{\mathbf{P}})\,
     \begin{tikzpicture}[baseline,transform shape,scale=0.7]

    \draw[thick,dotted] (0,0) circle (1cm); 
        \fill[thick,gray!100,opacity=0.4] (0,0) circle (0.99cm); 
              \draw[very thick, orange, opacity=0.9] (0,-1.8)  -- (0,1.8);    
                  \node[above left, orange] at (0,-1.8) {\large $\CD^{(p)}$};
                  
                    \node[OrangeRed!100,thick,font=\small,right] at (0,0.3) {$\widehat{\CO}(\hat{y}_2)$};
\filldraw[thick,OrangeRed!100,font=\small] (0,0) circle (0.03cm);

\end{tikzpicture}\ ,
\end{equation}
with $\widehat{\mathbf{P}}$ being translation generators of the defect conformal group. This expansion is nothing but the \textbf{defect OPE}, or the OPE for defect primaries:
\begin{align}\label{eq:OPE defect operator DCFT gen}
\widehat{\CO}_1(\hat{y}_1)\,\widehat{\CO}_2(\hat{y}_2)=\sum_{\widehat{\CO}}\,\widehat{C}_{\widehat{\CO}_1\widehat{\CO}_2\widehat{\CO}}(\hat{y}_{12},\partial/\partial \hat{y})\,\widehat{\CO}(\hat{y})\ .
\end{align}
Similarly, by expanding a state with a single insertion of a bulk local operator, we have:
\begin{equation}
\begin{tikzpicture}[baseline,transform shape,scale=0.7]

    \draw[thick,dotted] (0,0) circle (1cm); 
        \fill[thick,gray!100,opacity=0.4] (0,0) circle (0.99cm); 
                 \draw[very thick, orange, opacity=0.9] (0,-1.8)  -- (0,1.8);    
                  \node[above left, orange] at (0,-1.8) {\large $\CD^{(p)}$};

          \node[black!100,thick,font=\small] at (-0.2,-0.2) {$\hat{y}$};
\filldraw[thick,Black!100,font=\small] (0,0) circle (0.03cm);

      \node[Sepia!100,thick,font=\small,below left] at (-0.7,-0.3) {$\CO(x)$};
   \filldraw[thick,Sepia!100,font=\small] (-0.7,-0.3) circle (0.03cm);

\end{tikzpicture}
=\sum_{\widehat{\CO}}\,B_{\CO\widehat{\CO}}(\hat{x}-\hat{y},x_\perp,\widehat{\mathbf{P}})\,
     \begin{tikzpicture}[baseline,transform shape,scale=0.7]

    \draw[thick,dotted] (0,0) circle (1cm); 
        \fill[thick,gray!100,opacity=0.4] (0,0) circle (0.99cm); 
              \draw[very thick, orange, opacity=0.9] (0,-1.8)  -- (0,1.8);    
                  \node[above left, orange] at (0,-1.8) {\large $\CD^{(p)}$};
                  
                    \node[OrangeRed!100,thick,font=\small,right] at (0,0.3) {$\widehat{\CO}(\hat{y})$};
\filldraw[thick,OrangeRed!100,font=\small] (0,0) circle (0.03cm);

\end{tikzpicture}\ .
\end{equation}
This operator expansion can be regarded as an operator identity and called the \textbf{Defect Operator Expansion (DOE)}, or the \textbf{Boundary Operator Expansion (BOE)} in case of boundary:
\begin{align}\label{eq:DOE gen expression 1}
\CO(x)=\sum_{\widehat{\CO}}\,B_{\CO\widehat{\CO}}(\hat{x}-\hat{y},x_\perp,\partial/\partial \hat{y})\,\widehat{\CO}(\hat{y})\ .
\end{align}
In making DOEs, we often match the parallel coordinates of the bulk and the defect local operators just for convenience:
\begin{align}\label{eq:DOE gen expression 2}
\CO(x)=\sum_{\widehat{\CO}}\,B_{\CO\widehat{\CO}}(x_\perp,\partial/\partial \hat{x})\,\widehat{\CO}(\hat{x})\ .
\end{align}
One can fix the functions that appear in \eqref{eq:OPE defect operator DCFT gen} and \eqref{eq:DOE gen expression 2} by comparing the defect conformal transformation law for both sides of the equations or by applying these operator identities inside correlation functions:
\begin{align}
\langle\,\CO(x)\,\widehat{\CO}(\hat{y})\,\rangle&=\sum_{\widehat{\CO}}\,B_{\CO\widehat{\CO}}(x_\perp,\partial/\partial \hat{x})\,\langle\,\widehat{\CO}(\hat{x})\,\widehat{\CO}(\hat{y})\,\rangle\ ,\\
\langle \,\widehat{\CO}_1(\hat{y}_1)\,\widehat{\CO}_2(\hat{y}_2)\,\widehat{\CO}_3(\hat{y}_3)\rangle&=\sum_{\widehat{\CO}}\,\widehat{C}_{\widehat{\CO}_1\widehat{\CO}_2\widehat{\CO}}(\hat{y}_{12},\partial/\partial \hat{y}_2)\,\langle\,\widehat{\CO}(\hat{y}_2)\,\widehat{\CO}(\hat{y}_3)\,\rangle\ .
\end{align}
Hence, the two functions $B_{\CO\widehat{\CO}}(x_\perp,\partial/\partial \hat{x})$ and $\widehat{C}_{\widehat{\CO}_1\widehat{\CO}_2\widehat{\CO}}(\hat{y}_{12},\partial/\partial \hat{y})$ are proportional to the bulk-defect two-point coefficients of $\langle\,\CO\,\widehat{\CO}\,\rangle$ and the defect three-point coefficients of $\langle\,\widehat{\CO}_1\,\widehat{\CO}_2\,\widehat{\CO}\,\rangle$ in respective ways ($b(\CO,\widehat{\CO})$ in \eqref{eq:bulk-defect 2pt} and $c(\CO_1,\CO_2,\widehat{\CO})$ in \eqref{eq:defect three-point} for scalar cases, see the next section for details). In particular, for the identity channel DOE of a bulk primary $\CO$, the function $B_{\CO\bm{1}}(x_\perp,\partial/\partial \hat{x})$ is proportional to the bulk one-point coefficient $a(\CO)$ (see \eqref{eq:bulk one-point} for the definition for bulk scalars), since $b(\CO,\bm{1})$ can be identified with $a(\CO)$.

\section{Correlation functions}\label{sec:Correlation functions in DCFT}
We now use the embedding space formalism to investigate correlation functions in DCFT similarly to section \ref{sec:Correlation functions in CFT}. Within this formalism, we can treat bulk scalar primaries in the same manner as without defect, while we must make a little twist for defect local primaries with SO$(d-p)$ flavor indices. After setting the stage, we will systematically construct DCFT correlators, exploiting homogeneity conditions and defect conformal Ward identities in the embedding space coordinates.

\paragraph{Embedding space formalism for defect scalar primaries.}
We deal with defect scalar primaries with $\mathrm{SO}(d-p)$ flavor indices using the embedding space formalism combined with encoding polynomial techniques for symmetric traceless tensors. Let $\widehat{\CO}_{\widehat{\Delta},i_1\cdots i_s}(\hat{y})$ be a defect scalar primary operator with conformal dimension $\widehat{\Delta}$ and transverse spin of rank-$s$. We denote its embedding space encoding polynomial of degree-$s$ by $\widehat{\CO}_{\widehat{\Delta},s}(Q,W)$. We impose the following homogeneity condition on $\widehat{\CO}_{\widehat{\Delta},s}(Q,W)$:
\begin{align}\label{eq:homogeneity condition defect local}
         \widehat{\CO}_{\widehat{\Delta},s}(\lambda\,Q,\alpha\,W)= \lambda^{-\widehat{\Delta}}\,\alpha^{s}\cdot \widehat{\CO}_{\widehat{\Delta},s}(Q,W)\qquad \lambda,\alpha>0\ .
 \end{align}
Here, we denoted the embedding space coordinates on the defect by $Q^M=(Q^A,Q^I=0)$ being subject to the relation $Q^2=Q^A\,Q_A=0$.
We also introduced a polarization vector associated with the transverse rotation group $ W^M=(W^A=0,W^I)$ with $W^I\in\mathbb{C}^{d-p}$ and $W^2=W^I\,W_I=0$.
The generators of the defect conformal group act on the embedding space encoding polynomial $\widehat{\CO}_{\widehat{\Delta},s}(Q,W)$ as:
\begin{align}\label{eq:generator action embed defect local 1}
\begin{aligned}
       [ \mathbf{J}_{AB},\widehat{\CO}_{\widehat{\Delta},s}(Q,W)]&=\widehat{\CJ}_{AB}(Q,W)\,\widehat{\CO}_{\widehat{\Delta},s}(Q,W)\\
       &=-\left(Q_A\,\frac{\partial}{\partial Q_B}-Q_B\,\frac{\partial}{\partial Q_A}\right)\,\widehat{\CO}_{\widehat{\Delta},s}(Q,W)\ ,
\end{aligned}
\end{align}
and 
\begin{align}\label{eq:generator action embed defect local 2}
\begin{aligned}
             [ \mathbf{J}_{IJ},\widehat{\CO}_{\widehat{\Delta},s}(Q,W)]&=\widehat{\CJ}_{IJ}(Q,W)\,\widehat{\CO}_{\widehat{\Delta},s}(Q,W)\\
             & =-\left(W_I\,\frac{\partial}{\partial W_J}-W_J\,\frac{\partial}{\partial W_I}\right)\,\widehat{\CO}_{\widehat{\Delta},s}(Q,W)\ .
\end{aligned}
\end{align}
We recover the physical space encoding polynomial $\widehat{\CO}_{\widehat{\Delta},s}(\hat{y},w)$ by setting:
\begin{align}
    Q^A=(1,\hat{y}^2,\hat{y}^a)\ ,\qquad   W^i=w^i\ ,\qquad w^2=w^i\,w_i=0\ .
\end{align}
After stripping all polarization vector $w^i$ and subtracting traces, we recover $\widehat{\CO}_{\widehat{\Delta},i_1\cdots i_s}(\hat{y})$.

\paragraph{Defect conformal Ward identities.}
We define correlation functions of local primaries in DCFT $\langle\cdots\rangle_{\text{DCFT}}$ in relation to CFT correlators $\langle\cdots\rangle_{\text{CFT}}$ by:
\begin{align}\label{eq:def of DCFT correlator}
    \langle\,\CO_1(x_1)\,\cdots \widehat{\CO}_1(\hat{y}_1)\cdots\,\rangle_{\text{DCFT}}=\frac{  \langle\,\CD^{(p)}\,\CO_1(x_1)\,\cdots \widehat{\CO}_1(\hat{y}_1)\cdots\,\rangle_{\text{CFT}}}{\langle\,\CD^{(p)}\,\rangle_{\text{CFT}}}\ .
\end{align}
For the expectation value of the identity operator to be one $ \langle\,\bm{1}\,\rangle_{\text{DCFT}}=1$, the right-hand side of \eqref{eq:def of DCFT correlator} is divided by the one-point function of the conformal defect operator:
\begin{align}
    \langle\,\CD^{(p)}\,\rangle_{\text{CFT}}\equiv \langle\Omega|\,\CD^{(p)}\,|\Omega\rangle=\langle\CD^{(p)}|\Omega\rangle=\langle\widehat{\CD}^{(p)}|\widehat{\CD}^{(p)}\rangle\ .
\end{align}
One can expand the numerator in the right-hand side of \eqref{eq:def of DCFT correlator} in two ways by choosing radial quantization origin on and away from the defect (see the second and the third line, respectively):
\begin{align}
    \begin{aligned}
       &\langle\,\CD^{(p)}\,\CO_1(x_1)\,\cdots \widehat{\CO}_1(\hat{y}_1)\cdots\,\rangle_{\text{CFT}}\\
        &=\langle\widehat{\CD}^{(p)}|\,\mathrm{R}\{\CO_1(x_1)\,\cdots \widehat{\CO}_1(\hat{y}_1)\cdots\}\,|\widehat{\CD}^{(p)}\rangle\\
        &=\langle\CD^{(p)}|\,\mathrm{R}\{\CO_1(x_1)\,\cdots \widehat{\CO}_1(\hat{y}_1)\cdots\}\,|\Omega\rangle\ .
    \end{aligned}
\end{align}
In both cases, the similar manipulation to section \ref{sec:Correlation functions in CFT} by use of the embedding space formalism leads to the defect conformal Ward identities. In particular, for the correlation functions of bulk and defect scalar primaries, we have:
\begin{align}\label{eq:conformal inv of DCFT correlator }
\begin{aligned}
      &\left(\sum_{\alpha=1}^n\,\CJ_{AB}(P_\alpha)+\sum_{\beta=1}^m\,\widehat{\CJ}_{AB}(Q_\beta,W_\beta)\right)\\
      &\qquad\qquad\cdot \langle\, \CO_{\Delta_1}(P_1)\,\cdots \CO_{\Delta_n}(P_n)\,\widehat{\CO}_1(Q_1,W_1)\cdots\widehat{\CO}_m(Q_m,W_m)\,\rangle_{\text{DCFT}}=0 \ ,\\
      &\left(\sum_{\alpha=1}^n\,\CJ_{IJ}(P_\alpha)+\sum_{\beta=1}^m\,\widehat{\CJ}_{IJ}(Q_\beta,W_\beta)\right)\\
      &\qquad\qquad\cdot \langle\, \CO_{\Delta_1}(P_1)\,\cdots \CO_{\Delta_n}(P_n)\,\widehat{\CO}_1(Q_1,W_1)\cdots\widehat{\CO}_m(Q_m,W_m)\,\rangle_{\text{DCFT}}=0 \ .
\end{aligned}
\end{align}
Akin to the CFT case, we learn from these defect conformal Ward identities that DCFT correlators depend on the embedding space coordinates only through the defect conformal invariants such as:
\begin{align}
X\cdot X'= X^M\,X'_M\ ,\qquad  X\bullet X'=X^A\,X'_A\ ,\qquad  X\circ X'=X^I\,X'_I\ .
\end{align}
Though conformal symmetry seems less constraining than without defect, it is still powerful enough to fix correlation functions up to model-dependent coefficients as in ordinary CFTs. In what follows, we abbreviate the subscript in $\langle\cdots\rangle_{\text{DCFT}}$ to simplify the notations unless otherwise noted and construct scalar correlation functions in DCFT.

\paragraph{Correlation functions of defect primaries.}
With no bulk operator insertions, DCFT correlation functions behave in the same manner as $p$-dimensional CFT correlators with $\mathrm{SO}(d-p)$ flavor indices. All one-point functions vanish due to the conformal symmetry on the defect. Defect primaries in different irreducible representations are orthogonal. The two-point function of defect identical primaries in the embedding space is as follows:
\begin{align}
    \langle\, \widehat{\CO}_{\widehat{\Delta},s}(Q_1,W_1)\,\widehat{\CO}_{\widehat{\Delta},s}(Q_2,W_2) \,\rangle=c(\widehat{\CO}_{\widehat{\Delta},s},\widehat{\CO}_{\widehat{\Delta},s})\cdot\frac{(W_1\circ W_2)^s}{(-2Q_1\circ Q_2)^{\widehat{\Delta}}}\ ,
\end{align}
which, in physical space, reduces to:\footnote{For instance, the two-point function of the displacement operator is fixed by defect conformal symmetry to have the form:
\begin{align}
      \langle\, \widehat{D}^i(\hat{y}_1)\,\widehat{D}^j(\hat{y}_2) \,\rangle=C_{\widehat{D}}\cdot\frac{\delta^{ij}}{|\hat{y}_{12}|^{2(p+1)}}\ .
\end{align}
The coefficient $C_{\widehat{D}}$ is related to defect trace anomaly coefficients that are key ingredients characterizing the dynamics of the defect (see e.g., \cite{Chalabi:2021jud,Herzog:2017kkj,Jensen:2018rxu,Bianchi:2015liz}).}
\begin{align}\label{eq:defect 2pt}
      \langle\, \widehat{\CO}_{\widehat{\Delta}}^{i_1\cdots i_s}(\hat{y}_1)\,\widehat{\CO}_{\widehat{\Delta},j_1\cdots j_s}(\hat{y}_2) \,\rangle=c(\widehat{\CO}_{\widehat{\Delta},s},\widehat{\CO}_{\widehat{\Delta},s})\cdot\frac{\delta_{j_1}^{(i_1}\cdots\delta_{j_s}^{i_s)}}{|\hat{y}_{12}|^{2\widehat{\Delta}}}\ .
\end{align}

Three-point functions of defect scalar primaries are rather involved due to the transverse spin structures. Here, we only mention two cases of importance in this thesis. When all operators are singlet in the transverse rotation group, the three-point correlator takes the same form as the CFT scalar three-point function \eqref{eq:conformal inv of 3pt CFT correlator phy}:
\begin{align}\label{eq:defect three-point}
    \langle\, \widehat{\CO}_{\widehat{\Delta}_1}(\hat{y}_1)\,\widehat{\CO}_{\widehat{\Delta}_2}(\hat{y}_2)\,\widehat{\CO}_{\widehat{\Delta}_3}(\hat{y}_3)\,\rangle=\frac{c(\widehat{\CO}_{\widehat{\Delta}_1},\widehat{\CO}_{\widehat{\Delta}_2},\widehat{\CO}_{\widehat{\Delta}_3})}{|\hat{y}_{12}|^{\widehat{\Delta}_1+\widehat{\Delta}_2-\widehat{\Delta}_3}|\hat{y}_{23}|^{\widehat{\Delta}_2+\widehat{\Delta}_3-\widehat{\Delta}_1}|\hat{y}_{13}|^{\widehat{\Delta}_1+\widehat{\Delta}_3-\widehat{\Delta}_2}}\  .
\end{align}
Meanwhile, when only one of the three primary operators has transverse spin indices, their correlation functions must vanish identically, as one cannot make $\mathrm{SO}(d-p)$-invariants out of a single polarization vector $W^I$:
\begin{align}\label{eq:defect three spin}
    \langle\, \widehat{\CO}_{\widehat{\Delta}_1}(\hat{y}_1)\,\widehat{\CO}_{\widehat{\Delta}_2}(\hat{y}_2)\,\widehat{\CO}_{\widehat{\Delta},i_1\cdots i_s}(\hat{y}_3)\,\rangle=0\qquad\qquad \text{for}\quad s\neq 0\  .
\end{align}
As in the CFT case, four and higher-point functions depend non-trivially on cross ratios and cannot be fixed solely from defect conformal symmetry. Nonetheless, one can investigate them by utilizing the defect OPEs \eqref{eq:OPE defect operator DCFT gen} to reduce higher-point functions into the sum of lower-point ones.

\paragraph{Bulk one-point functions.}
Unlike the case without defects, we have a single building block $P\circ P(=-P\bullet P)$, and the bulk scalar one-point function does not vanish:\footnote{While we do not use in this thesis, spinning bulk primary can also have a non-zero one-point function except the case of co-dimension one defects. For instance, the one-point function of the stress tensor $T_{\mu\nu}$ turns out to be:
\begin{align}
    \begin{aligned}
    \langle\,T_{ab}(x)\,\rangle=\frac{d-p-1}{d}\cdot \frac{a(T)}{|x_\perp|^d}\,\delta_{ab}\ ,\quad
      \langle\,T_{ij}(x)\,\rangle=-\frac{a(T)}{|x_\perp|^d}\cdot\left(\frac{p+1}{d}\,\delta_{ij}-\frac{x_i x_j}{|x_\perp|^2}\right)\ ,\quad
            \langle\,T_{ai}(x)\,\rangle=0\ .
    \end{aligned}
\end{align}
The one-point coefficient $a(T)$ is related to the boundary trace anomaly coefficients via Ward identities and carries dynamical information of the defect (see e.g., \cite{Jensen:2018rxu,Bianchi:2019sxz}).}
\begin{align}
    \langle\,\CO_{\Delta}(P)\,\rangle =\frac{a(\CO_{\Delta})}{(P\circ P)^{\frac{\Delta}{2}}}\ .
\end{align}
Back to physical space, we have:
\begin{align}\label{eq:bulk one-point}
    \langle\,\CO_{\Delta}(x)\,\rangle =\frac{a(\CO_{\Delta})}{|x_\perp|^{\Delta}}\ .
\end{align}

\paragraph{Bulk-defect two-point functions.}
We can construct a bulk-defect two-point function out of three defect conformal invariants $P\circ P,P\bullet Q,P\circ W$:
\begin{align}
    \langle\,\CO_{\Delta}(P)\,\widehat{\CO}_{\widehat{\Delta},s}(Q,W)\,\rangle=b(\CO_{\Delta},\widehat{\CO}_{\widehat{\Delta},s})\cdot\frac{(P\circ W)^s}{(-2P\bullet Q)^{\widehat{\Delta}}\,(P\circ P)^{\frac{\Delta-\widehat{\Delta}+s}{2}}}\ .
\end{align}
Reduction to physical space follows from the following replacements combined with stripping off $w^i$'s and subtracting traces:
\begin{align}
    P\circ P\mapsto |x_\perp|^2\ ,\qquad -2P\bullet Q\mapsto |\hat{x}-\hat{y}|^2+|x_\perp|^2\ ,\qquad P\circ W\mapsto x_\perp\cdot w\ .
\end{align}
We end up with the following expression:
\begin{align}\label{eq:bulk-defect 2pt}
  \langle\,\CO_{\Delta}(x)\,\widehat{\CO}_{\widehat{\Delta}}^{i_1\cdots i_s}(\hat{y})\,\rangle
  =b(\CO_{\Delta},\widehat{\CO}_{\widehat{\Delta},s})\cdot \frac{ x_\perp^{(i_1}\cdots x_\perp^{i_s)}}{(|\hat{x}-\hat{y}|^2+|x_\perp|^2)^{\widehat{\Delta}}\,|x_\perp|^{\Delta-\widehat{\Delta}+s}}\ .
\end{align}

\paragraph{On higher-point functions: DCFT data and crossing constraints.}
We again cannot fix higher-point functions, as they depend on the defect conformal invariants. Nevertheless, one can compute any-point functions utilizing DOEs and bulk/defect OPEs. All we need to do that is a set of model-dependent DCFT data, {\it i.e.,} the bulk and defect operator spectra and bulk/defect three-point coefficients that govern the fusion rules of bulk and defect primaries, and bulk-defect two-point coefficients that control defect operator contents that appear in the DOEs of bulk primaries:
\begin{align}\label{eq:DCFT data list}
    \{\Delta,\widehat{\Delta},c(\CO_i,\CO_j,\CO_k),c(\widehat{\CO}_i,\widehat{\CO}_j,\widehat{\CO}_k),b(\CO,\widehat{\CO}),a(\CO)\}\ .
\end{align}
This DCFT data is redundant: One can evaluate any correlators in CFT only by using DOEs and defect OPEs, but one still has bulk OPEs. The simplest example to illustrate this redundancy would be the bulk scalar two-point function $\langle\,\CO_{\Delta_1}(x_1)\,\CO_{\Delta_2}(x_2)\,\rangle$, depending non-trivially on two defect conformal invariants:
\begin{align}
    \frac{P_1\circ P_2}{(P_1\circ P_1)^{\frac{1}{2}}(P_2\circ P_2)^{\frac{1}{2}}}&\xrightarrow[\text{space}]{\text{physical}}\frac{x_{1,\perp}\cdot x_{2,\perp}}{|x_{1,\perp}|\cdot|x_{2,\perp}|}\ ,\\
  \frac{-2 P_1\cdot P_2}{(P_1\circ P_1)^{\frac{1}{2}}(P_2\circ P_2)^{\frac{1}{2}}}&\xrightarrow[\text{space}]{\text{physical}}\frac{|x_{12}|^2}{|x_{1,\perp}|\cdot|x_{2,\perp}|}  \ .
\end{align}
Similarly to the CFT four-point functions, one can investigate $\langle\,\CO_{\Delta_1}(x_1)\,\CO_{\Delta_2}(x_2)\,\rangle$ by performing the bulk OPE $\CO_{\Delta_1}\times \CO_{\Delta_2}$ or taking the DOEs of two bulk primaries:
\begin{align}
\begin{aligned}
       \langle\,\CO_{\Delta_1}(x_1)\,\CO_{\Delta_2}(x_2)\,\rangle&=\sum_{\CO}\,C_{\CO_{\Delta_1}\CO_{\Delta_2}\CO}(x_{12},\partial/\partial x_2)\,\langle\,\CO(x_2)\,\rangle\ ,\\
     &= \sum_{\widehat{\CO}}\,B_{\CO_{\Delta_1}\widehat{\CO}}(x_{1,\perp},\partial/\partial \hat{x}_1)\,B_{\CO_{\Delta_2}\widehat{\CO}}(x_{2,\perp},\partial/\partial \hat{x}_2)\,\langle\,\widehat{\CO}(\hat{x}_1)\,\widehat{\CO}(\hat{x}_2)\,\rangle\ .
\end{aligned}
\end{align}
This identity is a DCFT version of crossing equation \cite{Liendo:2012hy,Billo:2016cpy} and expected to give enough constraints to eliminate the redundancy of the DCFT data \eqref{eq:DCFT data list}:
\begin{equation}\label{eq:defect crossing equation}
    \langle\,\CO_{\Delta_1}(x_1)\,\CO_{\Delta_2}(x_2)\,\rangle=\sum_{\CO}\,
\begin{tikzpicture}[baseline,transform shape,scale=0.7]
  \draw[black!100] (-0.5,0) -- (-1,0.5) node [above, font=\small] {$\CO_{\Delta_1}$}; 
  \draw[black!100] (-0.5,0) -- (-1,-0.5) node [below, font=\small] {$\CO_{\Delta_2}$};

    \draw[black!100,thick,decorate,decoration={snake,amplitude=0.4mm,segment length=2mm,post length=0.1mm}] (-0.5,0) -- (0.5,0) node [midway,above, font=\small] {$\CO$}; 

                 \draw[very thick, orange, opacity=0.9] (0.5,-1.2)  -- (0.5,1.2);                   \node[above right, orange] at (0.5,-1.2) {\large $\CD^{(p)}$};
 \filldraw[ultra thick,black!100,font=\small] (0.5,0) circle (0.03cm);  
\end{tikzpicture}
=\sum_{\widehat{\CO}}\,
     \begin{tikzpicture}[baseline,transform shape,scale=0.7]
  \draw[black!100] (0,0.5) -- (-1,0.5) node [above, font=\small] {$\CO_{\Delta_1}$}; 
  \draw[black!100] (0,-0.5) -- (-1,-0.5) node [below, font=\small] {$\CO_{\Delta_2}$};

                 \draw[very thick, orange, opacity=0.9] (0,-1.2)  -- (0,1.2);                
                 \node[above right, orange] at (0,-1.2) {\large $\CD^{(p)}$};
                   \draw[black!100,thick,decorate,decoration={snake,amplitude=0.4mm,segment length=2mm,post length=0.1mm}] (0,-0.5) -- (0,0.5); 
                             \node[black!100,thick,right] at (0,0.5) {\small $\widehat{\CO}$};      
         \node[black!100,thick,right] at (0,-0.5) {\small $\widehat{\CO}$};      

\filldraw[ultra thick,black!100,font=\small] (0,-0.5) circle (0.03cm);
\filldraw[ultra thick,black!100,font=\small] (0,0.5) circle (0.03cm);
         
\end{tikzpicture}\ .
\end{equation}

\paragraph{Bulk-defect-defect three-point functions.}
Lastly, we explain bulk-defect-defect three-point functions of scalar primaries that will be of great use in our analysis.
We cannot fix them just from defect conformal symmetry but can evaluate them with the aid of DOEs. Throughout this thesis, we restrict ourselves to the case where two defect scalar primaries have no transverse spin indices. We now summarize the conformal block expansion of the bulk-defect-defect three-point functions and relegate its complete but somewhat lengthy derivation to appendix \ref{app:Conformal block expansion of bulk-defect-defect three-point function}. 

\begin{figure}[ht!]
    \centering
       \begin{tikzpicture}
          \draw[very thick, orange, opacity=0.9] (0,-3)  -- (0,2.7);    
                  \node[below, orange] at (0,-3) {\large $\CD^{(p)}$};
        
        \coordinate[label=0:$\hat{y}_1$] (A) at (0,-2) {};
        \coordinate[label=180:$x$] (B) at (-2.6,-0.3) {};
        \coordinate[label=0:$\hat{y}_2$] (C) at (0,1.5) {};
        
        \path[draw] (A) -- (B) -- (C) -- cycle;
        \draw[fill=gray!15,opacity=0.8]    (A) -- (B) -- (C) -- cycle;
        \node[RoyalBlue!100,font=\large] at (-1,-0.3) {$S$};
        
        \path pic["$\theta$",draw,angle radius=7mm,angle eccentricity=1.3] {angle = A--B--C};
     \end{tikzpicture} 
    \caption{The triangle spanned by one point $x$ on the bulk and two points $\hat{y}_1,\hat{y}_2$ on the defect $\CD^{(p)}$. The angle $\theta$ does not change under defect conformal transformations and is related to the defect cross ratio \eqref{eq:defect cross ratio} by $\upsilon=(\sin\theta)^2$.}
    \label{fig:cross-ratio}
\end{figure}
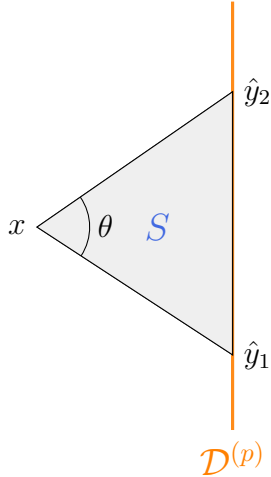

First of all, we make use of defect conformal symmetry to fix a bulk-defect-defect three-point function of scalar primaries up to some function of a single defect conformal invariant:
\begin{align}\label{eq:three-point embed}
\begin{aligned}
        \langle\, \CO_{\Delta}(P)&\,\widehat{\CO}_{\widehat{\Delta}_1}(Q_1)\,\widehat{\CO}_{\widehat{\Delta}_2}(Q_2) \,\rangle\\
        &=\frac{g(\upsilon)}{(P\circ P)^{\frac{\Delta}{2}} \,(-2P\bullet Q_1)^{\frac{\widehat{\Delta}^-_{12}}{2}}\,(-2P\bullet Q_2)^{\frac{\widehat{\Delta}^-_{21}}{2}}\,(-2Q_1\bullet Q_2)^{\frac{\widehat{\Delta}^+_{12}}{2}} }\ ,
\end{aligned}
\end{align}
where $\widehat{\Delta}^\pm_{ij}=\widehat{\Delta}_i\pm \widehat{\Delta}_j$ and the defect cross ratio $\upsilon$ is given by:
\begin{align}\label{eq:defect cross ration emb}
     \upsilon=\frac{(P\circ P)\,(-2Q_1\bullet Q_2)}{(-2P\bullet Q_1)\,(-2P\bullet Q_2)}\ .
\end{align}
In going back to physical space, we have:
\begin{align}\label{eq:bulk-defect-defect 3pt}
          \langle\, \CO_{\Delta}(x)\,\widehat{\CO}_{\widehat{\Delta}_1}(\hat{y}_1)\,\widehat{\CO}_{\widehat{\Delta}_2}(y_2) \,\rangle=T^{\widehat{\Delta}_1,\widehat{\Delta}_2}_{\Delta}(x,\hat{y}_1,\hat{y_2})\cdot g(\upsilon)\ .
\end{align}
Here, the function $T^{\widehat{\Delta}_1,\widehat{\Delta}_2}_{\Delta}(x,\hat{y}_1,\hat{y_2})$ behaves in the same way as the left-hand side of \eqref{eq:bulk-defect-defect 3pt} under defect conformal transformations. The cross ratio in the physical space is given by:
\begin{align}\label{eq:defect cross ratio}
\upsilon=\frac{|x_\perp|^2\,|\hat{y}_{12}|^2}{|x-\hat{y}_1|^2\,|x-\hat{y}_2|^2}
    \ ,
\end{align}
Notice that $\upsilon$ is related to an angle of the triangle formed by the three points $x,\hat{y}_1,\hat{y}_2$ (see figure \ref{fig:cross-ratio}) by $\upsilon=(\sin\theta)^2$. This geometrical interpretation follows immediately by considering the area of the triangle in two ways:
\begin{align}
        S=\frac{1}{2}\cdot |x_\perp|\cdot |\hat{y}_{12}|
        =\frac{1}{2}\cdot |\hat{x}-\hat{y}_1|\cdot |\hat{x}-\hat{y}_2|\cdot |\sin\theta|\ .
\end{align}
Hence, the defect cross ratio takes its value within the interval $[0,1]$:
\begin{align}
    0\leq \upsilon\leq 1\ .
\end{align}
The former equality is satisfied when the triangle collapses, whereas the latter holds whenever $\theta=\pi/2$.
The function of defect conformal invariant $g(\upsilon)$ can be evaluated by performing the DOE of the bulk operator $\CO_{\Delta}(x)$ to have the following expression:
\begin{align}
   g(\upsilon)=  \sum_{\widehat{\Delta}}\, \frac{b(\CO_{\Delta},\widehat{\CO}_{\widehat{\Delta}})\,c(\widehat{\CO}_{\widehat{\Delta}},\widehat{\CO}_{\widehat{\Delta}_1},\widehat{\CO}_{\widehat{\Delta}_2})}{c(\widehat{\CO}_{\widehat{\Delta}},\widehat{\CO}_{\widehat{\Delta}})}\cdot G^{\widehat{\Delta}^-_{12}}_{\widehat{\Delta}}(\upsilon)\ .
\end{align}
Here, we have denoted the conformal block in this expansion by $G^{\widehat{\Delta}^-_{12}}_{\widehat{\Delta}}(\upsilon)$ which collects all the contributions from a single DOE channel of $\CO_{\Delta}$ and has the following expression:
\begin{align}\label{eq:conformal block}
    G^{\widehat{\Delta}^-_{12}}_{\widehat{\Delta}}(\upsilon)=\upsilon^{\widehat{\Delta}/2}\cdot {}_2F_1\left(\frac{\widehat{\Delta}+\widehat{\Delta}^-_{12}}{2},\frac{\widehat{\Delta}-\widehat{\Delta}^-_{12}}{2};\widehat{\Delta}+1-\frac{p}{2};\upsilon\right)\ .
\end{align}
We have used the standard definitions of Gauss's hypergeometric function:
\begin{align}\label{eq:hypergeometric series expansion}
    {}_2F_1(\alpha,\beta;\gamma;z)=\sum_{n=0}^{\infty}\,\frac{(\alpha)_n(\beta)_n}{(\gamma)_n\,n!}\cdot z^n\ ,
\end{align}
where $(a)_n$ stands for the Pochhammer symbol defined through the relation:
\begin{align}
\begin{aligned}
        (a)_n&=a\cdot (a+1)\cdots (a+n-1)\\
        &=\frac{\Gamma(a+n)}{\Gamma(a)}\qquad  \text{for }\quad a\neq 0,-1,-2,\cdots \ .
\end{aligned}
\end{align}
Notably, only SO$(d-p)$ singlets contribute to the expansion because of \eqref{eq:defect three spin}.

Because the general form of the bulk-defect-defect three-point function is rather involved, for the sake of simplicity, it is convenient to utilize defect conformal transformations to set $(\hat{y}_1,\hat{y}_2)\mapsto(0,\infty)$, while defining the defect local primaries at the infinity by:
\begin{align}\label{eq:defect primary at infinity}
    \widehat{\CO}_{\widehat{\Delta}}(\infty)=\lim_{|\hat{y}|\to\infty}\,|\hat{y}|^{2\widehat{\Delta}}\cdot \widehat{\CO}_{\widehat{\Delta}}(\hat{y})\ .
\end{align}
Then, one ends up with the following simplified expressions:
\begin{align}\label{eq:conformal block expansion main}
    \begin{aligned}
              \langle\, \CO_{\Delta}(x)\,\widehat{\CO}_{\widehat{\Delta}_1}(0)\,\widehat{\CO}_{\widehat{\Delta}_2}(\infty) \,\rangle&=
                    \frac{1}{|x_\perp|^{\Delta} \,|x|^{\widehat{\Delta}^-_{12}}} \cdot\sum_{\widehat{\CO}}\, \frac{b(\CO_{\Delta},\widehat{\CO}_{\widehat{\Delta}})\,c(\widehat{\CO}_{\widehat{\Delta}},\widehat{\CO}_{\widehat{\Delta}_1},\widehat{\CO}_{\widehat{\Delta}_2})}{c(\widehat{\CO}_{\widehat{\Delta}},\widehat{\CO}_{\widehat{\Delta}})}\cdot G^{\widehat{\Delta}^-_{12}}_{\widehat{\Delta}}(\upsilon)\ ,\\
       G^{\widehat{\Delta}^-_{12}}_{\widehat{\Delta}}(\upsilon)&=\upsilon^{\widehat{\Delta}/2}\cdot {}_2F_1\left(\frac{\widehat{\Delta}+\widehat{\Delta}^-_{12}}{2},\frac{\widehat{\Delta}-\widehat{\Delta}^-_{12}}{2};\widehat{\Delta}+1-\frac{p}{2};\upsilon\right)\ ,             
    \end{aligned}
\end{align}
with $\upsilon=|x_\perp|^2/|x|^2$ being the defect cross ratio in this frame.

\section{Defect Operator Expansion of bulk scalar primary}\label{eq:DOE of bulk scalar primary}
We here record the explicit form of the DOE of a bulk scalar primary $\CO_{\Delta}(x)$ \cite[appendix B.1]{Billo:2016cpy}:
\begin{align}\label{eq:DOE of bulk scalar}
\begin{aligned}
        \CO_{\Delta}(x)&=\sum_{\widehat{\CO}} \,\frac{b(\CO_{\Delta},\widehat{\CO}_{\widehat{\Delta},s})/c(\widehat{\CO}_{\widehat{\Delta},s},\widehat{\CO}_{\widehat{\Delta},s})}{|x_\perp|^{\Delta-\widehat{\Delta}}}\\
        &\qquad\qquad\qquad\cdot\sum_{n=0}^\infty\,\frac{ x_\perp^{i_1}\cdots x_\perp^{i_s}\,(-1)^n\,|x_\perp|^{2n-s}}{2^{2n}\,(\widehat{\Delta}+1-p/2)_n\,n!}\cdot (\widehat{\partial}^{\,2}_x)^n\,\widehat{\CO}_{\widehat{\Delta},i_1\cdots i_s}(\hat{x})\ ,
\end{aligned}
\end{align}
where the coefficients $b(\CO_{\Delta},\widehat{\CO}_{\widehat{\Delta},s})$ and $c(\widehat{\CO}_{\widehat{\Delta},s},\widehat{\CO}_{\widehat{\Delta},s})$ are bulk-defect and defect two-point coefficients defined in \eqref{eq:bulk-defect 2pt} and \eqref{eq:defect 2pt}, respectively. The sum is taken over all possible defect scalar primaries with transverse spin indices. Notably, there are no contributions from defect primaries with parallel spin indices.\footnote{We can verify this fact similarly to footnote \ref{fot:symmetric and traceless argument} in section \ref{sec:Operator Product Expansion}.}

We can confirm \eqref{eq:DOE of bulk scalar} by using it inside the bulk-defect two-point function \eqref{eq:bulk-defect 2pt}:\footnote{In deriving this, we have used the following identities and \eqref{eq:defect 2pt}:
\begin{align}
    (\widehat{\partial}^{\,2}_x)^n\,\frac{1}{|\hat{x}-\hat{y}|^{2\widehat{\Delta}}}&=\frac{2^{2n}\,(\widehat{\Delta})_n\,(\widehat{\Delta}+1-p/2)_n}{|\hat{x}-\hat{y}|^{2\widehat{\Delta}+2n}}\ ,\\
    \sum_{n=0}^\infty\,\frac{(\widehat{\Delta})_n}{n!}\, \left(-\frac{|x_\perp|^{2}}{|\hat{x}-\hat{y}|^{2}}\right)^n&=\frac{1}{(1+|x_\perp|^{2}/|\hat{x}-\hat{y}|^{2})^{\widehat{\Delta}}}\ .
\end{align}}
\begin{align}
\begin{aligned}
  \langle\,\CO_{\Delta}(x)&\,\widehat{\CO}_{\widehat{\Delta}}^{i_1\cdots i_s}(\hat{y})\,\rangle
  =\frac{b(\CO_{\Delta},\widehat{\CO}_{\widehat{\Delta},s})/c(\widehat{\CO}_{\widehat{\Delta},s},\widehat{\CO}_{\widehat{\Delta},s})}{|x_\perp|^{\Delta-\widehat{\Delta}}}\\
  &\qquad\cdot \sum_{n=0}^\infty\,\frac{x^{j_1}_{\perp}\cdots x^{j_s}_{\perp}\,(-1)^n\,|x_\perp|^{2n-s}}{2^{2n}\,(\widehat{\Delta}+1-p/2)_n\,n!}\cdot (\widehat{\partial}^{\,2}_x)^n\,\langle\,\widehat{\CO}_{\widehat{\Delta},j_1\cdots j_s}(\hat{x})\,\widehat{\CO}_{\widehat{\Delta}}^{i_1\cdots i_s}(\hat{y})\,\rangle\\
  &=\frac{b(\CO_{\Delta},\widehat{\CO}_{\widehat{\Delta},s})}{|x_\perp|^{\Delta-\widehat{\Delta}}}\cdot\sum_{n=0}^\infty\,\frac{x^{j_1}_{\perp}\cdots x^{j_s}_{\perp}\,(-1)^n\,|x_\perp|^{2n-s}}{2^{2n}\,(\widehat{\Delta}+1-p/2)_n\,n!}\cdot (\widehat{\partial}^{\,2}_x)^n\,\frac{\delta_{j_1}^{(i_1}\cdots \delta_{j_s}^{i_s)}}{|\hat{x}-\hat{y}|^{2\widehat{\Delta}}}\\
  &=b(\CO_{\Delta},\widehat{\CO}_{\widehat{\Delta},s})\cdot\frac{x^{(i_1}_{\perp}\cdots x^{i_s)}_{\perp}}{|x_\perp|^{\Delta-\widehat{\Delta}+s}\,|\hat{x}-\hat{y}|^{2\widehat{\Delta}}}\cdot\sum_{n=0}^\infty\,\frac{(\widehat{\Delta})_n}{n!}\, \left(-\frac{|x_\perp|^{2}}{|\hat{x}-\hat{y}|^{2}}\right)^n\\  
  &=b(\CO_{\Delta},\widehat{\CO}_{\widehat{\Delta},s})\cdot\frac{x^{(i_1}_{\perp}\cdots x^{i_s)}_{\perp}}{(|\hat{x}-\hat{y}|^2+|x_\perp|^2)^{\widehat{\Delta}}\,|x_\perp|^{\Delta-\widehat{\Delta}+s}}\ .
\end{aligned}
\end{align}

In the case of boundary, the transverse rotational group becomes trivial, and we do not have to care about spin structures. Hence, the BOE of a bulk scalar takes the simpler form:
\begin{align}\label{eq:BOE of bulk scalar}
    \CO_{\Delta}(x)=\sum_{\widehat{\CO}} \,\frac{b(\CO_{\Delta},\widehat{\CO}_{\widehat{\Delta}})/c(\widehat{\CO}_{\widehat{\Delta}},\widehat{\CO}_{\widehat{\Delta}})}{|x_\perp|^{\Delta-\widehat{\Delta}}}\,\sum_{n=0}^\infty\,\frac{(-1)^n\,|x_\perp|^{2n}}{2^{2n}\,(\widehat{\Delta}+3/2-d/2)_n\,n!}\, (\widehat{\partial}^{\,2}_x)^n\,\widehat{\CO}_{\widehat{\Delta}}(\hat{x})\ .
\end{align}

\section{Defect Operator Expansion spectrum of free scalar field}\label{sec:Defect Operator Expansion spectrum of free scalar field}
In this section, we illustrate the defect operator contents that appear in the DOE of the free Klein-Gordon field in two- and more-dimensional spacetime for later use. We here do not deal with the case where the bulk Klein-Gordon field is not single-valued against the rotation around the defect, such as a monodromy defect. See refs. \cite{Lauria:2020emq,Soderberg:2017oaa} for readers who are interested in this point.

Let $\phi(x)$ be a Klein-Gorden field that satisfies the Klein-Gorden equation as an operator identity:
\begin{align}\label{eq:Klein-Gorden equation}
    \Box\,\phi(x)=0\ ,
\end{align}
with $\Box$ being the Laplacian in the flat $d$-dimensional spacetime. Owing to the defect conformal symmetry, we obtain the following DOE of $\phi(x)$ \eqref{eq:DOE of bulk scalar}:
\begin{align}\label{eq:DOE of free Klein Gordon field 0}
    \phi(x)\supset A\,\frac{x_{\perp}^{i_1}\cdots x_{\perp}^{i_s}}{|x_\perp|^{\Delta_{\phi}-\widehat{\Delta}+s}}\cdot\widehat{\CO}_{\widehat{\Delta},i_1\cdots i_s}(\hat{x})+(\text{descendants.})\ ,
\end{align}
with $A\equiv b(\phi,\widehat{\CO})/c(\widehat{\CO},\widehat{\CO})\neq 0$ and $\Delta_{\phi}=d/2-1$.
Because the Klein-Gorden field is subject to the Klein-Gorden equation \eqref{eq:Klein-Gorden equation}, the right-hand side of \eqref{eq:DOE of free Klein Gordon field 0} should vanish identically against the action of the Laplace differential operator $\Box_x\equiv \frac{\partial}{\partial x^{\mu}}\frac{\partial}{\partial x_{\mu}}$:
 \begin{align}\label{eq:RHS of free btd OPE}
 \begin{aligned}
       0&=\Box_x\,A\,\frac{x_{\perp}^{i_1}\cdots x_{\perp}^{i_s}}{|x_\perp|^{\Delta_{\phi}-\widehat{\Delta}+s}}\cdot\widehat{\CO}_{\widehat{\Delta},i_1\cdots i_s}(\hat{x})+(\text{descendants.})\\
          &=A\,(\widehat{\Delta}-s-\Delta_\phi)(\widehat{\Delta}+s-\Delta_\phi+d-p-2)\\
          &\qquad\cdot \frac{x_{\perp}^{i_1}\cdots x_{\perp}^{i_s}}{|x_\perp|^{\Delta_{\phi}-\widehat{\Delta}+s}}\cdot \widehat{\CO}_{\widehat{\Delta},i_1\cdots i_s}(\hat{x})+(\text{descendants.})\ ,
 \end{aligned}
 \end{align}
where we used the following identity:
\begin{align}\label{eq:action of Lap on spinning btd OPE}
       \Box_x\,\frac{x_\perp^{(i_1}\cdots x_\perp^{i_s)}}{|x_\perp|^\delta}=\delta\,(\delta+2-d+p-2s)\cdot \frac{x_\perp^{(i_1}\cdots x_\perp^{i_s)}}{|x_\perp|^{\delta+2}}\ .
\end{align}
For $A$ to be non-zero, one has the following equation:
\begin{align}
    (\widehat{\Delta}-s-\Delta_\phi)(\widehat{\Delta}+s-\Delta_\phi+d-p-2)=0\ .
\end{align}
There are two solutions to this equation:
\begin{align}\label{eq:EoM DOE free sol 1}
    \widehat{\Delta}=\frac{d-2}{2}+s\ ,
 \end{align}   
and
 \begin{align}\label{eq:EoM DOE free sol 2}
    \widehat{\Delta}=1+p-s-\frac{d}{2}\ .
\end{align}

\paragraph{Remarks on the first solution \eqref{eq:EoM DOE free sol 1}.}
The first case \eqref{eq:EoM DOE free sol 1} is compatible with the unitarity bound of the $p$-dimensional CFT on the defect:
\begin{align}\label{eq:defect local unitarity}
    \widehat{\Delta}\geq \begin{dcases}
    0& \text{for}\quad p=1,2\\
        \frac{p-2}{2}& \text{for}\quad p=3,\cdots,d-1
    \end{dcases}\ ,
\end{align}
where the equality is satisfied by the identity operator for $p=1,2$ and the Klein-Gordon field on the defect for $p=3,\cdots,d-1$.
One may identify the primary operator having the conformal dimension specified by the condition \eqref{eq:EoM DOE free sol 1} with the one defined through the limiting form:
\begin{align}
\widehat{\phi}_{i_1\cdots i_s}(\hat{x})=\lim_{x_\perp\to0}\,\partial_{(i_1}\cdots\partial_{i_s)} \phi(x)\ ,
\end{align}
To be more precise, the operator $\widehat{\phi}_{i_1\cdots i_s}(\hat{x})$ is defined inside correlators, if necessary, by subtracting divergences associated with the limit, so that they are finite and remain defect conformal invariant.

\paragraph{Remarks on the second solution \eqref{eq:EoM DOE free sol 2}.}
First of all, in the case of the co-dimension one defect $p=d-1$ where the transverse spin is trivial, the condition \eqref{eq:EoM DOE free sol 2} turns out to be $\widehat{\Delta}=d/2$, in line with the unitarity bound on the defect \eqref{eq:defect local unitarity}. The corresponding operator can be identified with $\widehat{\phi}{\,'}$ defined through the limiting form $\widehat{\phi}{\,'}(\hat{x})=\lim_{|x_\perp|\to0}\,|x_\perp|^{-1}\,\phi(x)$.

The identity operator with $\widehat{\Delta}=s=0$ can appear only when $p=d/2-1$. Relatedly, the bulk one-point functions of the Klein-Gordon field must vanish unless $p=d/2-1$. An example of this case is a line defect in four-dimensional spacetime $(d=4,p=1)$ that will be our main focus in section \ref{sec:line defect free}.

Let us investigate \eqref{eq:EoM DOE free sol 2} further combined with the unitarity bound on the defect \eqref{eq:defect local unitarity}, which implies:
\begin{align}
    \begin{dcases}
    s\leq 2-\frac{d}{2} & \text{for}\quad p=1\\
        s\leq 2-\frac{d-p}{2} &\text{for}\quad p=2,\cdots ,p-2\\
    \end{dcases}\ .
\end{align}
For one-dimensional defect ($p=1$), we cannot solve this inequality with finite $s$.\footnote{The only solution for $p=1$ seems $d=2,s=1$. In this case, however, the transverse rotational group becomes trivial, leading to a contradiction.}
For $p=2,\cdots,p-2$ the inequality is satisfied by the Klein-Gordon field on the co-dimension two defects with one transverse spin index ($p=d-2$, $\widehat{\Delta}=p/2-1$ and $s=1$). This situation is realized in vector coupled with co-dimension two matter (see \cite[the last example of section 5.4]{Billo:2016cpy}) but is irrelevant in this thesis.

\chapter{Review of Rychkov-Tan's approach to critical phenomena}\label{chap:Review of Rychkov-Tan}
This chapter is an introduction to the axiomatic approach to critical phenomena. After reviewing a conventional perturbative study of the critical O$(N)$ model without defects in $d=4-\epsilon$ dimensions (section \ref{sec:Critical O(N) vector model without defects}), we expand on Rychkov-Tan's axiomatic approach \cite{Rychkov:2015naa} (section \ref{sec:Axioms for homogeneous critical systems}). We then describe a generalization of the axiomatic approach in the presence of a defect proposed by \cite{Yamaguchi:2016pbj} (section \ref{sec:Axioms in the presence of a defect}).

\section{Critical O$(N)$ vector model without defects}\label{sec:Critical O(N) vector model without defects}
Here, we review a perturbative study of the O$(N)$ vector model without defects (see e.g., \cite{Wilson:1971dc,Wilson:1973jj,Collins:1984xc,Peskin:1995ev,Kleinert:2001ax,kardar_2007}, and also \cite{Henriksson:2022rnm} for a comprehensive review). 
The bare action of the O$(N)$ vector model in $d=4-\epsilon$ dimensions is given by:
\begin{align}\label{eq:action without defect}
    I = \int_{\mathbb{R}^{d}}\,\d^d x \,\left(\frac{1}{2}\, |\partial \Phi_{1} |^2\, +\,   \frac{\lambda_0}{4!}\, |\Phi_1|^4 \right) \, ,\qquad |\Phi_1|^4 \equiv (\Phi_{1}^{\alpha}\Phi_{1}^{\alpha})^2\ . 
\end{align}
We denote an O$(N)$ vector field by $\Phi_{1}^{\alpha}$ ($\alpha=1,\cdots, N$).
The two-point function of $\Phi_{1}^{\alpha}$ in the free limit $(\lambda_0=0)$ is normalized as follows:
\begin{align}\label{eq;2pt un-normalized}
    \langle\,\Phi_{1}^{\alpha}(x_1)\,\Phi_{1}^{\beta}(x_2)\,\rangle\big|_{\lambda_0=0}=\frac{1}{(d-2)\,\mathrm{Vol}(\mathbb{S}^{d-1})}\cdot \frac{\delta^{\alpha\beta}}{|x_{12}|^{d-2}}\ ,
\end{align}
with $\mathrm{Vol}(\mathbb{S}^{d-1})=2\pi^{d/2}/\Gamma(d/2)$ being the volume of a unit $(d-1)$-sphere.
We work in the massless renormalization in the minimal subtraction scheme by setting the bare and renormalized mass to zero, as we are only interested in the Wilson-Fisher fixed point.

We first introduce \textbf{wave-function renormalization} $Z_1$ and denote the \textbf{renormalized operator} by $W_{1}^{\alpha}(x)$ henceforth:
\begin{align}\label{eq:Phi1 bare renormalized rel}
    \Phi_{1}^{\alpha}(x)=Z_1\cdot  W_{1}^{\alpha}(x)\ .
\end{align}
Assuming $0<\epsilon\ll 1$, we expand the bare coupling $\lambda_0$ and the wave-function renormalization $Z_1$ in inverse powers of $\epsilon$ to cancel the divergences of the correlation functions of renormalized operators such as $\langle\,W_{1}^{\alpha}\,W_{1}^{\beta}  \,\rangle=Z_1^{-2}\,\langle\,\Phi_{1}^{\alpha}\,\Phi_{1}^{\beta}  \,\rangle$ and $\langle\,W_{1}^{\alpha}\,W_{1}^{\beta}\,W_{1}^{\gamma}  \,W_{1}^{\sigma} \,\rangle=Z_1^{-4}\,\langle\,\Phi_{1}^{\alpha}\,\Phi_{1}^{\beta}\,\Phi_{1}^{\gamma}  \,\Phi_{1}^{\sigma} \,\rangle$. Through standard diagrammatic calculations, one finds that:
\begin{align}
    Z_1&=1-\frac{\lambda^2}{(4\pi)^4}\cdot\frac{N+2}{36\epsilon}+O(\lambda^3)\ ,\\
    \lambda_0&=\mu^{\epsilon}\cdot\left(\lambda+\frac{N+8}{3\,\epsilon}\cdot \frac{\lambda^2}{(4\pi)^2}+O(\lambda^3)\right)\ ,\label{eq:lambda expanding invers order}
\end{align}
Here, we introduced a \textbf{momentum scale} $\mu$ for the \textbf{renormalized coupling constant} $\lambda$ to be dimensionless.
The scale dependence of $\lambda$ can be seen through the \textbf{beta function} $\beta_{\lambda}$:
 \begin{align}\label{eq:beta function}
     \beta_{\lambda}= \frac{\partial \lambda}{\partial \log \mu} =-\epsilon\, \lambda + \frac{N+8}{3}\cdot \frac{\lambda^2}{(4\pi)^2}+O(\lambda^3) \, .
 \end{align}
The theory becomes scale-independent and critical when the beta function vanishes $\beta_{\lambda}=0$. Its trivial solution is $\lambda=0$, which is uninteresting as the model is free and has no interacting degrees of freedom. One has a non-trivial zero of the beta function for $\lambda=\lambda_\ast$ with:
 \begin{align}\label{eq: Wilson-Fisher FP}
        \frac{\lambda_\ast}{(4\pi)^2}\equiv \frac{3}{N+8}\, \epsilon\, +\, O(\epsilon^2)\, .
\end{align}
This interacting fixed point on the theory space of coupling constants is called the \textbf{Wilson-Fisher fixed point}. The \textbf{anomalous dimension} of the O$(N)$ vector field by $\Phi_{1}^{\alpha}$ is denoted by $\gamma_1$ and can be computed as follows:
\begin{align}
\gamma_1=\left.\frac{\partial \log Z_1}{\partial \log \mu}\right|_{\lambda=\lambda_\ast}
=\frac{N+2}{4\,(N+8)^2}\,\epsilon^2+O(\epsilon^3)\ .
\end{align}

We now comment on the relation between conformal dimension $\Delta_1$ of the renormalized operator $W_{1}^{\alpha}(x)$ and anomalous dimension $\gamma_1$ of the bare operator $\Phi_{1}^{\alpha}$. From the relation \eqref{eq:Phi1 bare renormalized rel}, the two-point function of the bare operators is related to that of renormalized ones by:
\begin{align}\label{eq:2pt bare ren dem}
\langle\,\Phi_{1}^{\alpha}(x_1)\,\Phi_{1}^{\beta}(x_2)  \,\rangle=Z_1^2\cdot  \left.\langle\,W_{1}^{\alpha}(x_1)\,W_{1}^{\beta}(x_2)  \,\rangle\right|_{\lambda}
\end{align} 
Correlation functions of the bare operators should be independent of the momentum scale $\mu$. Hence, we have: 
\begin{align}
    \frac{\d}{\d\log \mu} \,\langle\,\Phi_{1}^{\alpha}(x_1)\,\Phi_{1}^{\beta}(x_2)  \,\rangle=0\ ,
\end{align}
leading to the \textbf{Callan-Symanzik equation} for general coupling constant $\lambda$:
\begin{align}\label{eq:2pt CS equation}
\left(\frac{\partial}{\partial \log\mu}+\beta_{\lambda}\,\frac{\partial}{\partial \lambda}+2\,\gamma_1 \right)\cdot \left.\langle\,W_{1}^{\alpha}(x_1)\,W_{1}^{\beta}(x_2)  \,\rangle\right|_{\lambda}=0\ .
\end{align}
At the critical point $\lambda=\lambda_{\ast}$, the beta function vanishes $\beta_{\lambda_\ast}=0$ and one finds that:
\begin{align}\label{eq:2pt CS equation ast}
\left(\frac{\partial}{\partial \log\mu}+2\,\gamma_1 \right)\cdot \left.\langle\,W_{1}^{\alpha}(x_1)\,W_{1}^{\beta}(x_2)  \,\rangle\right|_{\lambda=\lambda_{\ast}}=0\ .
\end{align}
Through performing the dimensional analysis and utilizing the Poincar\'e invariance of the theory for \eqref{eq:2pt bare ren dem}, we learn that the two-point function of the renormalized operators must behave as $|x_{12}|^{-2\Delta_1^{(0)}}\,f(|x_{12}|\,\mu)$ at criticality with $\Delta_1^{(0)}=d/2-1$ being the canonical (engineering) dimension of the bare operator $\Phi_{1}^{\alpha}(x)$. This observation leads to the following differential equation:
\begin{align}\label{eq:2pt CS equation ast 2}
\left( \frac{\partial}{\partial \log |x_{12}|}-\frac{\partial}{\partial \log\mu}+2\Delta_1^{(0)}\right)\cdot \left.\langle\,W_{1}^{\alpha}(x_1)\,W_{1}^{\beta}(x_2)  \,\rangle\right|_{\lambda=\lambda_{\ast}}=0\ .
\end{align}
From \eqref{eq:2pt CS equation ast} and \eqref{eq:2pt CS equation ast 2}, it turns out that $f(|x_{12}|\,\mu)\propto (|x_{12}|\,\mu)^{-2\gamma_1}$ and:
\begin{align}
    \left.\langle\,W_{1}^{\alpha}(x_1)\,W_{1}^{\beta}(x_2)  \,\rangle\right|_{\lambda=\lambda_{\ast}}\propto \frac{1}{|x_{12}|^{2(\Delta_1^{(0)}+\gamma_1)}}\ .
\end{align}
Comparing this expression with the general form of the scalar two-point functions in CFT \eqref{eq:conformal inv of 2pt CFT correlator phy}, we conclude that:
\begin{align}
    \Delta_1=\Delta_1^{(0)}+\gamma_1=1-\frac{1}{2}\,\epsilon+\frac{N+2}{4\,(N+8)^2}\,\epsilon^2+O(\epsilon^3)\ .
\end{align}

This argument holds even for composite operators and can be generalized to the case with several coupling constants and to the models with boundaries and defects. In general, the conformal dimension $\Delta (W)$ of a renormalized operator $W$ is the sum of the (engineering) canonical dimension $\Delta^{(0)}(\Phi)$ and the anomalous dimension $\gamma(\Phi)$ of the bare operator $\Phi$:
\begin{align}\label{eq:conf dim anom dim}
    \Delta (W)=\Delta^{(0)}(\Phi)+\gamma(\Phi)\ ,\qquad \gamma(\Phi)=\left.\frac{\partial \log Z(\Phi)}{\partial \log \mu}\right|_{\lambda=\lambda_\ast}\ ,
\end{align}
where $Z(\Phi)$ is a wave-function renormalization of $\Phi$ determined for the two-point function of renormalized operators $\langle\,W\,W\,\rangle=Z(\Phi)^{-2}\cdot \langle\,\Phi\,\Phi\,\rangle$ to be finite.

For instance, consider the lowest-lying composite operator $\Phi_{2}\equiv|\Phi_1|^{2}$ in free theory. Denoting its renormalized counterpart by $W_2=Z_2\cdot |\Phi_1|^{2}$, a standard perturbative calculation gives:
\begin{align}
    Z_2=1-\frac{N+2}{3\,\epsilon}\cdot\frac{\lambda}{(4\pi)^2}+O(\lambda^2)\ .
\end{align}
And the conformal dimension of $W_2$ turns out to be:
\begin{align}
    \Delta_2=d-2+\gamma_2=2-\epsilon+\frac{N+2}{N+8}\,\epsilon+O(\epsilon^2)\ ,\qquad \gamma_2=\left.\frac{\partial \log Z_2}{\partial \log \mu}\right|_{\lambda=\lambda_\ast}\ .
\end{align}
Critical exponents such as specific heat $C$, magnetization $|\vec{m}|$ and susceptibility $\chi$ exponent are related to $\Delta_1$ as well as $\Delta_2$ through the relations:
\begin{align}
     \alpha=2-\frac{d}{d-\Delta_2}\ ,\qquad \beta=\frac{\Delta_1}{d-\Delta_2}\ ,\qquad \gamma=\frac{d-2\Delta_1}{d-\Delta_2}\ ,
\end{align}
with 
\begin{align}
     C\sim |T-T_c|^{-\alpha}\ ,\qquad |\vec{m}|\sim |T-T_c|^{\beta} \ ,\qquad \chi\sim |T-T_c|^{-\gamma}\ , \qquad \text{as}\quad T\sim T_c\ ,
\end{align}
where $T_c$ is the critical temperature of the system.

\section{Axioms for homogeneous critical systems}\label{sec:Axioms for homogeneous critical systems}
We then move to the axiomatic approach to the critical O$(N)$ model \cite{Rychkov:2015naa}.
We here spell out their original axioms and methodology to study the homogeneous critical O$(N)$ model without resorting to the standard diagrammatic calculations. 

The first axiom states that the theory at the fixed point has conformal symmetry and is described by CFT:
\begin{itemize}\setlength{\leftskip}{8mm}
    \item[\textbf{Axiom \mylabel{cftaxiom1}{\RomanNumeralCaps{1}}.}] The theory at the Wilson-Fisher fixed point has conformal symmetry. 
\end{itemize}

The second axiom follows from the intuition that the Wilson-Fisher conformal field theory in $d=4-\epsilon$ dimensions $(\epsilon\neq0)$ is smoothly connected to the free field theory in four dimensions $(\epsilon=0)$. It postulates the one-to-one correspondence between local operators in both theories:
\begin{itemize}\setlength{\leftskip}{10mm}
    \item[\textbf{Axiom \mylabel{cftaxiom2}{\RomanNumeralCaps{2}}.}] For every local operator $\CO_{\text{free}}$ in the free theory ($\epsilon=0$), there exists a local
operator at the Wilson-Fisher fixed point ($\epsilon\neq0$), $\CO_{\text{WF}}$, which tends to $\CO_{\text{free}}$ in the free limit $\epsilon \rightarrow 0$: $\lim_{\epsilon \rightarrow 0}\CO_{\text{WF}}=\CO_{\text{free}}$. 
\end{itemize}
In what follows, we abuse notations to denote Wilson-Fisher theory counterparts of the free theory operators such as $\Phi_{1}^{\alpha},\Phi_{2p}\equiv|\Phi_1|^{2p}$ and $\Phi_{2p+1}^{\alpha}\equiv\Phi_1^\alpha|\Phi_1|^{2p}$ in the same expressions as their renormalized counterparts in perturbation theory. Then, axiom \ref{cftaxiom2} states that:
\begin{align}
    W_{1}^{\alpha}\xrightarrow[]{\epsilon\to0}\Phi_{1}^{\alpha}\ ,\qquad W_{2p}\xrightarrow[]{\epsilon\to0}\Phi_{2p}\ ,\qquad W_{2p+1}^{\alpha}\xrightarrow[]{\epsilon\to0}\Phi_{2p+1}^{\alpha}\qquad (p=1,2,\cdots)\ .
\end{align}

Both free and Wilson-Fisher theories at criticality in $(4-\epsilon)$ dimensions fulfill the above two axioms. So, we must add one extra axiom to focus only on the Wilson-Fisher theory:
\begin{itemize}\setlength{\leftskip}{12mm}
    \item[\textbf{Axiom \mylabel{cftaxiom3}{\RomanNumeralCaps{3}}.}]
At the Wilson-Fisher fixed point, $W_1^\alpha$ and $W_3^\alpha$ are related by the following equation of motion:
    \begin{align}\label{eq:classical EoM O(N) homogeneous CFT}
    \Box \, W_1^\alpha(x)=\kappa\cdot W_3^\alpha(x)\  ,
\end{align}
 where $\Box$ is the Laplacian in $d=4-\epsilon$ dimensions. 
\end{itemize}
The third axiom indicates that $W_3^\alpha$ is the descendant of $W_1^\alpha$ in the Wilson-Fisher theory. This recombination of conformal multiplet would not be the case in free theory in $(4-\epsilon)$ dimensions where these two operators are independent of each other. The coefficient $\kappa$ is implicitly assumed to be non-zero for finite $\epsilon$ to make a distinction from free theory, and to vanish in taking $\epsilon\to0$ to reproduce the Klein-Gordon equation in four dimensions.

\begin{figure}[ht!]
	\centering
    \begin{tikzpicture}[transform shape,scale=1]
    
 \draw[ultra thick,->] (0,0) to (7.5,0) node [right, font=\Large] {$\substack{\text{Spacetime}\\ \text{dimensions}}$};
\draw[ultra thick,->] (0,0) to (0,5) node [above, font=\Large, black] {$\substack{\text{Theory space of}\\ \text{coupling constants}}$};

 \draw[ultra thick,RoyalBlue] (0,4) to (7,4);

 \draw[ultra thick,PineGreen] (7,4) to[out=-140,in=5] (0,1.5);

 \draw[ultra thick,RoyalBlue] (7.5,5.5) to (8.5,5.5) node [right, font=\Large] {$:\substack{\text{Free theory}\\ \text{(free CFT)}}$};

 \draw[ultra thick,PineGreen] (7.5,4.5) to (8.5,4.5) node [right, font=\Large] {$:\substack{\text{Wilson-Fisher theory}\\ \text{(interacting CFT)}}$};

 \draw[ultra thick,Black,dotted] (7,4) to (7,0) node [below, font=\normalsize,Black] {$4$};

  \draw[ultra thick,Black,dotted] (4,4) to (4,0) node [below, font=\normalsize,Black] {$(4-\epsilon)$};

  \draw[ultra thick,OrangeRed,<-] (3.8,2.2) to (3.8,4);

    \draw[ultra thick,violet,<-] (4,2.05) to[out=20,in=-140] (7,3.8);

 \node[font=\large,very thick,OrangeRed] at (1.9,3.1) {$\substack{\text{Standard} \\ \text{perturbative calculation}\\ \text{+ RG analysis}}$}; 

 \node[font=\large,very thick,violet] at (5.5,1.5) {$\substack{\text{Rychkov-Tan's}\\ \text{axiomatic approach}}$}; 
 
    \end{tikzpicture}
	\caption{Illustrated is a schematic drawing to clarify the difference between the conventional perturbative approach and Rychkov-Tans's axiomatic approach. In the former approach, one starts from the free theory in $d=4-\epsilon$ dimensions and adds interactions perturbatively (vertical direction in this figure). By requiring the vanishing of the beta functions, one identifies the scale-invariant RG fixed point corresponding to the Wilson-Fisher theory and then explores various quantities there. On the other hand, in the axiomatic approach, we focus on the Wilson-Fisher theories on the green slope in this figure smoothly connected to the free theory in four dimensions. To investigate the Wilson-Fisher theory in $d=4-\epsilon$ dimensions, we expand every quantity in powers of $\epsilon$ along the green slope to make the most of the conformal symmetry.}
	\label{fig:perturbative vs axiomatic}
\end{figure}
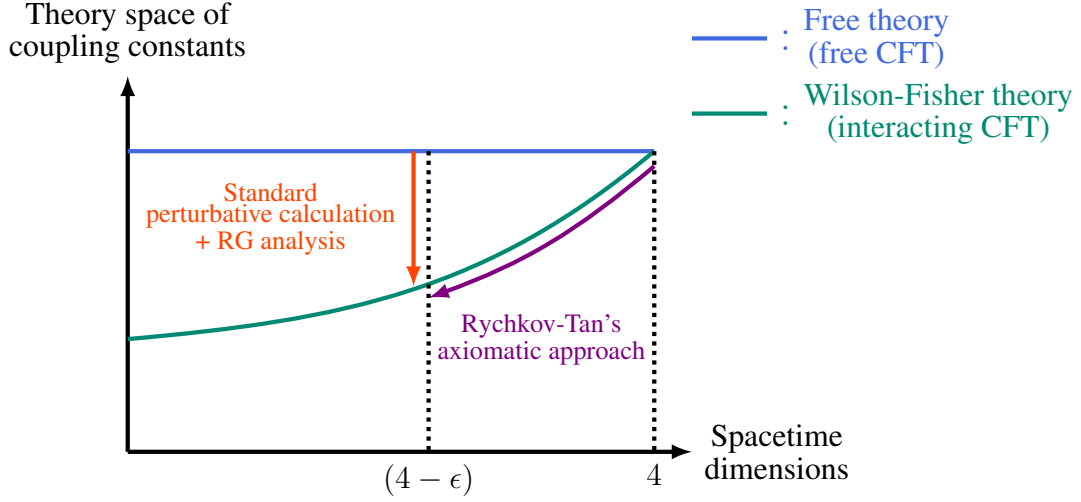

We now emphasize the difference between this axiomatic framework and the conventional perturbative calculations (see figure \ref{fig:perturbative vs axiomatic} for an illustration). Within the axiomatic framework, we compare the free theory in four dimensions with the Wilson-Fisher theory in $d=4-\epsilon$ dimensions, as we will explain in the rest of this section. On the other hand, the conventional approach performs perturbative expansion from the free theory in $d=4-\epsilon$ dimensions to access the Wilson-Fisher theory in $d=4-\epsilon$ dimensions as we described in section \ref{sec:Critical O(N) vector model without defects}.

\paragraph{Rough sketch of Rychkov-Tan's strategy for anomalous dimensions.}
We now take a brief look at the machinery of Rychkov-Tan's axiomatic framework. 
Denoting the conformal dimensions of $W_1^\alpha,W_{2p}$ and $W_{2p+1}^{\alpha}$ by $\Delta_1,\Delta_{2p}$ and $\Delta_{2p+1}$, we derive their leading non-trivial anomalous dimensions:
\begin{align}
    \Delta_1&=\frac{d-2}{2}+\gamma_1 \ , \qquad \gamma_1=\gamma_{1,1}\,\epsilon+\gamma_{1,2}\,\epsilon^2+\cdots  \ ,\\
    \Delta_n&=n\,\frac{d-2}{2}+\gamma_n \ , \qquad \gamma_n=\gamma_{n,1}\,\epsilon+\gamma_{n,2}\,\epsilon^2+\cdots\ .
\end{align}
To do away with technical difficulties, we cut corners by assuming that the anomalous dimension of $\Phi_{1}^{\alpha}$ starts from order $\epsilon^2$:
\begin{align}\label{eq:initial condition RT}
    \gamma_{1,1}=0\ ,\qquad \gamma_{1,2}\neq 0\ ,
\end{align}
while the others from order $\epsilon$.\footnote{One can show $\Phi_{1}^{\alpha}$ has no anomalous dimension at order $\epsilon$ only from the axioms. This assumption is, of course, unnecessary, but without that, the calculations would get so involved that one cannot grasp the whole picture.}
Whenever we work in the axiomatic framework, we use a re-defined version of $\Phi_{1}^{\alpha}$:
\begin{align}\label{eq:On vector redefinition}
    \left.\Phi_{1}^{\alpha}\right|_{\text{new}}=2\pi\cdot \left.\Phi_{1}^{\alpha}\right|_{\text{old}}\ ,
\end{align}
so that the two-point function of $\Phi_{1}^{\alpha}$ is unit-normalized in four dimensions (see \eqref{eq;2pt un-normalized} for canonical normalization under the action \eqref{eq:action without defect}):
\begin{align}
    \langle\,\Phi_{1}^{\alpha}(x_1)\,\Phi_{1}^{\beta}(x_2)\,\rangle\big|_{\lambda_0=0,d=4}=\frac{\delta^{\alpha\beta}}{|x_{12}|^2}\ .
\end{align}

We start by axiom \ref{cftaxiom1} and focus our attention on the identity operator channel of the OPE of $W_{1}^{\alpha}$:\footnote{See \eqref{eq:CFT OPE scalars} for the generic form of the OPEs for primary operators in CFT}
\begin{align}\label{eq:bulk W1W1 OPE}
    W_1^\alpha(x)\times W_1^\beta(0)\supset c_1\cdot\frac{\delta^{\alpha\beta}}{|x|^{2\Delta_1}}\cdot \bm{1} \ .
\end{align}
Here the coefficient should be $c_1=1+O(\epsilon)$, since the OPE \eqref{eq:bulk W1W1 OPE} reduces to the following expression when $\epsilon=0$ (axiom \ref{cftaxiom2}):
\begin{align}
    \Phi_1^\alpha(x)\times \Phi_1^\beta(0)\supset \frac{\delta^{\alpha\beta}}{|x|^2}\cdot \bm{1} \ .
\end{align}
From axiom \ref{cftaxiom3}, one obtains the OPE of $W_{3}^{\alpha}$ by acting the Laplacian on both sides of \eqref{eq:bulk W1W1 OPE}:
\begin{align}\label{eq:bulk W3W3 OPE}
\begin{aligned}
    W_3^\alpha(x)\times W_3^\beta(0)&\supset \frac{4\,\Delta_1(\Delta_1+1)(2\Delta_1+2-d)(2\Delta_1+4-d)}{\kappa^2}\cdot c_1\cdot \frac{\delta^{\alpha\beta}}{|x|^{2\Delta_1+4}}\cdot\bm{1}\\
    &= \frac{32\,\epsilon^2\,\gamma_{1,2}+O(\epsilon^3)}{\kappa^2}\cdot \frac{\delta^{\alpha\beta}}{|x|^{2\Delta_1+4}}\cdot \bm{1} \ ,
\end{aligned}
\end{align}
From Axiom \ref{cftaxiom2}, this OPE should be equated to the following in taking $\epsilon\to0$ limit:\footnote{One can calculate the free theory OPEs via Wick's theorem.}
\begin{align}\label{eq:bulk P3P3 OPE}
\Phi_3^\alpha(x)\times \Phi_3^\beta(0)\supset \frac{2\,(N+2)\,\delta^{\alpha\beta}}{|x|^6}\cdot \bm{1} \ .
\end{align}
As a result, we find that:
\begin{align}\label{eq:rel kappa gamma12}
    \kappa^2 =\frac{16}{N+2}\cdot \gamma_{1,2}\,\epsilon^2+O(\epsilon^3)\ .
\end{align}
We proceed to consider $W_1^\alpha(x)$ channel OPEs of the operators such as $W_{2p}$ and $W_{2p+1}^{\alpha}$. In particular, we pay attention to a few sub-leading terms and use axiom \ref{cftaxiom3} to obtain:
\begin{align}\label{eq:OPE W2p}
\begin{aligned}
       W_{2p}(x)\times W_{2p+1}^{\alpha}(0) & \supset \frac{c_{2p}}{|x|^{\Delta_{2p}+\Delta_{2p+1}-\Delta_1}}\cdot [1+\cdots+q_3^{(2p)}\,|x|^2\,\Box+\cdots]\cdot W_1^\alpha(0)\\
    &\supset \frac{c_{2p}}{|x|^{\Delta_{2p}+\Delta_{2p+1}-\Delta_1}}\cdot [W_1^\alpha(0)+\kappa\,q_3^{(2p)}\,|x|^2\,W_3^\alpha(0)] \ ,
\end{aligned}
\end{align}
and 
\begin{align}\label{eq:OPE W2p+1}
\begin{aligned}
       W_{2p+1}^{\alpha}(x)&\times W_{2p+2}(0) \\
   &\supset \frac{c_{2p+1}}{|x|^{\Delta_{2p+1}+\Delta_{2p+2}-\Delta_1}}\cdot [1+\cdots+q_3^{(2p+1)}\,|x|^2\,\Box+\cdots]\cdot W_1^\alpha(0)\\
    &\supset \frac{c_{2p+1}}{|x|^{\Delta_{2p+1}+\Delta_{2p+2}-\Delta_1}}\cdot [W_1^\alpha(0)+\kappa\,q_3^{(2p+1)}\,|x|^2\cdot W_3^\alpha(0)] \ ,
\end{aligned}
\end{align}
Here, two coefficients $q_3^{(2p)}$ and $q_3^{(2p+1)}$ are determined by conformal symmetry (see \eqref{eq:CFT OPE scalars} and \eqref{eq:coefficient q3}). At the leading order in $\epsilon$, we have:\footnote{Use \eqref{eq:rel kappa gamma12} to obtain these two equations. Precisely, one must pay special attention to the case with $p=1$. There, one cannot apply the formula \eqref{eq:CFT OPE scalars} designed for primary operators as one of the operators in the left-hand side of \eqref{eq:OPE W2p} (and also \eqref{eq:OPE W2p+1}) turns into a descendant operator. However, careful analysis reveals that the recursion relations \eqref{eq:CFT recursion relation} hold as they stand even for $p=1$. See the original paper \cite{Rychkov:2015naa} for details on this point.}
\begin{align}
   \kappa\, q_3^{(2p)}&=\frac{\gamma_{2p+1,2}-\gamma_{2p,1}}{(N+2)\,\kappa}\,\epsilon+O(\epsilon^0)\ ,\\
    \kappa\,   q_3^{(2p+1)}&=\frac{\gamma_{2p+2,2}-\gamma_{2p+1,1}}{(N+2)\,\kappa}\,\epsilon+O(\epsilon^0)\ .
\end{align}
These two OPEs (equation \eqref{eq:OPE W2p} and \eqref{eq:OPE W2p+1}) tend to the following ones as $\epsilon\to0$ in respective ways:
\begin{align}
        \Phi_{2p}(x)\times \Phi_{2p+1}^{\alpha}(0)&\supset \frac{2^{2p}\,p!\,(N/2)_p}{|x|^{4p}}\cdot \left(\Phi_1^\alpha(0)+\frac{3p}{N+2}\,|x|^2\cdot \Phi_3^\alpha(0)\right) \ ,
\end{align}
and 
\begin{align}
             \Phi_{2p+1}^{\alpha}(x)&\times \Phi_{2p+2}(0)\supset \frac{2^{2p+1}\,(p+1)!\,(N/2)_p}{|x|^{4p+2}}\cdot \left(\Phi_1^\alpha(0)+\frac{6p+N+2}{2\,(N+2)}\,|x|^2\cdot \Phi_3^\alpha(0)\right) \ .
\end{align}
Hence, we obtain the following recursion relations for anomalous dimensions:
\begin{align}\label{eq:CFT recursion relation}
\gamma_{2p+1,2}-\gamma_{2p,1}&=3p\cdot \frac{\kappa}{\epsilon}+O(\epsilon)\ ,\\
\gamma_{2p+2,2}-\gamma_{2p+1,1}&=\frac{6p+N+2}{2}\cdot \frac{\kappa}{\epsilon} +O(\epsilon)\ ,
\end{align}
which can be solved under the initial condition $\gamma_{1,1}=0$ \eqref{eq:initial condition RT} to give:
\begin{align}\label{eq:anomalous dimension rec CFT 1}
    \gamma_{2p,1}=\frac{p\,(N+6p-4)}{2}\,\cdot\frac{\kappa}{\epsilon}+O(\epsilon)\ ,\qquad \gamma_{2p+1,1}=\frac{p\,(N+6p+2)}{2}\cdot\frac{\kappa}{\epsilon}+O(\epsilon)\ .
\end{align}
Now, all we have left is to determine $\kappa$. The last piece of the puzzle comes from the equation of motion in axiom \ref{cftaxiom3} that implies:
\begin{align}
    \Delta_3=\Delta_1+2 \quad \longrightarrow \quad \gamma_{3,1}=1\ .
\end{align}
Combining this with \eqref{eq:anomalous dimension rec CFT 1}, we are able to determine $\kappa$ as:
\begin{align}\label{eq:value of kappa}
    \kappa= \frac{2}{N+8}\,\epsilon+O(\epsilon^2)\ .
\end{align}
Plugging this into \eqref{eq:rel kappa gamma12} and \eqref{eq:anomalous dimension rec CFT 1}, we finally find that:
\begin{align}
    \Delta_1&=\frac{d-2}{2}+\frac{N+2}{4\,(N+8)^2}\,\epsilon^2 +O(\epsilon^3)\ ,\label{eq:Delta1 axiomatic}\\
    \Delta_{2p}&=2p\cdot\frac{d-2}{2}+\frac{p\,(N+6p-4)}{N+8}\,\epsilon+O(\epsilon^2)\ ,\\
 \Delta_{2p+1}&=(2p+1)\cdot\frac{d-2}{2}+\frac{p\,(N+6p+2)}{N+8}\,\epsilon+O(\epsilon^2)\ .
\end{align}
These results agree with the perturbative results up to the first non-trivial order in $\epsilon$. Similarly, one can compute leading anomalous dimensions of more generic operators such as O$(N)$ symmetric traceless tensors (see \cite[section 4.2]{Rychkov:2015naa}).\footnote{According to the standard perturbative calculations, the anomalous dimensions of the broken conserved currents also start at the second order in $\epsilon$ \cite[section 9]{Wilson:1973jj}. One can deal with them within the axiomatic framework by considering the multiplet recombination phenomena just like $\Phi_{1}^{\alpha}$ \cite{Giombi:2016hkj}.}

\section{Axioms in the presence of a defect}\label{sec:Axioms in the presence of a defect}
In the presence of a defect, Rychkov-Tan's axioms are modified in the following manner \cite{Yamaguchi:2016pbj}:
\begin{itemize}\setlength{\leftskip}{11mm}
    \item[\textbf{Axiom \mylabel{dcftaxiom1}{\RomanNumeralCaps{1}'}.}] In the presence of a defect, the theory at the Wilson-Fisher fixed point has the \textbf{defect conformal symmetry}.
      \item[\textbf{Axiom \mylabel{dcftaxiom2}{\RomanNumeralCaps{2}'}.}] \textbf{For a bulk/defect local operator} $\CO_{\text{free}}/\widehat{\CO}_{\text{free}}$ in the free theory with a defect, there exists a local operator $\CO_{\text{WF}}/\widehat{\CO}_{\text{WF}}$ at the Wilson-Fisher fixed point, which tends to $\CO_{\text{free}}/\widehat{\CO}_{\text{free}}$ in the limit $\epsilon \to 0$.
          \item[\textbf{Axiom \mylabel{dcftaxiom3}{\RomanNumeralCaps{3}'}.}] At the Wilson-Fisher fixed point, \textbf{two bulk operators} $W_1^\alpha$ and $W_3^\alpha$ are related by the following equation of motion:
    \begin{align}\label{eq:classical EoM O(N)}
    \Box \, W_1^\alpha(x)=\kappa\cdot W_3^\alpha(x)\  ,
\end{align}
where $\Box$ is the Laplacian in $d=4-\epsilon$ dimensions.
\end{itemize}
In the presence of a defect, the assumption of conformal symmetry on the fixed point (axiom \ref{cftaxiom1}) turns into that of defect conformal symmetry (axiom \ref{dcftaxiom1}), making it possible to use DCFT techniques within this axiomatic framework. The second and the third axioms are almost unchanged. But take care that the one-to-one correspondence between operators in free $(\epsilon=0)$ and Wilson-Fisher theory $(\epsilon\neq0)$ is for both bulk and defect local operators. 
Because axiom \ref{cftaxiom1} allows to use the bulk OPEs, one can do the same analysis as the last section. Hence, we still have the same bulk operator spectrum and $\kappa$ as the case without defects (see e.g., \eqref{eq:value of kappa} and \eqref{eq:Delta1 axiomatic}).
The equation of motion \eqref{eq:classical EoM O(N)} in axiom \ref{dcftaxiom3} is only about bulk operators and says nothing about defect local operators. Nonetheless, our axioms are sufficient to extract information on the defect since the DOE bridges the bulk and defect CFT data.

The author of \cite{Yamaguchi:2016pbj} originally introduced these three modified axioms to explore a class of defect local non-composite operators on the co-dimension two monodromy defect of the critical Ising model in $(4-\epsilon)$ dimensions. It is straightforward to generalize his analysis to the O$(N)$ model \cite{Soderberg:2017oaa} and to the case with Neumann/Dirichlet boundary and interface \cite{Soderberg:2017oaa,Giombi:2020rmc,Dey:2020jlc}. These three axioms, in particular the equation of motion \eqref{eq:classical EoM O(N)}, can be combined with the analytic conformal bootstrap, giving fine analytic studies of critical phenomena even in the presence of a defect where ordinary numerical bootstrap does not work \cite{Herzog:2022jlx}. 

Let us clarify below what we will do in the following three chapters leveraging these three modified axioms. We will briefly review in section \ref{sec:Neu lowest RT} and \ref{sec:Dir lowest RT} how to study non-composite operators on Neumann and Dirichlet boundaries of the critical O$(N)$ model within the axiomatic framework. As one of the main results in this thesis, we will utilize modified axioms in section \ref{1st order: scalar} and \ref{1st order: transverse spin} to derive the leading anomalous dimensions of non-composite operators on the line defect and the critical defect coupling in the critical O$(N)$ model.

 As we argued in section \ref{sec:Axioms for homogeneous critical systems}, in the homogeneous critical systems, one can derive conformal dimensions of generic local operators, whichever they are composite or non-composite ones.
With a defect, however, we found no research that deals with conformal dimensions of generic defect local composite operators within the axiomatic framework. In the rest of this thesis, we also demonstrate in the critical O$(N)$ model with three kinds of defects that it is possible to uncover defect local composite operator spectra using three modified axioms by focusing on the constraints coming from the analyticity of correlators. We expect that our analysis applies to other models and is enough to show the validity of the axiomatic approach.

\chapter{The O$(N)$ model with line defect}\label{chap:line defect}
This chapter investigates the critical behavior of the O$(N)$ model with a line defect $(p=1)$ in $d=4-\epsilon$ dimensions using the axioms introduced in section \ref{sec:Axioms in the presence of a defect}. Section \ref{sec:line defect known} reviews known perturbative studies by the authors of \cite{Allais:2014fqa,Cuomo:2021kfm} (section \ref{sec:line defect known}). We then reveal the structure of the free O$(N)$ model with a line defect in four dimensions (section \ref{sec:line defect free}). Based on the results, we take the axiomatic approach and reproduce the results in known literature (section \ref{sec:DCFT data on the line defect}). In particular, we invent a methodology to study conformal dimensions of composite operators by focusing on the analyticity of the correlators. 
We also study the defect operator spectrum for Ising DCFT $(N=1)$ in section \ref{sec:N=1 anomalous dimension}.

\section{Review of perturbative results}\label{sec:line defect known}
We consider the O$(N)$ model in $d=4-\epsilon$ dimensions with a line defect $(p=1)$ extending along the first axis $\hat{x}_1$.\footnote{We here would like to stress that, even in $d=4-\epsilon$ dimensions, we fix the dimension of the defect to one, as we eventually aim to access the critical O$(N)$ in three dimensions with a line defect by taking the limit $\epsilon\to1$.} The total bare action of this model is the bulk action $I_{\text{bulk}}(=I$ in \eqref{eq:action without defect}) plus the defect localized action $I_{\text{defect}}$:\footnote{Notice that we put a minus sign in the localized magnetic field relatively to \cite{Cuomo:2021kfm} $h_0|_{\text{here}}=-h_0|_{\text{there}}$.}
\begin{align}\label{eq:magnetized model}
\begin{aligned}      I_{\text{tot}}&=I_{\text{bulk}}+I_{\text{defect}} \\
    &=\int \d^d x \,\left( \frac{1}{2}\,|\partial\Phi_1|^2 + \frac{\lambda_0}{4!}\, |\Phi_1|^4\right) - h_0\, \int \d\hat{x}^1\,\Phi^{1}_{1} \ .
\end{aligned}
\end{align}
We first present perturbative results according to \cite{Allais:2014fqa,Cuomo:2021kfm}. See also \cite[section 3.2]{Gimenez-Grau:2022czc} for a nice and concise review.

One renormalizes the bulk fields and the bulk coupling constant to have the same results as without defects (see section \ref{sec:Critical O(N) vector model without defects}):
\begin{align}
    \Phi_{1}^{\alpha}(x)&=Z_1\cdot W_{1}^{\alpha}(x)\ ,\\
    Z_1&=1-\frac{\lambda^2}{(4\pi)^4}\cdot\frac{N+2}{36\epsilon}+O(\lambda^3)\ ,\\
    \lambda_0&=\mu^{\epsilon}\cdot \left(\lambda+\frac{N+8}{3\,\epsilon}\cdot \frac{\lambda^2}{(4\pi)^2}+O(\lambda^3)\right)\ .
\end{align}
On the other hand, the defect coupling is renormalized so that the bulk one-point function of the bare operator $\Phi_{1}^{\alpha}(x)$ is finite:\footnote{As explained in the next section, the bulk one-point function of $\Phi_{1}^{\alpha}(x)$ does not vanish in free theory and is proportional to the defect coupling $h$ (see equation \eqref{eq:bulk one point phi}). Although we already know the bulk coupling $\lambda$ is of order $\epsilon$ at the Wilson-Fisher fixed point, we have no idea about the defect coupling $h$ at criticality. Hence, we should expand the defect coupling to all orders for every fixed order in $\lambda$.}
\begin{align}
    h_0=\mu^{\epsilon/2}\cdot \left(h+\frac{\lambda}{(4\pi)^2}\cdot\frac{h^3}{12\,\epsilon}+O(\lambda^2)\right)\ .
\end{align}
The beta functions for renormalized bulk and defect coupling constants are:
 \begin{align}
     \beta_{\lambda}&=\frac{\partial \lambda}{\partial \log \mu} =-\epsilon\, \lambda + \frac{N+8}{3}\cdot \frac{\lambda^2}{(4\pi)^2}+O(\lambda^3) \label{eq:beta function bulk}\ ,\\
         \beta_{h}&=\frac{\partial h}{\partial \log \mu} =-\frac{\epsilon}{2}\, h +\frac{\lambda^2}{(4\pi)^2}\cdot\frac{h^3}{6}+O(\lambda^2) \label{eq:beta function defect}\ .
 \end{align}
They vanish non-trivially when $(\lambda,h)=(\lambda_\ast,h_\ast)$ with:
 \begin{align}
 \frac{\lambda_\ast}{(4\pi)^2}&=\frac{3}{N+8}\,\epsilon+O(\epsilon^2)\ ,\label{eq:critical lambda}\\
 h_\ast^2&= N+8+O(\epsilon)\ .\label{eq:critical defect coupling}
 \end{align}
We refer to this point in the parameter space as the Wilson-Fisher fixed point in the presence of a line defect.

Next, let us focus on the anomalous dimensions of the local operators. Because the bulk operators are insensitive to the defect coupling, their anomalous dimensions take the same values as those without defects.
Meanwhile, the defect local operators are affected by the defect and behave differently from bulk ones. One naturally classifies defect local operators according to the following enhanced symmetry group on the critical point:
\begin{align}\label{eq:residual symmetry group on the defect}
    \mathrm{SL}(2,\mathbb{R})\times \mathrm{SO}(d-1)\times \mathrm{O}(N-1)\ .
\end{align}
The first two correspond to the defect conformal group, which is a direct product of the parallel conformal group and the transverse rotation group. The defect localized action $I_{\text{defect}}$ breaks the bulk O$(N)$ symmetry down to its subgroup O$(N-1)$. 
The authors of \cite{Cuomo:2021kfm} based on the representation theory of \eqref{eq:residual symmetry group on the defect} and derived conformal dimensions of the low-lying defect local operators through conventional perturbative calculations as specified below.

First, consider the lowest-lying defect local operators defined by the limiting form when no interactions are present $(\lambda_0=0)$:
\begin{align}
\widehat{\Phi}_1^{\,1}(\hat{x})\equiv\lim_{|x_\perp|\to0}\,\Phi_1^{\,1}(x) \ ,\qquad \widehat{\Phi}_1^{\,\hat{\alpha}}(\hat{x})\equiv\lim_{|x_\perp|\to0}\,\Phi_1^{\,\hat{\alpha}}(x)\ .
\end{align}
Here $\widehat{\Phi}_1^{\,1}$ is singlet both in $\mathrm{SO}(d-1)$ and $\mathrm{O}(N-1)$, whereas $\widehat{\Phi}_1^{\,\hat{\alpha}}$ is in the vector representation of $\mathrm{O}(N-1)$. Denoting their renormalized counterparts by $\widehat{W}_{1}^{\,1}=\widehat{Z}(\widehat{\Phi}_1^{\,1})\cdot \widehat{\Phi}_1^{\,1}$ and $\widehat{W}_{1}^{\,\hat{\alpha}}=\widehat{Z}(\widehat{\Phi}_{1}^{\,\hat{\alpha}})\cdot \widehat{\Phi}_{1}^{\,\hat{\alpha}}$, their wave function renormalizations are calculated as follows:
\begin{align}
    \widehat{Z}(\widehat{\Phi}_1^{\,1})=1-\frac{\lambda}{(4\pi)^2}\cdot\frac{h^2}{4\,\epsilon}+O(\lambda^2)\ ,\qquad \widehat{Z}(\widehat{\Phi}_{1}^{\,\hat{\alpha}})=1-\frac{\lambda}{(4\pi)^2}\cdot\frac{h^2}{12\,\epsilon}+O(\lambda^2)\ .
\end{align}
According to the formula \eqref{eq:conf dim anom dim}, these expressions lead to the following conformal dimensions of the renormalized operators:
\begin{align}
\widehat{\Delta}(\widehat{W}_1^{\,1})&=d/2-1+ \widehat{\gamma}(\widehat{\Phi}_1^{\,1})=1+\epsilon+O(\epsilon^2)\ ,\qquad \left.\widehat{\gamma}(\widehat{\Phi}_1^{\,1})=\frac{\partial\log \widehat{Z}(\widehat{\Phi}_1^{\,1})}{\partial \log \mu}\right|_{\substack{\lambda=\lambda_\ast\\ h=h_\ast}} \label{eq:conf dim low scalar cuomo}\\
\widehat{\Delta}(\widehat{W}_1^{\,\hat{\alpha}})&=d/2-1+ \widehat{\gamma}(\widehat{\Phi}_1^{\,\hat{\alpha}})=1+O(\epsilon^2) ,\qquad \left.\widehat{\gamma}(\widehat{\Phi}_1^{\,\hat{\alpha}})=\frac{\partial\log \widehat{Z}(\widehat{\Phi}_1^{\,\hat{\alpha}})}{\partial \log \mu}\right|_{\substack{\lambda=\lambda_\ast\\ h=h_\ast}}\ .\label{eq:conf dim tilt cuomo}
\end{align}
One can identify $\widehat{W}_1^{\,\hat{\alpha}}$ as the tilt operator \eqref{eq:def of tilt} associated with the symmetry breaking from O$(N)$ to O$(N-1)$.\footnote{In this case, symmetry breaking currents in free theory are $J^{1\hat{\alpha}}_\mu=\Phi_1^{1}\,\partial_\mu\,\Phi_1^{\hat{\alpha}}-\Phi_1^{\hat{\alpha}}\,\partial_\mu\,\Phi_1^{1}$.} The calculations performed by the authors of \cite{Cuomo:2021kfm} are at one-loop level, but we expect the conformal dimension of $\widehat{W}_1^{\,\hat{\alpha}}$ to remain unity at all orders in $\epsilon$.

Next, let us proceed to the defect local operators in the vector representation of SO$(d-1)$:
\begin{align}
        \widehat{U}_{i}^{\,1}(\hat{x})&\equiv\widehat{Z}(\widehat{\Phi}^{\,1}_{2,i})\cdot \widehat{\Phi}^{\,2}_{i}(\hat{x}) \ ,\qquad \widehat{\Phi}^{\,1}_{2,i}(\hat{x})\equiv\lim_{|x_\perp|\to0}\,\partial_{i}\,\widehat{\Phi}_{i}^{\,1}(x) \ ,\\
            \widehat{U}_{i}^{\,\hat{\alpha}}(\hat{x})&\equiv\widehat{Z}(\widehat{\Phi}^{\,\hat{\alpha}}_{2,i})\cdot \widehat{\Phi}^{\,\hat{\alpha}}_{2,i}(\hat{x})\ , \qquad \widehat{\Phi}^{\,\hat{\alpha}}_{2,i}(\hat{x})\equiv\lim_{|x_\perp|\to0}\,\partial_{i}\,\widehat{\Phi}^{\,\hat{\alpha}}(x) \ .
\end{align}
According to \cite{Cuomo:2021kfm}, their conformal dimensions are given by:
\begin{align}
    \widehat{\Delta}(\widehat{U}_{s=1}^{\,1})&=2+O(\epsilon^2)\ ,\\
    \widehat{\Delta}(\widehat{U}_{s=1}^{\,\hat{\alpha}})&=2-\frac{1}{3}\,\epsilon+O(\epsilon^2)\ .
\end{align}
The latter operator $\widehat{U}_{s=1}^{\,\hat{\alpha}}$ can be identified as the displacement operator having protected conformal dimension two (see equation \eqref{eq:anomalous conservation stress tensor}). In section \ref{1st order: transverse spin}, we will take one step further in axiomatic framework and consider operators in the symmetric and traceless representation of SO$(d-1)$ $\widehat{U}_{i_1\cdots i_s}^{\,1}$ and $\widehat{U}_{i_1\cdots i_s}^{\,\hat{\alpha}}$, tending to:
\begin{align}
    \widehat{\Phi}^{\,1}_{1+s,i_1\cdots i_s}(\hat{x})&\equiv\lim_{|x_\perp|\to0}\,\partial_{(i_1}\cdots \partial_{i_s)}\,\widehat{\Phi}_{i}^{\,1}(x)\ ,\\
    \widehat{\Phi}^{\,\hat{\alpha}}_{1+s,i_1\cdots i_s}(\hat{x})&\equiv\lim_{|x_\perp|\to0}\,\partial_{(i_1}\cdots \partial_{i_s)}\,\widehat{\Phi}_{i_1\cdots i_s}^{\,\hat{\alpha}}(x)\ ,
\end{align}
as $\epsilon\to0$.

The structure of the second-order defect composite operators is a little involved. According to the classification \eqref{eq:residual symmetry group on the defect}, their bare operators fall into the three classes:\footnote{Note that $\widehat{\Phi}_1^{\,1}\widehat{\Phi}_1^{\,\hat{\alpha}}$ and $\widehat{\Phi}_1^{\,(\hat{\alpha}}\widehat{\Phi}_1^{\,\hat{\beta})}$ make sense only for $N\geq2$ and $N\geq3$ respectively. Also, in case of $N=1$ (Ising model) we fail to define $|\widehat{\Phi}_1^{\,\hat{\gamma}}|^2$.}
\begin{align}\label{eq:second order composite list 1}
    \widehat{\Phi}_1^{\,1}\widehat{\Phi}_1^{\,\hat{\alpha}}(\hat{x})\equiv\lim_{|x_\perp|\to0}\,\Phi_1^1\Phi_1^{\hat{\gamma}}(x)\ ,\qquad \widehat{\Phi}_1^{\,(\hat{\alpha}}\widehat{\Phi}_1^{\,\hat{\beta})}(\hat{x})\equiv \lim_{|x_\perp|\to0}\,\Phi_1^{\,(\hat{\gamma}}\Phi_1^{\,\hat{\sigma})}(x) \ ,
\end{align}
and 
\begin{align}\label{eq:second order composite list 2}
    \begin{dcases}
    |\widehat{\Phi}_1^{\,1}|^2(\hat{x})\equiv\lim_{|x_\perp|\to0}\,\Phi_1^{1}\Phi_1^{1}(x)\\
    |\widehat{\Phi}_1^{\,\hat{\gamma}}|^2(\hat{x})\equiv\lim_{|x_\perp|\to0}\,\sum_{\hat{\gamma}=2}^N\,\Phi_1^{\,\hat{\gamma}}\Phi_1^{\,\hat{\gamma}}(x)
\end{dcases}\ .
\end{align}
The first operator in \eqref{eq:second order composite list 1} is in the vector representation of O$(N-1)$, whereas the second is a symmetric and traceless tensor of rank-two. The conformal dimensions of their renormalized counterparts $\widehat{V}^{\,\hat{\alpha}}$ and $\widehat{T}^{\,\hat{\alpha}\hat{\beta}}$ are given by:
\begin{align}
\widehat{\Delta}(\widehat{V})&=2+\frac{N+10}{N+8}\,\epsilon+O(\epsilon^2)\ ,\label{eq:conf dim V}\\
\widehat{\Delta}(\widehat{T})&=2+\frac{2}{N+8}\,\epsilon+O(\epsilon^2)\ .\label{eq:conf dim T}
\end{align}
Two O$(N-1)$ singlet operators \eqref{eq:second order composite list 2} are degenerate in free theory $(\lambda_0=0)$. However, once the interaction is turned on, such degeneracy is resolved due to the bulk interaction term to form a new set of operators $\widehat{S}^{\pm}$ having the following conformal dimensions:
\begin{align}\label{eq:conf dim Spm}
    \widehat{\Delta}(\widehat{S}^{\pm})=2+\frac{3N+20\pm\sqrt{N^2+40N+320}}{2\,(N+8)}\,\epsilon+O(\epsilon^2)\ .
\end{align}

\section{Structure of free O$(N)$ model in four dimensions with a line defect}\label{sec:line defect free}
We now perform a detailed analysis of the four-dimensional free O$(N)$ model with a line defect. The case with $N=1$ is partially analyzed in \cite{Kapustin:2005py} and \cite[section 5.4]{Billo:2016cpy} such as one- and two-point functions of non-composite fields. We extend their analysis to get all the information necessary for our purpose. 
We first study correlation functions in section \ref{sec:Correlation functions in free theory DCFT}. We then utilize the free DCFT correlators to uncover the DOEs of $\Phi_1^\alpha$ and $\Phi_3^\alpha$ in section \ref{eq:Bulk-to-defect operator product expansions free}.

\subsection{Correlation functions}\label{sec:Correlation functions in free theory DCFT}
With the redefined version of the O$(N)$ vector field \eqref{eq:On vector redefinition} $\left.\Phi_1^{\alpha}\right|_{\text{new}}=2\pi\cdot \left.\Phi_1^{\alpha}\right|_{\text{old}}$, we rewrite the action of the free model by:
\begin{align}\label{eq:magnetized model free}  
I_{\text{tot,free}}=\frac{1}{8\pi^2}\,\int \d^4 x \, |\partial\Phi_1|^2  - \frac{h}{2\pi}\, \int \d\hat{x}^1\,\Phi^{1}_{1} \ .
\end{align}
As the defect coupling $h$ is marginal, this model is described by DCFT at any value of $h$.
We denote the correlation functions in the absence of the defect by $\langle \,\cdots\,\rangle_0$, which we compute by the following path-integral (see also \eqref{eq:path integral correlation QFT}):
\begin{align}
    \langle \,\cdots\,\rangle_0
        \equiv
        \int \CD \Phi_1\, (\,\cdots\,)\,\exp\left( -\frac{1}{8\pi^2}\int\d^4 x\,|\partial \Phi_1|^2\right) \ .
\end{align}
The two-point function of $\Phi_1^{\alpha}$ are unit-normalized under this normalization:
\begin{align}\label{eq:free scalar propargator}
  \langle\,\Phi_1^\alpha(x_1)\,\Phi_1^\beta(x_2)\,\rangle_0=\frac{\delta^{\alpha\beta}}{|x_1-x_2|^{2}}\ .
\end{align}
Let us define the DCFT correlators $\langle \,\cdots\,\rangle$ by:
\begin{align}
    \langle\,\cdots\,\rangle \equiv \frac{\langle\,\cdots \,e^{\frac{h}{2\pi}\,\int \d \hat{y}^1\,\Phi_1^1}\,\rangle_0}{\langle\,e^{\frac{h}{2\pi}\,\int \d \hat{y}^1\,\Phi_1^1}\,\rangle_0}\ ,
\end{align}
by identifying $e^{-I_{\text{defect}}}=e^{\frac{h}{2\pi}\,\int \d \hat{y}^1\,\Phi_1^1}$ with the defect operator $\CD^{(p=1)}$ in line with \eqref{eq:def of DCFT correlator}.
The bulk one-point function of $\Phi_1^{\alpha}$ does not vanish in the presence of the defect and is computed as follows:
\begin{align}\label{eq:bulk one point phi}
\begin{aligned}
                  \langle\,\Phi_1^\alpha(x)\,\rangle
             &=\frac{\langle\,\Phi_1^\alpha(x)\,e^{\frac{h}{2\pi}\,\int \d \hat{y}^1\,\Phi_1^1}\,\rangle_0}{\langle\,e^{\frac{h}{2\pi}\,\int \d \hat{y}^1\,\Phi_1^1}\,\rangle_0}\\
        &=\frac{h}{2\pi}\,\int \d \hat{y}^1\,\langle\,\Phi_1^\alpha(x)\,\Phi_1^1(\hat{y})\,\rangle_0\\
                &=\frac{h}{2\pi}\,\int \d \hat{y}^1\,\frac{\delta^{\alpha 1}}{|x-\hat{y}|^2}\\
             &=
                \frac{\delta^{\alpha1}\,\hat{h}}{|x_\perp|}\ .
\end{aligned}
\end{align}
In deriving this, we contracted $\Phi_1^{\alpha}$ with the defect operator $e^{\frac{h}{2\pi}\,\int \d \hat{y}^1\,\Phi_1^1}$ using Wick's theorem, and introduced $\hat{h}$ to shorten equations:
\begin{align}\label{eq:defect coupling redefinition}
    \hat{h}\equiv\frac{h}{2}\ .
\end{align}
The bulk two-point functions of $\Phi_1^{\alpha}$ are calculated in the similar fashion:
\begin{align}\label{eq:bulk two point phi}
\begin{aligned}
                 \langle\,\Phi_1^\alpha(x_1)\,\Phi_1^\beta(x_2)\,\rangle
             &=\frac{\langle\,\Phi_1^\alpha(x_1)\,\Phi_1^\beta(x_2)\,e^{\frac{h}{2\pi}\,\int \d \hat{y}^1\,\Phi_1^1}\,\rangle_0}{\langle\,e^{\frac{h}{2\pi}\,\int \d \hat{y}^1\,\Phi_1^1}\,\rangle_0}\\
        &= \langle\,\Phi_1^\alpha(x)\,\rangle\cdot \langle\,\Phi_2^\beta(x)\,\rangle+ \langle\,\Phi_1^\alpha(x_1)\,\Phi_1^\beta(x_2)\,\rangle_0\\
                &=\frac{\delta^{\alpha1}\,\delta^{\beta1}\,\hat{h}^2}{|x_{1,\perp}|\,|x_{2,\perp}|}+\frac{\delta^{\alpha\beta}}{|x_1-x_2|^{2}}\ .
\end{aligned}
\end{align}
This immediately leads to the two-point functions involving $\widehat{\Phi}_1^{\,\alpha}=\lim_{|x_\perp|\to0}\,\Phi_1^\alpha(x)$:
\begin{align}\label{eq:fund 2pt}
    \langle\,\Phi_1^\alpha(x)\,\widehat{\Phi}_1^{\,\beta}(\hat{y})\,\rangle=\frac{\delta^{\alpha\beta}}{|x-\hat{y}|^2} \ ,\qquad
        \langle\,\widehat{\Phi}_1^{\,\alpha}(\hat{y}_1)\,\widehat{\Phi}_1^{\,\beta}(\hat{y}_2)\,\rangle=\frac{\delta^{\alpha\beta}}{|\hat{y}_{12}|^{2}} \ .
\end{align}
Note that, to compute the bulk-defect two-point function $  \langle\,\Phi_1^\alpha\,\widehat{\Phi}_1^{\,\beta}\,\rangle$, we have taken the limit $|x_\perp|\to0$ inside correlators so that they are finite and compatible with defect conformal symmetry:
\begin{align}
\begin{aligned}
            \langle\,\Phi_1^\alpha(x_1)\,\widehat{\Phi}_1^{\,\beta}(\hat{x_2})\,\rangle
            &=\lim_{|x_{2,\perp}|\to0}\,\left(\cancel{\frac{\delta^{\alpha1}\,\delta^{\beta1}\,\hat{h}^2}{|x_{1,\perp}|\,|x_{2,\perp}|}}+\frac{\delta^{\alpha\beta}}{|x_1-x_2|^{2}}\right)\\
            &=\frac{\delta^{\alpha\beta}}{|x_1-\hat{x_2}|^{2}}\ .
\end{aligned}
\end{align}

Consider defect local operators having transverse spin indices:
\begin{align}
    \widehat{\Phi}_{s+1}^{\,\alpha,i_1\cdots i_s}= \lim_{|x_\perp|\to0}\,\partial^{(i_1}\cdots\partial^{i_s)}\Phi_1^{\alpha}\ .
\end{align}
To compute their correlation functions, we start by acting derivatives to transverse directions on one of the operators in the left-hand side of \eqref{eq:bulk two point phi}:
\begin{align}
\begin{aligned}     \langle\,\partial^{(i_1}\cdots\partial^{i_s)}\Phi_1^\alpha(x_1)\,\Phi_1^\beta(x_2)\,\rangle&=(-2)^s\,s!\cdot \frac{x_{\perp,12}^{(i_1}\cdots x_{\perp,12}^{i_s)}}{|x_{12}|^{2(s+1)}}\,\delta^{\alpha\beta}\\
&\qquad+(\text{singular terms in $|x_{1,\perp}|$})\ ,
\end{aligned}
\end{align}
 By taking $|x_{1,\perp}|\to0$ and subtracting divergences, we find that:
\begin{align}\label{eq:transverse spin free 1}
               \langle\,\widehat{\Phi}_{s+1}^{\,\alpha,i_1\cdots i_s}(\hat{y})\,\Phi_1^\beta(x)\,\rangle=2^s\,s!\cdot \frac{x_{\perp}^{(i_1}\cdots x_{\perp}^{i_s)}}{|x-\hat{y}|^{2(s+1)}}\,\delta^{\alpha\beta}\ .
\end{align}
From this result, it is straightforward to derive the two-point function of $\widehat{\Phi}_{s+1}^{\,\alpha,i_1\cdots i_s}$:
\begin{align}\label{eq:transverse spin free 2}
               \langle\,\widehat{\Phi}_{s+1}^{\,\alpha,i_1\cdots i_s}(\hat{y}_1)\,\widehat{\Phi}_{s+1}^{\,\beta,j_1\cdots j_s}(\hat{y}_2)\,\rangle&=2^s\,(s!)^2\cdot \frac{\delta^{(i_1}_{j_1}\cdots \delta^{i_s)}_{j_s}}{|\hat{y}_{12}|^{2(s+1)}}\,\delta^{\alpha\beta}\ .
\end{align}

Wick's theorem allows us to calculate all the correlators of our interest from the ones derived up to this point.
For instance, the bulk one-point function $\langle\,\Phi_3^\alpha(x)\,\rangle$ and the bulk-defect two-point function $ \langle\,\Phi_3^\alpha(x)\,\widehat{\Phi}_1^{\,\beta}(\hat{y})\,\rangle$ are computed as follows:
\begin{align}\label{eq:bulk one point Phi3 line}
\begin{aligned}      \langle\,\Phi_3^\alpha(x)\,\rangle&=\langle\,\Phi_1^\alpha(x)\,\rangle\cdot\langle\,\Phi_1^\beta(x)\,\rangle\cdot\langle\,\Phi_1^\beta(x)\,\rangle\\
         &=\frac{\delta^{\alpha1}\,\hat{h}^3}{|x_\perp|^3}\ ,
\end{aligned}
\end{align}
and 
\begin{align}\label{eq:line defect correlator Phi3 hat Phi1}
\begin{aligned}
           \langle\,\Phi_3^\alpha(x)\,\widehat{\Phi}_1^{\,\beta}(\hat{y})\,\rangle&= \langle\,\Phi_1^\alpha(x)\,\widehat{\Phi}_1^{\,\beta}(\hat{y})\,\rangle\cdot\langle\,\Phi_1^\gamma(x)\,\rangle\cdot\langle\,\Phi_1^\gamma(x)\,\rangle\\
       &\qquad+2\, \langle\,\Phi_1^\alpha(x)\,\widehat{\Phi}_1^{\,\gamma}(\hat{y})\,\rangle\cdot \langle\,\Phi_1^\beta(x)\,\rangle\cdot\langle\,\Phi_1^\gamma(x)\,\rangle\\
       &= \frac{\hat{h}^2\,(1 + 2\,\delta^{\alpha1})\,\delta^{\alpha\beta}}{|x-\hat{y}|^2\,|x_\perp|^2} \ .
\end{aligned}
\end{align}
Provided \eqref{eq:transverse spin free 1} and \eqref{eq:bulk one point phi}, some calculations similar to the above result in the following bulk-defect two-point functions involving $\widehat{\Phi}_{s+1}^{\,\alpha,i_1\cdots i_s}$:
\begin{align}
              \langle\,\widehat{\Phi}_{s+1}^{\,\alpha,i_1\cdots i_s}(\hat{y})\,\Phi_3^\beta(x)\,\rangle&=\hat{h}^2\,(1 + 2\,\delta^{\alpha1})\,2^s\,s!\cdot\frac{x_{\perp}^{(i_1}\cdots x_{\perp}^{i_s)}}{|x-\hat{y}|^{2(s+1)}}\,\delta^{\alpha\beta}\ .
\end{align}

Our further interests are in the correlation functions involving the following operators:
\begin{align}
\widehat{\Phi}_3^{\,\alpha}(\hat{x})\equiv\lim_{|x_\perp|\to0}\,\Phi_3^\alpha(x)\ ,
\end{align}
and
\begin{align}\label{eq:second order defect local app}
\widehat{\Phi}_2 \in \left\{|\widehat{\Phi}_1^{\,1}|^{2}\ ,~ |\widehat{\Phi}_{1}^{\,\hat{\gamma}}|^{2}\ ,~ \widehat{\Phi}_{1}^{\,1}\widehat{\Phi}_{1}^{\,\hat{\gamma}}\ , ~ \widehat{\Phi}_{1}^{\,(\hat{\gamma}}\widehat{\Phi}_{1}^{\,\hat{\sigma})}\right\}\ .
\end{align}
We here note the following correlators including $\widehat{\Phi}_3^{\,\alpha}$:
\begin{align}\label{eq:two-point functions line}
           \langle\,\Phi_3^\alpha(x)\,\widehat{\Phi}_3^{\,\beta}(\hat{y})\,\rangle= \frac{2\,(N+2)}{|x-\hat{y}|^6}\,\delta^{\alpha\beta}\ ,\qquad
        \langle\,\widehat{\Phi}_3^{\,\alpha}(\hat{y}_1)\,\widehat{\Phi}_3^{\,\beta}(\hat{y}_2)\,\rangle= \frac{2\,(N+2)}{|\hat{y}_{12}|^6}\,\delta^{\alpha\beta}\ .
\end{align}
In addition, we will make heavy use of the following defect three-point functions and bulk-defect-defect three-point functions concerning $\widehat{\Phi}_2$ in our analysis:
\begin{align}\label{eq:defect three-point phi2}
\begin{aligned}
       \langle\,\widehat{\Phi}_1^{\alpha}(\hat{x})\,\widehat{\Phi}_1^{\,\beta}(\hat{y}_1)\,\widehat{\Phi}_2(\hat{y_2})\,\rangle&= \frac{c(\widehat{\Phi}_1^{\alpha},\widehat{\Phi}_1^{\,\beta},\widehat{\Phi}_2)}{|\hat{x}-\hat{y}_2|\,|\hat{y}_{12}|^2}\ ,\\
   \langle\,\widehat{\Phi}_3^{\alpha}(\hat{x})\,\widehat{\Phi}_1^{\,\beta}(\hat{y}_1)\,\widehat{\Phi}_2(\hat{y_2})\,\rangle&= \frac{c(\widehat{\Phi}_3^{\alpha},\widehat{\Phi}_1^{\,\beta},\widehat{\Phi}_2)}{|\hat{x}-\hat{y}_1|^2\,|\hat{x}-\hat{y}_2|^4}\ ,
\end{aligned}
\end{align}
and
 \begin{align}
      \langle\,  \Phi_1^{\alpha}(x)\, \widehat{\Phi}_1^{\,\beta}(\hat{y}_1)\,\widehat{\Phi}_2(\hat{y}_2) \,\rangle&=\frac{c(\widehat{\Phi}_1^{\alpha},\widehat{\Phi}_1^{\,\beta},\widehat{\Phi}_2)}{|x-\hat{y}_2|^2\,|\hat{y}_{12}|^2}\ ,\\
      \langle\,  \Phi_3^{\alpha}(x)\, \widehat{\Phi}_1^{\,\beta}(\hat{y}_1)\,\widehat{\Phi}_2(\hat{y}_2) \,\rangle&=\frac{(1+2\,\delta^{\alpha1})\,\hat{h}^2\,c(\widehat{\Phi}_3^{\alpha},\widehat{\Phi}_1^{\,\beta},\widehat{\Phi}_2)}{|x-\hat{y}_2|^2\,|\hat{y}_{12}|^2\,|x_\perp|^2}+ \frac{c(\Phi_3^{\alpha},\widehat{\Phi}_1^{\,\beta},\widehat{\Phi}_2)}{|x-\hat{y}_1|^2\,|x-\hat{y}_2|^4} \ ,\label{eq:bdd three point Phi2}
 \end{align}
Here, the defect three-point coefficients $c(\widehat{\Phi}_1^{\alpha},\widehat{\Phi}_1^{\,\beta},\widehat{\Phi}_2)$ and $c(\widehat{\Phi}_3^{\alpha},\widehat{\Phi}_1^{\,\beta},\widehat{\Phi}_2)$ are listed in table \ref{tab:list three-point coeff}.

\begin{table}[t]
\centering
\renewcommand{\arraystretch}{1.2}
\begin{tabular}{>{\centering}m{2.2cm}>{\centering}m{4cm}>{\centering\arraybackslash}m{4cm}}
\toprule
$\widehat{\Phi}_2$  & $c(\widehat{\Phi}_1^{\alpha},\widehat{\Phi}_1^{\,\beta},\widehat{\Phi}_2)$ & $c(\widehat{\Phi}_3^{\alpha},\widehat{\Phi}_1^{\,\beta},\widehat{\Phi}_2)$ \\ \midrule
$|\widehat{\Phi}^{\,1}|^{2}$  & $2\,\delta^{\alpha1}\delta^{\beta1}$  & $2\,\delta^{\alpha\beta}\,+4\,\delta^{\alpha1}\delta^{\beta1}$ \\
$|\widehat{\Phi}_{1}^{\, \hat{\alpha}}|^2$  & $2\,\delta^{\alpha\beta}-2\,\delta^{\alpha1}\delta^{\beta1}$  & $2\,(N+1)\,\delta^{\alpha\beta}-4\,\delta^{\alpha1}\delta^{\beta1}$  \\
 $\widehat{\Phi}^{\,1}\widehat{\Phi}^{\,\hat{\gamma}}$  &$\delta^{\alpha1}\delta^{\beta\hat{\gamma}}+\delta^{\beta1}\delta^{\alpha\hat{\gamma}}$  & $2\,(\delta^{\alpha1}\delta^{\beta\hat{\gamma}}+\delta^{\beta1}\delta^{\alpha\hat{\gamma}})$  \\
$\widehat{\Phi}^{\,(\hat{\gamma}}\widehat{\Phi}^{\,\hat{\sigma})}$  & $2\,\delta^{\alpha(\hat{\gamma}}\delta^{\hat{\sigma})\beta}$  & $4\,\delta^{\alpha(\hat{\gamma}}\delta^{\hat{\sigma})\beta}$ \\
\bottomrule
\end{tabular}
\caption{List of defect three-point coefficients involving second-order composite operators \eqref{eq:second order defect local app}. See equation \eqref{eq:defect three-point phi2} for the expressions of defect three-point functions.
}
\label{tab:list three-point coeff}
\end{table}

\subsection{Defect operator expansions}\label{eq:Bulk-to-defect operator product expansions free}
According to section \ref{eq:DOE of bulk scalar primary}, we now write down the DOEs of two bulk local operators $\Phi_1^\alpha$ and $\Phi_3^\alpha$ from the correlation functions derived in the last section. 

\paragraph{Defect operator expansion of $\Phi_1^\alpha$.}
The DOE of $\Phi_1^\alpha$ is given by the expression:
\begin{align}\label{eq:phi1 OPE Wilsonline}
         \Phi_1^\alpha(x)&=\frac{\delta^{\alpha1}\,\hat{h}}{|x_\perp|} \cdot\bm{1}+\widehat{\Phi}_1^{\alpha}(\hat{x})+\sum_{s=1}^{\infty}  \frac{1}{s!}\cdot x_\perp^{(i_1}\cdots x_\perp^{i_s)}\cdot \widehat{\Phi}_{s+1,i_1\cdots i_s}^{\,\alpha}(\hat{x}) +(\text{descendants.})\ .
\end{align}
Because the free O$(N)$ vector is just an $n$-copies of the Klein-Gordon field, the defect primaries appearing in its DOE are as anticipated in section \ref{sec:Defect Operator Expansion spectrum of free scalar field}.

\paragraph{Defect operator expansion of $\Phi_3^\alpha$.}
It is clear from the one- and two-point functions involving $\Phi^{\alpha}_3$ (see section \ref{sec:Correlation functions in free theory DCFT}) that the DOE of $\Phi^{\alpha}_3$ contains $\widehat{\Phi}^{\,\alpha}_3$, as well as the same defect primaries as those that appear in the DOE of $\Phi_1^\alpha$ \eqref{eq:phi1 OPE Wilsonline}.
But this is not the end of the story, and there are additional contributions to the DOE of $\Phi^{\alpha}_3$. Relevant terms in our analysis are:
\begin{align}\label{eq:phi3 OPE Wilsonline}
\begin{aligned}
                 \Phi_3^\alpha(x)&\supset\frac{\delta^{\alpha1}\,\hat{h}^3}{|x_\perp|^3} \cdot \bm{1}+\frac{(1+2\,\delta^{\alpha1})\,\hat{h}^2}{|x_\perp |^{2}}\cdot \widehat{\Phi}_1^{\,\alpha}(\hat{x})\\
      &\qquad+\frac{(1+2\,\delta^{\alpha1})\,\hat{h}^2}{|x_\perp|^2}\cdot \sum_{s=1}^{\infty}\, \frac{1}{s!}\cdot x_\perp^{(i_1}\cdots x_\perp^{i_s)}\cdot \widehat{\Phi}_{s+1,i_1\cdots i_s}^{\,\alpha}(\hat{x})\\
      &\qquad\qquad+\sum_{n=0}^{\infty}\,\frac{b(\Phi_3^\alpha,\widehat{\mathsf{O}}_{2n+3}^{\,\alpha})}{c(\widehat{\mathsf{O}}_{2n+3}^{\,\alpha},\widehat{\mathsf{O}}_{2n+3}^{\,\alpha})}\cdot |x_\perp|^{2n}\cdot \widehat{\mathsf{O}}_{2n+3}^{\,\alpha}(\hat{x})\ ,
\end{aligned}
\end{align}
where the coefficients appearing in the last line fulfill the conditions:
\begin{align}\label{eq:btd 2n+3 OPE coeff}
\frac{b(\Phi^{\alpha}_3,\widehat{\mathsf{O}}_{2n+3}^{\,\alpha})\,c(\widehat{\mathsf{O}}_{2n+3}^{\,\alpha},\widehat{\Phi}_1^{\,\beta},\widehat{\Phi}_2)}{c(\widehat{\mathsf{O}}_{2n+3}^{\,\alpha},\widehat{\mathsf{O}}_{2n+3}^{\,\alpha})}= c(\Phi_3^{\alpha},\widehat{\Phi}_1^{\,\beta},\widehat{\Phi}_2)\cdot \frac{(-1)^n\,(2)_n}{(n+5/2)_n} \ .
\end{align}

 One can confirm above expressions by looking at the bulk-defect-defect three-point functions $\langle\,  \Phi_3^{\alpha}(x)\, \widehat{\Phi}_1^{\,\beta}(\hat{y}_1)\,\widehat{\Phi}_2(\hat{y}_2) \,\rangle$ given in equation \eqref{eq:bdd three point Phi2}, as explained below. By setting $(\hat{y}_1,\hat{y}_2)\mapsto(0,\infty)$, we find the following conformal block expansions:\footnote{The defect local primaries at the infinity is defined in \eqref{eq:defect primary at infinity}.}
\begin{align}\label{eq:conformal block Phi3}
    \begin{aligned}
       \langle\,\Phi^{\alpha}_3(x)\,\widehat{\Phi}_1^{\,\beta}(0)\,\widehat{\Phi}_2(\infty) \,\rangle  
       &=\frac{(1+2\,\delta^{\alpha1})\,\hat{h}^2\,c(\widehat{\Phi}_3^{\alpha},\widehat{\Phi}_1^{\,\beta},\widehat{\Phi}_2)}{|x_\perp|^2}+ \frac{c(\Phi_3^{\alpha},\widehat{\Phi}_1^{\,\beta},\widehat{\Phi}_2)}{|x|^2} \\
       &=\frac{|x|}{|x_\perp|^3}\cdot \left[(1+2\,\delta^{\alpha1})\,\hat{h}^2\,c(\widehat{\Phi}_3^{\alpha},\widehat{\Phi}_1^{\,\beta},\widehat{\Phi}_2)\cdot G_1^{-1}(\upsilon)\right.\\
       &\qquad\qquad\left.+ c(\Phi_3^{\alpha},\widehat{\Phi}_1^{\,\beta},\widehat{\Phi}_2)\cdot  \sum_{n=0}^{\infty}\,\frac{(-1)^n\,(2)_n}{(n+5/2)_n}\cdot G_{2n+3}^{-1}(\upsilon)\right]  \ ,
    \end{aligned}
\end{align}
with $\upsilon=|x_\perp|^2/|x|^2$.
In deriving this, we employed the hypergeometric identity \eqref{eq:hypergeometric identity 1} and:
\begin{align}
    G^{-1}_{1}(\upsilon)=\upsilon^{1/2}\ ,\qquad G_{2n+3}^{-1}(\upsilon)=\upsilon^{n+3/2}\cdot {}_2F_1(1+n,2+n;7/2+2n;\upsilon)\ .
\end{align}
We learn from \eqref{eq:conformal block Phi3} that an infinite number of defect primaries having odd integer conformal dimensions $(2n+3)$ $(n=0,1,\cdots)$ appear in the DOE of $\Phi^{\alpha}_3$, which we denote by $\widehat{\mathsf{O}}_{2n+3}^{\,\alpha}$. It is worthwhile noting that $\widehat{\mathsf{O}}_{3}^{\,\alpha}$ can be identified with $\widehat{\Phi}^{\,\alpha}_3$, and $\widehat{\mathsf{O}}_{2n+3}^{\,\alpha}$ ($n\geq 1$) is some composite operator in the vector representation of O$(N)$ made out from three $\widehat{\Phi}_1^{\,\alpha}$'s and $2n$ parallel derivatives $\hat{\partial}^a$.
By comparing \eqref{eq:conformal block Phi3} with \eqref{eq:conformal block expansion main}, we conclude \eqref{eq:phi3 OPE Wilsonline} and \eqref{eq:btd 2n+3 OPE coeff}.\footnote{It is possible to perform a similar analysis for the bulk-defect-defect three-point function $\langle\,  \Phi_3^{\alpha}\, \widehat{\Phi}_1^{\,\beta}\,\widehat{\Phi}_1^{\,\gamma} \,\rangle$:
\begin{align}
\begin{aligned}
        \langle\,  &\Phi_3^{\alpha}(x)\, \widehat{\Phi}_1^{\,\beta}(0)\,\widehat{\Phi}_1^{\,\gamma}(\infty) \,\rangle=\frac{\delta^{\alpha1}\delta^{\beta\gamma}\,\hat{h}^3}{|x_\perp|^3}+\frac{2\,\hat{h}\,(\delta^{\alpha1}\delta^{\beta\gamma}+\delta^{\beta1}\delta^{\alpha\gamma}+\delta^{\gamma1}\delta^{\alpha\beta})}{|x_\perp|\cdot |x|^2}\\
        &=\frac{1}{|x_\perp|^3}\cdot \left[\delta^{\alpha1}\delta^{\beta\gamma}\,\hat{h}^3\cdot G_{0}^{0}(\upsilon)+2\,\hat{h}\,(\delta^{\alpha1}\delta^{\beta\gamma}+\delta^{\beta1}\delta^{\alpha\gamma}+\delta^{\gamma1}\delta^{\alpha\beta})\cdot \sum_{n=0}^{\infty}\,\frac{(-1)^n\,n!}{(n+3/2)_n}\cdot G_{2n+2}^{0}(\upsilon)\right]\ .
\end{aligned}
\end{align}
This conformal block expansion implies that there are further contributions to the DOE of $\Phi^{\alpha}_3$ coming from $\widehat{\Phi}_2\equiv |\widehat{\Phi}_1|^2$ and a tower of operators with even integer conformal dimensions, consisting of two $\widehat{\Phi}_1^{\,\alpha}$'s and even numbers of parallel derivatives $\hat{\partial}^a$. However, such operators are irrelevant in our analysis and ignored afterward, as they come into play at order $\epsilon^2$. \label{fot:infinite even line defect}
}

\section{DCFT data on the line defect}\label{sec:DCFT data on the line defect}
We make the most of the modified Rychkov-Tan's axioms introduced in section \ref{sec:Axioms in the presence of a defect} to reproduce the perturbative results reviewed in section \ref{sec:line defect known}. 
As a warm-up, we start with the critical defect coupling and conformal dimensions of non-composite operators in section \ref{1st order: scalar} and \ref{1st order: transverse spin}. We then move to composite ones. There are little twists and turns in the case of composite operators, but we demonstrate here that our framework works well when combined with the analyticity of Euclidean correlators (Schwinger functions) postulated a-priori in any Euclidean QFTs.

\subsection{Lowest-lying defect local operator and critical defect coupling}\label{1st order: scalar}
Let us derive the conformal dimensions of the lowest-lying defect local operators $(\widehat{W}_1^{\,1},\widehat{W}_1^{\,\hat{\alpha}})$ and the critical defect coupling $\hat{h}$. 
Owing to axiom \ref{dcftaxiom1}, the theory at the fixed point is described by DCFT. Hence, the DOE of $W^\alpha_1(x)$ is fixed by the defect conformal symmetry to have the form:
\begin{align}\label{eq:V1 OPE Wilsonline}
    W^\alpha_1(x)\supset C_0^\alpha\cdot\frac{1}{|x_\perp|^{ \Delta_1}} \cdot \bm{1}+C_{1}^\alpha\cdot\frac{1}{|x_\perp|^{\Delta_1- \widehat{\Delta}(\widehat{W}_1^\alpha)}}\cdot \widehat{W}_1^{\,\alpha}(\hat{x}) \ .
\end{align}
Axiom \ref{dcftaxiom2} states that this DOE \eqref{eq:V1 OPE Wilsonline} should reduce to \eqref{eq:phi1 OPE Wilsonline} as $\epsilon\to0$. As a result, we have:
\begin{align}
    C_{1}^\alpha=1+O(\epsilon)\ ,\qquad
     C_{0}^\alpha=\delta^{\alpha1}\,\hat{h}+O(\epsilon) \ .
\end{align}
Employing the equation of motion in axiom \ref{dcftaxiom3}, one can deduce the DOE of $W^\alpha_3$ from \eqref{eq:V1 OPE Wilsonline}:\footnote{Use \eqref{eq:action of Lap on spinning btd OPE} to evaluate the action of the Laplace differential operator.}
\begin{align}
\begin{aligned}
        W^\alpha_3(x)
        &\supset\frac{C_{0}^\alpha}{\kappa}\cdot\frac{ \Delta_1\,( \Delta_1-1+\epsilon)}{|x_\perp|^{ \Delta_1+2}} \cdot \bm{1}  \\
       &\qquad\qquad+\frac{C_{1}^\alpha}{\kappa}\cdot\frac{[\Delta_1- \widehat{\Delta}(\widehat{W}_1^\alpha)]\cdot[\Delta_1- \widehat{\Delta}(\widehat{W}_1^\alpha)-1+\epsilon]}{|x_\perp|^{ \Delta_1- \widehat{\Delta}(\widehat{W}_1^\alpha)+2}}\cdot \widehat{W}_1^{\,\alpha}(\hat{x})\ .
\end{aligned}
\end{align}
Because this DOE should be identical to \eqref{eq:phi3 OPE Wilsonline} from axiom \ref{dcftaxiom2}, one obtains the following equations:
\begin{align}
\hat{h}^3\,\kappa&=\hat{h}\, \Delta_1\,( \Delta_1-1+\epsilon)+O(\epsilon^2) \ ,\label{eq:equation for defect coupling}\\
(1+2\,\delta^{\alpha1})\,\hat{h}^2\,\kappa&=[\Delta_1- \widehat{\Delta}(\widehat{W}_1^\alpha)]\cdot[\Delta_1- \widehat{\Delta}(\widehat{W}_1^\alpha)-1+\epsilon]+O(\epsilon^2) \ .\label{eq:equation for hat W1}
\end{align}
We can solve these two equations by exploiting the values of two bulk quantities $\kappa$ \eqref{eq:value of kappa} and $\Delta_1$ \eqref{eq:Delta1 axiomatic} to find that:
\begin{align}\label{eq:aphi in O(N) model WFFP}
    \hat{h}^2=\frac{N+8}{4}+O(\epsilon)\quad \xrightarrow[\eqref{eq:defect coupling redefinition}]{\hat{h}=h/2}\quad  h^2=N+8+O(\epsilon) \ ,
\end{align}
and 
\begin{align}\label{conf dim: defect local scalar}
    \begin{aligned}
        \widehat{\Delta}(\widehat{W}_1^{\alpha})
            &=
                \Delta_1+\frac{1+2\,\delta^{\alpha1}}{2}\,\epsilon+O(\epsilon^2) \\
            &=
                1+\epsilon\,\delta^{\alpha1}+O(\epsilon^2)\ .
    \end{aligned}
\end{align}
These are in agreement with the perturbative results \eqref{eq:critical defect coupling}, \eqref{eq:conf dim low scalar cuomo} and \eqref{eq:conf dim tilt cuomo}.

For those who wonder if it is not sufficient to look at the leading behavior of the DOE for small $|x_{\perp}|$, we here verify our result \eqref{conf dim: defect local scalar} all orders in $|x_{\perp}|$ by considering the bulk-defect two-point function:
\begin{align}
  \langle\,W^\alpha_1(x)\,\widehat{W}_1^{\beta}(\hat{y})\,\rangle
  =\frac{b(W^\alpha_1,\widehat{W}_1^{\beta})}{|x-\hat{y}|^{2\widehat{\Delta}(\widehat{W}_1^{\alpha})}\,|x_\perp|^{\Delta_1-   \widehat{\Delta}(\widehat{W}_1^{\alpha})}}\ ,
\end{align}
with $b(W^\alpha_1,\widehat{W}_1^{\beta})=\delta^{\alpha\beta}+O(\epsilon)$.
Let us apply the equation of motion \eqref{eq:classical EoM O(N)} to this expression using the formula for the Laplace differential operator $\Box_x=\frac{\partial}{\partial x^\mu}\frac{\partial}{\partial x_\mu}$ in the presence of a $p$-dimensional defect:
\begin{align}\label{eq:Action of Laplacian most generic}
    \Box_x \,\left(\frac{1}{|x|^a\,|x_\perp|^b}\right)
        =
          \frac{  a\,(a + 2b + 2 - d)}{|x|^{a+2}\,|x_\perp|^b}  
                +
                \frac{b\,(b + 2 + p - d)}{|x|^a\,|x_\perp|^{b+2}}  \ .
\end{align}
Then we we have:
\begin{align}\label{eq:bd 2pt W3 W1}
\begin{aligned}
      \langle\,W^\alpha_3(x)\,\widehat{W}_1^{\beta}(\hat{y})\,\rangle&=\frac{b(W^\alpha_1,\widehat{W}_1^{\beta})}{\kappa}\cdot \frac{2\widehat{\Delta}(\widehat{W}_1^{\alpha})\cdot[2\Delta_1-2+\epsilon]}{|x-\hat{y}|^{2\widehat{\Delta}(\widehat{W}_1^{\alpha})+2}\,|x_\perp|^{\Delta_1-   \widehat{\Delta}(\widehat{W}_1^{\alpha})}}\\
      &
 +  \frac{b(W^\alpha_1,\widehat{W}_1^{\beta})}{\kappa}\cdot\frac{[\Delta_1-   \widehat{\Delta}(\widehat{W}_1^{\alpha})]\cdot [\Delta_1-   \widehat{\Delta}(\widehat{W}_1^{\alpha})-1+\epsilon]}{|x-\hat{y}|^{2\widehat{\Delta}(\widehat{W}_1^{\alpha})}\,|x_\perp|^{\Delta_1-   \widehat{\Delta}(\widehat{W}_1^{\alpha})+2}}\ .
\end{aligned}
\end{align}
The first term in the right-hand side of \eqref{eq:bd 2pt W3 W1} vanishes in taking $\epsilon\to0$ since its coefficient is of order $\epsilon$:
\begin{align}
 \frac{b(W^\alpha_1,\widehat{W}_1^{\beta})}{\kappa}\cdot    2\,\widehat{\Delta}(\widehat{W}_1^{\alpha})\cdot[2\Delta_1-2+\epsilon]=O(\epsilon)\ .
\end{align}
On the other hand, the coefficient in the second term is finite even after setting $\epsilon\to0$:
\begin{align}
\begin{aligned}
      \frac{b(W^\alpha_1,\widehat{W}_1^{\beta})}{\kappa}&\cdot   [\Delta_1-   \widehat{\Delta}(\widehat{W}_1^{\alpha})]\cdot [\Delta_1-   \widehat{\Delta}(\widehat{W}_1^{\alpha})-1+\epsilon]\\
      &=\delta^{\alpha\beta}\cdot\frac{N+8}{4}\cdot (1+2\,\delta^{\alpha1})+O(\epsilon)\ .
\end{aligned}
\end{align}
Hence, when $\epsilon=0$, the bulk-defect two-point function \eqref{eq:bd 2pt W3 W1} reduces to:
\begin{align}
    \langle\,\Phi_3^\alpha(x)\,\widehat{\Phi}_1^{\,\beta}(\hat{y})\,\rangle= \frac{\hat{h}^2\,(1 + 2\,\delta^{\alpha1})\,\delta^{\alpha\beta}}{|x-\hat{y}|^2\,|x_\perp|^2} \quad\text{with}\quad  \hat{h}^2=\frac{N+8}{4}\ ,
\end{align}
in agreement with the free theory analysis \eqref{eq:line defect correlator Phi3 hat Phi1}. Therefore, our result is consistent with defect conformal symmetry to all orders in $|x_\perp|$.

\subsection{Defect local operators with transverse spin indices}\label{1st order: transverse spin}
The defect local operators with transverse spin indices $\widehat{U}_{i_1\cdots i_s}^{\,1}$ and $\widehat{U}_{i_1\cdots i_s}^{\,\hat{\alpha}}$ appear in the DOE of $W^\alpha_1$ in the following manner (axiom \ref{dcftaxiom1}):\footnote{The result in this section has some overlaps with \cite[Appendix C]{Giombi:2022vnz} that appeared shortly before our original paper \cite{Nishioka:2022qmj}.}
\begin{align}
        W^\alpha_1(x)\supset C_s^\alpha\cdot \frac{x_\perp^{(i_1}\cdots x_\perp^{i_s)}}{|x_\perp|^{ \Delta_1- \widehat{\Delta}(\widehat{U}^{\alpha}_s)+s}}\cdot\widehat{U}_{i_1\cdots i_s}^{\,\alpha}(\hat{x})\ ,
\end{align}
with $C_s^\alpha=1/s!+O(\epsilon)$ from \eqref{eq:phi1 OPE Wilsonline} (axiom \ref{dcftaxiom2}).
From axiom \ref{dcftaxiom3} and \eqref{eq:action of Lap on spinning btd OPE}, one has:
\begin{align}
\begin{aligned}
W^\alpha_3(x)&\supset\frac{C_s^\alpha}{\kappa}\cdot [\Delta_1- \widehat{\Delta}(\widehat{U}^{\alpha}_s)+s]\cdot [\Delta_1- \widehat{\Delta}(\widehat{U}^{\alpha}_s)-s-1+\epsilon]\\
&\qquad\qquad\qquad\qquad\qquad\qquad\cdot 
              \frac{x_\perp^{(i_1}\cdots x_\perp^{i_s)}}{|x_\perp|^{ \Delta_1- \widehat{\Delta}(\widehat{U}^{\alpha}_s)+s+2}}\cdot\widehat{U}_{i_1\cdots i_s}^{\,\alpha}(\hat{x})\ .    
\end{aligned}
\end{align}
Comparing this with \eqref{eq:phi3 OPE Wilsonline} in taking $\epsilon\to0$ limit (axiom \ref{dcftaxiom2}), we see that:
\begin{align}\label{eq:conf dim:transverse spin}
    \begin{aligned}
       \widehat{\Delta}(\widehat{U}^{\alpha}_s)
            &= 
            \Delta_1+s+\frac{1+2\,\delta^{\alpha1}}{2\,(s+2)}\,\epsilon+O(\epsilon^2) \\
            &=
            s+1+\frac{2\,\delta^{\alpha1}-s-1}{2\,(s+2)}\,\epsilon+O(\epsilon^2)\ .
    \end{aligned}
\end{align}
This result agrees with the perturbative calculations for $s=1$ and also with the limiting behavior of defect local operators $ \widehat{\Delta}(\widehat{U}^{\alpha}_s)\xrightarrow[]{s\to\infty}\Delta_1+s$ predicted by the analytic defect conformal bootstrap \cite[section 2.1]{Lemos:2017vnx}.

\subsection{Composite defect local operators}\label{sec:defect composite operator}
We now derive the conformal dimensions of the following defect composite operators:
\begin{align}\label{eq:second order operators with no transverse spin indices}
 \widehat{W}_2 \in \left\{\widehat{V}^{\hat{\alpha}}\ ,~ \widehat{T}^{\hat{\alpha}\hat{\beta}}\ ,~ \widehat{S}_+\ , ~ \widehat{S}_-\right\} \ ,
\end{align}
tending to free theory ones listed in \eqref{eq:second order defect local app} as $\epsilon\to0$.
As a remainder, we note that in taking the limit $\epsilon\to0$ $\widehat{V}^{\hat{\alpha}}$ and $\widehat{T}^{\hat{\alpha}\hat{\beta}}$ become $\widehat{\Phi}_{1}^{\,1}\widehat{\Phi}_{1}^{\,\hat{\alpha}}$ and $\widehat{\Phi}_{1}^{\,(\hat{\alpha}}\widehat{\Phi}_{1}^{\,\hat{\beta})}$ respectively, while $\widehat{S}^\pm$ turn into some linear combinations of $|\widehat{\Phi}_1^{\,1}|^{2}$ and $|\widehat{\Phi}_{1}^{\,\hat{\gamma}}|^{2}$. Without loss of generality, we make such that $\widehat{S}^\pm$ are unit-normalized and orthogonal to each other in the limit $\epsilon\to0$ by introducing the mixing angle $\theta$:
\begin{align}\label{eq:scalar mixing}
    \lim_{\epsilon\to0}\begin{pmatrix}
    \widehat{S}_+\\ \widehat{S}_-
    \end{pmatrix}
    =
    \begin{pmatrix}
    \cos\theta & -\sin\theta\\
    \sin\theta &\cos\theta
    \end{pmatrix}\,
    \begin{pmatrix}
    \frac{1}{\sqrt{2}}\cdot |\widehat{\Phi}_1^{\,1}|^2\\ \frac{1}{\sqrt{2\,(N-1)}}\cdot |\widehat{\Phi}_1^{\,\hat{\gamma}}|^2
    \end{pmatrix}\ .
\end{align}

Because composite operators do not appear in the DOE of $\Phi_1^\alpha(x)$ in the free theory $(\epsilon=0)$, we cannot use the same approach as the last two subsections. Instead, we take the following strategy to tackle the problem:
\begin{itemize}
    \item We calculate the DOE of $W_1$ up to the first order in $\epsilon$ utilizing the equation of motion \eqref{eq:classical EoM O(N)}. 
    \item Thereby, we investigate the following bulk-defect-defect three-point function:
        \begin{align}
           \langle\, W^{\alpha}_1(x)\,\widehat{W}_1^{\,\beta}(0)\,\widehat{W}_2(\infty) \,\rangle
        \end{align}
    \item The correlator turns out to be non-analytic along $|\hat{x}|=0$ where any pairs of operators do not coincide, in contradiction to the analyticity of Euclidean correlators away from the coincidence of points \cite{Osterwalder:1973dx,Osterwalder:1974tc}. Requiring non-analytic terms to vanish at order $\epsilon$, we obtain constraints on the conformal dimensions.\footnote{We comment that the analyticity of bulk-defect-defect three-point correlators are firstly used \cite{Lauria:2020emq} to argue the triviality of defect operator spectrum when the bulk scalar field is free, and subsequently for free Maxwell theory in \cite{Herzog:2022jqv}. The authors of \cite{Behan:2020nsf,Behan:2021tcn} applied a similar idea for bulk-bulk-boundary three-point functions. See also \cite{Lauria:2023uca} for a related subject.}
\end{itemize}

\paragraph{Defect operator expansion of $W_1^\alpha$.}
In the free theory $(\epsilon=0)$, the Klein-Gordon equation severely constrains the DOE of $W_1^\alpha=\Phi_1^\alpha$ to take the form given in \eqref{eq:phi1 OPE Wilsonline}. In the Wilson-Fisher theory $(\epsilon\neq0)$, however, other operators can appear in the DOE of $W_1^\alpha$, whose contributions relevant in our analysis at order $\epsilon$ are as follows:\footnote{Similarly, a tower of operators having even integer conformal dimensions when $\epsilon=0$ should appear according to footnote \ref{fot:infinite even line defect} in section \ref{eq:Bulk-to-defect operator product expansions free}. But we here neglect them, as they are irrelevant in our analysis restricted to the first order in $\epsilon$.}
\begin{align}\label{eq:OPE of W1 all order}
    \begin{aligned}
        W_1^\alpha(x)
            &\supset
                \frac{C_1^\alpha}{|x_\perp|^{\Delta_1-\widehat{\Delta}(\widehat{W}_1^{\,\alpha})}}\cdot\widehat{W}_1^{\,\alpha}(\hat{x})
                \\
                &\qquad\qquad+                \sum_{n=0}^{\infty}\,\frac{b(W_1^\alpha,\widehat{\mathsf{O}}_{2n+3}^{\prime\,\alpha})/c(\widehat{\mathsf{O}}_{2n+3}^{\prime\,\alpha},\widehat{\mathsf{O}}_{2n+3}^{\prime\,\alpha})}{|x_\perp|^{\Delta_1-\widehat{\Delta}(\widehat{\mathsf{O}}_{2n+3}^{\prime\,\alpha})}}\cdot\widehat{\mathsf{O}}_{2n+3}^{\prime\,\alpha}(\hat{x})\ .
    \end{aligned}
\end{align}
Here, the defect local operator appearing in above expression $\widehat{\mathsf{O}}_{2n+3}^{\prime\,\alpha}$ ($n=0,1,\cdots$) has conformal dimension $\widehat{\Delta}(\widehat{\mathsf{O}}_{2n+3}^{\prime\,\alpha})=2n+3+O(\epsilon)$ and is identical with $\widehat{\mathsf{O}}_{2n+3}^{\,\alpha}$ that appears in the DOE of $\Phi_3^\alpha$ \eqref{eq:phi3 OPE Wilsonline} when $\epsilon=0$. The DOE coefficients are subject to the following relations:
\begin{align}\label{eq:W1 to higher order OPE coeff}
    \frac{b(W_1^\alpha,\widehat{\mathsf{O}}_{2n+3}^{\prime\,\alpha})}{c(\widehat{\mathsf{O}}_{2n+3}^{\prime\,\alpha},\widehat{\mathsf{O}}_{2n+3}^{\prime\,\alpha})}=\frac{\epsilon}{(N+8)\,(n+1)(2n+3)}\cdot \frac{b(\Phi_3^\alpha,\widehat{\mathsf{O}}_{2n+3}^{\,\alpha})}{c(\widehat{\mathsf{O}}_{2n+3}^{\,\alpha},\widehat{\mathsf{O}}_{2n+3}^{\,\alpha})}+O(\epsilon^2)\ .
\end{align}
To derive the relation for DOE coefficients \eqref{eq:W1 to higher order OPE coeff}, we first apply the equation of motion \eqref{eq:classical EoM O(N)} to the DOE of $W_1^\alpha$ \eqref{eq:OPE of W1 all order}:
\begin{align}\label{eq:OPE of W3 all order line}
\begin{aligned}
            W_3^\alpha(x)
            &\supset
                \frac{b(W_1^\alpha,\widehat{\mathsf{O}}_{2n+3}^{\prime\,\alpha})}{\kappa\cdot c(\widehat{\mathsf{O}}_{2n+3}^{\prime\,\alpha},\widehat{\mathsf{O}}_{2n+3}^{\prime\,\alpha})}\\
                &\qquad\qquad\cdot\frac{[\Delta_1-\widehat{\Delta}(\widehat{\mathsf{O}}_{2n+3}^{\prime\,\alpha})]\cdot [\Delta_1-\widehat{\Delta}(\widehat{\mathsf{O}}_{2n+3}^{\prime\,\alpha})-1+\epsilon]}{|x_\perp|^{\Delta_1-\widehat{\Delta}(\widehat{\mathsf{O}}_{2n+3}^{\prime\,\alpha})+2}}\cdot\widehat{\mathsf{O}}_{2n+3}^{\prime\,\alpha}(\hat{x}) \ .
\end{aligned}
\end{align}
Due to axiom \ref{dcftaxiom3}, this DOE should reduce to the expression given in \eqref{eq:phi3 OPE Wilsonline} as $\epsilon\to0$. Then, we compare \eqref{eq:OPE of W3 all order line} with \eqref{eq:phi3 OPE Wilsonline}, while substituting $\kappa$ \eqref{eq:value of kappa}, to conclude \eqref{eq:W1 to higher order OPE coeff}.

\paragraph{Study of bulk-defect-defect three-point functions involving $W_1^\alpha$.}
It follows from the DOE of $W_1^\alpha$ \eqref{eq:OPE of W1 all order} and the generic form of the conformal block expansion \eqref{eq:defect primary at infinity} that the three-point function $\langle\, W_1^{\,\alpha}\,\widehat{W}_1^{\,\beta}\,\widehat{W}_2 \,\rangle$ takes the form:
\begin{align}\label{eq:3pt app W1}
    \begin{aligned}
    \langle\, & W_1^\alpha(x)\, \widehat{W}_1^{\,\beta}(0)\,\widehat{W}_{2}(\infty)\,\rangle\\
       & =
            \frac{1}{|x_\perp|^{\Delta_1}\,|x|^{\widehat{\Delta}(\widehat{W}_{1}^{\, \beta})-\widehat{\Delta}(\widehat{W}_2)}}\cdot
            \left[C_1^\alpha\cdot c(\widehat{W}_1^{\,\alpha},\widehat{W}_1^{\,\beta},\widehat{W}_2)\cdot G^{\widehat{\Delta}(\widehat{W}_{1}^{\, \beta})-\widehat{\Delta}(\widehat{W}_2)}_{\widehat{\Delta}(\widehat{W}_{1}^{\, \alpha})}(\upsilon)\right.\\
        &\qquad\qquad\qquad\left.     
            +  \sum_{n=0}^{\infty}\,\frac{b(W_1^\alpha,\widehat{\mathsf{O}}_{2n+3}^{\prime\,\alpha})\,c(\widehat{\mathsf{O}}_{2n+3}^{\prime\,\alpha},\widehat{W}_1^{\,\beta},\widehat{W}_2)}{c(\widehat{\mathsf{O}}_{2n+3}^{\prime\,\alpha},\widehat{\mathsf{O}}_{2n+3}^{\prime\,\alpha})}\cdot G_{\widehat{\Delta}(\widehat{\mathsf{O}}_{2n+3}^{\prime\,\alpha})}^{\widehat{\Delta}(\widehat{W}_{1}^{\, \beta})-\widehat{\Delta}(\widehat{W}_2)}(\upsilon)\right]\ ,
    \end{aligned}
\end{align}
with $\upsilon=|x_\perp|^2/|x|^2$.
One can expand the first term in the parenthesis in the following manner:
\begin{align}\label{eq: G}
\begin{aligned}
  &  G^{\widehat{\Delta}(\widehat{W}_{1}^{\, \beta})-\widehat{\Delta}(\widehat{W}_2)}_{\widehat{\Delta}(\widehat{W}_{1}^{\, \alpha})}(\upsilon)\\
    &\qquad=\upsilon^{\widehat{\Delta}(\widehat{W}_{1}^{\, \alpha})/2}\cdot{}_2F_1\left(\tfrac{\delta^{\alpha1}+\delta^{\beta1}-\widehat{\Gamma}(\widehat{W}_2)}{2}\,\epsilon,1;\frac{3}{2};\upsilon\right)+O(\epsilon^2)\\
    &\qquad=\upsilon^{\widehat{\Delta}(\widehat{W}_{1}^{\, \alpha})/2}+\frac{\delta^{\alpha1}+\delta^{\beta1}-\widehat{\Gamma}(\widehat{W}_2)}{3}\,\epsilon\cdot \upsilon^{3/2}\cdot {}_2F_1(1,1;5/2;\upsilon)+O(\epsilon^2)\ ,
\end{aligned}
\end{align}
where we used \eqref{conf dim: defect local scalar} and \eqref{eq:hypergeometric series expansion}, and introduced $\widehat{\Gamma}(\widehat{W}_2)$ by the relation: 
\begin{align}\label{eq:Delta W2 redef}
\widehat{\Delta}(\widehat{W}_2)=2+\widehat{\Gamma}(\widehat{W}_2)\cdot \epsilon+O(\epsilon^2)\ .    
\end{align}
Meanwhile, with the aid of the hypergeometric summation formula \eqref{eq:hypergeometric identity 2} and two identities concerning DOE coefficients \eqref{eq:W1 to higher order OPE coeff} and \eqref{eq:btd 2n+3 OPE coeff}, the second term becomes:
\begin{align}\label{eq: sum G}
    \begin{aligned}   \sum_{n=0}^{\infty}&\,\frac{b(W_1^\alpha,\widehat{\mathsf{O}}_{2n+3}^{\prime\,\alpha})\,c(\widehat{\mathsf{O}}_{2n+3}^{\prime\,\alpha},\widehat{W}_1^{\,\beta},\widehat{W}_2)}{c(\widehat{\mathsf{O}}_{2n+3}^{\prime\,\alpha},\widehat{\mathsf{O}}_{2n+3}^{\prime\,\alpha})}\cdot G_{\widehat{\Delta}(\widehat{\mathsf{O}}_{2n+3}^{\prime\,\alpha})}^{\widehat{\Delta}(\widehat{W}_{1}^{\, \beta})-\widehat{\Delta}(\widehat{W}_2)}(\upsilon)\\
     &\qquad= \frac{c(\Phi_3^{\alpha},\widehat{\Phi}_1^{\,\beta},\widehat{\Phi}_2)}{3\,(N+8)}\, \epsilon\cdot \sum_{n=0}^{\infty}\,\frac{(-1)^n\,n!}{(2n/3+1)\,(n+5/2)_n} \cdot \upsilon^{n+3/2}\\
     &\quad\qquad\qquad\qquad\qquad\qquad\qquad\cdot{}_2F_1\left({n+1,n+2\atop 2n+7/2};\upsilon\right)+O(\epsilon^2)\\
 &\qquad= \frac{c(\Phi_3^{\alpha},\widehat{\Phi}_1^{\,\beta},\widehat{\Phi}_2)}{3\,(N+8)}\,\epsilon\cdot \upsilon^{3/2}\cdot {}_2F_1(1,1;5/2;\upsilon)+O(\epsilon^2)\ .
    \end{aligned}
\end{align}
Combining these two expressions with the relation $C_1^\alpha\cdot c(\widehat{W}_1^{\,\alpha},\widehat{W}_1^{\,\beta},\widehat{W}_2)=c(\widehat{\Phi}_1^{\,\alpha},\widehat{\Phi}_1^{\,\beta},\widehat{\Phi}_2)+O(\epsilon)$, we are able to evaluate the three-point function $\langle\, W_1^{\,\alpha}\,\widehat{W}_1^{\,\beta}\,\widehat{W}_2 \,\rangle$ up to the first order in $\epsilon$:
\begin{align}\label{eq:3pt gen}
\begin{aligned}
     \langle\, &W^{\alpha}_1(x)\,\widehat{W}_1^{\,\beta}(0)\,\widehat{W}_2(\infty) \,\rangle\\
     &= C_1^\alpha\cdot c(\widehat{W}_1^{\,\alpha},\widehat{W}_1^{\,\beta},\widehat{W}_2)\cdot\frac{1}{|x_\perp|^{\Delta_1-\widehat{\Delta}(\widehat{W}_1^{\,\beta})} \,|x|^{\widehat{\Delta}(\widehat{W}_{1}^{\,\alpha})+\widehat{\Delta}(\widehat{W}_1^{\,\beta})-\widehat{\Delta}(\widehat{W}_2)}}\\
     &\quad + \frac{\epsilon}{3\,(N+8)}\cdot\left\{(N+8)\cdot [\delta^{\alpha1}+\delta^{\beta1}-\widehat{\Gamma}(\widehat{W}_2)]\cdot c(\widehat{\Phi}_1^{\,\alpha},\widehat{\Phi}_1^{\,\beta},\widehat{\Phi}_2)+c(\widehat{\Phi}_3^{\,\alpha},\widehat{\Phi}_1^{\,\beta},\widehat{\Phi}_2)\right\}\\
     &\qquad\qquad\qquad\qquad\qquad\qquad\qquad\qquad\cdot \frac{|x_\perp|^2}{|x|^2}\cdot{}_2F_1\left(1,1;\frac{5}{2};\frac{|x_\perp|^2}{|x|^2}\right)+O(\epsilon^2)\ .
\end{aligned}
\end{align}

\paragraph{Constraint from analyticity.}
According to Osterwalder-Schrader axioms \cite{Osterwalder:1973dx,Osterwalder:1974tc} that one postulates in any Euclidean QFTs, Euclidean QFT correlators must be analytic if any pairs of operators are away from each other.
In taking the limit $\hat{x}\to0$, the second term in three-point function $\langle\, W_1^{\,\alpha}\,\widehat{W}_1^{\,\beta}\,\widehat{W}_2 \,\rangle$ \eqref{eq:3pt gen} turns out to be non-analytic due to odd integer powers of $|\hat{x}|$:
\begin{align}
    {}_2F_1\left(1,1;\frac{5}{2};\frac{|x_\perp|^2}{|x|^2}\right)\xrightarrow[\hat{x}\sim0]{}\frac{3\,\pi}{2}\cdot \frac{|\hat{x}|}{|x|}+\cdots\ ,
\end{align}
as clear from Kummer's connection formula for Gauss's hypergeometric function \eqref{eq:Kummer's connection formula}.
The only way to resolve this non-analyticity is to require that:
\begin{align}\label{eq:master equation line defect}
    (N+8)\cdot [\delta^{\alpha1}+\delta^{\beta1}-\widehat{\Gamma}(\widehat{W}_2)]\cdot c(\widehat{\Phi}_1^{\,\alpha},\widehat{\Phi}_1^{\,\beta},\widehat{\Phi}_2)+c(\widehat{\Phi}_3^{\,\alpha},\widehat{\Phi}_1^{\,\beta},\widehat{\Phi}_2)=0\ ,
\end{align}
with $c(\widehat{\Phi}_1^{\,\alpha},\widehat{\Phi}_1^{\,\beta},\widehat{\Phi}_2)$ and $c(\widehat{\Phi}_3^{\,\alpha},\widehat{\Phi}_1^{\,\beta},\widehat{\Phi}_2)$ being the defect three-point coefficients listed in table \ref{tab:list three-point coeff}.
By solving this constraint, one immediately finds that the conformal dimensions of the O$(N)$ vector field $\widehat{V}^{\hat{\alpha}}$ and the rank-two symmetric and traceless tensor $\widehat{T}^{\hat{\alpha}\hat{\beta}}$ are given as follows:
\begin{align}
\widehat{\Delta}_{\widehat{V}}&=2+\frac{N+10}{N+8}\,\epsilon+O(\epsilon^2)\ ,\label{conf dim: vector}\\
\widehat{\Delta}_{\widehat{T}}&=2+\frac{2}{N+8}\,\epsilon+O(\epsilon^2)\ ,\label{conf dim: tensor}
\end{align}
in agreement with perturbative results \eqref{eq:conf dim V} and \eqref{eq:conf dim T}.
For the conformal dimensions of $\widehat{S}^\pm$, we have to solve the following simultaneous linear equations with three unknowns $\widehat{\Gamma}(\widehat{S}^{\pm}),\theta$ with $\Gamma_{\widehat{S}_+}\geq \Gamma_{\widehat{S}_-}$:
\begin{align}
\begin{aligned}
 (N+8)\cdot (\Gamma_{\widehat{S}_+}&-\delta^{\alpha1}-\delta^{\beta1})\cdot [\sqrt{N-1}\cdot c(\widehat{\Phi}_1^{\,\alpha},\widehat{\Phi}_1^{\,\beta},|\widehat{\Phi}_1^{\,1}|^2)-\tan\theta\cdot c(\widehat{\Phi}_1^{\,\alpha},\widehat{\Phi}_1^{\,\beta},|\widehat{\Phi}_1^{\,\hat{\gamma}}|^2)]\\
  &=\sqrt{N-1}\cdot c(\widehat{\Phi}_3^{\,\alpha},\widehat{\Phi}_1^{\,\beta},|\widehat{\Phi}^{\,1}|^2)-\tan\theta\cdot c(\widehat{\Phi}_3^{\,\alpha},\widehat{\Phi}_1^{\,\beta},|\widehat{\Phi}_1^{\,\hat{\gamma}}|^2)   \ ,\\
 (N+8)\cdot  (\Gamma_{\widehat{S}_-}&-\delta^{\alpha1}-\delta^{\beta1})\cdot [\sqrt{N-1}\cdot \tan\theta\cdot c(\widehat{\Phi}^{\,\alpha},\widehat{\Phi}_1^{\,\beta},|\widehat{\Phi}^{\,1}|^2)+c(\widehat{\Phi}^{\,\alpha},\widehat{\Phi}_1^{\,\beta},|\widehat{\Phi}^{\,\hat{\gamma}}|^2)]\\
  &=\sqrt{N-1}\cdot \tan\theta\cdot c(\widehat{\Phi}_3^{\,\alpha},\widehat{\Phi}_1^{\,\beta},|\widehat{\Phi}^{\,1}|^2)+c(\widehat{\Phi}_3^{\,\alpha},\widehat{\Phi}_1^{\,\beta},|\widehat{\Phi}^{\,\hat{\gamma}}|^2)    \ .
\end{aligned}
\end{align}
One can readily solve them to find that:
\begin{align}
\widehat{\Delta}_{\widehat{S}^\pm}&=2+\frac{3N+20\pm \sqrt{N^2+40N+320}}{2\,(N+8)}\,\epsilon+O(\epsilon^2)\ ,\label{conf dim: scalars}\\
\tan\theta&=\frac{N+18+\sqrt{N^2+40N+320}}{2\,\sqrt{N-1}}\ .
\end{align}
These expressions are compatible with perturbative results \eqref{eq:conf dim Spm}.

\section{Defect operator spectrum for $N=1$ (Ising DCFT)}\label{sec:N=1 anomalous dimension}
In the previous section, we have derived the conformal dimensions of the lowest-lying composite operators within the axiomatic framework. It would be possible to do the same for higher-order composite operators. However, additional complexity comes in due to the multiplet mixing by the bulk interaction term, which has already happened in our analysis for $\widehat{S}^{\pm}$. Nevertheless, we do not have to worry about this affair for $N=1$. Thus, we here go one step further and study higher-order composite operators $\widehat{W}_p$ $(p=1,2,\cdots)$, tending to $\widehat{\Phi}_{p}$ as $\epsilon\to0$:
\begin{align}\label{eq: def of p-order compodite op}
    \widehat{\Phi}_{p}(\hat{x})\equiv \lim_{|x_\perp|\to0} \,|\Phi_{1}|^p(x) \qquad (\text{free theory with } N=1)\ .
\end{align}

One can deduce one- and two-point functions necessary for our analysis from the results presented in section \ref{sec:Correlation functions in free theory DCFT} by setting $N=1$. We need three-point correlators involving $\widehat{\Phi}_{p}$ to access the information about higher-order composite operators. We are particularly interested in the following defect three-point functions:
\begin{align}
\langle\, \widehat{\Phi}_1(\hat{x})\,\widehat{\Phi}_p(\hat{y}_1) \,\widehat{\Phi}_{p+1}(\hat{y}_2)\,\rangle&=\frac{c(\widehat{\Phi}_1,\widehat{\Phi}_p,\widehat{\Phi}_{p+1})}{|\hat{x}-\hat{y}_2|^2\,|\hat{y}_{12}|^{2p}}\ ,\\
\langle\, \widehat{\Phi}_3(\hat{x})\,\widehat{\Phi}_p(\hat{y}_1) \,\widehat{\Phi}_{p+1}(\hat{y}_2)\,\rangle&=\frac{c(\widehat{\Phi}_1,\widehat{\Phi}_p,\widehat{\Phi}_{p+1})}{|\hat{x}-\hat{y}_2|^2\,|\hat{x}-\hat{y}_2|^4\,|\hat{y}_{12}|^{2p-2}}\ ,
\end{align}
which can be computed straightforwardly via Wick's theorem.
The defect three-point constants appearing in the above expressions are given by:
\begin{align}\label{eq:N=1 line defect 3pt}
    c(\widehat{\Phi}_1,\widehat{\Phi}_p,\widehat{\Phi}_{p+1})=(p+1)!\ ,\qquad c(\widehat{\Phi}_3,\widehat{\Phi}_p,\widehat{\Phi}_{p+1})=3p\,(p+1)!\ .
\end{align}
The following bulk-defect-defect three-point functions are also important in our analysis:
\begin{align}
\langle\, \Phi_1(x)\,\widehat{\Phi}_p(\hat{y}_1) \,\widehat{\Phi}_{p+1}(\hat{y}_2)\,\rangle&=\frac{c(\widehat{\Phi}_1,\widehat{\Phi}_p,\widehat{\Phi}_{p+1})}{|x-y_2|^2\,|\hat{y}_{12}|^{2p}}\ ,\\
\langle\, \Phi_3(x)\,\widehat{\Phi}_p(\hat{y}_1) \,\widehat{\Phi}_{p+1}(\hat{y}_2)\,\rangle&=\frac{3\,\hat{h}\,c(\widehat{\Phi}_1,\widehat{\Phi}_p,\widehat{\Phi}_{p+1})}{|x-y_2|^2\,|\hat{y}_{12}|^{2p}\,|x_\perp|^2}+\frac{c(\widehat{\Phi}_1,\widehat{\Phi}_p,\widehat{\Phi}_{p+1})}{|x-y_2|^2\,|x-y_2|^4\,|\hat{y}_{12}|^{2p-2}}\ .
\end{align}

Following the prescription presented in section \ref{sec:defect composite operator}, we can derive the following recursive constraint for the conformal dimension of $\widehat{W}_p$:
\begin{align}\label{eq:master equation N=1}
    9\cdot [\widehat{\Gamma}(\widehat{W}_1)+\widehat{\Gamma}(\widehat{W}_{p})-\widehat{\Gamma}(\widehat{W}_{p+1})]\cdot c(\widehat{\Phi}_1,\widehat{\Phi}_p,\widehat{\Phi}_{p+1})+c(\widehat{\Phi}_3,\widehat{\Phi}_p,\widehat{\Phi}_{p+1})=0\ ,
\end{align}
where we introduced $\widehat{\Gamma}(\widehat{W}_{p})$ through the relation:
\begin{align}
    \widehat{\Delta}(\widehat{W}_{p})=p+\widehat{\Gamma}(\widehat{W}_p)\cdot\epsilon+O(\epsilon^2)\ .
\end{align}
Provided the initial condition $\widehat{\Gamma}(\widehat{W}_{1})=1$, that follows from \eqref{conf dim: defect local scalar}, and the defect three-point coefficients given in \eqref{eq:N=1 line defect 3pt}, the constraint \eqref{eq:master equation N=1} simplifies to the following expression:
\begin{align}
\Gamma_{\widehat{W}_{p+1}}=\Gamma_{\widehat{W}_p}+1+\frac{p}{3}\ .
\end{align}
This recursion relation is solved to give:
\begin{align}\label{eq:Ising DCFT composite}
   \widehat{\Delta}(\widehat{W}_{p})=p+\frac{p\,(p+5)}{6}\,\epsilon+O(\epsilon^2)\ .
\end{align}
We remark that, at the first order in $\epsilon$, we can identify $\widehat{W}_{2}$ with $\widehat{S}^+$ as their conformal dimensions are identical:
\begin{align}
  \widehat{\Delta}(\widehat{W}_{2})=\widehat{\Delta}(\widehat{S}^+)=2+\frac{7}{3}\,\epsilon+O(\epsilon^2)  \ .
\end{align}

\chapter{The O$(N)$ model with Neumann boundary}\label{chap:Neumann boundary}
This chapter is on the O$(N)$ model in $d=4-\epsilon$ dimensions with Neumann boundary.\footnote{We can regard the Neumann boundary as a co-dimension one (or ($3-\epsilon$)-dimensional) defect. Recall that the goal of the perturbative expansion in powers of $\epsilon$ is to obtain some theoretical predictions for the O$(N)$ model in three dimensions with the Neumann boundary (two-dimensional defect) eventually by setting $\epsilon\to1$.} The action of the model is given by the same expression as \eqref{eq:action without defect} but restricted to the half-spacetime $\mathbb{R}_{+}^{d}\equiv \mathbb{R}^{d-1}\times \mathbb{R}_{\geq 0}$, on whose boundary Neumann boundary condition is imposed:
\begin{align}\label{eq:action Neumann}
    I = \int_{\mathbb{R}_{+}^{d}}\,\d^d x \,\left(\frac{1}{2}\, |\partial \Phi_{1} |^2\, +\,   \frac{\lambda_0}{4!}\, |\Phi_1|^4 \right) \ ,\qquad \lim_{|x_\perp|\to0}\,\frac{\partial}{\partial x_\perp}\,\Phi_{1}^{\alpha}(x)=0\ .
\end{align}
We here describe results found in existing literature while setting up our notations. Similarly to the case with a line defect, the critical coupling constant and bulk operator spectrum are the same as those without boundaries.\footnote{See \eqref{eq:critical lambda} for the critical value of bulk coupling constant.}
Unlike the line defect case, the Neumann boundary condition preserves the O$(N)$ symmetry. Hence, there is no multiplet mixing on the boundary, and one can readily tackle the higher-order composite operators in this model.
We are interested in the conformal dimensions of the boundary local operators that take the following forms in the free theory:
\begin{align}\label{eq:Neumann lowest-lying defect local def}
    \widehat{\Phi}_{1}^{\,\alpha}(\hat{x})\equiv\lim_{|x_\perp|\to0}\,\Phi_{1}^{\,\alpha}(x)\ ,
\end{align}
and 
\begin{align}\label{eq:Neumann composite defect local def}
\widehat{\Phi}_{2p}(\hat{x})\equiv\lim_{|x_\perp|\to0}\,|\Phi_{1}|^{2p}(x)\ ,\qquad \widehat{\Phi}_{2p+1}^{\,\alpha}(\hat{x})\equiv\lim_{|x_\perp|\to0}\,\Phi_{1}^{\,\alpha}|\Phi_{1}|^{2p}(x) \qquad p=1,2,\cdots\ .
\end{align}
We denote their renormalized operators in conventional perturbative calculations (section \ref{sec:Neumann diagrammatic}) as well as their Wilson-Fisher counterparts in axiomatic approach (section \ref{sec:Neumann Axiomatic approach}) as follows:
\begin{align}\label{eq:list of boundary composite Neumann}
\widehat{W}_{1}^{\,\alpha}(\hat{y})\ ,\qquad \widehat{W}_{2p}(\hat{y})\ ,\qquad \widehat{W}_{2p+1}^{\,\alpha}(\hat{y})\ .
\end{align}
We also denote their conformal dimensions by $\widehat{\Delta}_1,\widehat{\Delta}_{2p}$ and $\widehat{\Delta}_{2p+1}$, which are, in the standard perturbative framework, computed as follows:
\begin{align}\label{eq:Neumann boundary composite wfr}
\begin{aligned}
    \widehat{\Delta}_{1}&=\frac{d-2}{2}+\widehat{\gamma}_1\ ,\qquad \widehat{\gamma}_1=\left.\frac{\partial\log \widehat{Z}_{1}}{\partial\log \mu}\right|_{\lambda=\lambda_\ast}\ ,\\
    \widehat{\Delta}_{2p}&=2p\cdot \frac{d-2}{2}+\widehat{\gamma}_{2p}\ ,\qquad \widehat{\gamma}_{2p}=\left.\frac{\partial\log \widehat{Z}_{2p}}{\partial\log \mu}\right|_{\lambda=\lambda_\ast}\ ,\\
    \widehat{\Delta}_{2p+1}&=(2p+1)\cdot \frac{d-2}{2}+\widehat{\gamma}_{2p+1}\ ,\qquad \widehat{\gamma}_{2p+1}=\left.\frac{\partial\log \widehat{Z}_{2p+1}}{\partial\log \mu}\right|_{\lambda=\lambda_\ast}\ ,
\end{aligned}
\end{align}
with $\widehat{Z}_1$, $Z_{2p}$ and $Z_{2p+1}$ being wave-function renormalizations defined through the relations:
\begin{align}\label{eq:Neumann wave-function renormalizations def}
\widehat{\Phi}_{1}^{\,\alpha}=\widehat{Z}_1\cdot \widehat{W}_{1}^{\,\alpha}\ ,\qquad \widehat{\Phi}_{2p}=\widehat{Z}_{2p} \cdot \widehat{W}_{2p}\ ,\qquad \widehat{\Phi}_{2p+1}^{\,\alpha}=\widehat{Z}_{2p+1}\cdot \widehat{W}_{2p+1}^{\,\alpha}\ ,
\end{align}

The conformal dimensions of $\widehat{W}_{1}^{\,\alpha}$ and $\widehat{W}_{2}$ have been computed from the standard perturbative calculations (see e.g., \cite{McAvity:1993ue,McAvity:1995zd,Reeve1980}):
\begin{align}
    \widehat{\Delta}_1&=1-\frac{N+5}{N+8}\,\epsilon+O(\epsilon^2)\ ,\label{eq:Neumann boundary lowest pert}\\
    \widehat{\Delta}_2&=2-\frac{6}{N+8}\,\epsilon+O(\epsilon^2)\ .\label{eq:Neumann boundary second lowest pert}
\end{align}
Also, the conformal dimension of $\widehat{W}_{1}^{\,\alpha}$ has already derived from the axiomatic perspective in \cite{Dey:2020jlc,Giombi:2020rmc}.

We first perform a detailed analysis of the free theory $(\lambda_0=0)$ in $d$-dimensions (section \ref{sec:Neumann free}). Using the results, we investigate the boundary operator spectrum in the standard perturbative framework (section \ref{sec:Neumann diagrammatic}) and the axiomatic framework (section \ref{sec:Neumann Axiomatic approach}) to confirm the consistency of these two approaches. Throughout this chapter, for convenience, we redefine the O$(N)$ vector field in the following manner:
\begin{align}\label{eq:BCFT bulk field normalization}
   \left.\Phi_{1}^{\alpha}\right|_{\text{new}}=\left(\frac{2\,(d-2)\,\pi^{d/2}}{\Gamma(d/2)}\right)^{1/2}\cdot \left.\Phi_{1}^{\alpha}\right|_{\text{old}}\ .
\end{align}
Notice that this expression tends to $ \left.\Phi_{1}^{\alpha}\right|_{\text{new}}=2\pi \cdot \left.\Phi_{1}^{\alpha}\right|_{\text{old}}$ in four dimensions, which is the same as the standard normalization to work in the axiomatic framework \eqref{eq:On vector redefinition}. Under this redefinition \eqref{eq:BCFT bulk field normalization}, the two-point function of $\Phi_{1}^{\alpha}$ behaves as follows:
\begin{align}
   \langle\,\Phi_{1}^\alpha(x_1)\,\Phi_{1}^\beta(x_2)\,\rangle
      \xrightarrow[]{}
        \frac{\delta^{\alpha\beta}}{|x_1-x_2|^{d-2}}+\cdots \qquad \text{as}\quad x_1\to x_2\ ,
\end{align}
and all equations simplify a great deal.

\section{Structure of free O$(N)$ model with Neumann boundary}\label{sec:Neumann free}
In this section, we work out the free O$(N)$ model with the Neumann boundary condition. In section \ref{sec:Correlation functions in free theory BCFT Neumann}, we calculate various correlators of the model in $d$-dimensions. We set the spacetime dimensions to be generic in studying correlation functions because they are necessary for the standard perturbative calculations in section \ref{sec:Neumann diagrammatic}.\footnote{See figure \ref{fig:perturbative vs axiomatic} to recall the difference between the standard perturbative approach and the axiomatic approach.} In section \ref{eq:Boundary-to-defect operator product expansions free Neumann}, similarly to the line defect case, we explore BOEs in four dimensions using correlation functions derived in section \ref{sec:Correlation functions in free theory BCFT Neumann}, setting the stage for the axiomatic approach in section \ref{sec:Neumann Axiomatic approach}.

\subsection{Correlation functions}\label{sec:Correlation functions in free theory BCFT Neumann}
Under the redefinition of the O$(N)$ vector field \eqref{eq:BCFT bulk field normalization}, the differential equation for the Green's function of $\Phi_{1}^{\alpha}$ reads:
\begin{align}\label{eq:4d free scalar propagator EoM}
    \Box_{x_1}\,\langle\,\Phi_{1}^\alpha(x_1)\,\Phi_{1}^\beta(x_2)\,\rangle
        =
            \frac{4\pi^{d/2}}{\Gamma(d/2)}\,\delta^{\alpha\beta}\,\delta^d(x_1-x_2)\ .
\end{align}
This equation can be solved under Neumann boundary condition $\lim_{x_\perp\to0}\,\frac{\partial}{\partial x_\perp}\,\Phi_1^\alpha(x)=0$ to give the following two-point function of $\Phi_{1}^{\alpha}$:
\begin{align}\label{eq:4d free scalar propagator Neumann}
    \langle\,\Phi_{1}^\alpha(x_1)\,\Phi_{1}^\beta(x_2)\,\rangle
        =
        \frac{\delta^{\alpha\beta}}{|x_1-x_2|^{d-2}}+ \frac{\delta^{\alpha\beta}}{|x_1-\bar{x}_2|^{d-2}}\ ,\qquad 
    \bar{x}^\mu
        =
            (\hat{x}^a,-x_\perp)\ .
\end{align}
One can derive the bulk one-point functions of composite operators such as $ \langle\,\Phi_{1}^\alpha\Phi_{1}^\beta\,\rangle$ and $\langle\,|\Phi_1|^{2}\,\rangle$ from \eqref{eq:4d free scalar propagator Neumann} by taking the coincident limit in \eqref{eq:4d free scalar propagator Neumann}:
\begin{align}\label{eq:N bulk 1pt}
  \langle\,\Phi_{1}^\alpha\Phi_{1}^\beta(x)\,\rangle
    =
        \frac{\delta^{\alpha\beta}}{2^{d-2}\,|x_\perp|^{d-2}}\ ,\qquad   \langle\,|\Phi_1|^{2}(x)\,\rangle
    =
        \frac{N}{2^{d-2}\,|x_\perp|^{d-2}}\ .
\end{align}
We then take either one or both bulk operators in \eqref{eq:4d free scalar propagator Neumann} to the boundary to find the two-point functions involving the defect local operator $\widehat{\Phi}_{1}^{\,\alpha}$ defined in \eqref{eq:Neumann lowest-lying defect local def}:
\begin{align}\label{eq:2pt Phi Neu}
    \langle\,\Phi_1^\alpha(x)\,\widehat{\Phi}_{1}^{\,\beta}(\hat{y})\,\rangle
        =
            \frac{2\, \delta^{\alpha\beta}}{|x-\hat{y}|^{d-2}} \ , \qquad
       \langle\,\widehat{\Phi}_{1}^{\,\alpha}(\hat{y}_1)\,\widehat{\Phi}_{1}^{\,\beta}(\hat{y}_2)\,\rangle
        =
            \frac{2\, \delta^{\alpha\beta}}{|\hat{y}_{12}|^{d-2}} \ .
\end{align}
In the same way as section \ref{sec:Correlation functions in free theory DCFT}, we can derive all other correlators needed for our analysis by utilizing Wick's theorem. First of all, the bulk-defect two-point functions involving $\Phi_3^\alpha$ take the following forms:
\begin{align}\label{eq:Neu btb 2pt Phi3}
      \langle\,\Phi_3^\alpha(x)\,\widehat{\Phi}_{1}^{\,\beta}(\hat{y})\,\rangle
        =
            \frac{(N/2+1)\,\delta^{\alpha\beta}}{2^{d-4}\,|x-\hat{y}|^{d-2}\,|x_\perp|^{d-2}}\ ,\qquad
            \langle\,\Phi_3^\alpha(x)\,\widehat{\Phi}_{3}^{\,\beta}(\hat{y})\,\rangle
                =
                \frac{  32\,(N/2+1)\,\delta^{\alpha\beta}}{|x-\hat{y}|^{3(d-2)}}\ .
\end{align}
The boundary two-point functions of composite operators turn out to be:
\begin{align}
                           \langle\, \widehat{\Phi}_{2p}(\hat{y}_1)\,\widehat{\Phi}_{2p}(\hat{y}_2) \,\rangle
                       & =
                        \frac{N\, g_{p-1}}{|\hat{y}_{12}|^{2p(d-2)}}\ ,\label{eq:Boundary two-point functions Neumann 1} \\
   \langle\, \widehat{\Phi}_{2p+1}^{\,\alpha}(\hat{y}_1)\,\widehat{\Phi}_{2p+1}^{\,\beta}(\hat{y}_2) \, \rangle
                        &=
                        \frac{f_p\, \delta^{\alpha\beta}}{|\hat{y}_{12}|^{(2p+1)(d-2)}}\ .   \label{eq:Boundary two-point functions Neumann 2}      
\end{align}
One can derive these two correlators by taking Wick's contraction once and making recursion relations between them.
We have introduced two combinatorial factors $f_p$ and $g_p$:
\begin{align}\label{eq:Neumann 3pf coeff f g}
    f_p=2^{4p+1}\,p!\, (N/2+1)_p\ ,\qquad g_p=2^{4p+3}\,(p+1)!\, (N/2+1)_p\ ,
\end{align}
which also appear in the boundary three-point functions:
    \begin{align}
     \langle\,\widehat{\Phi}_1^\alpha(\hat{x})\, \widehat{\Phi}_{2p}(\hat{y}_1)\,\widehat{\Phi}_{2p+1}^{\,\beta}(\hat{y}_2)\,\rangle&=\frac{f_p\, \delta^{\alpha\beta}}{|\hat{x}-\hat{y}_2|^{d-2}\,|\hat{y}_{12}|^{2p(d-2)}}\ ,\label{eq:3pt ddd Neumann 1}\\
     \langle\,\widehat{\Phi}_1^\alpha(\hat{x})\, \widehat{\Phi}_{2p+1}^{\beta}(\hat{y}_1)\,\widehat{\Phi}_{2p+2}(\hat{y}_2)\,\rangle
    &=   \frac{g_p\, \delta^{\alpha\beta}}{|\hat{x}-\hat{y}_2|^{d-2}\,|\hat{y}_{12}|^{(2p+1)(d-2)}}\ ,\label{eq:3pt ddd Neumann 2}\\
       \langle\,\widehat{\Phi}_3^\alpha(\hat{x})\, \widehat{\Phi}_{2p}(\hat{y}_1)\,\widehat{\Phi}_{2p+1}^{\,\beta}(\hat{y}_2)\,\rangle&=\frac{ 12p\,f_p\, \delta^{\alpha\beta}}{|\hat{x}-\hat{y}_1|^{d-2}\,|\hat{x}-\hat{y}_2|^{2(d-2)}\,|\hat{y}_{12}|^{(2p-1)(d-2)}} \label{eq:3pt ddd Neumann 3} \ ,\\
     \langle\,\widehat{\Phi}_3^\alpha(\hat{x})\, \widehat{\Phi}_{2p+1}^{\,\beta}(\hat{y}_1)\,\widehat{\Phi}_{2p+2}(\hat{y}_2)\,\rangle&=\frac{2\,(N+6p+2)\,g_p\, \delta^{\alpha\beta}}{|\hat{x}-\hat{y}_1|^{d-2}\,|x-\hat{y}_2|^{2(d-2)}\,|\hat{y}_{12}|^{2p(d-2)}}\ .\label{eq:3pt ddd Neumann 4}
\end{align}
It is straightforward to compute boundary three-point functions given the results for two-point functions \eqref{eq:Boundary two-point functions Neumann 1} and \eqref{eq:Boundary two-point functions Neumann 2}.
On the other hand, it requires a little care for the bulk-boundary-boundary three-point functions, as we have to bear in mind the non-vanishing bulk one-point functions \eqref{eq:N bulk 1pt}. The results are:
\begin{align}
    \langle\,\Phi_1^\alpha(x)\, \widehat{\Phi}_{2p}(\hat{y}_1)\,\widehat{\Phi}_{2p+1}^{\,\beta}(\hat{y}_2)\,\rangle
        &= 
            \frac{f_p\, \delta^{\alpha\beta}}{|x-\hat{y}_2|^{d-2}\,|\hat{y}_{12}|^{2p(d-2)}}\ ,\label{eq:3pt bdd Neumann 1}\\
    \langle\,\Phi_1^\alpha(x)\, \widehat{\Phi}_{2p+1}^{\beta}(\hat{y}_1)\,\widehat{\Phi}_{2p+2}(\hat{y}_2)\,\rangle
        &=  
            \frac{g_p\, \delta^{\alpha\beta}}{|x-\hat{y}_2|^{d-2}\,|\hat{y}_{12}|^{(2p+1)(d-2)}}\ ,\label{eq:3pt bdd Neumann 2}
    \end{align}
    and
     \begin{align}
    &\langle\,\Phi_3^\alpha(x)\, \widehat{\Phi}_{2p}(\hat{y}_1)\,\widehat{\Phi}_{2p+1}^{\,\beta}(\hat{y}_2)\,\rangle
         \nonumber\\ 
         & \quad 
         = 
                \frac{N+2}{2^{d-2}}\cdot \frac{f_p\, \delta^{\alpha\beta}}{|x-\hat{y}_2|^{d-2}\,|\hat{y}_{12}|^{2p(d-2)}\,|x_\perp|^{d-2}}
                 +
                 \frac{12\,p\, f_p\,\delta^{\alpha\beta}}{|x-\hat{y}_1|^{d-2}\,|x-\hat{y}_2|^{2(d-2)}\,|\hat{y}_{12}|^{(2p-1)(d-2)}}
                   \ , \label{eq:3pt bdd Neumann 3}\\
    &\langle\,\Phi_3^\alpha(x)\, \widehat{\Phi}_{2p+1}^{\,\beta}(\hat{y}_1)\,\widehat{\Phi}_{2p+2}(\hat{y}_2)\,\rangle
         \nonumber \\
        & \quad 
        =\frac{N+2}{2^{d-2}}\cdot
            \frac{g_p\, \delta^{\alpha\beta}}{|x-\hat{y}_2|^{d-2}\,|\hat{y}_{12}|^{(2p+1)(d-2)}\,|x_\perp|^{d-2}}
             +
          \frac{2\,(N+6p+2)\,g_p\, \delta^{\alpha\beta}}{|x-\hat{y}_1|^{d-2}\,|x-\hat{y}_2|^{2(d-2)}\,|\hat{y}_{12}|^{2p(d-2)                               }}\ .
                \label{eq:3pt bdd Neumann 4}
\end{align}

\subsection{Boundary operator expansions in four dimensions}\label{eq:Boundary-to-defect operator product expansions free Neumann}
In the same spirit as section \ref{eq:Bulk-to-defect operator product expansions free}, we here make use of the results we have derived so far to write down the BOEs of $\Phi_1^\alpha$ and $\Phi_3^\alpha$ in four dimensions.

\paragraph{Boundary operator expansion of $\Phi_1^\alpha$.}
According to the analysis in section \ref{sec:Defect Operator Expansion spectrum of free scalar field}, the only boundary local primaries allowed to be in the BOE of $\Phi_1^\alpha$ are $\widehat{\Phi}_{1}^{\,\alpha}(\hat{x})=\lim_{|x_\perp|\to0}\,\Phi_{1}^{\,\alpha}(x)$ and $  \widehat{\Psi}_{2}^{\,\alpha}(\hat{x})=\lim_{|x_\perp|\to0}\,|x_\perp|^{-1}\,\Phi_{1}^{\,\alpha}(x)$ having conformal dimensions one and two respectively. As the latter does not survive under the Neumann boundary condition, the BOE of $\Phi_1^\alpha$ takes the form:
\begin{align}\label{eq:BOE of Phi1 Neumann}
    \Phi_1^\alpha(x) =\widehat{\Phi}_{1}^{\,\alpha}(\hat{x})+(\text{descendants.})\ .
\end{align}
Here, we have matched the BOE coefficient from the bulk-defect and defect two-point functions \eqref{eq:2pt Phi Neu}.

\paragraph{Boundary operator expansion of $\Phi_3^\alpha$.}
Similarly to the line defect case, there exists an infinite number of defect local operators in the BOE of $\Phi_3^\alpha$:
\begin{align}\label{eq:btb OPE of phi3 all order}
  \Phi_3^{\alpha}(x)
    =
        \frac{N+2}{4\,|x_\perp|^2}\cdot \widehat{\Phi}_1^{\,\alpha}(\hat{x}) +\sum_{n=0}^{\infty}\,\frac{b(\Phi_3^\alpha,\widehat{\mathsf{P}}_{2n+3}^\alpha)}{c(\widehat{\mathsf{P}}_{2n+3}^\alpha,\widehat{\mathsf{P}}_{2n+3}^\alpha)}\cdot |x_\perp|^{2n}\cdot \widehat{\mathsf{P}}_{2n+3}^\alpha(\hat{x})+(\text{descendants.})\ .
\end{align}
The BOE coefficients $b(\Phi_3^\alpha,\widehat{\mathsf{P}}_{2n+3}^\alpha)/c(\widehat{\mathsf{P}}_{2n+3}^\alpha,\widehat{\mathsf{P}}_{2n+3}^\alpha)$ satisfy the following relations:
\begin{align}    \frac{b(\Phi_3^\alpha,\widehat{\mathsf{P}}_{2n+3}^\alpha)\,c(\widehat{\mathsf{P}}_{2n+3}^\alpha,\widehat{\Phi}_{2p},\widehat{\Phi}_{2p+1}^{\,\beta})}{c(\widehat{\mathsf{P}}_{2n+3}^\alpha,\widehat{\mathsf{P}}_{2n+3}^\alpha)}
        &=
            12\,p\,f_p\,\delta^{\alpha\beta} \cdot \frac{(-1)^n\,(n+1)!}{(n+3/2)_n}\ ,\label{eq:b/c 2n+3 Neumann 1}\\
            \frac{b(\Phi_3^\alpha,\widehat{\mathsf{P}}_{2n+3}^\alpha)\,c(\widehat{\mathsf{P}}_{2n+3}^\alpha,\widehat{\Phi}_{2p+1}^{\beta},\widehat{\Phi}_{2p+2})}{c(\widehat{\mathsf{P}}_{2n+3}^\alpha,\widehat{\mathsf{P}}_{2n+3}^\alpha)}
       & =
            2\,(N+6p+2)\,g_p\,\delta^{\alpha\beta}\cdot\frac{(-1)^n\,(n+1)!}{(n+3/2)_n}\ . \label{eq:b/c 2n+3 Neumann 2}
\end{align}
A few comments on this BOE \eqref{eq:btb OPE of phi3 all order} are in order. The first term is manifest from the two-point functions involving $\widehat{\Phi}_1^{\,\alpha}$ and $\widehat{\mathsf{P}}_{3}^\alpha$ is nothing but $\widehat{\Phi}_3^{\,\alpha}$ (see \eqref{eq:Neu btb 2pt Phi3}). The higher-order operator $\widehat{\mathsf{P}}_{2n+3}^\alpha$ ($n=1,2,\cdots$) having odd conformal dimension $\widehat{\Delta}(\widehat{\mathsf{P}}_{2n+3}^\alpha)=2n+3$ is again made up of three $\widehat{\Phi}_1^{\,\alpha}$'s and $2n$ parallel derivatives $\hat{\partial}^a$. One can confirm these facts by performing the conformal block expansions of bulk-boundary-boundary three-point functions (equation \eqref{eq:3pt bdd Neumann 3} and \eqref{eq:3pt bdd Neumann 4}):
\begin{align}\label{eq:3pt bdd Neumann 3 exp}
\begin{aligned}
     \langle\,&\Phi_3^\alpha(x)\, \widehat{\Phi}_{2p}(0)\,\widehat{\Phi}_{2p+1}^{\,\beta}(\infty)\,\rangle 
        \\
        &=   \frac{N+2}{4}\cdot \frac{f_p\, \delta^{\alpha\beta}}{|x_\perp|^{2}}
                 +
                 \frac{12\,p\, f_p\,\delta^{\alpha\beta}}{|x|^2}\\
            &=f_p\, \delta^{\alpha\beta}\cdot \frac{|x|}{|x_\perp|^3}\cdot\left[\frac{N+2}{4}\cdot G^{-1}_{1}(\upsilon)+ 12\,p\cdot \sum_{n=0}^{\infty}\,\frac{(-1)^n\,(n+1)!}{(n+3/2)_n}\cdot G^{-1}_{2n+3}(\upsilon)\right]   \ ,
\end{aligned}
\end{align}
and
\begin{align}\label{eq:3pt bdd Neumann 4 exp}
\begin{aligned}
    \langle\,&\Phi_3^\alpha(x)\, \widehat{\Phi}_{2p+1}^{\beta}(0)\,\widehat{\Phi}_{2p+2}(\infty)\,\rangle
        \\
        &= \frac{N+2}{4}\cdot
            \frac{g_p\, \delta^{\alpha\beta}}{|x_\perp|^2}
             +
          \frac{2\,(N+6p+2)\,g_p\, \delta^{\alpha\beta}}{|x|^2}\\
        &=g_p\,\delta^{\alpha\beta}\cdot\frac{|x|}{|x_\perp|^3}
            \cdot\left[ \frac{N+2}{4}\cdot G^{-1}_{1}(\upsilon)\right.\\
            &\qquad\qquad\qquad\qquad\qquad\left.+2\,(N+6p+2)\cdot\sum_{n=0}^{\infty}\,\frac{(-1)^n\,(n+1)!}{(n+3/2)_n}\cdot G^{-1}_{2n+3}(\upsilon)\right]\ ,
\end{aligned}
\end{align}
with the cross ratio $\upsilon=|x_\perp|^2/|x|^2$.
To reach the last line both in \eqref{eq:3pt bdd Neumann 3 exp} and \eqref{eq:3pt bdd Neumann 4 exp}, we used the hypergeometric identity \eqref{eq:hypergeometric identity 1} and the explicit forms of conformal blocks in the presence of a boundary ($p=d-1=3-\epsilon$) given in \eqref{eq:conformal block expansion main}, namely:
\begin{align}
    G^{-1}_{1}(\upsilon)=\upsilon^{1/2}\ ,\qquad G_{2n+3}^{-1}(\upsilon)=\upsilon^{n+3/2}\,{}_2F_1(1+n,2+n;5/2+2n;\upsilon)\ .
\end{align}
We compare these two conformal block expansions \eqref{eq:3pt bdd Neumann 3 exp} and \eqref{eq:3pt bdd Neumann 4 exp} with the general expression \eqref{eq:conformal block expansion main} to obtain the relations for BOE coefficients \eqref{eq:b/c 2n+3 Neumann 1} and \eqref{eq:b/c 2n+3 Neumann 2}. Lastly, we would like to emphasize that there are no further contributions to the BOE of $\Phi_3^\alpha$, and the expression \eqref{eq:btb OPE of phi3 all order} is exact.

\section{Standard perturbative approach}\label{sec:Neumann diagrammatic}
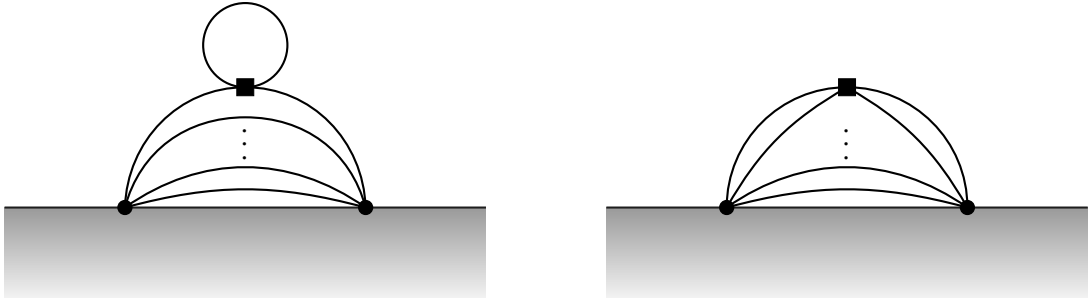
\begin{figure}[t]
    \centering
       \begin{tikzpicture}[transform shape,scale=0.8]
        \begin{scope}[shift={(-5,0)}]        
        \fill[top color=gray!80,bottom color=gray!10] (-4,0) rectangle (4,-1.5);

            \draw[thick, black!100,opacity=0.8] (-4,0)  -- (4,0);              
            \coordinate (A) at (-2,0) {};
            \coordinate (B) at (2,0) {};
            \coordinate (C) at (0,2) {};  
            \coordinate (D) at (0,1.1) {};
                  
            \filldraw[black,very thick] (A) circle (0.1);
                                           
            \filldraw[black,very thick] (B) circle (0.1);                                        
            \node[rectangle,draw,fill] (r) at (C) {};
            
            \draw[black!100,thick] (A) to[out=15,in=165] (B) ;
            \draw[black!100,thick] (A) to[out=35,in=145] (B) ;
            
            \draw[black!100,thick] (A) to[out=75,in=180] (0,1.5) to[out=0,in=105]  (B) ;
            \draw[black!100,thick] (A) to[out=88,in=180] (C) to[out=0,in=92]  (B) ;

            \draw[thick] (C) arc[
                        start angle=-90,
                        end angle=270,
                        x radius=0.7cm,
                        y radius=0.7cm
                    ] ;

            \node[Black!100,font=\large,rotate around={90:(0,0)}] at (D) {$\cdots$};
      
        \end{scope}

        \begin{scope}[shift={(5,0)}]          
        \fill[top color=gray!80,bottom color=gray!10] (-4,0) rectangle (4,-1.5);
            \draw[thick, black!100,opacity=0.8] (-4,0)  -- (4,0);    
                        
            \coordinate (A) at (-2,0) {};
            \coordinate (B) at (2,0) {};
            \coordinate (C) at (0,2) {};  
            \coordinate (D) at (0,1.1) {};  
            \coordinate (E) at (0,2) {};             
                  
            \filldraw[black,very thick] (A) circle (0.1);
                                       
            \filldraw[black,very thick] (B) circle (0.1);        
        
            \node[rectangle,draw,fill] (r) at (C) {};

            \draw[black!100,thick] (A) to[out=15,in=165] (B) ;
            \draw[black!100,thick] (A) to[out=35,in=145] (B) ;
            \draw[black!100,thick] (A) to[out=90,in=180] (C) to[out=0,in=90]  (B) ;

            \draw[black!100,thick] (C) to[out=-30,in=120] (B);
            \draw[black!100,thick] (C) to[out=210,in=60] (A);
            
            \node[Black!100,font=\large,rotate around={90:(0,0)}] at (D) {$\cdots$};
        
        \end{scope}

     \end{tikzpicture} 
\caption{We illustrated the first-order perturbative contributions to the two-point functions of the composite operators $\widehat{W}_{2p}$, $\widehat{W}_{2p+1}^{\, \alpha}$. Black circles ($\bullet$) represent boundary composite operators, whereas black squares ({\tiny $\blacksquare$}) stand for the bulk quartic interaction vertex.}
\label{Fig:Quantum correction}
\end{figure}

We now perform standard perturbative calculations to derive leading anomalous dimensions of boundary local operators $\widehat{\Phi}_{2p},\widehat{\Phi}_{2p+1}^{\,\alpha}$ $(p=0,1,\cdots)$. In this section, We work in the minimal subtraction scheme up to one-loop level (see figure \ref{Fig:Quantum correction}).\footnote{For $p=0$, $\widehat{\Phi}_{2p}$ and $\widehat{\Phi}_{2p+1}^{\,\alpha}$ reduces to the identity operator $\bm{1}$ and $\widehat{\Phi}_{1}^{\,\alpha}$, respectively.}
Before getting into perturbative calculations, we introduce several notations and lay the groundwork to make the main calculations transparent.
Denoting the bare correlators by:
\begin{align}
    I_{2p}=\langle\,\widehat{\Phi}_{2p}(\hat{y})\,\widehat{\Phi}_{2p}(0)\,\rangle\ ,\qquad \delta^{\alpha\beta}\,I_{2p+1}=\langle\,\widehat{\Phi}_{2p+1}^{\,\alpha}(\hat{y})\,\widehat{\Phi}_{2p+1}^{\,\beta}(0)\,\rangle\ ,
\end{align}
we expand them in inverse powers of $\epsilon$ together with wave-function renormalizations (see \eqref{eq:Neumann wave-function renormalizations def} for their definitions):
\begin{align}\label{eq:Neumann I exp}
I_{2p}&=I_{2p,0}+\delta I_{2p}\ ,\qquad \delta I_{2p}=\frac{\delta_1 I_{2p}}{\epsilon}+\frac{\delta_2 I_{2p}}{\epsilon^2}+O\left(\frac{1}{\epsilon^3}\right)\ ,\\
I_{2p+1}&=I_{2p+1,0}+\delta I_{2p+1}\ ,\qquad \delta I_{2p+1}=\frac{\delta_1 I_{2p+1}}{\epsilon}+\frac{\delta_2 I_{2p+1}}{\epsilon^2}+O\left(\frac{1}{\epsilon^3}\right)\ ,
\end{align}
and 
\begin{align}\label{eq:Neumann Z exp}
\widehat{Z}_{2p}&=1+\delta \widehat{Z}_{2p}\ ,\qquad \delta \widehat{Z}_{2p}=\frac{\delta_1 \widehat{Z}_{2p}}{\epsilon}+\frac{\delta_2 \widehat{Z}_{2p}}{\epsilon^2}+O\left(\frac{1}{\epsilon^3}\right)\ ,\\
\widehat{Z}_{2p+1}&=1+\delta\widehat{Z}_{2p+1}\ ,\qquad \delta\widehat{Z}_{2p+1} =\frac{\delta_1 \widehat{Z}_{2p+1}}{\epsilon}+\frac{\delta_2 \widehat{Z}_{2p+1}}{\epsilon^2}+O\left(\frac{1}{\epsilon^3}\right)\ .
\end{align}
Here, the two symbols $I_{2p,0}$ and $I_{2p+1,0}$ stand for free theory two-point functions in the presence of a boundary:\footnote{The explicit forms of these two correlators are taken from \eqref{eq:Boundary two-point functions Neumann 1} and \eqref{eq:Boundary two-point functions Neumann 2}.}
\begin{align}
    I_{2p,0}&=\langle\,\widehat{\Phi}_{2p}(\hat{y})\,\widehat{\Phi}_{2p}(0)\,\rangle_{\text{free}}=\frac{N\, g_{p-1}}{|\hat{y}|^{2p(d-2)}}\ ,\label{eq:I2p,0 explicit form}\\
     \delta^{\alpha\beta}\cdot I_{2p+1,0}&=\langle\,\widehat{\Phi}_{2p+1}^{\,\alpha}(\hat{y})\,\widehat{\Phi}_{2p+1}^{\,\beta}(0)\,\rangle_{\text{free}}=\frac{f_p\, \delta^{\alpha\beta}}{|\hat{y}|^{(2p+1)(d-2)}}\ .
\end{align}
It is worthwhile recording their recursion relations:
\begin{align}\label{eq:2pt recursion Neumann}
    I_{2p+1,0}=\frac{2\,(N+2p)}{N}\cdot\frac{I_{2p,0}}{|\hat{y}|^{d-2}}\ ,\qquad I_{2p,0}=4pN \cdot\frac{I_{2p-1,0}}{|\hat{y}|^{d-2}}\ .
\end{align}

Under the standard perturbative framework in the minimal subtraction scheme, one can compute the conformal/anomalous dimensions of the operators from their wave-function renormalizations through the relations \eqref{eq:Neumann boundary composite wfr}.
The wave functions renormalizations are to be determined to cancel the divergences of two-point functions of renormalized operators:
 \begin{align}\label{eq:Neumann W and Phi}
\begin{aligned}
        \langle\,\widehat{W}_{2p}(\hat{y})\,\widehat{W}_{2p}(0)\,\rangle&=\widehat{Z}_{2p}^{-2}\cdot\langle\,\widehat{\Phi}_{2p}(\hat{y})\,\widehat{\Phi}_{2p}(0)\,\rangle\ ,\\
            \langle\,\widehat{W}_{2p+1}^{\,\alpha}(\hat{y})\,\widehat{W}_{2p+1}^{\,\beta}(0)\,\rangle&=\widehat{Z}_{2p+1}^{-2}\cdot\langle\,\widehat{\Phi}_{2p+1}^{\,\alpha}(\hat{y})\,\widehat{\Phi}_{2p+1}^{\,\beta}(0)\,\rangle\ .
\end{aligned}
\end{align}
Plugging \eqref{eq:Neumann I exp} and \eqref{eq:Neumann Z exp} into \eqref{eq:Neumann W and Phi}, one finds that:
 \begin{align}\label{eq:Neumann W and Phi ast}
\begin{aligned}
        \langle\,\widehat{W}_{2p}(\hat{y})\,\widehat{W}_{2p}(0)\,\rangle&=I_{2p,0}+\frac{\delta_1 I_{2p}-2\,I_{2p,0}\cdot\delta_1 Z_{2p}}{\epsilon}+O\left(\frac{1}{\epsilon^2}\right)\ ,\\
            \langle\,\widehat{W}_{2p+1}^{\,\alpha}(\hat{y})\,\widehat{W}_{2p+1}^{\,\beta}(0)\,\rangle&=I_{2p+1,0}+\frac{\delta_1 I_{2p+1}-2\,I_{2p+1,0}\cdot\delta_1 Z_{2p+1}}{\epsilon}+O\left(\frac{1}{\epsilon^2}\right)\ .
\end{aligned}
\end{align}
To cancel the divergences in the right-hand side of the above equations, we have:
\begin{align}\label{eq:relation Z and I Neumann}
  \delta_1 Z_{2p}=\frac{\delta_1 I_{2p}}{2\,I_{2p,0}}  \ ,\qquad   \delta_1 Z_{2p+1}=\frac{\delta_1 I_{2p+1}}{2\,I_{2p+1,0}} \ .
\end{align}

Let us now tackle perturbative calculations.
First of all, as $\widehat{\Phi}_{0}$ stands for the identity operator $\bm{1}$, it receives no perturbative corrections at any orders in perturbation theory if appropriately renormalized:
\begin{align}\label{eq:Neumann Z initial condition}
    Z_0=1 \ ,\qquad \delta I_0=0 \qquad (\text{exact})\ .
\end{align}
This fact is also clear from the relation between the correlators in free and perturbation theory:
\begin{align}
    \langle\,\cdots\,\rangle=\frac{\left\langle\,\cdots\,\exp\left(-  \frac{\lambda'_0}{4!}\,\int_{\mathbb{R}_{+}^{d}}\d^d x \,  |\Phi_1|^4\right)\,\right\rangle_{\text{free}}}{\left\langle\,\exp\left(-  \frac{\lambda'_0}{4!}\,\int_{\mathbb{R}_{+}^{d}}\d^d x \,  |\Phi_1|^4\right)\,\right\rangle_{\text{free}}}\ ,
\end{align}
where we redefined the coupling constant correspondingly to the change of the normalization of the O$(N)$ vector field \eqref{eq:BCFT bulk field normalization}:
\begin{align}\label{eq:change of normalization lambda0}
    \lambda'_0=\left(\frac{\Gamma(d/2)}{2\,(d-2)\,\pi^{d/2}}\right)^{2}\cdot\lambda_0=\frac{\lambda_0}{(4\pi^2)^2}+O(\epsilon)\ .
\end{align}
Let us focus our attention to the perturbative corrections to $\delta I_{2p}$ at the first order in $\lambda_0$:
\begin{align}
\begin{aligned}
          \delta I_{2p}&=-  \frac{\lambda'_0}{4!}\,\int_{\mathbb{R}_{+}^{d}}\d^d x \,\langle\,|\Phi_1|^4(x)\,\widehat{\Phi}_{2p}(\hat{y})\,\widehat{\Phi}_{2p}(0)\,\rangle_{\text{free}}\\
     &\qquad\qquad\qquad+ I_{2p,0}\cdot \frac{\lambda'_0}{4!}\,\int_{\mathbb{R}_{+}^{d}}\d^d x \,  \langle\,|\Phi_1|^4(x)\,\rangle_{\text{free}}+O(\lambda_0^2)\ .
\end{aligned}
\end{align}
Applying Wick contraction once, we find that:
\begin{align}\label{eq:Neumann delta I 2p}
\begin{aligned}
          \delta I_{2p}&=-  \frac{\lambda'_0}{3!}\,\int_{\mathbb{R}_{+}^{d}}\d^d x \,\langle\,\Phi_{1}^{\,\alpha}(x)\,\widehat{\Phi}_{1}^{\beta}(\hat{y})\,\rangle_{\text{free}}\cdot \langle\,\Phi_3^{\alpha}(x)\,\widehat{\Phi}_{2p-1}^{\,\beta}(\hat{y})\,\widehat{\Phi}_{2p}(0)\,\rangle_{\text{free}}\\
          &\qquad+2p\cdot\langle\,\widehat{\Phi}_{1}^{\,\alpha}(\hat{y})\,\widehat{\Phi}_{1}^{\beta}(0)\,\rangle_{\text{free}}\cdot \left[-  \frac{\lambda'_0}{4!}\,\int_{\mathbb{R}_{+}^{d}}\d^d x \,\langle\,|\Phi_1|^4(x)\,\widehat{\Phi}_{2p-1}^{\,\alpha}(\hat{y})\,\widehat{\Phi}_{2p-1}^{\,\beta}(0)\,\rangle_{\text{free}}\right]\\
     &\qquad\qquad\qquad+ I_{2p,0}\cdot \frac{\lambda'_0}{4!}\,\int_{\mathbb{R}_{+}^{d}}\d^d x \,  \langle\,|\Phi_1|^4(x)\,\rangle_{\text{free}}+O(\lambda_0^2)\ ,
\end{aligned}
\end{align}
After utilizing the free theory results in the last section, one can perform the integrals in the first lines of \eqref{eq:Neumann delta I 2p} owing to the integration formula \eqref{eq:Integral over half-space} to give:
\begin{align}
\begin{aligned}
     &  (\text{First line of }\eqref{eq:Neumann delta I 2p})\\
&=-  \frac{g_{p-1}\,N\,\lambda'_0}{3!\,|\hat{y}|^{2p}}\,\int_{\mathbb{R}_{+}^{d}}\d^d x \,  \left[
            \frac{2^{3-d}\,(N+2)\,|\hat{y}|^{d-2}}{|x|^{d-2}\,|x_\perp|^{d-2}\,|x-\hat{y}|^{d-2}}
             +
          \frac{4\,(N+6p-4)\,|\hat{y}|^{2(d-2)}}{|x-\hat{y}|^{2(d-2)}\,|x|^{2(d-2)}}\right]\\
          &=-  \frac{N+8p-6}{\epsilon}\cdot \frac{\lambda}{(4\pi)^2}\cdot I_{2p,0}+O(\lambda^2,\epsilon^0)\ ,
\end{aligned}
\end{align}
where we have used \eqref{eq:change of normalization lambda0}, \eqref{eq:lambda expanding invers order} and \eqref{eq:I2p,0 explicit form} to reach the last line.
The parenthesis in second line of \eqref{eq:Neumann delta I 2p} can be associated with $\delta^{\alpha\beta}\cdot I_{2p-1,0}$ due to the relation:
\begin{align}
\begin{aligned}
    \delta^{\alpha\beta}\cdot \delta I_{2p-1}&=\left[-  \frac{\lambda'_0}{4!}\,\int_{\mathbb{R}_{+}^{d}}\d^d x \,\langle\,|\Phi_1|^4(x)\,\widehat{\Phi}_{2p-1}^{\,\alpha}(\hat{y})\,\widehat{\Phi}_{2p-1}^{\,\beta}(0)\,\rangle_{\text{free}}\right]\\
    &\qquad\qquad+\delta^{\alpha\beta}\cdot I_{2p-1,0}\cdot \frac{\lambda'_0}{4!}\,\int_{\mathbb{R}_{+}^{d}}\d^d x \,  \langle\,|\Phi_1|^4(x)\,\rangle_{\text{free}}+O(\lambda_0^2)\ .
\end{aligned}
\end{align}
Plugging this into the second line of \eqref{eq:Neumann delta I 2p}, one finds that:
\begin{align}\label{eq:Neumann delta I 2p ast}
\begin{aligned}
       (\text{Last two lines of }&\eqref{eq:Neumann delta I 2p})=\frac{4pN}{|\hat{y}|^{2(d-2)}}\cdot \delta I_{2p-1,0} \\
       &+\cancel{\left(I_{2p,0}-\frac{4pN}{|\hat{y}|^{2(d-2)}}\cdot I_{2p-1,0}\right)}\cdot \frac{\lambda'_0}{4!}\,\int_{\mathbb{R}_{+}^{d}}\d^d x \,  \langle\,|\Phi_1|^4(x)\,\rangle_{\text{free}}+O(\lambda_0^2)\ .
\end{aligned}
\end{align}
The second term in \eqref{eq:Neumann delta I 2p ast} cancels out due to \eqref{eq:2pt recursion Neumann}, resulting in the following recursion relation:
\begin{align}
     \delta I_{2p}=-  \frac{N+8p-6}{\epsilon}\cdot  \frac{\lambda}{(4\pi)^2}\cdot I_{2p,0}+ \frac{I_{2p,0}}{I_{2p-1,0}}\cdot\delta I_{2p-1,0}+O(\lambda^2,\epsilon^0)\ .
\end{align}
With the aid of \eqref{eq:relation Z and I Neumann}, one can rephrase this recursion relation in the language of the wave-function renormalizations:
\begin{align}\label{eq:Neumann Z recursion 1}
    \delta \widehat{Z}_{2p}-\delta \widehat{Z}_{2p-1}=-\frac{N+8p-6}{2\,\epsilon}\cdot\frac{\lambda}{(4\pi)^2}+O(\lambda^{2}, \epsilon^0)\ .
\end{align}
Similar analysis for $\delta I_{2p+1}$ leads to the following recursion relation:
\begin{align}\label{eq:Neumann Z recursion 2}
    \delta \widehat{Z}_{2p+1}-\delta \widehat{Z}_{2p}=-\frac{24p-N-2}{6\,\epsilon}\cdot \frac{\lambda}{(4\pi)^2} +O(\lambda^2, \epsilon^0)\, .
\end{align}
We can solve these two recursion relations \eqref{eq:Neumann Z recursion 1} and \eqref{eq:Neumann Z recursion 2} under the initial condition \eqref{eq:Neumann Z initial condition} to find that:
\begin{align}
   \delta \widehat{Z}_{2p}&=-\frac{p\, (N+12p-10)}{3\,\epsilon}\cdot\frac{\lambda}{(4\pi)^2} +O(\lambda^2)\ , \\
    \delta \widehat{Z}_{2p+1}&=-\frac{(2p-1)\,N+2\,(12p^2+2p-1)}{6\,\epsilon}\cdot\frac{\lambda}{(4\pi)^2} +O(\lambda^2) \ .
\end{align}
From these two expressions and the formula \eqref{eq:Neumann boundary composite wfr}, we obtain the leading anomalous dimensions of $\widehat{\Phi}_{2p}$ and $\widehat{\Phi}_{2p+1}^{\,\alpha}$:
\begin{align}\label{eq: Diagrammatic result, Neumann}
\begin{aligned}
        \widehat{\gamma}_{2p}&=\left.\frac{\partial\log \widehat{Z}_{r}}{\partial\log \mu}\right|_{\lambda=\lambda_\ast}=\frac{p\, (N+12p-10)}{N+8}\,\epsilon+O(\epsilon^2) \ ,\\
        \widehat{\gamma}_{2p+1}&=\left.\frac{\partial\log \widehat{Z}_{r}}{\partial\log \mu}\right|_{\lambda=\lambda_\ast}= \frac{(2p-1)\,N+2\,(12p^2+2p-1)}{2\,(N+8)}\,\epsilon+O(\epsilon^2)\ .    
\end{aligned}
\end{align}
And the conformal dimensions of renormalized operators $\widehat{W}_{2p}$ and $\widehat{W}_{2p+1}^{\,\alpha}$ are given by:
\begin{align}
\begin{aligned}
        \widehat{\Delta}_{2p}&=2p\cdot\frac{d-2}{2}+\widehat{\gamma}_{2p}=2p +\frac{6p\, (2p-3)}{N+8}\,\epsilon+O(\epsilon^2)\ ,\\
    \widehat{\Delta}_{2p+1}&=(2p+1)\cdot\frac{d-2}{2}+\widehat{\gamma}_{2p+1}=2p+1-\frac{N+6p\,(1-2p)+5}{N+8}\,\epsilon+O(\epsilon^2)\ .
\end{aligned}
\end{align}

\section{Axiomatic approach}\label{sec:Neumann Axiomatic approach}
We now take the axiomatic approach to study the critical O$(N)$ model in $d=4-\epsilon$ dimensions similarly to the line defect case. We start by deriving the conformal dimension of the lowest-lying boundary local operator $\widehat{W}_{1}^\alpha(\hat{x})$ focusing on the BOE of $W_1^\alpha$ (section \ref{sec:Neu lowest RT}) as a warm-up and proceed to the composite operators $\widehat{W}_{2p}$ and $\widehat{W}_{2p+1}^{\,\alpha}$ (section \ref{sec:Neu higher order}).

\subsection{Lowest-lying boundary local operator}\label{sec:Neu lowest RT}
Owing to the boundary conformal symmetry (axiom \ref{dcftaxiom1}), we have the following BOE of $W^\alpha_1$:
\begin{align}\label{eq:W1_bOPE_Neumann}
    W_1^\alpha(x)
        \supset
             D\cdot\frac{1}{|x_\perp|^{\Delta_1-\widehat{\Delta}_{1}}}\cdot\widehat{W}_{1}^\alpha(\hat{x}) \ ,
\end{align}
where the coefficient is given by $D = 1+O(\epsilon)$ as this expression should reduce to \eqref{eq:BOE of Phi1 Neumann} in taking $\epsilon\to0$ (axiom \ref{dcftaxiom2}). Employing the equation of motion \eqref{eq:classical EoM O(N)} (axiom \ref{dcftaxiom3}) with the help of \eqref{eq:action of Lap on spinning btd OPE}, one finds that:
\begin{align}
\begin{aligned}
     W_3^\alpha(x)
        &=
            \frac{1}{\kappa}\,\Box\,W_1^\alpha(x)\\
        &\supset
    \frac{D}{\kappa}\cdot \frac{(\Delta_1-\widehat{\Delta}_{1})(\Delta_1-\widehat{\Delta}_{1}+1)}{|x_\perp|^{\Delta_1-\widehat{\Delta}_{1}+2}}\cdot \widehat{W}_{1}^\alpha(\hat{x})\ .     
\end{aligned}
\end{align}
Making use of axiom \ref{dcftaxiom2} and comparing this expression with \eqref{eq:btb OPE of phi3 all order} in the limit $\epsilon\to0$, we obtain the following equation:
\begin{align}
    \frac{N+2}{4}
        =
            \frac{(\Delta_1 - \widehat{\Delta}_{1})(\Delta_1 - \widehat{\Delta}_{1}+1)}{\kappa} + O(\epsilon) \ .
\end{align}
Substituting bulk variables $\kappa$ \eqref{eq:value of kappa} and $\Delta_1$ \eqref{eq:Delta1 axiomatic} into this equation, one obtains the conformal dimension of the lowers-lying boundary local operator $\widehat{W}_{1}^\alpha$:
\begin{align}\label{eq:anomalous dim lowest Neu}
    \begin{aligned}
           \widehat{\Delta}_{1}
                &=
               \frac{d-2}{2}-\frac{N+2}{2\,(N+8)}\,\epsilon+O(\epsilon^2) \\
               &=1-\frac{N+5}{N+8}\,\epsilon+O(\epsilon^2)\ ,
    \end{aligned}
\end{align}
in agreement with the known perturbative result \eqref{eq:Neumann boundary lowest pert}.

\subsection{Boundary composite operators}\label{sec:Neu higher order}
We here compute conformal dimensions of boundary composite operators $\widehat{W}_{2p}$ and $\widehat{W}_{2p+1}^{\,\alpha}$ by taking the same steps as in section \ref{sec:defect composite operator}. We start by evaluating the DOE of $W_1^\alpha$ from the equation of motion \eqref{eq:classical EoM O(N)} combined with the free theory DOE of $\Phi_3^{\alpha}$ \eqref{eq:btb OPE of phi3 all order}. Armed with the DOE of $W_1^\alpha$, we exploit the generic forms of the conformal block expansions \eqref{eq:conformal block expansion main} and some hypergeometric identities to obtain the closed-form expressions of the bulk-boundary-boundary three-point functions involving $W_1^\alpha$ at order $\epsilon$. Requiring the removal of their unphysical singularities, we derive the conformal dimensions of a class of boundary composite operators.

\paragraph{Boundary operator expansion of $W_1^\alpha$.}
It turns out that the BOE of $W_1^\alpha$ takes the form:
\begin{align}\label{eq:OPE of W1 all order Neumann}
    \begin{aligned}
        W_1^\alpha(x)
            &=
                \frac{D}{|x_\perp|^{\Delta_1-\widehat{\Delta}_1}}\cdot \widehat{W}_1^{\,\alpha}(\hat{x})\\
            &\qquad
                +
   \sum_{n=0}^{\infty}\,\frac{b(W_1^\alpha,\widehat{\mathsf{P}}_{2n+3}^{\prime\,\alpha})/c(\widehat{\mathsf{P}}_{2n+3}^{\prime\,\alpha},\widehat{\mathsf{P}}_{2n+3}^{\prime\,\alpha})}{|x_\perp|^{\Delta_1-\widehat{\Delta}(\widehat{\mathsf{P}}_{2n+3}^{\prime\,\alpha})}}\cdot \widehat{\mathsf{P}}_{2n+3}^{\prime\,\alpha}(\hat{x})+(\text{descendants})\ .
    \end{aligned}
\end{align}
Here, the BOE coefficient $D=1+O(\epsilon)$ is the same as in \eqref{eq:W1_bOPE_Neumann}. As $\epsilon\to0$, the boundary primary $\widehat{\mathsf{P}}_{2n+3}^{\prime\,\alpha}$ ($n=0,1,\cdots$) turns into $\widehat{\mathsf{P}}_{2n+3}^{\,\alpha}$ with conformal dimension $(2n+3)$ that appears in the BOE of $\Phi_3^\alpha$ \eqref{eq:btb OPE of phi3 all order}, and its BOE coefficient fulfills the condition:
\begin{align}\label{eq:W1 to higher order OPE coeff Neumann}
    \frac{b(W_1^\alpha,\widehat{\mathsf{P}}_{2n+3}^{\prime\,\alpha})}{c(\widehat{\mathsf{P}}_{2n+3}^{\prime\,\alpha},\widehat{\mathsf{P}}_{2n+3}^{\prime\,\alpha})}
        =
            \frac{\epsilon}{(N+8)\,(n+1)(2n+1)}\cdot\frac{b(\Phi_3^\alpha,\widehat{\mathsf{P}}_{2n+3}^\alpha)}{c(\widehat{\mathsf{P}}_{2n+3}^\alpha,\widehat{\mathsf{P}}_{2n+3}^\alpha)}+O(\epsilon^2)\ .
\end{align}
One way to confirm these relations is to apply the equation of motion \eqref{eq:classical EoM O(N)} with the value of $\kappa$ given in \eqref{eq:value of kappa} to the BOE of $W_1^\alpha$ \eqref{eq:OPE of W1 all order Neumann}:
\begin{align}\label{eq:Neumann DOE epsion limit}
    \begin{aligned}
        W_3^\alpha(x)
            \supset
                &\frac{b(W_1^\alpha,\widehat{\mathsf{P}}_{2n+3}^{\prime\,\alpha})}{\kappa\cdot c(\widehat{\mathsf{P}}_{2n+3}^{\prime\,\alpha},\widehat{\mathsf{P}}_{2n+3}^{\prime\,\alpha})}\\
                &\qquad\cdot\frac{[\Delta_1-\widehat{\Delta}(\widehat{\mathsf{P}}_{2n+3}^{\prime\,\alpha})]\cdot [\Delta_1-\widehat{\Delta}(\widehat{\mathsf{P}}_{2n+3}^{\prime\,\alpha})+1]}{|x_\perp|^{\Delta_1-\widehat{\Delta}(\widehat{\mathsf{P}}_{2n+3}^{\prime\,\alpha})+2}}\cdot \widehat{\mathsf{P}}_{2n+3}^{\prime\,\alpha}(\hat{x})\\
                \xrightarrow[\epsilon\to0]{}&\frac{b(W_1^\alpha,\widehat{\mathsf{P}}_{2n+3}^{\prime\,\alpha})}{\epsilon\cdot  c(\widehat{\mathsf{P}}_{2n+3}^{\prime\,\alpha},\widehat{\mathsf{P}}_{2n+3}^{\prime\,\alpha})}\cdot (N+8)\,(n+1)(2n+1)\cdot |x_\perp|^{2n}\cdot \widehat{\mathsf{P}}_{2n+3}^{\,\alpha}(\hat{x})\ .
    \end{aligned}
\end{align}
Comparing the last line of \eqref{eq:Neumann DOE epsion limit} with \eqref{eq:btb OPE of phi3 all order}, we end up with \eqref{eq:W1 to higher order OPE coeff Neumann}.

\paragraph{Study of bulk-boundary-boundary three-point functions involving $W_1^\alpha$.}
Implementing the general formula for the conformal block expansion \eqref{eq:conformal block expansion main} combined with the BOE \eqref{eq:OPE of W1 all order Neumann}, we can express the bulk-boundary-boundary three-point function $\langle\, W_1^\alpha\, \widehat{W}_{2p}\,\widehat{W}_{2p+1}^{\,\beta}\,\rangle$ as follows:
\begin{align}\label{eq:Neumann bbb 3pt 1}
    \begin{aligned}
    \langle\,& W_1^\alpha(x)\, \widehat{W}_{2p}(0)\,\widehat{W}_{2p+1}^{\,\beta}(\infty)\,\rangle
        =
            \frac{1}{|x_\perp|^{\Delta_1}\,|x|^{\widehat{\Delta}_{2p}-\widehat{\Delta}_{2p+1}}}\\
        &\qquad\cdot
            \left[ D\cdot c(\widehat{W}_1^{\,\alpha},\widehat{W}_{2p},\widehat{W}_{2p+1}^{\,\beta})\cdot G^{\widehat{\Delta}_{2p}-\widehat{\Delta}_{2p+1}}_{\widehat{\Delta}_1}(\upsilon)\right.\\
        &\qquad\qquad\left.     
            +
            \sum_{n=0}^{\infty}\,\frac{b(W_1^\alpha,\widehat{\mathsf{O}}_{2n+3}^{\prime\,\alpha})\,c(\widehat{\mathsf{O}}_{2n+3}^{\prime\,\alpha},\widehat{W}_{2p+1}^{\,\beta},\widehat{W}_{2p+2})}{c(\widehat{\mathsf{O}}_{2n+3}^{\prime\,\alpha},\widehat{\mathsf{O}}_{2n+3}^{\prime\,\alpha})}\cdot G^{\widehat{\Delta}_{2p}-\widehat{\Delta}_{2p+1}}_{\widehat{\Delta}'_{2n+3}}(\upsilon)\right]\ ,
    \end{aligned}
\end{align}
with $\upsilon=|x_{\perp}|^2/|x|^2$ being the cross ratio.
We now expand the first and the second terms in the parenthesis in the following manner:\footnote{Use the series expansion of the Gauss's hypergeometric functions \eqref{eq:hypergeometric series expansion} in going from the first to the second line.}
\begin{align}
    \begin{aligned}
        G^{\widehat{\Delta}_{2p}-\widehat{\Delta}_{2p+1}}_{\widehat{\Delta}_1}(\upsilon)&=\upsilon^{\widehat{\Delta}_1/2}\cdot {}_2F_1\left(\frac{\widehat\gamma_{1,1}+\widehat\gamma_{2p,1}-\widehat\gamma_{2p+1,1}}{2}\,\epsilon,1;\frac{1}{2};\upsilon\right)+O(\epsilon^2)\\
            &=
                \upsilon^{\widehat{\Delta}_1/2}+(\widehat\gamma_{1,1}+\widehat\gamma_{2p,1}-\widehat\gamma_{2p+1,1})\,\epsilon\cdot \upsilon^{3/2}\cdot {}_2F_1(1,1;3/2;\upsilon)+O(\epsilon^2)\ .
    \end{aligned}
\end{align}
and\footnote{Recall the relations among BOE coefficients \eqref{eq:W1 to higher order OPE coeff Neumann} and \eqref{eq:b/c 2n+3 Neumann 1}. We obtain the last line of \eqref{eq:Neumann resummation 0} by applying the summation formula for Gauss's hypergeometric functions \eqref{eq:hypergeometric identity 2}.} 
\begin{align}\label{eq:Neumann resummation 0}
    \begin{aligned}
&\sum_{n=0}^{\infty}\,\frac{b(W_1^\alpha,\widehat{\mathsf{O}}_{2n+3}^{\prime\,\alpha})\,c(\widehat{\mathsf{O}}_{2n+3}^{\prime\,\alpha},\widehat{W}_{2p+1}^{\,\beta},\widehat{W}_{2p+2})}{c(\widehat{\mathsf{O}}_{2n+3}^{\prime\,\alpha},\widehat{\mathsf{O}}_{2n+3}^{\prime\,\alpha})}\,G^{\widehat{\Delta}_{2p}-\widehat{\Delta}_{2p+1}}_{\widehat{\Delta}'_{2n+3}}(\upsilon)\\
       &= \frac{12\,p\,f_p \, \delta^{\alpha\beta}}{N+8}\,\epsilon\cdot \sum_{n=0}^{\infty}\,\frac{(-1)^n\,n!}{(2n+1)\,(n+3/2)_n} \cdot \upsilon^{n+3/2}\cdot{}_2F_1\left({n+1,n+2\atop 2n+5/2};\upsilon\right)+O(\epsilon^2) \\
       &=\frac{12\,p\,f_p \, \delta^{\alpha\beta}}{N+8}\,\epsilon\cdot \upsilon^{3/2}\cdot {}_2F_1(1,1;3/2;\upsilon)+O(\epsilon^2)\ .
    \end{aligned}
\end{align}
We have expanded anomalous dimensions of composite operators in powers of $\epsilon$: 
\begin{align}\label{eq:Neumann anomalous dim expansion}   \widehat{\gamma}_{2p}=\widehat{\gamma}_{2p,1}\,\epsilon+\widehat{\gamma}_{2p,2}\,\epsilon^2+\cdots\ , \qquad \widehat{\gamma}_{2p+1}=\widehat{\gamma}_{2p+1,1}\,\epsilon+\widehat{\gamma}_{2p+1,2}\,\epsilon^2+\cdots\ ,
\end{align}
and denoted their leading coefficients by $\widehat{\gamma}_{r,1}$. We further notice from \eqref{eq:3pt ddd Neumann 1} that:
\begin{align}
      D\cdot c(\widehat{W}_1^{\,\alpha},\widehat{W}_{2p},\widehat{W}_{2p+1}^{\,\beta})=c(\widehat{\Phi}_1^{\,\alpha},\widehat{\Phi}_{2p},\widehat{\Phi}_{2p+1}^{\,\beta})+O(\epsilon)=f_p\,\delta^{\alpha\beta}+O(\epsilon)\ .
\end{align}
Substituting these expressions into \eqref{eq:Neumann bbb 3pt 1}, we can eventually evaluate the bulk-boundary-boundary three-point correlator $\langle\, W_1^\alpha\, \widehat{W}_{2p}\,\widehat{W}_{2p+1}^{\,\beta}\,\rangle$ up to the first order in $\epsilon$:
\begin{align}\label{eq:W1 3pt expansion 1}
    \begin{aligned}
    \langle\, W_1^\alpha(x)\, &\widehat{W}_{2p}(0)\,\widehat{W}_{2p+1}^{\,\beta}(\infty)\,\rangle\\
        &=
            D\cdot c(\widehat{W}_1^{\,\alpha},\widehat{W}_{2p},\widehat{W}_{2p+1}^{\,\beta})\cdot \frac{1}{|x_\perp|^{\Delta_1-\widehat{\Delta}_1}\,|x|^{\widehat{\Delta}_1+\widehat{\Delta}_{2p}-\widehat{\Delta}_{2p+1}}}  \\
        &\qquad\qquad 
            +
                \frac{f_p\, \delta^{\alpha\beta}}{N+8}\,\epsilon\cdot \left[ (N+8)\cdot (\widehat\gamma_{1,1}+\widehat\gamma_{2p,1}-\widehat\gamma_{2p+1,1})+ 12\,p\right]\\
                &\qquad\qquad\qquad\qquad\qquad\cdot \frac{|x_\perp|^2}{|x|^2} \cdot {}_2F_1\left(1,1;\frac{3}{2};\frac{|x_\perp|^2}{|x|^2}\right)+O(\epsilon^2)\ .
    \end{aligned}
\end{align}
One performs a similar manipulation for the other bulk-boundary-boundary three-point correlator $\langle\, W_1^\alpha\,\widehat{W}_{2p+1}^{\,\beta}\,\widehat{W}_{2p+2}\,\rangle$ to find that:
\begin{align}\label{eq:W1 3pt expansion 2}
    \begin{aligned}
    \langle\, W_1^\alpha(x)\, &\widehat{W}_{2p+1}^{\,\beta}(0)\,\widehat{W}_{2p+2}(\infty)\,\rangle\\
        &=
            D\cdot c(\widehat{W}_1^{\,\alpha},\widehat{W}_{2p}^{\,\beta},\widehat{W}_{2p+2})\cdot \frac{1}{|x_\perp|^{\Delta_1-\widehat{\Delta}_1}\,|x|^{\widehat{\Delta}_1+\widehat{\Delta}_{2p+1}-\widehat{\Delta}_{2p+2}}}\\
        & \qquad\qquad  +
            \frac{g_p\,\delta^{\alpha\beta}}{N+8}\,\epsilon\cdot\left[ (N+8)\,(\widehat\gamma_{1,1}+\widehat\gamma_{2p+1,1}-\widehat\gamma_{2p+2,1})+ 2\,(N+6p+2)\right]\\
            &\qquad\qquad\qquad\qquad\qquad\cdot \frac{|x_\perp|^2}{|x|^2} \cdot  {}_2F_1\left(1,1;\frac{3}{2};\frac{|x_\perp|^2}{|x|^2}\right)+O(\epsilon^2)\ .
    \end{aligned}
\end{align}

\paragraph{Constraint from analyticity.}
It is clear from the asymptotic behavior of Gauss's hypergeometric functions in the last lines of \eqref{eq:W1 3pt expansion 1} and \eqref{eq:W1 3pt expansion 2}:
\begin{align}
    {}_2F_1\left(1,1;\frac{3}{2};\frac{|x_\perp|^2}{|x|^2}\right)\xrightarrow[\text{with \eqref{eq:Kummer's connection formula}}]{|\hat{x}|\sim 0} \frac{\pi}{2}\cdot \frac{|x_\perp|}{|\hat{x}|}+\cdots\ ,
\end{align}
that the bulk-boundary-boundary three-point functions $\langle\, W_1^\alpha\, \widehat{W}_{2p}\,\widehat{W}_{2p+1}^{\,\beta}\,\rangle$ and $\langle\, W_1^\alpha\,\widehat{W}_{2p+1}^{\,\beta}\,\widehat{W}_{2p+2}\,\rangle$ are not analytic as $|\hat{x}|\sim 0$.
These non-analytic behaviors are similar to the line defect case but more manifest here as the bulk-boundary-boundary three-point correlators involving $W_1^\alpha$ become singular as $|\hat{x}|\to0$. They again contradict the analyticity of correlators for non-coincident configurations of points. To resolve this issue, we demand that:
\begin{align}
    (N+8)\cdot (\widehat\gamma_{1,1}+\widehat\gamma_{2p,1}-\widehat\gamma_{2p+1,1})+ 12\,p&=0\ ,\\
    (N+8)\,(\widehat\gamma_{1,1}+\widehat\gamma_{2p+1,1}-\widehat\gamma_{2p+2,1})+ 2\,(N+6p+2)&=0\ .
\end{align}
Given the result derived in the previous subsection \eqref{eq:anomalous dim lowest Neu} (and also \eqref{eq:Neumann anomalous dim expansion}), we find that these two constraints result in the following recursion relations with the initial condition $\widehat{\gamma}_{1,1}=-\frac{N+2}{2\,(N+8)}$:
\begin{align}\label{eq:Neu recursion rel}
    \widehat{\gamma}_{2p+1,1}
        =
            \widehat{\gamma}_{2p,1}-\frac{N-24p+2}{2\,(N+8)}\ ,\qquad 
    \widehat{\gamma}_{2p+2,1}
        =
            \widehat{\gamma}_{2p+1,1}+\frac{3\,(N+8p+2)}{2\,(N+8)}\ ,
\end{align}
by solving which, we can reproduce the perturbative results \eqref{eq: Diagrammatic result, Neumann}.

\chapter{The O$(N)$ model with Dirichlet boundary}\label{chap:Dirichlet}
This chapter deals with the O$(N)$ model in $d=4-\epsilon$ dimensions with Dirichlet boundary. The action of the model reads:\footnote{In this chapter, we consider the model in $d=4-\epsilon$ dimensions while fixing the co-dimensions of the defect (Dirichlet boundary) to one.}
\begin{align}\label{eq:action Dirichlet}
    I = \int_{\mathbb{R}_{+}^{d}}\,\d^d x \,\left(\frac{1}{2}\, |\partial \Phi_{1} |^2\, +\,   \frac{\lambda_0}{4!}\, |\Phi_1|^4 \right) \ ,\qquad \lim_{|x_\perp|\to0}\,\Phi_{1}^{\alpha}(x)=0\ .
\end{align}
The critical O$(N)$ model with the Dirichlet boundary condition describes the so-called ordinary transition of the statistical Ising model. On the other hand, the Neumann boundary condition (chapter \ref{chap:Neumann boundary}) corresponds to the special transition. We have another boundary universality class for the bulk Wilson-Fisher fixed point associated with the extraordinary transition. However, we do not deal with that case throughout this thesis since we cannot study it within our axiomatic framework. See footnote \ref{fot:on extraordinary} in chapter \ref{chap:conclusion} and the paragraph involved for a detailed description.

In what follows, we explore the conformal dimensions of the defect local operators tending to the following ones in the free theory:
\begin{align}\label{eq:Dirichlet lowest-lying defect local def}
    \widehat{\Psi}_{2}^{\,\alpha}(\hat{x})\equiv\lim_{|x_\perp|\to0}\,|x_\perp|^{-1}\,\Phi_{1}^{\,\alpha}(x)=\lim_{|x_\perp|\to0}\,\frac{\partial}{\partial x_\perp}\,\Phi_{1}^{\,\alpha}(x)\ ,
\end{align}
and 
\begin{align}\label{eq:Dirichlet composite defect local def}
\begin{aligned}
    \widehat{\Psi}_{4p}(\hat{x})&\equiv\lim_{|x_\perp|\to0}\,|x_\perp|^{-2p}|\,\Phi_{1}|^{2p}(x)\ ,\\
\widehat{\Psi}_{4p+1}^{\,\alpha}(\hat{x})&\equiv\lim_{|x_\perp|\to0}\,|x_\perp|^{-2p-1}\,\Phi_{1}^{\,\alpha}|\Phi_{1}|^{2p}(x) \ ,
\end{aligned}
\end{align}
for $p=1,2,\cdots$. Here, the subscripts for free theory operators represent their conformal dimensions in four dimensions.
We express their renormalized operators or Wilson-Fisher counterparts as follows:
\begin{align}\label{eq:Dirichlet operators interest}
\widehat{\CW}_{2}^{\,\alpha}(\hat{y})\ ,\qquad \widehat{\CW}_{4p}(\hat{y})\ ,\qquad \widehat{\CW}_{4p+2}^{\,\alpha}(\hat{y})\ .
\end{align}
We also denote their conformal dimensions by $\widehat{\Delta}_{2},\widehat{\Delta}_{4p},\widehat{\Delta}_{4p+2}$, respectively.
The conformal dimension of $\widehat{\CW}_{2}^{\,\alpha}$ is derived from standard perturbative calculations (see e.g., \cite{McAvity:1993ue,McAvity:1995zd}) and also from the axiomatic approach (see \cite{Dey:2020jlc,Giombi:2020rmc}):
\begin{align}\label{eq:Dirichlet pert lowest known}
    \widehat{\Delta}_2=2-\frac{N+5}{N+8}\,\epsilon +O(\epsilon^2)\ .
\end{align}
We note that the conformal dimension of $\widehat{\CW}_{2}^{\,\alpha}$ is related to the critical exponent for the surface magnetization that can be measured experimentally (see table \ref{tab:on surface exponent} in chapter \ref{chap:Introduction} for a comparison between the theoretical prediction and the experimental data):
\begin{align}\label{eq:surface magnetization definition}
    \widehat{m}\sim |T_c-T|^{\widehat{\beta}}\ ,\qquad \widehat{\beta}=\frac{\widehat{\Delta}_2}{d-\Delta_2} \ , \qquad \text{as}\quad T\sim T_c\ .
\end{align}

We use the same normalization of the O$(N)$ vector field as the Neumann case \eqref{eq:BCFT bulk field normalization} and take the same path as chapter \ref{chap:Neumann boundary} with attention to the differences in boundary conditions. Section \ref{sec:Dirichlet free} reveals the structure of the free O$(N)$ model with Dirichlet boundary. Using the results, we work out the conformal dimensions of the boundary local operators listed in \eqref{eq:Dirichlet operators interest} both in the standard perturbative approach and the axiomatic approach (section \ref{sec:Dirichlet diagrammatic} and \ref{sec:Dirichlet Axiomatic approach}).

\section{Structure of free O$(N)$ model with Dirichlet boundary}\label{sec:Dirichlet free}
We now investigate the free O$(N)$ model in $d$-dimensions with Neumann boundary condition. After computing correlation functions in $d$-dimensions (section \ref{sec:Correlation functions in free theory BCFT Dirichlet}), we specify spacetime dimensions to four and spell out the BOEs of two bulk primaries $\Phi_1^\alpha$ and $\Phi_3^\alpha$, that are closely tied to the multiplet recombination phenomena at the Wilson-Fisher fixed point and play the central roles in our axiomatic analysis (section \ref{sec:Dirichlet Axiomatic approach}). We will use the expressions of correlation functions in $d$-dimensions in the standard perturbative calculations (section \ref{sec:Dirichlet diagrammatic}). On the other hand, we will need the free theory BOEs in four dimensions in the axiomatic approach (section \ref{sec:Dirichlet Axiomatic approach}).

\subsection{Correlation functions}\label{sec:Correlation functions in free theory BCFT Dirichlet}
Under Dirichlet boundary condition $\lim_{|x_\perp|\to0}\,\Phi_{1}^{\alpha}(x)=0$, the differential equation for the two-point function of $\Phi_{1}^{\alpha}$ \eqref{eq:4d free scalar propagator EoM} is solved by:
\begin{align}\label{eq:4d free scalar propagator Dirichlet}
    \langle\,\Phi_{1}^\alpha(x_1)\,\Phi_{1}^\beta(x_2)\,\rangle
        =
        \frac{\delta^{\alpha\beta}}{|x_1-x_2|^{d-2}}- \frac{\delta^{\alpha\beta}}{|x_1-\bar{x}_2|^{d-2}}\ ,\qquad 
    \bar{x}^\mu
        =
            (\hat{x}^a,-x_\perp) .
\end{align}
Similarly to the Neumann case, we take the coincident limit of the above expression to find bulk one-point functions of composite operators $\Phi_{1}^\alpha\Phi_{1}^\beta$ and $|\Phi_1|^{2}$:
\begin{align}\label{eq:Dirichlet composite bulk 1pt}
  \langle\,\Phi_{1}^\alpha\Phi_{1}^\beta(x)\,\rangle=-\frac{\delta^{\alpha\beta}}{2^{d-2}\,|x_\perp|^{d-2}}\ ,\qquad   \langle\,|\Phi_1|^{\,2}(x)\,\rangle=-\frac{N}{2^{d-2}\,|x_\perp|^{d-2}}\ ,
\end{align}
Performing the same manipulation as the equation \eqref{eq:2pt Phi Neu}, one obtains the two-point functions involving the lowest-lying boundary local operator $\widehat{\Psi}_{2}^{\,\alpha}$ \eqref{eq:Dirichlet lowest-lying defect local def}:
\begin{align}\label{eq:2pt Phi Dir}
    \langle\,\Phi_1^\alpha(x)\,\widehat{\Psi}_{2}^{\,\beta}(\hat{y})\,\rangle
        =
            \frac{ 2(d-2)\,  \delta^{\alpha\beta}\,x_{\perp}}{|x-\hat{y}|^d}\ ,\qquad
       \langle\,\widehat{\Psi}_{2}^{\,\alpha}(\hat{y}_1)\,\widehat{\Psi}_{2}^{\,\beta}(\hat{y}_2)\,\rangle
        =
            \frac{2(d-2)\, \delta^{\alpha\beta}}{|\hat{y}_{12}|^d} \ .
\end{align}
Wick's theorem allows us to calculate all the other correlators of our interest. The bulk-boundary two-point functions concerning $\Phi_3^\alpha$ turn out to take the following forms:
\begin{align}\label{eq:Dir btb 2pt Phi3}
\begin{aligned}
      \langle\,\Phi_3^\alpha(x)\,\widehat{\Psi}_{2}^{\,\beta}(\hat{y})\,\rangle
        &=
            -\frac{(d-2)(N/2+1)\,\delta^{\alpha\beta}}{2^{d-4}\,|x-\hat{y}|^{d}\,|x_\perp|^{d-3}}\ ,\\
            \langle\,\Phi_3^\alpha(x)\,\widehat{\Psi}_{6}^{\,\beta}(\hat{y})\,\rangle
                &=
                \frac{  32(d-2)^3\,(N/2+1)\,\delta^{\alpha\beta}\,x_\perp^3}{|x-\hat{y}|^{3d}}\ .
\end{aligned}
\end{align}
It follows from the principle of mathematical induction combined with Wick's contraction that the boundary two-point functions of composite operators \eqref{eq:Dirichlet composite defect local def} have the following expressions:
\begin{align}\label{eq:Neumann free boundary 2pt expression}
            \langle\, \widehat{\Psi}_{4p}(\hat{y}_1)\,\widehat{\Psi}_{4p}(\hat{y}_2) \,\rangle = \frac{N\, b_{p-1}}{|\hat{y}_{12}|^{2\, p\, d}} \, ,\qquad
             \langle\,  \widehat{\Psi}_{4p+2}^{\,\alpha}(\hat{y}_1)\, \widehat{\Psi}_{4p+2}^{\,\beta}(\hat{y}_2) \,\rangle = \frac{a_p\, \delta^{\alpha\beta}}{|\hat{y}_{12}|^{(2p+1)\, d}}\, .  
    \end{align}
The two symbols $a_p$ and $b_p$ are defined by:
\begin{align}\label{eq:a and b coeff}
    a_{p}\equiv 2^{6p+2}\, p!\, (N/2+1)_{p}\ ,\qquad  b_{p}\equiv 2^{6p+5}\, (p+1)!\, (N/2+1)_{p}\  .
\end{align}
They are also related to the boundary three-point coefficients in the following manner:
    \begin{align}
     \langle\,\widehat{\Psi}_{2}^\alpha(\hat{x})\, \widehat{\Psi}_{4p}(\hat{y}_1)\,\widehat{\Psi}_{4p+2}^{\,\beta}(\hat{y}_2)\,\rangle&=\frac{a_{p}\, \delta^{\alpha\beta}}{|\hat{x}-\hat{y}_{2}|^{d}|\hat{y}_{12}|^{2\, p\, d}}\ ,\label{eq:3pt ddd Dirichlet 1}\\
     \langle\,\widehat{\Psi}_{2}^\alpha(\hat{x})\, \widehat{\Psi}_{4p+2}^{\beta}(\hat{y}_1)\,\widehat{\Psi}_{4p+4}(\hat{y}_2)\,\rangle
    &=   \frac{b_{p}\, \delta^{\alpha\beta}}{|\hat{x}-\hat{y}_{2}|^{d}|\hat{y}_{12}|^{(2p+1)d}}\ ,\label{eq:3pt ddd Dirichlet 2}\\
      \langle\,\widehat{\Psi}_{6}^\alpha(\hat{x})\, \widehat{\Psi}_{4p+2}^{\,\beta}(\hat{y}_1)\,\widehat{\Psi}_{4p+4}(\hat{y}_2)\,\rangle&=\frac{4(N+6p+2)\, b_{p}\, \delta^{\alpha\beta}}{|\hat{x}-\hat{y}_{1}|^d |\hat{x}-\hat{y}_{2}|^{2d} |\hat{y}_{12}|^{2pd}}\ ,\label{eq:3pt ddd Dirichlet 3}\\
       \langle\,\widehat{\Psi}_{6}^\alpha(\hat{x})\, \widehat{\Psi}_{4p}(\hat{y}_1)\,\widehat{\Psi}_{ 4p+2}^{\,\beta}(\hat{y}_2)\,\rangle&=\frac{24p\, a_{p}\, \delta^{\alpha\beta}}{|\hat{x}-\hat{y}_{1}|^d |\hat{x}-\hat{y}_{2}|^{2d} |\hat{y}_{12}|^{(2p-1)d}} \label{eq:3pt ddd Dirichlet 4} \ .
\end{align}
We further record the following bulk-boundary-boundary three-point functions that will be used in section \ref{eq:Boundary-to-defect operator product expansions free Dirichlet} to uncover the BOE structures of $\Phi_{1}^\alpha$ and $\Phi_{3}^\alpha$ in four dimensions:
\begin{align}
              \langle\,\Phi_{1}^\alpha(x)\, \widehat{\Psi}_{4p}(\hat{y}_1)\,\widehat{\Psi}_{4p+2}^{\,\beta}(\hat{y}_2)\,\rangle&= \frac{d-2}{2}\cdot 
              \frac{a_{p}\,  \delta^{\alpha\beta}\, x_{\perp}}{|x-\hat{y}_2|^d \,|\hat{y}_{12}|^{2pd}}\ ,\label{eq:3pt bdd Dirichlet 1}\\
                \langle\,\Phi_{1}^\alpha(x)\, \widehat{\Psi}_{4p+2}^{\beta}(\hat{y}_1)\,\widehat{\Psi}_{4p+4}(\hat{y}_2)\,\rangle
     &=  \frac{d-2}{2}\cdot \frac{b_{p}\, \delta^{\alpha\beta}\, x_{\perp}}{|x-\hat{y}_2|^d \,|\hat{y}_{12}|^{(2p+1)d}}\ ,\label{eq:3pt bdd Dirichlet 2}
     \end{align}
          and
     \begin{align}
     \begin{aligned}
     &\langle\,\Phi_{3}^\alpha(x)\, \widehat{\Psi}_{4p}(\hat{y}_1)\,\widehat{\Psi}_{4p+2}^{\,\beta}(\hat{y}_2)\,\rangle& \\
     &\quad=-\frac{N+2}{2^{d-2}}\cdot\frac{(d/2-1)\cdot a_{p}\,\delta^{\alpha\beta}}{|x-\hat{y}_2|^d\,|\hat{y}_{12}|^{2pd}\,x_\perp^{d-3}}+\frac{(d/2-1)^3\cdot 24\,p\,  a_{p}\, \delta^{\alpha\beta}\, x_{\perp}^3}{|x-\hat{y}_1|^d \,|x-\hat{y}_2|^{2d}\,|\hat{y}_{12}|^{(2p-1)d}}\ , &\label{eq:3pt bdd Dirichlet 3} 
     \end{aligned}
     \end{align}
     \begin{align}
     \begin{aligned}
     &\langle\,\Phi_{3}^\alpha(x)\, \widehat{\Psi}_{4p+2}^{\,\beta}(\hat{y}_1)\,\widehat{\Psi}_{4p+4}(\hat{y}_2)\,\rangle& \\
     &\quad=-\frac{N+2}{2^{d-2}}\cdot\frac{(d/2-1)\cdot b_{p}\,\delta^{\alpha\beta}}{|x-\hat{y}_2|^d\,|\hat{y}_{12}|^{(2p+1)d}\,x_\perp^{d-3}}+\frac{(d/2-1)^3\cdot 4\,(N+6p+2)\,  b_{p}\, \delta^{\alpha\beta}\, x_{\perp}^3}{|x-\hat{y}_1|^d \,|x-\hat{y}_2|^{2d}\,|\hat{y}_{12}|^{2pd}}\, .&\label{eq:3pt bdd Dirichlet 4}
     \end{aligned}
\end{align}
One way to compute these four bulk-boundary-boundary three-point correlators is to isolate bulk operators by applying Wick's contractions and to exploit the results for the boundary two-point functions \eqref{eq:Neumann free boundary 2pt expression}.
Notice that all these correlators we have enumerated so far are in line with the generic forms of the BCFT correlators (see section \ref{sec:Correlation functions in DCFT}) and thus are compatible with boundary conformal symmetry.

\subsection{Boundary operator expansions in four dimensions}\label{eq:Boundary-to-defect operator product expansions free Dirichlet}
Let us investigate the BOEs of $\Phi_1^\alpha$ and $\Phi_3^\alpha$ in four dimensions from the results presented in the last subsection. We extract BOE coefficients by comparing the correlation functions with the generic forms of BCFT correlators (section \ref{sec:Correlation functions in DCFT}). For the BOE of $\Phi_3^\alpha$, we also implement the conformal block expansions of the bulk-boundary-boundary three-point functions involving $\Phi_3^\alpha$ to see its BOE content inaccessible from lower-point functions.

\paragraph{Boundary operator expansion of $\Phi_1^\alpha$.}
According to section \ref{sec:Defect Operator Expansion spectrum of free scalar field}, the Klein-Gordon equation requires that the boundary local operators that contribute to the BOE of $\Phi_1^\alpha$ have conformal dimension one or two, and all other operators are prohibited.
Unlike the Neumann case, the only boundary local operator that shows up in the BOE of $\Phi_1^\alpha$ is $\widehat{\Psi}_{2}^{\,\alpha}(\hat{x})$ (see \eqref{eq:Dirichlet lowest-lying defect local def} for its definition) having conformal dimension two:
\begin{align}\label{eq:OPE of phi1 O(N) Dirichlet}
    \Phi_1^\alpha(x) = x_\perp\cdot \widehat{\Psi}_{2}^\alpha(\hat{x})+(\text{descendants})\ .
\end{align}

\paragraph{Boundary operator expansion of $\Phi_3^\alpha$.} 
The BOE of $\Phi_3^\alpha$ takes the form:
\begin{align}\label{eq:btb OPE of phi3 all order; Dirichlet}
\begin{aligned}
  \Phi_3^{\alpha}(x) =  -\frac{N+2}{4\,|x_\perp|}\cdot \widehat{\Psi}_2^\alpha(\hat{x})+
    \sum_{n=0}^{\infty}\,\frac{b(\Phi_3^\alpha,\widehat{\mathsf{Q}}_{2n+6}^\alpha)}{c(\widehat{\mathsf{Q}}_{2n+6}^\alpha,\widehat{\mathsf{Q}}_{2n+6}^\alpha)}\, x_\perp^{2n+3}\cdot \widehat{\mathsf{Q}}_{2n+6}^\alpha(\hat{x})+(\text{descendants.}) \ .
\end{aligned}
\end{align}
We find a series of composite operators $\widehat{\mathsf{Q}}_{2n+6}^\alpha$ with $\widehat{\Delta}(\widehat{\mathsf{Q}}_{2n+6}^\alpha)=2n+6$ involving three $\widehat{\Psi}_{2}^\alpha$'s and $2n$ parallel derivatives $\hat{\partial}^a$. We notice that their BOE coefficients satisfy the following identities:
\begin{align}\label{eq:bc_over_c_Dirichlet}
\begin{aligned}
    \frac{b(\Phi_3^\alpha,\widehat{\mathsf{Q}}_{2n+6}^\alpha)\,c(\widehat{\mathsf{Q}}_{2n+6}^\alpha,\widehat{\Psi}_{4p},\widehat{\Psi}_{4p+2}^{\,\beta})}{c(\widehat{\mathsf{Q}}_{2n+6}^\alpha,\widehat{\mathsf{Q}}_{2n+6}^{\alpha})}&=24p\, a_p\cdot  \frac{(-1)^n\,(2)_n (4)_n}{(n+9/2)_n \, n!}\, \delta^{\alpha\beta}\ ,\\
    \frac{b(\Phi_3^\alpha,\widehat{\mathsf{Q}}_{2n+6}^\alpha)\,c(\widehat{\mathsf{Q}}_{2n+6}^\alpha,\widehat{\Psi}_{4p+2}^{\,\beta},\widehat{\Psi}_{4p+4})}{c(\widehat{\mathsf{Q}}_{2n+6}^\alpha,\widehat{\mathsf{Q}}_{2n+6}^{\alpha})}&=4(N+6p+2)\, b_p\cdot \frac{(-1)^n\,(2)_n (4)_n}{(n+9/2)_n \, n!}\, \delta^{\alpha\beta}\ .
\end{aligned}
\end{align}
One can confirm this BOE \eqref{eq:btb OPE of phi3 all order; Dirichlet} immediately from the following conformal block expansions of the bulk-boundary-boundary three-point functions \eqref{eq:3pt bdd Dirichlet 3} and \eqref{eq:3pt bdd Dirichlet 4}:
\begin{align}\label{eq:3pt bdd Dirichlet 3 exp ast}
\begin{aligned}
     \langle\,&\Phi_3^\alpha(x)\, \widehat{\Psi}_{4p}(0)\,\widehat{\Psi}_{4p+2}^{\,\beta}(\infty)\,\rangle
     \\
     &=-\frac{N+2}{4}\cdot \frac{a_p\, \delta^{\alpha\beta}}{|x_\perp|}+\frac{24p\, a_p\, |x_\perp|^3}{|x|^4}\\
     &=a_p\, \delta^{\alpha\beta}\cdot\frac{|x|^2}{x_\perp^3}\cdot \left[
            -\frac{N+2}{4}\cdot G^{-2}_{2}(\upsilon)
            +        24\,p\cdot  \sum_{n=0}^{\infty}\,\frac{(-1)^n\,(n+1)\,(4)_n}{(n+9/2)_n}\cdot G^{-2}_{2n+6}(\upsilon)\right]\ ,
\end{aligned}
\end{align}
and 
\begin{align}\label{eq:3pt bdd Dirichlet 4 exp ast}
\begin{aligned}
     \langle\,&\Phi_3^\alpha(x)\, \widehat{\Psi}_{4p+2}^{\,\beta}(0)\,\widehat{\Psi}_{4p+4}(\infty)\,\rangle\\
     &=-\frac{N+2}{4}\cdot \frac{b_p\, \delta^{\alpha\beta}}{|x_\perp|}+4(N+6p+2)\cdot  \frac{b_p\, \delta^{\alpha\beta}\,|x_\perp|^3}{|x|^4}\\
     &=b_p\, \delta^{\alpha\beta}\cdot \frac{|x|^2}{x_\perp^3}\cdot \left[
            -\frac{N+2}{4}\cdot G^{-2}_{2}(\upsilon)\right.\\
            &\qquad\qquad\qquad\qquad\left.
            +        4(N+6p+2)\cdot  \sum_{n=0}^{\infty}\, \frac{(-1)^n\,(n+1)\, (4)_n}{(n+9/2)_n}\cdot G^{-2}_{2n+6}(\upsilon)\right]\ ,
\end{aligned}
\end{align}
with $\upsilon=|x_\perp|^2/|x|^2$ being the cross ratio. In deriving these conformal block expansions, we employed the hypergeometric identity \eqref{eq:hypergeometric identity 1} and the explicit form of the conformal block \eqref{eq:conformal block expansion main}. More specifically, we used:
\begin{align}
G_2^{-2}(\upsilon)=\upsilon^2\ ,\qquad G^{-2}_{2n+6}(\upsilon)=\upsilon^{n+3}\cdot {}_2F_1(n+2,n+4;2n+11/2;\upsilon)\ .   
\end{align}
One can read off the BOE relations \eqref{eq:bc_over_c_Dirichlet} by comparing \eqref{eq:3pt bdd Dirichlet 3 exp ast} and \eqref{eq:3pt bdd Dirichlet 4 exp ast} with \eqref{eq:conformal block expansion main}.

\section{Standard perturbative approach}\label{sec:Dirichlet diagrammatic}
In the section, we derive anomalous dimensions of $\widehat{\Psi}_{4p}$ and $\widehat{\Psi}_{4p+2}^{\,\alpha}$ ($p=0,1,\cdots$) from standard perturbative calculations. Since the calculations are straightforward and almost identical to the Neumann case (section \ref{sec:Neumann diagrammatic}), we omit the details and only leave the rough sketch of the derivations. 

We first introduce wave-function renormalizations of the boundary local operators of our interest through the relations:
\begin{align}
    \widehat{\Psi}_{4p}=\widehat{Z}_{4p}\cdot \widehat{\CW}_{4p}\ ,\qquad \widehat{\Psi}_{4p+2}^{\,\alpha}=\widehat{Z}_{4p+2}\cdot\widehat{\CW}_{4p+2}^{\,\alpha}\ .
\end{align}
which are to be determined to cancel the divergence of $\langle\,\widehat{\CW}_{4p}\,\widehat{\CW}_{4p}\,\rangle$ and $\langle\,\widehat{\CW}_{4p+2}^{\,\alpha}\,\widehat{\CW}_{4p+2}^{\,\beta}\,\rangle$. By calculating these two-point functions up to the one-loop level in a similar manner to the Neumann case, one arrives at the recursion relations:
\begin{align} \label{eq:recursion relation Dirichlet}
\begin{aligned}
    \delta \widehat{Z}_{4p}-\delta \widehat{Z}_{4p-2}=-\frac{N+12p-10}{6\,\epsilon}\cdot  \frac{\lambda}{(4\pi)^2} + O(\lambda^2, \epsilon^0) \ , \\
     \delta \widehat{Z}_{4p+2}-\delta \widehat{Z}_{4p}=-\frac{12p-N-2}{6\,\epsilon}\cdot  \frac{\lambda}{(4\pi)^2} +O(\lambda^2, \epsilon^0)\ ,
\end{aligned}
\end{align}
where $\widehat{Z}_{4p}=1+ \delta \widehat{Z}_{4p}$ and $\widehat{Z}_{4p+2}=1+ \delta \widehat{Z}_{4p+2}$.
These recursion relations are solved under initial condition $\delta\widehat{Z}_0=0$, which follows from the fact that there are no quantum corrections to the identity operator $\widehat{\Psi}_{0}=\bm{1}$ to give the following expressions:
\begin{align}
         Z_{4p}&=1-\frac{2\, p\, (p-1)}{\epsilon}\cdot \frac{\lambda}{(4\pi)^2}\, +O(\lambda^2)\, , \\
     Z_{4p+2}&=1-\frac{12p^2 -N-2}{6\,\epsilon}\cdot  \frac{\lambda}{(4\pi)^2}+O(\lambda^2)\, .
\end{align}
Then, one can calculate the anomalous dimensions of $\widehat{\Psi}_{4p}$ and $\widehat{\Psi}_{4p+2}^{\,\alpha}$:
\begin{align}\label{eq:Dir diagram cal anomalous dimensions}
\begin{aligned}
        \widehat{\gamma}_{4p}&=\left.\frac{\partial\log \widehat{Z}_{4p}}{\partial\log \mu}\right|_{\lambda=\lambda_\ast}=\frac{6p\,(p-1)}{N+8}\,\epsilon+O(\epsilon^2)\ ,\\
        \widehat{\gamma}_{4p+2}&=\left.\frac{\partial\log \widehat{Z}_{4p+2}}{\partial\log \mu}\right|_{\lambda=\lambda_\ast}=\frac{12\, p^2 -N-2}{2(N+8)}\,\epsilon+O(\epsilon^2)\ .
\end{aligned}
\end{align}
Because the canonical dimensions of $\widehat{\Psi}_{4p}$ and $\widehat{\Psi}_{4p+2}^{\,\alpha}$ are given by $4p\cdot d/2$ and $(4p+2)\cdot d/2$ respectively, the conformal dimension of the renormalized operators turn out to be:
\begin{align}
\begin{aligned}
        \widehat{\Delta}_{4p}&=4p\cdot\frac{d}{2}+\widehat{\gamma}_{4p}=4p-\frac{p\, (N-6p+14)}{N+8}\,\epsilon+O(\epsilon^2)\ ,\\
        \widehat{\Delta}_{4p+2}&=(4p+2)\cdot\frac{d}{2}+\widehat{\gamma}_{4p+2}=4p+2-\frac{N-6p^2 +p\,(N+8)+5}{N+8}\,\epsilon+O(\epsilon^2)\ .
\end{aligned}
\end{align}

We remark that the conformal dimension of $\widehat{\CW}_{4p}$ is the same as spacetime dimensions $\widehat{\Delta}_{4}=4-\epsilon=d$, making it possible to identify $\widehat{\CW}_{4p}$ with the displacement operator $\widehat{D}$ (see section \ref{sec:Bulk and defect local primaries} for details) up to the first order in $\epsilon$.\footnote{This statement is true in free theory in $d$ dimensions as below.
The (improved) stress tensor of the free $O(N)$-model is given by (see e.g., \cite[equation (1.34)]{Ammon:2015wua}):
\begin{align}
    T_{\mu\nu}(x)\propto\partial_{\mu}\Phi_{1}^{\alpha}\partial_{\nu}\Phi_{1}^{\alpha}-\frac{1}{2}\,\delta_{\mu\nu}\,|\partial\Phi_{1}|^2-\frac{d-2}{4(d-1)}\,(\partial_{\mu}\partial_{\nu}-\delta_{\mu\nu}\,\Box)\,|\Phi_{1}|^2\qquad (\text{free})\ ,
\end{align}
and the displacement operator in BCFT is related to the transverse component of the stress tensor by the relation \eqref{eq:displacement BCFT}. Hence, under Dirichlet boundary condition $\lim_{|x_\perp|\to0}\,\Phi_{1}^{\alpha}(x)=0$, we have:
\begin{align}
   \widehat{D}(\hat{x})= \lim_{|x_\perp|\to0}\,T_{\perp\perp}(x)\propto \partial_{\perp}\Phi_{1}^{\alpha}\partial_{\perp}\Phi_{1}^{\alpha}= \widehat{\Psi}_4\ .
\end{align}
} It may be interesting to check whether its higher-order anomalous dimensions vanish as expected.

\section{Axiomatic approach}\label{sec:Dirichlet Axiomatic approach}
Finally, we apply the axiomatic framework to the O$(N)$ model with a Dirichlet boundary. It is straightforward to derive the conformal dimensions of the lowest-lying boundary local operator (section \ref{sec:Dir lowest RT}). Similarly to the previous two chapters, we derive conformal dimensions of composite operators by requiring the analyticity of correlators away from the coincidence of points at order $\epsilon$.

\subsection{Lowest-lying boundary local operator}\label{sec:Dir lowest RT}
Thanks to axiom \ref{dcftaxiom1}, the theory is described by BCFT and the BOE of $W_1^\alpha$ takes the form:
\begin{align}
    W_1^\alpha(x)
        \supset D'\cdot\frac{1}{|x_\perp|^{\Delta_1-\widehat{\Delta}_2}}\cdot \widehat{\CW}_{2}^{\,\alpha}(\hat{x})  \ .
\end{align}
Here, we have $D'=1+O(\epsilon)$ from axiom \ref{dcftaxiom2} combined with \eqref{eq:OPE of phi1 O(N) Dirichlet}. We then apply the equation of motion \eqref{eq:classical EoM O(N)} (axiom \ref{dcftaxiom2}) to have:
\begin{align}
\begin{aligned}
     W_3^\alpha(x)
        &=
        \frac{1}{\kappa}\,\Box\,W_1^\alpha(x)\\
        &\supset \frac{D'}{\kappa}\cdot \frac{(\Delta_1-\widehat{\Delta}_2)(\Delta_1-\widehat{\Delta}_2+1)}{|x_\perp|^{\Delta_1-\widehat{\Delta}_2+2}}\cdot\widehat{\CW}_{2}^{\,\alpha}(\hat{x}) \ .  
\end{aligned}
\end{align}
For this BOE to match \eqref{eq:btb OPE of phi3 all order; Dirichlet} as $\epsilon\to0$ (axiom \ref{dcftaxiom2}), we obtain the equation that determines $\widehat{\Delta}_2$:
\begin{align}
- \frac{N+2}{4}\,\kappa=(\widehat{\Delta}_2-\Delta_1)(\widehat{\Delta}_2-\Delta_1-1) +O(\epsilon^2) \ .
\end{align}
Provided that the conformal dimension of $\widehat{\CW}_{2}^{\,\alpha}$ should be $\widehat{\Delta}_2=2+O(\epsilon)$, this equation is solved by:
\begin{align}\label{eq:anomalous dim lowest Dir}
    \begin{aligned}
    \widehat{\Delta}_2
        &=
        \frac{d}{2}-\frac{N+2}{2\,(N+8)}\,\epsilon+O(\epsilon^2) \\
        &=
        2 -\frac{N+5}{N+8}\,\epsilon+O(\epsilon^2)\ ,
    \end{aligned}
\end{align}
reproducing the perturbative result \eqref{eq:Dirichlet pert lowest known}.

\subsection{Boundary composite operators}\label{sec:Dir higher order}
We now focus on the boundary composite operators $\widehat{\CW}_{4p}$ and $\widehat{\CW}_{4p+2}^{\,\alpha}$, tending to $\widehat{\Psi}_{4p}$ and $\widehat{\Psi}_{4p+2}^{\,\alpha}$ as $\epsilon\to0$ (see \eqref{eq:Dirichlet composite defect local def} for the definitions for free theory operators). Our ultimate goal is to find the closed-form expressions of the bulk-boundary-boundary three-point functions such as $ \langle\, W_1^\alpha, \widehat{\CW}_{4p}\,\widehat{\CW}_{4p+2}^{\,\beta}\,\rangle$ and $\langle\, W_1^\alpha\, \widehat{\CW}^{\, \beta}_{4p+2}\,\widehat{\CW}_{4p+4}\,\rangle$, and remove their unphysical singularity to obtain the constraints for anomalous dimensions of the operators involved at order $\epsilon$. To this end, we derive the BOE of $W_1^\alpha$, plug it into the generic form of the conformal block expansions of bulk-defect-defect three-point functions given in \eqref{eq:conformal block expansion main}, and apply some resummation techniques to get closed-form expressions.

\paragraph{Boundary operator expansion of $W_1^\alpha$.}
The final expression for the BOE of $W_1^\alpha$ is:
\begin{align}\label{eq:OPE of W1 all order;Dirichlet}
    \begin{aligned}
        W_1^\alpha(x)
            &=
                \frac{D'}{|x_\perp|^{\Delta_1-\widehat{\Delta}_2}}\cdot\widehat{\CW}_2^{\,\alpha}(\hat{x}) 
          \\
            &\qquad    +
  \sum_{n=0}^{\infty}\,\frac{b(W_1^\alpha,\widehat{\mathsf{Q}}_{2n+6}^{\prime\,\alpha})/c(\widehat{\mathsf{Q}}_{2n+6}^{\prime\,\alpha},\widehat{\mathsf{Q}}_{2n+6}^{\prime\,\alpha})}{|x_\perp|^{\Delta_1-\widehat{\Delta}(\widehat{\mathsf{Q}}_{2n+6}^{\prime\,\alpha})}}\cdot \widehat{\mathsf{Q}}_{2n+6}^{\prime\,\alpha}(\hat{x})+(\text{descendants})\ ,
    \end{aligned}
\end{align}
where $\lim_{\epsilon\rightarrow 0}\,\widehat{\mathsf{Q}}_{2n+6}^{\prime\,\alpha}=\widehat{\mathsf{Q}}_{2n+6}^\alpha$ (the same boundary composite operator that appeared in section \ref{eq:Boundary-to-defect operator product expansions free Dirichlet}) with $\widehat{\Delta}(\widehat{\mathsf{Q}}_{2n+6}^{\prime\,\alpha})=2n+6+O(\epsilon)$. We further note an identity for BOE coefficients:
\begin{align}\label{eq:W1 to higher order OPE coeff;Dirichlet}
\frac{b(W_1^\alpha,\widehat{\mathsf{Q}}_{2n+6}^{\prime\,\alpha})}{c(\widehat{\mathsf{Q}}_{2n+6}^{\prime\,\alpha},\widehat{\mathsf{Q}}_{2n+6}^{\prime\,\alpha})}=\frac{\epsilon}{(N+8)\,(n+2)(2n+5)}\cdot \frac{b(\Phi_3^\alpha,\widehat{\mathsf{Q}}_{2n+6}^\alpha)}{c(\widehat{\mathsf{Q}}_{2n+6}^\alpha,\widehat{\mathsf{Q}}_{2n+6}^\alpha)}+O(\epsilon^2)\ .
\end{align}
Given the value of $\kappa$ \eqref{eq:value of kappa}, one can check these relations by comparing \eqref{eq:btb OPE of phi3 all order; Dirichlet} with the following BOE in taking $\epsilon\to0$ limit (axiom \ref{dcftaxiom2}):
\begin{align}
    \begin{aligned}
         W_3^\alpha(x)&\overset{\eqref{eq:classical EoM O(N)}}{=}\frac{1}{\kappa}\,\Box W_1^\alpha(x)\\
         &\overset{\eqref{eq:OPE of W1 all order;Dirichlet}}{\supset} \frac{b(W_1^\alpha,\widehat{\mathsf{Q}}_{2n+6}^{\prime\,\alpha})}{\kappa\cdot c(\widehat{\mathsf{Q}}_{2n+6}^{\prime\,\alpha},\widehat{\mathsf{Q}}_{2n+6}^{\prime\,\alpha})}\cdot \frac{[\Delta_1-\widehat{\Delta}(\widehat{\mathsf{Q}}_{2n+6}^{\prime\,\alpha})]\cdot [\Delta_1-\widehat{\Delta}(\widehat{\mathsf{Q}}_{2n+6}^{\prime\,\alpha})+1]}{|x_\perp|^{\Delta_1-\widehat{\Delta}(\widehat{\mathsf{Q}}_{2n+6}^{\prime\,\alpha})+2}}\cdot \widehat{\mathsf{Q}}_{2n+6}^{\prime\,\alpha}(\hat{x})\\
         &  \xrightarrow[\epsilon\to0]{}  \frac{b(W_1^\alpha,\widehat{\mathsf{Q}}_{2n+6}^{\prime\,\alpha})}{\epsilon\cdot c(\widehat{\mathsf{Q}}_{2n+6}^{\prime\,\alpha},\widehat{\mathsf{Q}}_{2n+6}^{\prime\,\alpha})}\cdot (N+8)\, (n+2)(2n+5)\cdot x_\perp^{2n+3} \cdot  \widehat{\mathsf{Q}}_{2n+6}^{\,\alpha}(\hat{x})\ .
    \end{aligned}
\end{align}

\paragraph{Study of bulk-boundary-boundary three-point functions involving $W_1^\alpha$.}
With the BOE of $W_1^\alpha$ \eqref{eq:OPE of W1 all order;Dirichlet} and the expression for the conformal block expansions \eqref{eq:conformal block expansion main}, the bulk-boundary-boundary three-point function $\langle\, W_1^\alpha\, \widehat{\CW}_{4p}\,\widehat{\CW}_{4p+2}^{\,\beta}\,\rangle$ becomes:
\begin{align}\label{eq: 3pt bdd, WilsonFisher;Dirichlet}
    \begin{aligned}
    \langle\,& W_1^\alpha(x)\, \widehat{\CW}_{4p}(0)\,\widehat{\CW}_{4p+2}^{\,\beta}(\infty)\,\rangle=\frac{1}{|x_\perp|^{\Delta_1}\, |x|^{\widehat{\Delta}_{4p}-\widehat{\Delta}_{4p+2}}}\\
    &\qquad\cdot\left[ D'\cdot c(\widehat{\CW}_2^{\,\alpha},\widehat{\CW}_{4p},\widehat{\CW}_{4p+2}^{\,\beta})\cdot G^{\widehat{\Delta}_{4p}-\widehat{\Delta}_{4p+2}}_{\widehat{\Delta}_2}(\upsilon)\right.\\
    &\qquad\qquad\left.  +\sum_{n=0}^{\infty}\,\frac{b(W_1^\alpha,\widehat{\mathsf{Q}}_{2n+6}^{\prime\,\alpha})\,c(\widehat{\mathsf{Q}}_{2n+6}^{\prime\,\alpha},\widehat{\CW}_{4p},\widehat{\CW}_{4p+2}^{\, \beta})}{c(\widehat{\mathsf{Q}}_{2n+6}^{\prime\,\alpha},\widehat{\mathsf{Q}}_{2n+6}^{\prime\,\alpha})}\cdot G^{\widehat{\Delta}_{4p}-\widehat{\Delta}_{4p+2}}_{\widehat{\Delta}(\widehat{\mathsf{Q}}_{2n+6}^{\prime\,\alpha})}(\upsilon)\right]\ .
    \end{aligned}
\end{align}
In what follows, we expand \eqref{eq: 3pt bdd, WilsonFisher;Dirichlet} at the first order in $\epsilon$ to obtain the closed-form expression.
 For the first term in the parenthesis of \eqref{eq: 3pt bdd, WilsonFisher;Dirichlet}, we have:
\begin{align}
    D'\cdot c(\widehat{\CW}_2^{\,\alpha},\widehat{\CW}_{4p},\widehat{\CW}_{4p+2}^{\,\beta})=c(\widehat{\Psi}_2^{\,\alpha},\widehat{\Psi}_{4p},\widehat{\Psi}_{4p+2}^{\,\beta})+O(\epsilon)=a_{p}\,  \delta^{\alpha\beta}+O(\epsilon)\ ,
\end{align}
and
\begin{align}
\begin{aligned}
    G^{\widehat{\Delta}_{4p}-\widehat{\Delta}_{4p+2}}_{\widehat{\Delta}_2}(\upsilon)&=\upsilon^{\widehat{\Delta}_2/2}\cdot {}_2F_1\left(\frac{\widehat{\gamma}_{2,1}+\widehat{\gamma}_{4p,1}-\widehat{\gamma}_{4p+2,1}}{2}\,\epsilon\, ,2; 3/2;\upsilon\right)+O(\epsilon^2) \\
    &\overset{\eqref{eq:hypergeometric series expansion}}{=} \upsilon^{\widehat{\Delta}_2/2} + \frac{\widehat{\gamma}_{2,1}+\widehat{\gamma}_{4p,1}-\widehat{\gamma}_{4p+2,1}}{3} \,\epsilon\cdot \upsilon^2\cdot h(\upsilon)\, +O(\epsilon^2)\ .
\end{aligned}
\end{align}
We have expanded the anomalous dimensions as follows:
\begin{align}
\widehat{\gamma}_{4p}=\widehat{\gamma}_{4p,1}\,\epsilon+\widehat{\gamma}_{4p,2}\,\epsilon^2+\cdots\ , \qquad \widehat{\gamma}_{4p+2}=\widehat{\gamma}_{4p+2,1}\,\epsilon+\widehat{\gamma}_{4p+2,2}\,\epsilon^2+\cdots\ .
\end{align}
We also denoted by $h(\upsilon)$ the following sum of two Gauss's hypergeometric functions:
\begin{align}\label{eq:function h def}
    h(\upsilon)={}_2F_1(1,2;5/2;\upsilon)+{}_2F_1(1,1;5/2;\upsilon)\overset{\eqref{eq:hypergeometric series expansion}}{=}\sum_{n=0}^{\infty}\,\frac{(1)_n\,(n+2)}{(5/2)_n}\cdot \upsilon^n\ .
\end{align}
Exploiting two relations for BOE coefficients \eqref{eq:bc_over_c_Dirichlet} and \eqref{eq:W1 to higher order OPE coeff;Dirichlet} as well as the hypergeometric identity:\footnote{This can be shown by expanding in powers of $\upsilon$ order by order.}
\begin{align}
 \sum_{n=0}^{\infty}\,\frac{(-1)^n\, (n+1)\,(4)_n}{(n+2)(2n+5)(n+9/2)_n} \cdot \upsilon^{n+1}\cdot {}_2F_1\left({n+2,n+4\atop 2n+11/2};\upsilon\right)=\frac{h(\upsilon) -2}{12}\ ,
\end{align}
one can verify that the second term of the parenthesis of \eqref{eq: 3pt bdd, WilsonFisher;Dirichlet} reduces to the following expression:
\begin{align} 
\begin{aligned}    \sum_{n=0}^{\infty}\,&\frac{b(W_1^\alpha,\widehat{\mathsf{Q}}_{2n+6}^{\prime\,\alpha})\,c(\widehat{\mathsf{Q}}_{2n+6}^{\prime\,\alpha},\widehat{\CW}_{4p},\widehat{\CW}_{4p+2}^{\, \beta})}{c(\widehat{\mathsf{Q}}_{2n+6}^{\prime\,\alpha},\widehat{\mathsf{Q}}_{2n+6}^{\prime\,\alpha})}\cdot G^{\widehat{\Delta}_{4p}-\widehat{\Delta}_{4p+2}}_{\widehat{\Delta}(\widehat{\mathsf{Q}}_{2n+6}^{\prime\,\alpha})}(\upsilon)
    \\
    &\qquad\qquad\qquad\qquad\qquad =
    \frac{2\,p \, a_p \, \delta^{\alpha\beta} }{N+8}\,\epsilon \cdot   \upsilon^2\cdot \left[h(\upsilon)-2\right]+O(\epsilon^2)\ .
\end{aligned}
\end{align}
Putting these all together, we end up with:
\begin{align}\label{eq:W1 3pt expansion 1; Dirichlet}
    \begin{aligned}
    \langle\,& W_1^\alpha(x)\, \widehat{\CW}_{4p}(0)\,\widehat{\CW}_{4p+2}^{\,\beta}(\infty)\,\rangle\\
    &=D'\cdot c(\widehat{\CW}_2^{\,\alpha},\widehat{\CW}_{4p},\widehat{\CW}_{4p+2}^{\,\beta})\cdot \frac{1}{|x_\perp|^{\Delta_1 -\widehat{\Delta}_2}|x|^{\widehat{\Delta}_2+\widehat{\Delta}_{4p}-\widehat{\Delta}_{4p+2}}}-\frac{4p\,a_p\,\delta^{\alpha\beta}}{N+8}\,\epsilon\cdot\frac{|x_\perp|^3}{|x|^2}  \\
     &\qquad +\frac{a_p\,\delta^{\alpha\beta}}{3\,(N+8)}\,\epsilon\cdot [(N+8)\,(\widehat{\gamma}_{2,1}+\widehat{\gamma}_{4p,1}-\widehat{\gamma}_{4p+2,1})+6p]\cdot \frac{|x_\perp|^3}{|x|^2}\cdot h(\upsilon)
+O(\epsilon^2)\ .
    \end{aligned}
\end{align}
It is straightforward to do a similarly analysis for $\langle\, W_1^\alpha\, \widehat{\CW}^{\, \beta}_{4p+2}\,\widehat{\CW}_{4p+4}\,\rangle$. The final result is:
\begin{align}\label{eq:W1 3pt expansion 2; Dirichlet}
    \begin{aligned}
    \langle\,& W_1^\alpha(x)\, \widehat{\CW}^{\, \beta}_{4p+2}(0)\,\widehat{\CW}_{4p+4}(\infty)\,\rangle\\
    &=D'\cdot c(\widehat{\CW}_2^{\,\alpha},\widehat{\CW}^{\, \beta}_{4p+2},\widehat{\CW}_{4p+4})\cdot \frac{1}{x_\perp^{\Delta_1 -\widehat{\Delta}_2}|x|^{\widehat{\Delta}_2+\widehat{\Delta}_{4p+2}-\widehat{\Delta}_{4p+4}}} \\
     &\quad +\frac{b_p\, \delta^{\alpha\beta}}{3\,(N+8)}\,\epsilon\cdot 
[(N+8)\,(\widehat{\gamma}_{2,1}+\widehat{\gamma}_{4p+2,1}-\widehat{\gamma}_{4p+4,1})\,+(N+6p+2) ]\\
&\qquad\qquad\qquad\qquad\cdot \frac{|x_\perp|^3}{|x|^2}\cdot h(\upsilon) -\frac{2\,(N+6p+2)\,b_p\, \delta^{\alpha\beta}}{3\,(N+8)}\,\epsilon\cdot\frac{|x_\perp|^3}{|x|^2} +O(\epsilon^2)\ .
    \end{aligned}
\end{align}

\paragraph{Constraint from analyticity.}
The two correlation functions we have derived so far $ \langle\, W_1^\alpha, \widehat{\CW}_{4p}\,\widehat{\CW}_{4p+2}^{\,\beta}\,\rangle$ \eqref{eq:W1 3pt expansion 1; Dirichlet} and $\langle\, W_1^\alpha\, \widehat{\CW}^{\, \beta}_{4p+2}\,\widehat{\CW}_{4p+4}\,\rangle$ \eqref{eq:W1 3pt expansion 2; Dirichlet} exhibit unphysical singularities due to the asymptotic form of $h(\upsilon)$ introduced in \eqref{eq:function h def}:
\begin{align}
    h(\upsilon) \xrightarrow[\text{with \eqref{eq:Kummer's connection formula}}]{|\hat{x}|\sim 0}  \frac{3\pi}{4}\cdot\frac{|x_{\perp}|}{|\hat{x}|} +\cdots\ .
\end{align}
All Euclidean correlators should be analytic away from the coincidence of points owing to the Euclidean QFT axioms \cite{Osterwalder:1973dx,Osterwalder:1974tc}.
All operators in \eqref{eq:W1 3pt expansion 1; Dirichlet} and \eqref{eq:W1 3pt expansion 2; Dirichlet} are still distant from each other when $|\hat{x}|=0$, but we have singularities, resulting in a contradiction.
To remove these unphysical singularities associated with the limiting behaviors as $|\hat{x}|\to0$, we require that:
\begin{align}
    (N+8)\,(\widehat{\gamma}_{2,1}+\widehat{\gamma}_{4p,1}-\widehat{\gamma}_{4p+2,1})+6p&=0\ ,\label{eq:Dirichlet axiom rec 1}\\
    (N+8)\,(\widehat{\gamma}_{2,1}+\widehat{\gamma}_{4p+2,1}-\widehat{\gamma}_{4p+4,1})\,+(N+6p+2)&=0\ .\label{eq:Dirichlet axiom rec 2}
\end{align}
Using \eqref{eq:anomalous dim lowest Dir} as the initial condition $\widehat{\gamma}_{2,1}=-\frac{N+2}{2\,(N+8)}$, one can solve these two recursion relations \eqref{eq:Dirichlet axiom rec 1} and \eqref{eq:Dirichlet axiom rec 2} to give the same results as in the standard perturbative approach \eqref{eq:Dir diagram cal anomalous dimensions}.

\chapter{Conclusion}\label{chap:conclusion}
In this thesis, we first described the uses of conformal symmetry in QFTs with and without defects (chapter \ref{chap:Foundations of conformal symmetry without defects} and \ref{chap:Elements of Defect Conformal Field Theory}). In chapter \ref{chap:Review of Rychkov-Tan}, we introduced the axiomatic framework to explore the critical phenomena while comparing it with the standard perturbative approach. We then studied the critical O$(N)$ model near four dimensions in the presence of three types of defects: line defect, Neumann and Dirichlet boundary (chapter \ref{chap:line defect}, \ref{chap:Neumann boundary} and \ref{chap:Dirichlet}). In particular, based on the axiomatic framework, we analyzed the model leveraging DCFT techniques to confirm agreement with the standard perturbative calculations. Below, we summarize the main result of the thesis and then give an outlook for the future.

For the model with a line defect (chapter \ref{chap:line defect}), we have shown that the critical defect coupling can be fixed uniquely from our axioms. Though the equation of motion only specifies the relation among bulk operators, one can exploit DOEs to propagate that constraint onto the defect coupling, reproducing the same result as the perturbative calculations \cite{Cuomo:2021kfm}. It is effortless to derive the conformal dimensions of the non-composite operators in the axiomatic approach because they appear in the DOE of the free O$(N)$ vector field $\Phi_1^\alpha$. However, this strategy does not apply to composite operators absent in the DOE of $\Phi_1^\alpha$. Nevertheless, we pointed out that it is still possible to determine conformal dimensions of composite operators within the axiomatic framework by requiring the resolution of unphysical non-analyticities of bulk-defect-defect three-point functions, giving consistent results with the known research.

We did a similar analysis for the case with Neumann and Dirichlet boundaries in chapter \ref{chap:Neumann boundary} and \ref{chap:Dirichlet}. As we could not find the conformal dimensions of the higher boundary composite operators in existing literature, we also performed standard perturbative calculations in both cases.
The perfect matching between the results of the two different approaches at the leading order in $\epsilon$ confirms the overall utility of the axioms, including the defect conformal symmetry and the bulk equation of motion \eqref{eq:classical EoM O(N)} in $(4-\epsilon)$ dimensions.

Relevant defect local operators having conformal dimensions smaller than the dimension of the defect have dynamical information about the defect since their conformal dimensions are related to defect critical exponents and are also to the stability of critical points under the defect RG flow in the theory space of coupling constants \cite{Kobayashi:2018lil}. The higher-order defect composite operators we derived in this thesis may seem unimportant as they are irrelevant defect local operators insensitive to defect critical exponents and the defect RG flow. Nevertheless, one may use them as input data for solving crossing equations to refine the conformal bootstrap approach in the presence of a defect. We hope our findings contribute to more precise predictions of critical exponents and a more conceptual understanding of extended objects in our world.

Roughly speaking, one of the advantages of the axiomatic approach over the conventional perturbative framework combined with RG analysis is that one can reduce the calculations to combinatorial problems. We only need some DCFT techniques and free DCFT data that are tractable via Wick's theorem without resorting to the standard diagrammatic calculations, which get more painstaking when the defect breaks translational invariance. We are sure that our approach is much simpler than the standard perturbative calculations in studying defect local non-composite operators and critical defect coupling constants. For defect local composite operators, it depends on personal preference whether to choose our approach or the conventional perturbative approach. Nevertheless, the axiomatic approach should still serve as a nice consistency check. It is worthwhile noting that our analysis has extended the potential applications of the analyticity of correlators, though it is a rather technical development. No one has succeeded in extending Rychkov-Tan's axiomatic framework to higher orders in $\epsilon$, even for the cases without defects. Despite this disadvantage, our axiomatic framework is still quite a shortcut to studying critical phenomena and has the potential application to various models with defects.

Unlike the line defect made by smearing the first component of the O$(N)$ vector along one spacetime direction, we imposed Neumann and Dirichlet boundary conditions uniformly for the O$(N)$ vector field in this thesis, and the models end up with no internal symmetry breaking on the boundary. Let us consider the case in which we impose the Neumann boundary condition for the first $m$ components of the O$(N)$ vector field $\Phi_{1}^{\alpha'}$ ($\alpha'=1,\cdots, m$) and Dirichlet boundary condition for the rest $\Phi_{1}^{\alpha''}$ ($\alpha''=m+1,\cdots, N$):
\begin{align}
\begin{aligned}
        \lim_{|x_\perp|\to0}\,\frac{\partial}{\partial x_\perp}\,\Phi_{1}^{\alpha'}(x)&=0\qquad \text{for }\alpha'=1,\cdots, m\ ,\\
           \lim_{|x_\perp|\to0}\,\Phi_{1}^{\alpha''}(x)&=0\qquad \text{for }\alpha''=m+1,\cdots, N\ .
\end{aligned}    
\end{align}
This mixed boundary condition causes the following internal symmetry breaking on the boundary:
\begin{align}
    \mathrm{O}(N)\to \mathrm{O}(m)\times \mathrm{O}(N-m)\ .
\end{align}
In such a case, one can identify the tilt operators associated with this symmetry breaking with the transverse component of the broken currents, according to the argument in section \ref{sec:Bulk and defect local primaries}. Recall that the conserved current of the O$(N)$ global symmetry is given by $J^{\alpha\beta}_\mu=\Phi_1^{\alpha}\,\partial_\mu\,\Phi_1^{\beta}-\Phi_1^{\beta}\,\partial_\mu\,\Phi_1^{\alpha}$ for $\alpha,\beta=1,\cdots,N$ in free theory. Then, the tilt operators should be expressed by $\hat{t}^{\alpha'\alpha''}=\widehat{\Phi}_1^{\alpha'}\,\partial_\perp\,\widehat{\Phi}_1^{\alpha''}$ when no interactions are present. Even with interactions, the conformal dimensions of the tilt operators should always be $(d-1)$ from the anomalous conservation law \eqref{eq:def of tilt} and receive no quantum corrections as the stress tensor and the displacement operator. It will be interesting to do the standard perturbative calculations or apply the axiomatic framework for the model to check this speculation.

As a byproduct of our analysis, we invented a systematic way to uncover the structure of free DCFTs. We studied various correlation functions of free DCFTs via Wick's theorem and extracted several DOE spectra. In particular, we demonstrated that, unlike Klein-Gordon fields, the DOE contents of generic operators are complicated but still accessible by looking at the conformal block expansion of bulk-defect-defect three-point functions. We expect this methodology and knowledge will be instructive for future readers eager to know more about the generic structures of DCFTs. We did not focus on other correlators that depend on cross ratios, such as bulk two-point functions, bulk-bulk-defect three-point functions, and defect four-point functions. It will be interesting to apply our results to the analysis of these correlators for further extraction of the free DCFT data.

One can employ our axiomatic approach to critical phenomena with defects if and only if there exist corresponding free DCFTs.
Hence, our axiomatic framework does not apply to the extraordinary transition that is also supposed to be described by BCFT, as it fails to have corresponding free theories in four dimensions.\footnote{There are three kinds of boundary critical phenomena depending on whether boundary or bulk orders first: ordinary, special, and extraordinary transition. The critical O$(N)$ model with Dirichlet and Neumann boundary, which we studied in the main text, describes the first two cases. On the other hand, the extraordinary transition breaks the $\mathbb{Z}_2$ symmetry $\Phi_1^\alpha\leftrightarrow -\Phi_1^\alpha$, and the bulk one-point function of the O$(N)$ vector field acquires non-zero expectation value. As argued in section \ref{sec:Defect Operator Expansion spectrum of free scalar field}, this contradicts the Klein-Gordon equation in four dimensions. Therefore, we have no free BCFTs in four-dimensional spacetime corresponding to the extraordinary transition. \label{fot:on extraordinary}} Likewise, one cannot apply our axiomatic framework directly to the recently proposed surface defect in the critical O$(N)$ model \cite{Trepanier:2023tvb,Giombi:2023dqs}.
Nevertheless, even in such cases, the concept of the equation of motion due to the multiplet recombination is still available \cite[appendix B.4]{Liendo:2012hy}, and one can combine the equation of motion with conformal bootstrap techniques to study DCFT as demonstrated in \cite{Herzog:2022jlx}, instead of conventional perturbative approach.
It will be interesting to pursue this direction further with the analytical constraints from correlation functions in mind.

\chapter*{Acknowledgment}
Firstly, I am deeply indebted to my supervisor Tatsuma Nishioka. He was an assistant professor when I joined the theoretical particle physics group at Tokyo University. He won rapid promotion and became an associate professor at Yukawa Institute for Theoretical Physics when I was about to get a master's degree and got a position as a professor at Osaka University a year later.
It was a precious experience for me, as a young graduate student, to get involved in three different environments.
 I could learn a lot in his company and through collaboration with him, such as proper deportment and what it means to work as a professional scientist in academia. My scientific interest had gradually deviated from what it was when I entered graduate school. Besides, I am the kind of person who does not listen to others and rarely goes to the office.
So, I suppose it was not easy for him to handle me.
It is his patience and kindness that have made it possible to complete this dissertation.

Going through three different research groups: Tokyo University, Yukawa Institute for Theoretical Physics, and Osaka University, I got acquainted with many people there. I would also like to express my gratitude to all of them. In particular, I would like to thank my collaborators Nozomu Kobayashi, Kento Watanabe, Kohki Kawabata, Soichiro Shimamori, and Tatsuma Nishioka. I am also grateful to my formal supervisor Yutaka Matsuo, for his continuous support and encouragement, though I was distant from him except for the first year in graduate school.

I also thank all financial support and grants both from our government and a private company: Forefront Physics and Mathematics Program to Drive Transformation (FoPM), a World-leading Innovative Graduate Study (WINGS) Program, the University of Tokyo, JSPS fellowship for young students, MEXT, and JSR fellowship, the University of Tokyo. All these add up to the foundations of my learning and inspired me to become more innovative in my search for knowledge throughout my Ph.D. journey.

Last but not least, I would like to thank my family, friends, and all the others involved who have decorated my private life. It is too hard to name just a few, and I would be rude if I put them in order or emphasize a particular person. What I owe them is beyond evaluation.

\appendix

\renewcommand{\appendixname}{Appendix}
\chapter{Useful identities}\label{app:Useful identities}
We here list several identities used in the main text.

\paragraph{Kummer's connection formula for Gauss's hypergeometric function.}
\begin{align}\label{eq:Kummer's connection formula}
    \begin{aligned}
    {}_2F_1(\alpha,&\beta;\gamma;z)=\frac{\Gamma(\gamma)\Gamma(\gamma-\alpha-\beta)}{\Gamma(\gamma-\alpha)\Gamma(\gamma-\beta)}\cdot{}_2F_1(\alpha,\beta;\alpha+\beta-\gamma+1;1-z)\\
    &+\frac{\Gamma(\gamma)\Gamma(\alpha+\beta-\gamma)}{\Gamma(\alpha)\Gamma(\beta)}\cdot (1-z)^{\gamma-\alpha-\beta}\cdot {}_2F_1(\gamma-\alpha,\gamma-\beta;\gamma-\alpha-\beta+1;1-z)\ .
    \end{aligned}
\end{align}

\paragraph{Decomposition of unity in terms of Gauss's hypergeometric functions.}
\begin{align}\label{eq:hypergeometric identity 1}
    1=\sum_{n=0}^{\infty}\,\frac{(-1)^n\,(\alpha)_n(\beta)_n}{(n+\lambda)_n\,n!}\cdot z^n\cdot {}_2F_1\left({\alpha+n,\beta+n\atop\lambda+1+2n};z\right)\ .
\end{align} 
This follows immediately from \cite[equation (9.1.32)]{luke1969special} by setting $t=2,u=0$:
\begin{align}
    \begin{aligned}
        1&=\sum_{n=0}^{\infty}\,\frac{(-1)^n\,(\gamma_1)_n\cdots (\gamma_t)_n}{(\sigma_1)_n\cdots (\sigma_u)_n\,(n+\lambda)_n\,n!}\\
        &\qquad\cdot z^n\cdot{}_tF_{u+1}\left({\gamma_1+n,\cdots,\gamma_t+n\atop \lambda+1+2n,\sigma_1+n,\cdots,\sigma_u+n};z\right)\ ,
    \end{aligned}
\end{align}
where $z\in\mathbb{C}$ for $1\leq t<u+1$ and $z\neq1,|\arg(1-z)|<\pi$ for $1\leq t=u+2$.
Here, ${}_pF_q$ is a generalized hypergeometric function defined by the power series:\footnote{The series expansion of the generalized hypergeometric function ${}_pF_q$ converges in the entire complex $z$-plain for $p\leq q$. When $q=p+1$, this is convergent as a power series only for $|z|<1$, but is analytically continued for $|\arg(1-z)|<\pi$. The convergence on the unit circle $|z|=1$ is conditional depending on $\chi_p\equiv\mathrm{Re}[(\beta_1+\cdots +\beta_{p+1})-(\alpha_1+\cdots+\alpha_p)]$: absolutely convergent for $\chi_p>0$, convergent except at $z=1$ for $-1<\chi_p\leq 0$ and divergent for $\chi_p\leq -1$.
}
\begin{align}
{}_pF_q\left({\alpha_1,\cdots ,\alpha_p\atop \beta_1,\cdots ,\beta_q};z\right)=\sum_{n=0}^{\infty}\,\frac{(\alpha_1)_n\cdots (\alpha_p)_n}{(\beta_1)_n\cdots (\beta_q)_n\, n!}\cdot z^n\ .
\end{align}

\paragraph{Decomposition of Gauss's hypergeometric function in terms of Gauss's hypergeometric functions.}
\begin{align}\label{eq:hypergeometric identity 2}
    \begin{aligned}
        {}_{2}F_{1}(\alpha,\beta;\gamma;z)&=\sum_{n=0}^{\infty}\,\frac{(-1)^n\, (\alpha)_n\,(\lambda-\gamma+1)_n(\gamma-\beta)_n}{(\gamma)_n\, (n+\lambda)_n\,n!}\cdot z^n\\
        &\qquad\qquad\cdot {}_{2}F_{1}\left({\alpha+n,\beta+\lambda-\gamma+1+n\atop\lambda+1+2n};z\right)\ .
    \end{aligned}
\end{align}
We start with equation (9.1.13) in \cite{luke1969special}:
\begin{align}\label{eq:gen hyp gen hyp}
    \begin{aligned}
        {}_{p+r}F_{q+s}&\left({a_1,\cdots ,a_p, c_1,\cdots ,c_r \atop b_1,\cdots b_q, d_1,\cdots ,d_s };z \omega\right)\\
        &\qquad\qquad=\sum_{n=0}^{\infty}\,\frac{(-1)^n\, (\alpha_1)_n\cdots (\alpha_t)_n \,(a_1)_n\cdots (a_p)_n }{ (\beta_1)_n\cdots (\beta_u)_n \,(b_1)_n\cdots (b_q)_n\,(n+\lambda)_n\,n!}\cdot z^n\\
        &\cdot  {}_{p+t}F_{q+u+1}\left({\alpha_1+n,\cdots ,\alpha_t+n, a_1+n,\cdots ,a_p+n \atop \lambda+1+2n , \beta_1+n,\cdots, \beta_u+n, b_1+n,\cdots ,b_q+n };z \right)\\
        &\qquad\qquad\cdot {}_{r+u+2}F_{s+t}\left({-n,n+\lambda,c_1,\cdots ,c_r, \beta_1,\cdots ,\beta_u \atop \alpha_1 , \cdots,\alpha_t, d_1,\cdots ,d_s };\omega \right)\ .
    \end{aligned}
\end{align}
By setting $p=t=r=s=1$, $q=u=0$ and $\omega=1$, we have:
\begin{align}
    \begin{aligned}
        {}_{2}F_{1}(\alpha,\beta;\gamma;z)&=\sum_{n=0}^{\infty}\,\frac{(-1)^n\, (\alpha)_n\,(\rho)_n}{ (n+\lambda)_n\,n!}\cdot  {}_{3}F_{2}\left({-n,n+\lambda,\beta \atop \rho,\gamma };1 \right)\\
        &\cdot z^n\cdot {}_{2}F_{1}(\alpha+n,\rho+n;\lambda+1+2n;z)\ .
    \end{aligned}
\end{align}
When $\rho=\lambda+\beta-\gamma+1$, we can make used of Saalschütz's theorem \cite[page 9]{bailey1935generalized}
\begin{align}
    {}_3F_2\left({a,b,-n\atop c,1+a+b-c-n};1\right)=\frac{(c-a)_n(c-b)_n}{(c)_n(c-a-b)_n}\ ,\qquad n=0,1,2,\cdots \ ,
\end{align}
 to simplify ${}_3F_2$ as follows:
 \begin{align}
     {}_{3}F_{2}\left({-n,n+\lambda,\beta \atop \lambda+\beta-\gamma+1,\gamma };1 \right)=\frac{(\lambda-\gamma+1)_n(\gamma-\beta)_n}{(\lambda+\beta-\gamma+1)_n(\gamma)_n}\ .
 \end{align}
Employing the relation $(z-n)_n=(-1)^n\,(1-z)_n$ that follows immediately from Euler's reflection formula $\Gamma(z)\Gamma(1-z)=\pi/\sin\pi z$, we conclude \eqref{eq:hypergeometric identity 2}.

An alternative way to show \eqref{eq:hypergeometric identity 2} is to set $p=q=u=0,s=1,r=t=2,\omega=1$ in \eqref{eq:gen hyp gen hyp}:
\begin{align}
    \begin{aligned}
        {}_{2}F_{1}(\alpha,\beta;\gamma;z)&=\sum_{n=0}^{\infty}\,\frac{(-1)^n\, (\sigma)_n\,(\rho)_n}{ (n+\lambda)_n\,n!}\cdot{}_{4}F_{3}\left({-n,n+\lambda,\alpha,\beta \atop \sigma,\rho,\gamma };1 \right)\\
        &\cdot z^n\,{}_{2}F_{1}(\sigma+n,\rho+n;\lambda+1+2n;z)\ ,
    \end{aligned}
\end{align}
and to use appropriate reduction formulas for ${}_{4}F_{3}$ after specifying arguments to particular values.

\paragraph{Integral over half-space.}
\begin{align}\label{eq:Integral over half-space}
\begin{aligned}
          & \int_{\mathbb{R}^{d}_+} \d^d x\, \frac{1}{|x_\perp|^{2\alpha}\, |x|^{2\beta}\, |x-\hat{y}|^{2\gamma}}\\
           &\quad= \frac{\pi^{\frac{d-1}{2}}\,\Gamma(\frac{1}{2}-\alpha)\, \Gamma(\alpha+\beta+\gamma-\frac{d}{2})\, \Gamma(\frac{d}{2}-\alpha-\gamma)\, \Gamma(\frac{d}{2}-\alpha-\beta)}{2\, \Gamma(\beta)\, \Gamma(\gamma)\, \Gamma(d-2\alpha-\beta-\gamma)}\cdot\frac{1}{|\hat{y}|^{2\alpha+2\beta+2\gamma-d}} \ ,
\end{aligned}
\end{align}
where $\beta\neq0,\gamma\neq0$ and the integral is over $d$-dimensional half-space:
\begin{align}
   \int_{\mathbb{R}^{d}_+} \d^d x= \int_{\mathbb{R}^{d-1}} \d^{d-1} \hat{x}\,\int_{0}^{\infty}\d x_{\perp}\ .
\end{align}
Our starting point to derive \eqref{eq:Integral over half-space} is to use Schwinger parametrization:
\begin{align}\label{eq:Schwinger parametrization}
    \frac{1}{x^{\Delta}}=\frac{1}{\Gamma(\Delta)}\,\int_0^\infty\frac{\d t}{t}\,t^{\Delta}\,e^{-t x}\ ,
\end{align}
which allows to rewrite the left-hand side of \eqref{eq:Integral over half-space} as follows:
\begin{align}
\begin{aligned}
        \int_{\mathbb{R}^{d}_+} \d^d x&\, \frac{1}{|x_\perp|^{2\alpha}\, |x|^{2\beta}\, |x-\hat{y}|^{2\gamma}}\\
        &=\frac{1}{\Gamma(\beta)\,\Gamma(\gamma)}\cdot\int_0^\infty\frac{\d s}{s}\,s^{\beta}\,\int_0^\infty\frac{\d t}{t}\,t^{\gamma}\, e^{-\frac{st}{s+t}\,|\hat{y}|^2}\\
        &\qquad\cdot\int_{\mathbb{R}^{d-1}} \d^{d-1} \hat{x}\,e^{-(s+t)\,\left|\hat{x}+\frac{t}{s+t}\,\hat{y}\right|^2}\,\int_{0}^{\infty}\d x_{\perp}\, \frac{1}{|x_\perp|^{2\alpha}}\,e^{-(s+t)|x_\perp|^{2}}\\
           &=\frac{\pi^{\frac{d-1}{2}}\,\Gamma\left(1/2-\alpha\right)}{2\,\Gamma(\beta)\,\Gamma(\gamma)}\cdot\int_0^\infty\frac{\d s}{s}\int_0^\infty\frac{\d t}{t}\cdot\frac{s^{\beta}\,t^{\gamma}\,e^{-\frac{st}{s+t}\,|\hat{y}|^2}}{(s+t)^{d/2-\alpha}}  \ .
\end{aligned}
\end{align}
The remaining integrals are doable by the change of integration variables $s=ru,t=r(1-u)$ as follows:
\begin{align}
\begin{aligned}
\int_0^\infty\frac{\d s}{s}&\,\int_0^\infty\frac{\d t}{t}\cdot \frac{s^{\beta}\,t^{\gamma}\,e^{-\frac{st}{s+t}\,|\hat{y}|^2}}{(s+t)^{d/2-\alpha}} \\
&=\int_0^1\d u\,u^{\beta-1}(1-u)^{\gamma-1}\,\int_0^\infty \frac{\d r}{r} \, r^{\alpha+\beta+\gamma-d/2}\, e^{-ru(1-u)\,|\hat{y}|^2}\\
&=\frac{\Gamma\left(\alpha+\beta+\gamma-\frac{d}{2}\right)}{|\hat{y}|^{2\alpha+2\beta+2\gamma-d}}\,\int_0^1\d u\,u^{d/2-\alpha-\gamma-1}(1-u)^{d/2-\alpha-\beta-1}\\
&=\frac{\Gamma\left(\alpha+\beta+\gamma-\frac{d}{2}\right)\,\Gamma\left(\frac{d}{2}-\alpha-\beta\right)\,\Gamma\left(\frac{d}{2}-\alpha-\gamma\right)}{\Gamma(d-2\alpha-\beta-\gamma)}\cdot\frac{1}{|\hat{y}|^{2\alpha+2\beta+2\gamma-d}}\ .
\end{aligned}
\end{align}

\chapter{A review of unitarity bound}\label{app:unitatiry bound}
We here leave a comprehensive review of unitarity bound from a Lorentzian perspective, which attracts much less attention since the original derivation due to \cite{Mack:1975je,Dobrev:1977qv} except a partial review by \cite{Kologlu:2019bco}. Interested readers are referred to \cite{Minwalla:1997ka} in which the author derives unitarity bound from the reflection positivity in Euclidean signature combined with the state/operator correspondence. Our approach relies only on the Wightman positivity of Lorentzian correlators. Hence, given the forms of correlation functions in CFT, we do not need conformal symmetry further to prove unitarity bound.\footnote{One can apply our method to obtain positivity constraints in a class of quantum field theories that are scale invariant but not conformal invariant. Let us consider scalar two-point correlators as an illustration. Though one fails to have the orthogonality of scalar operators with different conformal dimensions without special conformal invariance, scale invariance is enough to fix the scalar two-point correlation function to have the following form in Euclidean signature:
\begin{align}
    \langle\,\CO_{\Delta'_1}(x_1)\,\CO_{\Delta'_2}(x_2)\,\rangle=\frac{c(\CO_{\Delta'_1},\CO_{\Delta'_2})}{|x_{12}|^{\Delta'_1+\Delta'_2}}\ ,
\end{align}
where we have denoted the scaling dimension of the scalar operator $\CO_{\Delta'_i}$ $(i=1,2)$ by $\Delta'_i$.
By utilizing Wightman positivity, we end up with the same unitarity bound for the scaling dimensions $\Delta'\geq d/2-1$ as in CFTs \eqref{eq:scalar unitarity bound d>2}.
} 

We first explain the analytic properties of correlation functions in Euclidean and Lorentzian QFTs following \cite{Haag:1992hx,Streater:1989vi} (section \ref{app:Remarks on analytic properties of correlation functions in QFT}). We then develop some group theoretical technologies in section \ref{app:Harmonic function and symmetric traceless tensor} and \ref{app:projector onto SO(d-1)}. In section \ref{app:Fourier transformation of two-point Wightman distributions}, we perform Fourier transformation of two-point Lorentzian correlators. Section \ref{app:Unitarity bound from Wightman positivity} is for derivations of unitary bound for scalar primaries and symmetric traceless tensors.

\section{Remarks on analytic properties of correlation functions in quantum field theory}\label{app:Remarks on analytic properties of correlation functions in QFT}

\paragraph{Lorentzian correlators (Wightman distributions).}
Let us consider $d$-dimensional Minkowski spacetime $\mathbb{R}^{1,d-1}$ whose coordinates are denoted by $\tilde{x}^{\tilde{\mu}}$ ($\tilde{\mu}=0,1,\cdots, d-1$):
\begin{align}
\tilde{x}^{\tilde{\mu}}=(\tilde{x}^0,\tilde{x}^1,\cdots \tilde{x}^{d-1})=(t,\vec{x})\ , \qquad \vec{x}\in\mathbb{R}^{d-1}\ ,
\end{align}
with the metric:
\begin{align}
    \d^2\tilde{s}=\eta_{\tilde{\mu}\tilde{\nu}}\,\d\tilde{x}^{\tilde{\mu}}\d\tilde{x}^{\tilde{\nu}}=-\d t^2+\d \vec{x}^{\,2}\ .
\end{align}
We denote an $n$-point scalar correlation function on $\mathbb{R}^{1,d-1}$ by:
\begin{align}
    W_n(\tilde{x}_1,\cdots,\tilde{x}_n)=\langle\Omega| \,\CO_{1}(\tilde{x}_1)\,\CO_{2}(\tilde{x}_2)\,\cdots\, \CO_{n}(\tilde{x}_n)\,|\Omega\rangle \ .
\end{align}
We require the vacuum of QFT $|\Omega\rangle$ to be invariant under Poincar\'e transformations, a semi-direct product of translations and Lorentz transformations: $\mathbb{R}^{1,d-1}\rtimes\mathrm{SO}(1,d-1)$. That is, the QFT vacuum $|\Omega\rangle$ is annihilated by Poincar\'e generators $\widetilde{\mathbf{P}}_{\tilde{\mu}}$ and $\widetilde{\mathbf{M}}_{\tilde{\mu}\tilde{\nu}}$:
\begin{align}
\widetilde{\mathbf{P}}_{\tilde{\mu}}\,|\Omega\rangle=0\ ,\qquad \widetilde{\mathbf{M}}_{\tilde{\mu}\tilde{\nu}}\,|\Omega\rangle=0\ ,
\end{align}
These generators are Hermitian: $(\widetilde{\mathbf{P}}^{\tilde{\mu}})^\dagger=\widetilde{\mathbf{P}}^{\tilde{\mu}},(\widetilde{\mathbf{M}}^{\tilde{\mu}\tilde{\nu}})^\dagger=\widetilde{\mathbf{M}}^{\tilde{\mu}\tilde{\nu}}$ and the zero-th component of $\widetilde{\mathbf{P}}^{\tilde{\mu}}$ is nothing but the Hamiltonian of the theory: $\widetilde{\mathbf{P}}^0=H$.
We further require the spectrum condition of the translation generators $\widetilde{\mathbf{P}}^{\tilde{\mu}}$ on any physical state $|\Psi\rangle$:
\begin{align}\label{eq:spectrum condition for generator}
\langle\Psi|\,\widetilde{\mathbf{P}}^{\tilde{\mu}}\,|\Psi\rangle\geq 0\ .
\end{align}
Here, we have introduced an order between two Minkowski coordinates: $\tilde{x}_1^{\tilde{\mu}}\geq \tilde{x}_2^{\tilde{\mu}}$, meaning that $\tilde{x}_1^{\tilde{\mu}}$ is in the forward light-cone of $\tilde{x}_2^{\tilde{\mu}}$:
\begin{align}
t_1\geq t_2\ ,\qquad -\tilde{x}_{12}^2= t_{12}^2-|\vec{x}_{12}|^2\geq 0\ .
\end{align}

Lorentzian correlators are oscillatory and not convergent in general and are termed the Wightman or tempered distributions. However, one can regard them as boundary values of holomorphic functions. To see this, let us complexify the spacetime coordinates $\tilde{x}^{\tilde{\mu}}$ as:
\begin{align}
    \tilde{x}^{\tilde{\mu}}= \xi^{\tilde{\mu}}-\i\, \zeta^{\tilde{\mu}}\ ,\qquad \xi^{\tilde{\mu}},\zeta^{\tilde{\mu}}\in\mathbb{R}^{1,d-1}\ .
\end{align}
From the transformation properties of fields $ \CO(x)=e^{-\i\,\widetilde{\mathbf{P}}\cdot \tilde{x}} \,\CO(0)\, e^{+\i\,\widetilde{\mathbf{P}}\cdot \tilde{x}}$, the $n$-point scalar correlation function $W_n(\tilde{x}_1,\cdots,\tilde{x}_n)$ can be expanded as:
\begin{align}\label{eq:n-pt wightman gen}
\begin{aligned}
           W_n(\tilde{x}_1,\cdots,\tilde{x}_n)&=\langle\Omega|  \,\CO_1(0) \,e^{\i\, \widetilde{\mathbf{P}}\cdot(\xi_1-\xi_{2})+\widetilde{\mathbf{P}}\cdot (\zeta_1-\zeta_{2})}\,\CO_2(0)\,e^{\i\, \widetilde{\mathbf{P}}\cdot(\xi_2-\xi_{3})+\widetilde{\mathbf{P}}\cdot (\zeta_2-\zeta_{3})}\\
           &\qquad\qquad\qquad\qquad\qquad\,\cdots\, e^{\i\, \widetilde{\mathbf{P}}\cdot(\xi_{n-1}-\xi_{in})+\widetilde{\mathbf{P}}\cdot (\zeta_{n-1}-\zeta_{n})} \,\CO_n(0) \,|\Omega\rangle\ .
\end{aligned}
\end{align}
Then, the $n$-point Wightman distribution turns out to be convergent and holomorphic on the domain:
\begin{align}\label{eq:Wn analytic dom}
    \zeta_1>\zeta_2>\cdots > \zeta_n\ ,
\end{align}
due to the exponentially dumping factors and the spectrum condition for $\widetilde{\mathbf{P}}^{\tilde{\mu}}$ \eqref{eq:spectrum condition for generator}. Here, for coordinates on Minkowski space, $\tilde{x}_1>\tilde{x}_2$ means that $t_1> t_2$ and $-\tilde{x}_{12}^2= t_{12}^2-|\vec{x}_{12}|^2> 0$. As we recover the un-complexified correlator by setting $\zeta_i$'s to zero keeping the condition \eqref{eq:Wn analytic dom}, we can regard the Lorentzian correlator $W_n(\tilde{x}_1,\cdots,\tilde{x}_n)$ as the boundary value of the holomorphic function. 

To evaluate the tempered distributions, we must smear them with some test functions called Schwartz functions to get finite answers. Let $f(u)$ be a Schwartz function with one variable. Then, it is well-behaved functions satisfying the following property:
\begin{align}
    u^n\,\frac{\partial^m}{\partial u^m}\,f(u) \quad \,\text{is bounded above for arbitrary }n,m\in\mathbb{Z}_{\geq0}\ .
\end{align}
Gaussian $f(u)=e^{-u^2}$ is one of the examples of such functions.
Consider a general finite linear combination of quantum states created by acting operators on the vacuum and integrating them against Schwartz functions:
\begin{align}    |\Phi\rangle=\sum_{n}\,\int_{\mathbb{R}^{1,d-1}}\,\d\tilde{x}_1\cdots \int_{\mathbb{R}^{1,d-1}}\,\d\tilde{x}_n\,f_1(\tilde{x}_1)\,\cdots \,f_n(\tilde{x}_n)\,\CO_1(\tilde{x}_1)\,\cdots \,\CO_n(\tilde{x}_n)\,|\Omega\rangle\ .
\end{align}
One can define the state conjugate to $|\Phi\rangle$ through the relation:
\begin{align}
\begin{aligned}
        \langle\Phi|&=[|\Phi\rangle]^\dagger\\     &=\sum_{n}\,\int_{\mathbb{R}^{1,d-1}}\,\d\tilde{x}_1\cdots \int_{\mathbb{R}^{1,d-1}}\,\d\tilde{x}_n\,[f_1(\tilde{x}_1)]^\ast\,\cdots \,[f_n(\tilde{x}_n)]^\ast\,\langle\Omega|\,\CO_1(\tilde{x}_1)\,\cdots \,\CO_n(\tilde{x}_n) .
\end{aligned}
\end{align}
Wightman positivity requires that the norm of any quantum state be positive. Hence, one has the following positivity condition:
\begin{align}\label{eq:Wightman positivity most gen}
    \langle\Phi|\Phi\rangle\geq 0\ ,
\end{align}
for arbitrary Schwartz functions $f_n(\tilde{x})$ $(n=1,2,\cdots)$.
In particular, for a state created by inserting a single operator on the vacuum, one has:
\begin{align}\label{eq:Wightman positivity gen}
    \int_{\mathbb{R}^{1,d-1}}\d^d \tilde{x}_1\,\int_{\mathbb{R}^{1,d-1}}\d^d \tilde{x}_2\, [f(\tilde{x}_1)]^\ast\, f(\tilde{x}_2)\cdot \langle\,\CO_1(\tilde{x}_1)\,\CO_2(\tilde{x}_2)\,\rangle\geq 0\ .
\end{align}

\paragraph{Analytic continuation to Euclidean signature: Schwinger functions.}
We also impose microcausality of the fields: any two local fields that are spacelike separated from each other must commute (boson) or anti-commute (fermion) inside physical correlators. For scalar fields, we have:
\begin{align}\label{eq:microcausality of the field}
    [\CO_1(\tilde{x}_1),\CO_2(\tilde{x}_2)]=0 \qquad \text{for} \quad \tilde{x}_1\approx \tilde{x}_2\quad (\text{spacelike separated})\ .
\end{align}
Microcausality combined with (complex) Lorentz transformation and the edge-of-the-wedge theorem helps to extend the analytic domain \eqref{eq:Wn analytic dom} to the wider one including the Euclidean configuration $\tilde{x}=\tilde{x}^{\mathrm{E}}=(-\i\, \tau,\vec{x})$ \cite{Osterwalder:1973dx,Osterwalder:1974tc}. This allows to analytically continue the Wightman distribution $W_n(\tilde{x}_1,\cdots,\tilde{x}_n)$ to its Euclidean counterpart (Schwinger function):
\begin{align}
\begin{aligned}
   & S_n(x_1,x_2,\cdots,x_n)\equiv W_n(\tilde{x}^{\mathrm{E}}_1,\cdots,\tilde{x}^{\mathrm{E}}_n)=\langle\,\CO_1(x_1)\,\CO_2(x_2)\,\cdots \,\CO_n(x_n)\,\rangle\ ,\\
   &x_i^\mu=(\vec{x}_i,\tau_i)\in\mathbb{R}^{d}\ ,\qquad  \mu=1,\cdots,d\ ,\qquad \d s^2=\d\vec{x}^{\,2}+\d\tau^2\ ,
\end{aligned}
\end{align}
which can be computed through the Euclidean path integral and enjoys the required properties of the Osterwalder-Schrader axioms \cite{Osterwalder:1973dx,Osterwalder:1974tc}.

Let us enumerate some features of Schwinger functions. Schwinger functions in $d\geq2$ spacetime dimension are invariant under any permutation of their arguments:
\begin{align}\label{eq:Schwinger permutation inv}
S_n(x_1,x_2,\cdots,x_n)=S_n(x_2,x_1,\cdots,x_n)=\cdots =S_n(x_n,x_{n-1},\cdots,x_1)\ ,
\end{align}
and are analytic away from the coincidence of points.
The permutation invariance \eqref{eq:Schwinger permutation inv} is inherited from the microcausality of the Wightman axiom. 
But, in the operator language, all operators should be Euclidean-time-ordered in line with the condition \eqref{eq:Wn analytic dom}:\footnote{See \eqref{eq:def of time ordering} for the definition of the symbol $\mathrm{T}_{\text{E}}$ that stands for Euclidean time ordering. We note that using the Euclidean Poincar\'e invariance $\mathbb{R}^d\rtimes\mathrm{SO}(d)$ of the Schwinger functions, we can rotate the Euclidean time axis. More precisely, because flat Euclidean spacetime $\mathbb{R}^{d}$ has no preferred directions, we can choose any direction as the Euclidean time and align operators in arbitrary order.
One should be aware that the situation is quite different in one-dimensional spacetime where the permutation invariance \eqref{eq:Schwinger permutation inv} does not hold due to the absence of spatial direction in Lorentzian spacetime.}
\begin{align}
    \langle\,\CO_1(x_1)\cdots \CO_n(x_n)\,\rangle= \langle\Omega| \,\mathrm{T}_{\text{E}}\,\{\CO_1(x_1)\cdots \CO_n(x_n)\}\,|\Omega\rangle\ .
\end{align}

\paragraph{Osterwalder-Schrader reconstruction theorem.}
We have seen how to get Euclidean QFT correlators (Schwinger functions) from Lorentzian QFT correlators (Wightman distributions). There is indeed an inverse of this manipulation: reconstruction of Wightman distributions from Schwinger functions due to the Osterwalder-Schrader reconstruction theorem \cite{Osterwalder:1973dx,Osterwalder:1974tc}. According to this theorem, one can calculate a Wightman distribution with a particular ordering:
\begin{align}
    \langle\Omega|\,\CO_1(x_1)\cdots \CO_n(x_n)\,|\Omega \rangle\ ,
\end{align}
in the following way: Start with Euclidean configuration with fixed ordering:
\begin{align}
\langle\Omega|\,\CO_1(\tilde{x}_{\mathrm{E},1})\cdots \CO_n(\tilde{x}_{\mathrm{E},n})\,|\Omega \rangle\ , \qquad \tilde{x}_{\mathrm{E},i}^{\tilde{\mu}}=(-\i \,\varepsilon_i,\vec{x}_{i,0})\ , 
\end{align}
with $\varepsilon_1>\varepsilon_2>\cdots >\varepsilon_n$ and arbitrary $\vec{x}_{i,0}$'s. Then, analytically continue the real parts of $\tilde{x}_{\mathrm{E},i}$'s to the given values: $\tilde{x}_{\mathrm{E},i}\to (t_i-\i\, \varepsilon_i,\vec{x}_{i})$ and take $\varepsilon_i\to 0$ limit, keeping the order of $\varepsilon_i$'s.

\section{Harmonic function and symmetric traceless tensor}\label{app:Harmonic function and symmetric traceless tensor}
Consider an $\mathrm{SO}(n)$ symmetric and traceless tensor of rank-$J$: $f_{i_1\cdots i_J}$. We define its encoding polynomial by contracting it with auxiliary vectors $\vec{r}\in\mathbb{C}^n$:
\begin{align}\label{eq:encoding polynomial}
    f_J(\vec{r})=f_{i_1\cdots i_J}\,r^{i_1}\cdots r^{i_J}\ .
\end{align}
Without loss of generality, we can restrict ourselves to the case $\vec{r}\in\mathbb{R}^n$, from which one can recover the argument for $\vec{r}\in\mathbb{C}^n$ by analytic continuation. 
This encoding polynomial is the homogeneous polynomial of degree-$J$ in $\vec{r}$: $f_J(\lambda \,\vec{r})= \lambda^J\cdot f_J(\vec{r})$.
The tracelessness of the original tensor $f_{i_1\cdots i_J}$ turns into the absence of the terms that are proportional to $r^2=\vec{r}\cdot\vec{r}$ in $f_J(\vec{r})$, implying that $f_J(\vec{r})$ must be some harmonic function in $d$-dimensional flat spacetime:
\begin{align}\label{eq:encoding polynomial harmonic}
    \frac{\partial}{\partial \vec{r}}\cdot  \frac{\partial}{\partial \vec{r}}\, f_J(\vec{r})=\Box_{\mathbb{R}^n}\,f_J(\vec{r})=0\ .
\end{align}
Here, we denoted the Laplace operator in $n$-dimensional flat spacetime by $\Box_{\mathbb{R}^n}$.
We remark that according to the theorem presented in section 2 of \cite{Bargmann:1977gy} and appendix A of \cite{Dobrev:1977qv}, there is a one-to-one correspondence between harmonic functions in $n$-dimension with homogeneity of degree-$J$ and rank-$J$ $\mathrm{SO}(n)$ symmetric and traceless tensors. 

Using the homogeneity of the encoding polynomials, we now define the re-scaled function depending only on the coordinate in $(n-1)$-dimensional unit sphere $\vec{\Omega}_{n-1}\in\mathbb{S}^{n-1}$:
\begin{align}\label{eq:encoding polynomial rescale}
    \hat{f}_J(\vec{\Omega}_{n-1})=r^{-J}\cdot f_J(\vec{r})\ ,\qquad \vec{r}=r\cdot \vec{\Omega}_{n-1}\ ,\qquad |\vec{\Omega}_{n-1}|=1\ .
\end{align}
Then, the Laplace equation \eqref{eq:encoding polynomial harmonic} reduces to:\footnote{Recall that the Laplacian of the metric $g_{\mu\nu}$ acts on some scalar function $f(\vec{r})$ as:
\begin{align}\label{eq:Laplacian general on scalar function}
    \Box_g\,f=\frac{1}{|\det g|^{1/2}}\cdot \partial_\mu\left(|\det g|^{1/2}\cdot g^{\mu\nu}\partial_\nu f\right)\ .
\end{align}
Because the flat space metric in the radial coordinate system is given by $\d s^2_{\mathbb{R}^n}=\d r^2+r^2\,\d s^2_{\mathbb{S}^{n-1}}$, we have:
\begin{align}\label{eq:d-dim Lap}
    \Box_{\mathbb{R}^n}\,f=\frac{1}{r^{n-1}}\cdot\partial_r(r^{n-1}\,\partial_r\,f)+\frac{1}{r^2}\cdot\Box_{\mathbb{S}^{n-1}}\,f \ .
\end{align}
}
\begin{align}\label{eq:sphere harmonic equation}
    \left[\Box_{\mathbb{S}^{n-1}}+J(J+n-2)\right]\,\hat{f}_J(\vec{\Omega}_{n-1})=0\ ,
\end{align}
with $\Box_{\mathbb{S}^{d-1}}$ being the Laplacian of $\mathbb{S}^{d-1}$.

Let us consider the simplest case where the re-scaled encoding polynomial $\hat{f}_J(\vec{\Omega}_{n-1})$ depends only on the single angular variable $\cos\theta$. Then, the differential equation \eqref{eq:sphere harmonic equation} takes the form of the Gegenbauer differential equation:\footnote{We used the fact that $\Box_{\mathbb{S}^{n-1}}$ is the Laplacian associated with the metric $\d s^2_{\mathbb{S}^{n-1}}=\d\theta^2+(\sin\theta)^2\,\d s^2_{\mathbb{S}^{n-2}}$ whose action on $f(\vec{r})$ can be computed recursively using the formula \eqref{eq:Laplacian general on scalar function}:
\begin{align}
    \Box_{\mathbb{S}^{n-1}}\,f=\frac{1}{(\sin\theta)^{n-2}}\,\partial_\theta \cdot[(\sin\theta)^{n-2}\,\partial_\theta\,f]+\frac{1}{(\sin\theta)^2}\cdot \Box_{\mathbb{S}^{n-2}}\,f\ .
\end{align}}
\begin{align}
   \left[ (1-t^2)\cdot\frac{\partial^2}{\partial t^2}-(n-1)\cdot\frac{\partial}{\partial t}+J(J+n-2)\right]\cdot \hat{f}_J(t)=0\ ,
\end{align}
with $t=\cos\theta$.
The solution regular on the interval $-1\leq \cos\theta\leq 1$ is given by:
\begin{align}
    \hat{f}_J(\cos\theta)\propto C_J^{n/2-1}(\cos\theta)\ ,
\end{align}
Here, $C_n^\nu(x)$ $(n=0,1,2,\cdots)$ is the Gegenbauer polynomial:
\begin{align}\label{eq:def of Gegenbauer polynomial}
 C_n^\nu(x)=\frac{(2\nu)_n}{n!}\cdot {}_2F_1\left(-n,2\nu+n;\nu+\frac{1}{2};\frac{1-x}{2}\right)= \frac{2^n\,(\nu)_n}{n!}\cdot x^n + \cdots\ .
\end{align}
satisfying the following orthogonality condition:
\begin{align}\label{eq:Gegenbauer integral 3 orthogonality}
           \int_0^\pi\d\theta \,(\sin\theta)^{2\nu}\cdot  C_n^\nu(\cos\theta)\, C_m^\nu(\cos\theta)=\frac{\pi\,\Gamma(n+2\nu)}{2^{2\nu-1}(n+\nu)\,\Gamma^2(\nu)\,
           n!}\,\delta_{n,m}\ .
\end{align}

\section{Decomposition of $\mathrm{SO}(1,d-1)$ symmetric and traceless projector}\label{app:projector onto SO(d-1)}
In this section, we argue a way to decompose $\mathrm{SO}(1,d-1)$ symmetric and traceless tensor onto $\mathrm{SO}(d-1)$ symmetric and traceless substructures that stabilize one future-directed time-like vector $\tilde{p}>0$ ($\tilde{p}\in\mathbb{R}^{1,d-1}$) \cite{Dobrev:1977qv}. We here do this decomposition focusing on the properties of orthogonal polynomials. See \cite[section 2.A]{Dobrev:1977qv} for more on the group-theoretical argument on the validity of this decomposition.

First of all, let us summarize the formal properties of the $\mathrm{SO}(1,d-1)$ symmetric and traceless projector of rank-$J$:
\begin{align}
   \text{Symmetry }:&\qquad   \Pi_{\tilde{\mu}_1\cdots\tilde{\mu}_J;\tilde{\nu}_1\cdots\tilde{\nu}_J}=\Pi_{\{\tilde{\mu}_1\cdots\tilde{\mu}_J\};\{\tilde{\nu}_1\cdots\tilde{\nu}_J\}}\ ,\\
   \text{Tracelessness }:&  \qquad \eta^{\tilde{\mu}_1\tilde{\mu}_2}\, \Pi_{\tilde{\mu}_1\cdots\tilde{\mu}_J;\tilde{\nu}_1\cdots\tilde{\nu}_J}=0\ ,\\
 \text{Idempotence }:&\qquad        \Pi_{\tilde{\mu}_1\cdots\tilde{\mu}_J;\tilde{\rho}_1\cdots\tilde{\rho}_J}\,\Pi^{\tilde{\rho}_1\cdots\tilde{\rho}_J}_{\tilde{\nu}_1\cdots\tilde{\nu}_J}= \Pi_{\tilde{\mu}_1\cdots\tilde{\mu}_J;\tilde{\nu}_1\cdots\tilde{\nu}_J}\ ,
\end{align}
where the indices in the braces $\{\bullet,\cdots,\bullet\}$ are symmetrized.
For the $\mathrm{SO}(1,d-1)$ symmetric traceless projector $\Pi_{\tilde{\mu}_1\cdots\tilde{\mu}_J;\tilde{\nu}_1\cdots\tilde{\nu}_J}$ of rank-$J$, we are interested in the following decomposition:
\begin{align}\label{eq:decomposition of projector}
    \begin{aligned}
     \Pi_{\tilde{\mu}_1\cdots\tilde{\mu}_J;\tilde{\nu}_1\cdots\tilde{\nu}_J}=\sum_{l=0}^{J}\, \widehat{\Pi}^{(J,l)}_{\tilde{\mu}_1\cdots\tilde{\mu}_J;\tilde{\nu}_1\cdots\tilde{\nu}_J}(\hat{p})\ .
    \end{aligned}
\end{align}
Here, $\widehat{\Pi}^{(J,l)}(\hat{p})$ $(l=0,1,\cdots,J)$ is a projector onto $\mathrm{SO}(d-1)$ symmetric and traceless subspace that leaves $\tilde{p}$ invariant.
We introduced the unit-normalized future-directed vector $\hat{p}$:
\begin{align}
    \hat{p}^{\tilde{\mu}}=\frac{\tilde{p}^{\tilde{\mu}}}{\sqrt{-\tilde{p}^2}}\ ,\qquad  \hat{p}^2=-1\ .
\end{align}
The projectors $\widehat{\Pi}^{(J,l)}(\hat{p})$ are subject to the similar properties to $\Pi_{\tilde{\mu}_1\cdots\tilde{\mu}_J;\tilde{\nu}_1\cdots\tilde{\nu}_J}$:
\begin{align}
   \text{Symmetry }:&\qquad     \widehat{\Pi}^{(J,l)}_{\tilde{\mu}_1\cdots\tilde{\mu}_J;\tilde{\nu}_1\cdots\tilde{\nu}_J}(\hat{p})=\widehat{\Pi}^{(J,l)}_{\{\tilde{\mu}_1\cdots\tilde{\mu}_J\};\{\tilde{\nu}_1\cdots\tilde{\nu}_J\}}(\hat{p})\ ,\\
   \text{Tracelessness }:&  \qquad \bm{\eta}^{\tilde{\mu}_1\tilde{\mu}_2}\,  \widehat{\Pi}^{(J,l)}_{\tilde{\mu}_1\cdots\tilde{\mu}_J;\tilde{\nu}_1\cdots\tilde{\nu}_J}(\hat{p})=0\ ,\label{eq:SO(d-1) projector Tracelessness}\\
    \text{Idempotence }:&\qquad        \widehat{\Pi}^{(J,l)}_{\tilde{\mu}_1\cdots\tilde{\mu}_J;\tilde{\rho}_1\cdots\tilde{\rho}_J}(\hat{p})\,\widehat{\Pi}^{(J,l);\tilde{\rho}_1\cdots\tilde{\rho}_J}_{\tilde{\nu}_1\cdots\tilde{\nu}_J}(\hat{p})= \widehat{\Pi}^{(J,l)}_{\tilde{\mu}_1\cdots\tilde{\mu}_J;\tilde{\nu}_1\cdots\tilde{\nu}_J}(\hat{p})\ ,\label{eq:SO(d-1) projector Idempotence}
\end{align}
but the tracelessness is restated against the reduced metric onto the directions orthogonal to $\tilde{p}$:
\begin{align}
    \bm{\eta}_{\tilde{\mu}\tilde{\nu}}=\eta_{\tilde{\mu}\tilde{\nu}}-\frac{\tilde{p}_{\tilde{\mu}} \tilde{p}_{\tilde{\nu}}}{p^2}=\eta_{\tilde{\mu}\tilde{\nu}}+\hat{p}_{\tilde{\mu}}\hat{p}_{\tilde{\nu}}\ ,\qquad \tilde{p}^{\tilde{\mu}}\,  \bm{\eta}_{\tilde{\mu}\tilde{\nu}}=0\ .
\end{align}
In addition to this, we require the orthogonality of the projectors having different degrees:
\begin{align}
 \text{Orthogonality }:&\qquad           \widehat{\Pi}^{(J,l)}_{\tilde{\mu}_1\cdots\tilde{\mu}_J;\tilde{\rho}_1\cdots\tilde{\rho}_J}(\hat{p})\,\widehat{\Pi}^{(J,l');\tilde{\rho}_1\cdots\tilde{\rho}_J}_{\tilde{\nu}_1\cdots\tilde{\nu}_J}(\hat{p})=0\quad\text{for}\quad l\neq l'\ .\label{eq:SO(d-1) projector Orthogonality}
\end{align}
We also make $\widehat{\Pi}^{(J,l)}_{\tilde{\mu}_1\cdots\tilde{\mu}_J;\tilde{\rho}_1\cdots\tilde{\rho}_J}(\hat{p})$ such that, for $1\leq l\leq J$, it is annihilated when contracted with $\hat{p}$ $(J-l+1)$-times or more:
\begin{align}\label{eq:SO(d-1) projector Nilpotency}
 \text{Nilpotency }:&\qquad      \hat{p}^{\tilde{\mu}_l}\cdots \hat{p}^{\tilde{\mu}_J} \,\widehat{\Pi}^{(J,l)}_{\tilde{\mu}_1\cdots\tilde{\mu}_J;\tilde{\nu}_1\cdots\tilde{\nu}_J}(\hat{p})=0\ \quad\text{for}\quad 1\leq l\leq J\ .
\end{align}

When $J=l$, the nilpotency condition \eqref{eq:SO(d-1) projector Nilpotency} takes the much simpler form to give:
\begin{align}\label{eq:projector J=l component}    \hat{p}^{\tilde{\mu}_1}\,\widehat{\Pi}^{(l,l)}_{\tilde{\mu}_1\cdots\tilde{\mu}_l;\tilde{\nu}_1\cdots\tilde{\nu}_l}(\hat{p})=0\ .
\end{align}
 Hence, we can regard $\widehat{\Pi}^{(l,l)}_{\tilde{\mu}_1\cdots\tilde{\mu}_l;\tilde{\nu}_1\cdots\tilde{\nu}_l}(\hat{p})$ as an SO$(d-1)$ symmetric and traceless projector of rank-$l$ in the effectively $(d-1)$-dimensional spacelike subspace orthogonal to the timelike direction $\hat{p}$.
Bearing this in mind, it is indeed possible to construct $\widehat{\Pi}^{(J,l)}_{\tilde{\mu}_1\cdots\tilde{\mu}_J;\tilde{\nu}_1\cdots\tilde{\nu}_J}(\hat{p})$ satisfying all the requirements above up to a proportionality constant $c_{J,l}$, to be determined from \eqref{eq:decomposition of projector} (or equivalently \eqref{eq:SO(d-1) projector Idempotence}):
\begin{align}\label{eq:SO(d-1) projector def}
    \begin{aligned}
         \widehat{\Pi}^{(J,l);\tilde{\nu}_1\cdots\tilde{\nu}_J}_{\tilde{\mu}_1\cdots\tilde{\mu}_J}(\hat{p})=c_{J,l}\cdot  \hat{p}_{\{\tilde{\mu}_{l+1}}\cdots \hat{p}_{\tilde{\mu}_{J}}\cdot \widehat{\Pi}^{(l,l);\{\tilde{\nu}_1\cdots\tilde{\nu}_l}_{\tilde{\mu}_1\cdots\tilde{\mu}_l\}}(\hat{p})\cdot \hat{p}^{\tilde{\nu}_{l+1}}\cdots \hat{p}^{\tilde{\nu}_{J}\}} \ .
    \end{aligned}
\end{align}
To fix the coefficient $c_{J,l}$, we contract each set of indices in the above expression with two null vectors $z_{\alpha=1,2}\in\mathbb{R}^{1,d-1}$ with $z_{\alpha}^2=0$ and translate them into the language of encoding polynomials. Firstly, the decomposition \eqref{eq:decomposition of projector} turns into:
\begin{align}\label{eq:decomposition of projector poly}
    (z_1\cdot z_2)^J=\sum_{l=0}^J\, f^{(J,l)}(\hat{p},z_1,z_2) \ ,
\end{align}
where $f^{J,l}(\hat{p},z_1,z_2)$ is a homogeneous polynomial of degree-$J$ both in $z_1$ and $z_2$:
\begin{align}\label{eq:def of fJs}
          f^{(J,l)}(\hat{p},z_1,z_2)&=z_1^{\mu_1}\cdots  z_1^{\mu_J} \,\widehat{\Pi}^{(J,l)}_{\tilde{\mu}_1\cdots\tilde{\mu}_J;\tilde{\nu}_1\cdots \tilde{\nu}_J}(\hat{p})\,z_2^{\nu_1}\cdots  z_2^{\nu_J} \ .
\end{align}

Let us first focus on the simplest case $J=l$. From \eqref{eq:SO(d-1) projector Tracelessness} and \eqref{eq:projector J=l component}, the encoding polynomial $f^{l,l}(\hat{p},z_1,z_2)$ should behave as a harmonic function of degree-$l$ both in $\bm{z}_{1;\mu}$ and in $\bm{z}_{2;\mu}$ with $\bm{z}_{i;\mu}=\bm{\eta}_{\mu\nu}\,z^\nu_i$ ($i=1,2$) on the effectively $(d-1)$-dimensional spacetime orthogonal to $\hat{p}$. Besides, other dependence comes only from the angle $\theta$ between $\bm{z}_{1;\mu}$ and $\bm{z}_{2;\mu}$ with $\cos\theta=\frac{\bm{z}_1\cdot \bm{z}_2}{|\bm{z}_1||\bm{z}_2|}$. Therefore, $f^{(l,l)}(\hat{p},z_1,z_2)$ can be expressed as follows (see appendix \ref{app:Harmonic function and symmetric traceless tensor} for details on harmonic functions):\footnote{One can fix the overall coefficient to match the leading behavior in $(\bm{z}_1\cdot \bm{z}_2)^s$. See \eqref{eq:def of Gegenbauer polynomial} for the definition of the Gegenbauer polynomial. 
The last line follows from the following identities:
\begin{align}
       z^\mu_{i}=-(z_i\cdot \hat{p})\,\hat{p}^\mu+\bm{z}^{\mu}_i\ ,\qquad |\bm{z}_i|=(-z_i\cdot \hat{p})\ ,\qquad
     \bm{z}_1\cdot \bm{z}_2=(z_1\cdot z_2)+(z_1\cdot \hat{p})(z_2\cdot \hat{p})\ .
\end{align}
}
\begin{align}\label{eq:explicit form of of fll}
\begin{aligned}
    f^{(l,l)}(\hat{p},z_1,z_2)&=c_{l,l}\cdot\frac{l!\,(|\bm{z}_1| |\bm{z}_2|)^l}{2^l\, \left(\frac{d-3}{2}\right)_l}\cdot C_l^{(\frac{d-3}{2})}\left(\frac{\bm{z}_1\cdot \bm{z}_2}{|\bm{z}_1||\bm{z}_2|}\right)\\
     &=c_{l,l}\cdot\frac{l!\,(-z_1\cdot \hat{p})^l(-z_2\cdot \hat{p})^l}{2^l\, \left(\frac{d-3}{2}\right)_l}\cdot C_l^{(\frac{d-3}{2})}\left(1-\frac{-z_1\cdot z_2}{(-z_1\cdot \hat{p})(-z_2\cdot \hat{p})}\right)   \ ,
\end{aligned}
\end{align}
Employing the explicit form of \eqref{eq:explicit form of of fll} together with \eqref{eq:SO(d-1) projector def} and \eqref{eq:def of fJs}, we arrive at:
\begin{align}\label{eq:f of J l}
\begin{aligned}
        f^{(J,l)}(\hat{p},z_1,z_2)&=c_{J,l}\cdot(-z_1\cdot \hat{p})^{J-l}(-z_2\cdot \hat{p})^{J-l}\cdot f^{(l,l)}(\hat{p},z_1,z_2)\\
        &=c_{J,l}\cdot \frac{l!\,(-z_1\cdot \hat{p})^J(-z_2\cdot \hat{p})^J}{2^l\, \left(\frac{d-3}{2}\right)_l}\cdot C_l^{(\frac{d-3}{2})}\left(1-\frac{-z_1\cdot z_2}{(-z_1\cdot \hat{p})(-z_2\cdot \hat{p})}\right) \ .
\end{aligned}
\end{align}
Then, our remaining task is to determine the normalization constant $c_{J,l}$. Setting $\hat{p}=(1,\vec{0}),z_i=(1,\vec{n}_i)$ with $\vec{n}_1\cdot \vec{n}_2=\cos\theta$, the completeness relation in terms of encoding polynomials \eqref{eq:decomposition of projector poly} becomes:
\begin{align}\label{eq:decomposition of projector poly ast}
   (-1)^J\cdot (1-\cos\theta)^J=\sum_{l=0}^J\, c_{J,l}\cdot\frac{l!}{2^l\, \left(\frac{d-3}{2}\right)_l}\cdot C_l^{(\frac{d-3}{2})}(\cos\theta) \ .
\end{align}
Comparing this expression with the identity:\footnote{We can prove this identity by using integration formula such as \cite[equation (7.391.4)]{zwillinger2014table}
combined with \cite[equation (8.962.4)]{zwillinger2014table} and the orthogonality of the Gegenbauer polynomials \eqref{eq:Gegenbauer integral 3 orthogonality}.
}
\begin{align}\label{eq:Gegenbauer expansion of 1-cos}
  (1-\cos\theta)^{m}=\sum_{n=0}^{m}\, \frac{(-1)^n\,2^{m}\,(2n+2\nu)\, (\nu+1/2)_m\,m!}{(2\nu)_{m+n+1}\,(m-n)!}\cdot C_n^\nu(\cos\theta)\ ,
\end{align}
we eventually find that:
\begin{align}\label{eq;coefficient cJl}
    c_{J,l}=(-1)^{J-l}\cdot\frac{2^{J}\,(d-3+2l)\, (d/2-1)_J\,J!}{(d-3)_{J+l+1}\,(J-l)!}\cdot \frac{2^l\,\left(\frac{d-3}{2}\right)_l}{l!}\ .
\end{align}

\section{Fourier transformation of two-point Wightman distributions}\label{app:Fourier transformation of two-point Wightman distributions}
We here review how to perform Fourier transformations of two-point Wightman distributions. As a warm-up, we start with the scalar case and move on to spinning operators.

\paragraph{Scalar primary.}
We now derive the momentum space expression of the two-point function of scalar primaries:
\begin{align}
    \langle\Omega|\,\CO_{\Delta}(\tilde{p}_1)\, \CO_{\Delta}(\tilde{p}_2)\, |\Omega\rangle\ .
\end{align}
Here, the momentum space operator $\CO_{\Delta}(\tilde{p})$ is related to $\CO_{\Delta}(\tilde{x})$ by the following Fourier transformation:
\begin{align}
    \CO_{\Delta}(\tilde{p})=\int_{\mathbb{R}^{1,d-1}}\d^d \tilde{x}\,e^{\i\,\tilde{p}\cdot \tilde{x}}\,\CO_{\Delta}(\tilde{x})\ .
\end{align}
Because of the translational invariance of the vacuum, we can extract the momentum conservation factor from the scalar two-point function:
\begin{align}
\begin{aligned}
      \langle\Omega|\,\CO_{\Delta}(\tilde{p}_1)\, \CO_{\Delta}(\tilde{p}_2)\, |\Omega\rangle=  &(2\pi)^d\,\delta^d(\tilde{p}_1+\tilde{p}_2)\cdot \int_{\mathbb{R}^{1,d-1}}\d^d x \,e^{-\i\,\tilde{p}_2\cdot \tilde{x}}\, \langle\Omega| \,\CO_{\Delta}(x) \,\CO_{\Delta}(0) \,|\Omega\rangle\ .
\end{aligned}
\end{align}
Then, we are left to calculate the following integral:\footnote{Throughout this chapter, we unit-normalize two-point functions of primary operators \eqref{eq:conformal inv of 2pt CFT correlator phy}. We used the Osterwalder-Schrader reconstruction theorem (see appendix \ref{app:Remarks on analytic properties of correlation functions in QFT}) to obtain the Lorentzian correlator in position space.}
\begin{align}
    \int_{\mathbb{R}^{1,d-1}}\d^d x \,e^{-\i\,\tilde{p}\cdot \tilde{x}}\,\langle\Omega| \,\CO_{\Delta}(x)\, \CO_{\Delta}(0)\, |\Omega\rangle= \int_{\mathbb{R}^{1,d-1}}\d^d x \, \frac{e^{-\i\,\tilde{p}\cdot \tilde{x}}}{[-(t-\i\,\varepsilon)^2+|\vec{x}|^2]^{\Delta}}\ .
\end{align}
In what follows, we compute the Lorentzian correlator from Euclidean counterpart from analytic continuation by introducing infinitesimal positive number $\varepsilon$.

We start with the identity:\footnote{
One can confirm this identity as below:
\begin{align}
\begin{aligned}
\int_{\mathbb{R}^d}&\d^d x \,\frac{e^{-\i\,x\cdot q}}{(x^2)^{\Delta}}=\mathrm{Vol}(\mathbb{S}^{d-2})\cdot\int_0^\pi\d\theta\,\sin^{d-2}\theta\,\int_{0}^{\infty}\d r\, r^{d-1-2\Delta}\, e^{-\i\,|q| r \cos\theta}\\
    &=\mathrm{Vol}(\mathbb{S}^{d-2}) \cdot \int_0^{\pi/2}\d\theta\,\sin^{d-2}\theta\,\int_{0}^{\infty}\d r\, r^{d-1-2\Delta}\cdot (e^{+\i\,|q| r\cos\theta}+e^{-\i\,|q| r\cos\theta})\\
    &=\frac{ 2^{d-2\Delta}\pi^{d/2}\,\Gamma(d/2-\Delta)}{\Gamma(\Delta)}\cdot (q^2)^{\Delta-d/2}\ ,
\end{aligned}
\end{align}}
\begin{align}\label{eq:Euclidean scalar 2pt gen}
\int_{\mathbb{R}^d}&\d^d x \,\frac{e^{-\i\,x\cdot q}}{(x^2)^{\Delta}}=\frac{2^{d-2\Delta}\pi^{d/2}\,\Gamma(d/2-\Delta)}{\Gamma(\Delta)}\,(q^2)^{\Delta-d/2}\ ,\qquad q\in\mathbb{R}^d\ .
\end{align}
Inverting this, we find that:
\begin{align}\label{eq:scalar 2pt 1}
    \frac{1}{(x^2)^{\Delta}}=\frac{ 2^{d-2\Delta}\pi^{d/2}\,\Gamma(d/2-\Delta)}{\Gamma(\Delta)}\,\int_{\mathbb{R}^d}\, \frac{\d^d q}{(2\pi)^d}\,e^{\i\,q\cdot x}\,(q^2)^{\Delta-d/2}\ .
\end{align}
We give a positive imaginary part to the $d$-th component of $x$: $x^\mu=(\vec{x},\varepsilon+\i\, t)$ with $\varepsilon>0$, and deform the integration contour of \eqref{eq:scalar 2pt 1} in the complex $p^{d}$-plain to pick up the discontinuity on the positive imaginary axis. Using the relation:\footnote{The symbol $\Theta(x)$ stands for the Heaviside step function with $\Theta(x)=1$ for $x>0$ and $\Theta(x)=0$ for $x<0$.}
\begin{align}
    (x+\i\,0)^\lambda-(x-\i\,0)^\lambda=   2\,\i\,\sin\pi\lambda\cdot (-x)^\lambda  \cdot \Theta(-x)\ ,\qquad x\in\mathbb{R}\ ,
\end{align}
we find that:
\begin{align}
\begin{aligned}
  \frac{1}{[-(t-\i\,\varepsilon)^2+ |\vec{x}|^2]^{\Delta}}=\frac{\pi^{d/2+1}2^{d-2\Delta+1}}{\Gamma(\Delta)\Gamma\left(\Delta-\frac{d-2}{2}\right)}\,\int_{\mathbb{R}^{1,d-1}}\frac{\mathrm{d}^{d}\tilde{p}}{(2\pi)^{d}}\,e^{\i\,\tilde{p}\cdot \tilde{x}}\, (-\tilde{p}^2)^{\Delta-d/2} \,\Theta(p)\ ,
\end{aligned}
\end{align}
where $\Theta(p)=\Theta(-p^2)\,\Theta(p^0)$ is a generalized function that returns one if $p>0$ and otherwise zero.
Inverting this again, one ends up with the following expression:
\begin{align}
\begin{aligned}
      \langle\Omega|\,\CO_{\Delta}(\tilde{p}_1)\, \CO_{\Delta}(\tilde{p}_2)\, |\Omega\rangle=(2\pi)^d\,\delta^d(\tilde{p}_1+\tilde{p}_2)\cdot  \frac{2^{d+1-2\Delta}\,\pi^{d/2+1}}{\Gamma(\Delta)\Gamma\left(\Delta-\frac{d-2}{2}\right)}\cdot (-\tilde{p}^2)^{\Delta-d/2}\,\Theta(\tilde{p})\ .
\end{aligned}
\end{align}

In one-dimensional Minkowski spacetime, employing analytic continuation, one finds that the two-point function of scalar primaries is given by:
 \begin{align}
     \langle\Omega| \,\CO_{\Delta}(t_1) \,\CO_{\Delta}(t_2) \,|\Omega\rangle=\frac{1}{(\varepsilon+\i\, t_{12})^{2\Delta}}\ ,
 \end{align}
Performing appropriate contour deformation, we see that:
\begin{align}\label{eq:one dimension scalar 2pt}
\begin{aligned}
      \langle\Omega|\,\CO_{\Delta}(E_1)\, \CO_{\Delta}(E_2)\, |\Omega\rangle&=  2\pi\,\delta(E_1+E_2)\cdot \int\d t \,e^{\i\,E_2\cdot t}\, \langle\Omega| \,\CO_{\Delta}(t) \,\CO_{\Delta}(0) \,|\Omega\rangle\\
    &=2\pi\,\delta(E_1+E_2)\cdot \frac{2\pi}{\Gamma(2\Delta)}\cdot E_2^{2\Delta-1}\,\Theta(E_2)\ .
\end{aligned}
\end{align}

\paragraph{Symmetric and traceless tensor.}
We unit-normalize the two-point Wightman function of traceless symmetric tensors of rank-$J$ \eqref{eq:2pt CFT spinning} and contract it with polarization vectors $z_i\in\mathbb{R}^{1,d-1}$ with $z_i^2=0,z_i\geq 0$ for $i=1,2$:
\begin{align}\label{eq:2pt spinning contracted}
    \begin{aligned}
    \langle\Omega|\,\CO_{\Delta,J} (\tilde{x}_1,z_1)\,\CO_{\Delta,J} (\tilde{x}_2,z_2)\, |\Omega\rangle=\frac{\left[\tilde{x}_{12}^2(z_1\cdot z_2)-2(x_{12}\cdot z_1)(\tilde{x}_{12}\cdot z_2)\right]^J}{[-(t_{12}-\i\,\varepsilon)^2+|\vec{x}_{12}|^2]^{\Delta+J}}\ .
    \end{aligned}
\end{align}
It is in general possible to decompose $\tilde{x}\in\mathbb{R}^{1,d-1}$ as:
\begin{align}\label{eq:decoomposition of x into two polarizations}
    \tilde{x}=\frac{\tilde{x}\cdot z_2}{z_1\cdot z_2} \,z_1+\frac{\tilde{x}\cdot z_1}{z_1\cdot z_2} \, z_2+\bm{x} \ ,\qquad \bm{x}\cdot z_1=\bm{x}\cdot z_2=0\ .
\end{align}
Then, \eqref{eq:2pt spinning contracted} takes the form:
\begin{align}
    \begin{aligned}
\langle\Omega|\,\CO_{\Delta,J} (\tilde{x}_1,z_1)\,\CO_{\Delta,J} (\tilde{x}_2,z_2)\, |\Omega\rangle=\frac{(-1)^J\, (-z_1\cdot z_2)^J\,(\bm{x}^2)^{J}}{[-(t_{12}-\i\,\epsilon_{12})^2+|\vec{x}_{12}|^2]^{\Delta+J}}\ ,
    \end{aligned}
\end{align}
whose Fourier transformation is given by:
\begin{align}\label{eq:2pt spinning contracted 2}
    \begin{aligned}
 &\langle\Omega|\,\CO_{\Delta,J} (\tilde{x}_1,z_1)\,\CO_{\Delta,J} (\tilde{x}_2,z_2)\, |\Omega\rangle\\
      &=(2\pi)^d\, \delta^d(p_1+p_2)\,\cdot (-1)^J\, (-z_1\cdot z_2)^J\cdot \int_{\mathbb{R}^{1,d-1}}\d^d x \, e^{-\i\, \tilde{p}_2 \cdot\tilde{x}}\,\frac{(\bm{x}^2)^{J}}{[-(t-\i\,\epsilon)^2+|\vec{x}|^2]^{\Delta+J}}\ .
    \end{aligned}
\end{align}
To perform the integral, similarly to the case for scalar primaries, we begin by calculating its Euclidean counterpart:
\begin{align}
    \begin{aligned}
   \int_{\mathbb{R}^{d}}& \d^d x \, e^{-\i\,q\cdot x }\,\frac{(\bm{x}^2)^{J}}{(x^2)^{\Delta+J}}=\left(-\frac{\partial}{\partial \vec{\bm{q}}}\cdot \frac{\partial}{\partial \vec{\bm{q}}}\right)^J\, \int_{\mathbb{R}^{d}} \d^d x \, e^{-\i\,q\cdot x }\,\frac{1}{(x^2)^{\Delta+J}}\\
   &=\frac{2^{d-2\Delta-2J}\pi^{d/2} \,\Gamma(d/2-\Delta-J)}{\Gamma(\Delta+J)}\cdot (-1)^J\cdot\left(\frac{\partial^2}{\partial |\bm{q}|^2}+\frac{d-3}{|\bm{q}|}\,\frac{\partial}{\partial |\bm{q}|}\right)^J\,(q^2)^{\Delta+J-d/2}\\
            &=\frac{2^{d-2\Delta}\pi^{d/2}\,\Gamma(d/2-\Delta-J)\,\left(\Delta+1-\frac{d}{2}\right)_J\left(\frac{d}{2}-\Delta\right)_J}{\Gamma(\Delta+J)}\\
      &\qquad\qquad\qquad\cdot  (\bm{q}^2)^{J}\,(q^2)^{\Delta-J-d/2}\cdot{}_2F_1\left({-J,1-J-\frac{d-2}{2}\atop\Delta-J-\frac{d-2}{2}};\frac{q^2}{\bm{q}^2}\right)\ .
    \end{aligned}
\end{align}
We used \eqref{eq:Euclidean scalar 2pt gen} and \eqref{eq:d-dim Lap} in going from the first to the second line. One can verify the last line from the principle of mathematical induction.
Inverting this Fourier integral and performing contour deformation on the complex $q^d$-plain leads:\footnote{We remark that, as $J$ is an integer, the hypergeometric function here is a polynomial and has no branch cuts. The following identities will be useful to derive \eqref{eq:spinninf FT interm}:
\begin{align}
    \frac{2\pi\,(-1)^J}{\Gamma(\Delta-J-\frac{d-2}{2})}&=2\sin\pi(d/2-\Delta+J)\,\Gamma(d/2-\Delta-J)\,\left(\Delta+1-\frac{d}{2}\right)_J\left(\frac{d}{2}-\Delta\right)_J\ ,\\
    {}_2F_1(\alpha,\beta;\gamma;z)&=(1-z)^{-\alpha}\cdot {}_2F_1\left(\alpha,\gamma-\beta;\gamma;\frac{z}{z-1}\right)\ .
\end{align}
}
\begin{align}\label{eq:spinninf FT interm}
\begin{aligned}
        &\frac{(\bm{x}^2)^{J}}{\left[-(t-\i\,\epsilon)^2+|\vec{x}|^2\right]^{\Delta+J}}= \frac{2^{d+1-2\Delta} \pi^{d/2+1} }{\Gamma\left(\Delta-J-\frac{d-2}{2}\right)\Gamma(\Delta+J)}\\
 &\quad\cdot \int_{\mathbb{R}^{1,d-1}}\d^d \tilde{p}\, e^{\i \,\tilde{p}\cdot \tilde{x}}\,\Theta(\tilde{p})\,(-\tilde{p}^2)^{\Delta-J-d/2}\,(\bm{p}^2-\tilde{p}^2)^J\cdot {}_2F_1\left({-J,\Delta-1\atop\Delta-J-\frac{d-2}{2}};\frac{\tilde{p}^2}{\tilde{p}^2-\bm{p}^2}\right)\ .
\end{aligned}
\end{align}
We then plug the above expression into \eqref{eq:2pt spinning contracted 2} and do the following replacement:
\begin{align}
    \bm{p}^2-\tilde{p}^2\mapsto  -\frac{2(-\tilde{p}\cdot z_1)(-\tilde{p}\cdot z_2)}{(-z_1\cdot z_2)}\ ,
\end{align}
After inverting the Fourier transformation, we find that:
\begin{align}\label{eq:FT of spinning Wightman}
    \begin{aligned}
    \langle\Omega|&\,\CO_{\Delta,J} (\tilde{p}_1,z_1)\,\CO_{\Delta,J} (\tilde{p}_2,z_2)\, |\Omega\rangle\\
   &=(2\pi)^d\, \delta^d(\tilde{p}_1+\tilde{p}_2)\,\Theta(\tilde{p}_2)\, (-\tilde{p}^2_2)^{\Delta-J-d/2}\cdot \frac{\pi^{d/2+1} 2^{d+1-2\Delta+J}\,(-\tilde{p}_2\cdot z_1)^J(-\tilde{p}_2\cdot z_2)^J}{\Gamma\left(\Delta-J-\frac{d-2}{2}\right)\Gamma(\Delta+J)}\\
   &\qquad\qquad\qquad\qquad\cdot {}_2F_1\left({-J,\Delta-1\atop \Delta-J-\frac{d-2}{2}};\frac{(-\tilde{p}^2_2)\,(-z_1\cdot z_2)}{2(-\tilde{p}_2\cdot z_1)(-\tilde{p}_2\cdot z_2)}\right)\ .
    \end{aligned}
\end{align}

We are now in a position to focus on the tensor structure of the two-point correlator of $\mathrm{SO
}(1,d-1)$ symmetric traceless tensors of rank-$J$ \eqref{eq:FT of spinning Wightman}, where the dependence on the future directed time-like vector $\tilde{p}_2>0$ breaks $\mathrm{SO
}(1,d-1)$ symmetry down to $\mathrm{SO}(d-1)$. 
From this perspective, it would be nicer to project out the tensor structure onto $\mathrm{SO}(d-1)$ symmetric and traceless sub-structures that keep $\hat{p}_2=\tilde{p}_2/\sqrt{-\tilde{p}_2^2}$ invariant (see appendix \ref{app:projector onto SO(d-1)} for details of this projection):
\begin{align}\label{eq:FT of spinning Wightman phys expansion}
    \begin{aligned}
   \langle\Omega|&\,\CO_{\Delta,\tilde{\mu}_1\cdots\tilde{\mu}_J} (\tilde{p}_1)\,\CO_{\Delta,\tilde{\mu}_1\cdots\tilde{\mu}_J} (\tilde{p}_2)\, |\Omega\rangle\\
   &=(2\pi)^d\, \delta^d(\tilde{p}_1+\tilde{p}_2)\,\Theta(-\tilde{p}^2_2)\,\Theta(\tilde{p}_2)\, (-\tilde{p}^2_2)^{\Delta-d/2}\,\sum_{l=0}^{J}\,\mathcal{A}_{l}(\Delta,J)\cdot \widehat{\Pi}^{(J,l)}_{\tilde{\mu}_1\cdots\tilde{\mu}_J;\tilde{\nu}_1\cdots \tilde{\nu}_J}(\hat{p}_2)\ .
    \end{aligned}
\end{align}
Equivalently:
{\small\begin{align}\label{eq:FT of spinning Wightman expansion}
    \begin{aligned}
\langle\Omega|\,\CO_{\Delta,J} (\tilde{x}_1,z_1)\,\CO_{\Delta,J} (\tilde{x}_2,z_2)\, |\Omega\rangle
   &=(2\pi)^d\, \delta^d(\tilde{p}_1+\tilde{p}_2)\,\Theta(-\tilde{p}^2_2)\,\Theta(\tilde{p}_2)\, (-\tilde{p}^2_2)^{\Delta-d/2}\\
   &\qquad\cdot \sum_{l=0}^{J}\,\mathcal{A}_{l}(\Delta,J)\cdot\frac{(-1)^{J-l}\,2^{J}\,(d-3+2l)\, (d/2-1)_J\,J!}{(d-3)_{J+l+1}\,(J-l)!}\\
   &\qquad\qquad\cdot C_l^{(\frac{d-3}{2})}\left(1-\frac{-z_1\cdot z_2}{(-z_1\cdot \hat{p})(-z_2\cdot \hat{p})}\right)\ .
    \end{aligned}
\end{align}}
To fix the expansion coefficients $\mathcal{A}_{l}(\Delta,J)$ $(l=0,1,\cdots, J)$ in \eqref{eq:FT of spinning Wightman phys expansion}, we make the following specification of parameters:
\begin{align}\label{eq:reduction to SO(d-1) specification}
    \tilde{p}_2=-\tilde{p}_1=(1,\vec{0})\ ,\qquad z_1=(1,\vec{n}_1)\ ,\qquad z_2=(1,\vec{n}_2)\ ,\qquad \vec{n}_1\cdot\vec{n}_2=\cos\theta\ .
\end{align}
Under this specification, the hypergeometric function in the second line of \eqref{eq:FT of spinning Wightman} turns into a polynomial of degree-$J$ in $\cos\theta$. Moreover, one can expand it in terms of Gegenbauer polynomials of lower degrees:\footnote{One has the following expressions for the polynomial that appears in the second line of \eqref{eq:FT of spinning Wightman expansion} and the Gegenbauer polynomial \eqref{eq:def of Gegenbauer polynomial} in terms of the Jacobi polynomials:
\begin{align}\label{eq:Jacobi integral 00}
        {}_2F_1\left({-J,\Delta-1\atop\Delta-J-\frac{d-2}{2}};\frac{1-\cos\theta}{2}\right)&=\frac{J!}{\left(\Delta-J-\frac{d-2}{2}\right)_J}\cdot P_J^{(\Delta-J-d/2,d/2-2)}(\cos\theta)\ ,\\
 C_n^\nu(x)&=\frac{(2\nu)_n}{(\nu+1/2)_n}\cdot P_n^{(\nu-1/2,\nu-1/2)}(x)\ ,
\end{align}
where we used the standard definition of the Jacobi polynomial \cite[equation (8.962.1)]{zwillinger2014table}:
\begin{align}\label{eq:Jacobi integral 000}
P_n^{(\alpha,\beta)}(x)=\frac{(\alpha+1)_n}{n!}\cdot {}_2F_1\left({-n,n+\alpha+\beta+1\atop 1+\alpha};\frac{1-x}{2}\right)\ .
\end{align}
Use the orthogonality of the Gegenbauer polynomials \eqref{eq:Gegenbauer integral 3 orthogonality} to reach the last line of \eqref{eq:FT of spinning Wightman expansion}.
The following integral formula will be of great use \cite[equation (7.391.9)]{zwillinger2014table}:
\begin{align}\label{eq:Jacobi integral 2}
    \begin{aligned}
           \int_{-1}^{1}\d x&\, (1-x)^\rho (1+x)^\beta \, P_n^{(\alpha,\beta)}(x)P_m^{(\rho,\beta)}(x)\\          &=\frac{2^{\beta+\rho+1}\,\Gamma(\alpha+\beta+m+n+1)\Gamma(\beta+n+1)\Gamma(\rho+m+1)\Gamma(\alpha-\rho-m+n)}{\Gamma(\alpha+\beta+n+1)\Gamma(\beta+\rho+m+n+2)\Gamma(\alpha-\rho)\,m!(n-m)!}\ .
           \end{aligned}
\end{align}
}
\begin{align}\label{eq:FT of spinning Wightman ast}
    \begin{aligned}
    &(\text{second line of \eqref{eq:FT of spinning Wightman}})\\
    &\xrightarrow[\text{under \eqref{eq:reduction to SO(d-1) specification}}]{}
\frac{2^{d+1-2\Delta+J}\pi^{d/2+1} }{\Gamma\left(\Delta-J-\frac{d-2}{2}\right)\Gamma(\Delta+J)}\cdot  {}_2F_1\left({-J,\Delta-1\atop\Delta-J-\frac{d-2}{2}};\frac{1-\cos\theta}{2}\right)\\
&=\frac{2^{d+1-2\Delta+J}\pi^{d/2+1} }{\Gamma\left(\Delta-J-\frac{d-2}{2}\right)\Gamma(\Delta+J)}\\
&\qquad\times\sum_{l=0}^{J}\,   \frac{ (d-3+2l)\,(d/2-1)_J\,(\Delta-1)_l(\Delta-J-d+2)_{J-l}\,J!}{(d-3)_{J+l+1}\left(\Delta-J-\frac{d-2}{2}\right)_J\, (J-l)!}\cdot C_l^{(\frac{d-3}{2})}(\cos\theta)\ .
    \end{aligned}
\end{align}
Comparing \eqref{eq:FT of spinning Wightman expansion} with \eqref{eq:FT of spinning Wightman ast}, one concludes that:
\begin{align}\label{eq:coeff of spinning Wightman decomp}
    \mathcal{A}_{l}(\Delta,J)=\frac{2^{d+1-2\Delta}\pi^{d/2+1}}{\Gamma\left(\Delta-\frac{d-2}{2}\right)\Gamma(\Delta+J)}\cdot (-1)^{J-l}\,(\Delta-1)_l(\Delta-J-d+2)_{J-l}\ .
\end{align}

\section{Unitarity bound from Wightman positivity}\label{app:Unitarity bound from Wightman positivity}
Using the results we obtained, we derive unitarity bound from Wightman positivity.

\paragraph{Scalar primary.}
Wightman positivity \eqref{eq:Wightman positivity gen} requires that:
\begin{align}\label{eq:Wightman positivity for 2pt scalar}
    &\int_{\mathbb{R}^{1,d-1}}\d^d \tilde{x}_1\,\int_{\mathbb{R}^{1,d-1}}\d^d \tilde{x}_2\, [f(\tilde{x}_1)]^\ast f(\tilde{x}_2)\, \langle\Omega| \,\CO_{\Delta}(\tilde{x}_1)\, \CO_{\Delta}(\tilde{x}_2) \,|\Omega\rangle\geq 0\ ,
\end{align}
with $f(\tilde{x})$ being a Schwartz function.
It is rather convenient to see the positivity condition \eqref{eq:Wightman positivity for 2pt scalar} in momentum space:
\begin{align}\label{eq:Wightman positivity for 2pt scalar mom}
    \begin{aligned}
     \frac{\pi^{d/2+1}2^{d+1-2\Delta}}{\Gamma(\Delta)\Gamma\left(\Delta-\frac{d-2}{2}\right)}\,\int_{\mathbb{R}^{1,d-1}}\frac{d^d \tilde{p}}{(2\pi)^d}\,|\widetilde{f}(\tilde{p})|^2 \, (-\tilde{p}^2)^{\Delta-d/2}\, \Theta(\tilde{p}^0)\,\Theta(-\tilde{p}^2) \geq0\ .
    \end{aligned}
\end{align}
We denote by $\widetilde{f}(\tilde{p})$ the momentum space expression of the Schwartz function defined through the relation:\footnote{Note that the Fourier-transformed Schwartz functions are also Schwartz functions. }
\begin{align}
    f(\tilde{x})=\int_{\mathbb{R}^{1,d-1}}\,\frac{\d^d \tilde{p}}{(2\pi)^d}\,e^{-\i\,\tilde{p}\cdot \tilde{x}}\,\widetilde{f}(\tilde{p})\ .
\end{align}

When $d>2$, this inequality \eqref{eq:Wightman positivity for 2pt scalar mom} is achieved for $\Delta>\frac{d-2}{2}$, and even for $\Delta=\frac{d-2}{2}$ with special attention to the identity for tempered distributions $\lim_{\epsilon\to 0}\,\epsilon \, x^{\epsilon-1}\,\Theta(x)=\delta(x)$:\footnote{This distributional identity can be shown by performing the Fourier transform for both sides.}
 \begin{align}
     \lim_{\Delta\to\frac{d-2}{2}}\,(\text{left-hand side of \eqref{eq:Wightman positivity for 2pt scalar mom}})=\frac{2^3\pi^{d/2+1}}{\Gamma\left(\frac{d-2}{2}\right)}\,\int_{\mathbb{R}^{1,d-1}}\frac{d^d p}{(2\pi)^d}\,|\widetilde{f}(\tilde{p})|^2 \, \delta(-\tilde{p}^2)\, \Theta(\tilde{p}^0) \geq 0\ .
 \end{align}
This inequality implies that scalar primary operators having conformal dimensions equal to the canonical (engineering) ones $(d/2-1)$ satisfy the massless free Klein-Gordon equations.

Scalar primaries must not have conformal dimensions less than the canonical ones, in which the momentum integral \eqref{eq:Wightman positivity for 2pt scalar mom} becomes singular near the origin $|p^2|\to 0$ and indefinite, resulting in a contradiction with Wightman positivity. Hence, the necessary and sufficient condition for Wightman positivity is:
\begin{align}\label{eq:scalar unitarity bound d>2}
    \Delta\geq\frac{d-2}{2}\qquad\text{for scalar primaries in } d>2 \text{ dimensions}\ ,
\end{align}
and the equality is satisfied by Klein-Gordon fields, whose momenta are on the forward light cone.

In two-dimensions ($d=2$), Wightman positivity requires $\Delta\geq0$:
\begin{align}\label{eq:Wightman positivity for 2pt scalar mom 2d}
    \begin{aligned}
     \frac{\pi^{2}2^{3-2\Delta}}{\Gamma(\Delta)^2}\,\int_{\mathbb{R}^{1,d-1}}\frac{d^2 \tilde{p}}{(2\pi)^2}\,|\widetilde{f}(\tilde{p})|^2 \, (\tilde{p}^+\tilde{p}^-)^{\Delta-1}\, \Theta(\tilde{p}^+)\,\Theta(\tilde{p}^+) \geq0\ ,
    \end{aligned}
\end{align}
where we have introduced light-cone momentum $\tilde{p}^\pm=\tilde{p}^0\pm \tilde{p}^1$.
When $\Delta=0$, we have:
\begin{align}\label{eq:Wightman positivity for 2pt scalar mom 2d 0}
    \begin{aligned}
  \lim_{\Delta\to0}\,(\text{left-hand side of \eqref{eq:Wightman positivity for 2pt scalar mom 2d}}) = \pi^{2}2^{3-2\Delta}\,\int_{\mathbb{R}^{1,d-1}}\frac{d^2 \tilde{p}}{(2\pi)^2}\,|\widetilde{f}(\tilde{p})|^2 \, \delta(\tilde{p}^+)\,\delta(\tilde{p}^-) \geq0\ .
    \end{aligned}
\end{align}
Hence, the corresponding operator has zero energy and can be identified with the identity operator $\bm{1}$.
Similarly in one-dimension ($d=1$), Wightman positivity for scalar two-point function \eqref{eq:one dimension scalar 2pt} gives:
 \begin{align}
     \frac{2\pi}{\Gamma(2\Delta)}\,\int\frac{\d E}{2\pi}\,|\widetilde{f}(E)|^2\,E^{2\Delta-1}\,\Theta(E)\geq 0\ ,
 \end{align}
tending to the following expression in the limit $\Delta\to0$:
  \begin{align}
     2\pi\,\int\frac{\d E}{2\pi}\,|\widetilde{f}(E)|^2\,\delta(E)\geq 0\qquad  \text{when }\quad\Delta=0\ .
 \end{align}
 Thus, we have 
 \begin{align}\label{eq:scalar unitarity bound d=1,2}
    \Delta\geq0\qquad\text{for scalar primaries in } d=1,2 \text{ dimensions}\ .
\end{align}
and the equality is satisfied by the identity operator $\bm{1}$.

\paragraph{Symmetric and traceless tensor.}
In the same way as the scalar case, Wightman positivity demands that:
\begin{align}\label{eq:Wightman positivity for 2pt spinning}
\begin{aligned}
        \int_{\mathbb{R}^{1,d-1}}\d^d \tilde{x}_1\,\int_{\mathbb{R}^{1,d-1}}&\d^d \tilde{x}_2\, [f^{\tilde{\mu}_1\cdots\tilde{\mu}_J}(\tilde{x}_1)]^\ast\, f^{\tilde{\nu}_1\cdots\tilde{\nu}_J}(\tilde{x}_2)\\
    &\cdot \langle\Omega| \,\CO_{\Delta,\tilde{\mu}_1\cdots\tilde{\mu}_J}(\tilde{x}_1) \,\CO_{\Delta,\tilde{\nu}_1\cdots\tilde{\nu}_J}(\tilde{x}_2) \,|\Omega\rangle\geq 0 \ ,
\end{aligned}
\end{align}
in position space.
In momentum space, this inequality takes the following form:
\begin{align}\label{eq:Wightman positivity for 2pt mom}
\begin{aligned}
     \sum_{l=0}^{J}\,(-1)^{J-l}&\,\mathcal{A}_{l}(\Delta,J)\cdot \int_{\mathbb{R}^{1,d-1}}\frac{\d^d \tilde{p}}{(2\pi)^d}\, \Theta(-\tilde{p}^2)\,\Theta(\tilde{p}^0)\, (-\tilde{p}^2)^{\Delta-d/2}\\
     &\qquad\cdot (-1)^{J-l}\, [f^{\tilde{\mu}_1\cdots\tilde{\mu}_J}(\tilde{p})]^\ast \,f^{\tilde{\nu}_1\cdots \tilde{\nu}_J}(\tilde{p})\cdot \widehat{\Pi}^{(J,l)}_{\tilde{\mu}_1\cdots\tilde{\mu}_J;\tilde{\nu}_1\cdots \tilde{\nu}_J}(\hat{p})\geq 0 \ ,
\end{aligned}
\end{align}
where the coefficients $\mathcal{A}_{l}(\Delta,J)$ ($l=0,1,\cdots,J$) are given in \eqref{eq:coeff of spinning Wightman decomp}.
We introduced the Schwartz function carrying spin indices $f^{\tilde{\mu}_1\cdots\tilde{\mu}_J}(\tilde{x})$. 
Plugging the explicit form of $\widehat{\Pi}^{(J,l)}_{\tilde{\mu}_1\cdots\tilde{\mu}_J;\tilde{\nu}_1\cdots \tilde{\nu}_J}(\hat{p})$ given in \eqref{eq:SO(d-1) projector def} and \eqref{eq;coefficient cJl} into this expression \eqref{eq:Wightman positivity for 2pt mom}, we see that:
\begin{align}\label{eq:integrand of the spinning positivity condition}
\begin{aligned}
    &(\text{second line of \eqref{eq:Wightman positivity for 2pt mom}})=\frac{2^{J}\,(d-3+2l)\, (d/2-1)_J\,J!}{(d-3)_{J+l+1}\,(J-l)!}\cdot \frac{2^l\,\left(\frac{d-3}{2}\right)_l}{l!}\\
    &\qquad\cdot 
 [f^{\{\tilde{\mu}_1\cdots\tilde{\mu}_J\}}(\tilde{p})  \cdot   \hat{p}_{\tilde{\mu}_{l+1}}\cdots \hat{p}_{\tilde{\mu}_{J}}]^\ast\cdot \widehat{\Pi}^{(l,l)}_{\tilde{\mu}_1\cdots\tilde{\mu}_l;\tilde{\nu}_1\cdots\tilde{\nu}_l}(\hat{p})\cdot [f^{\{\tilde{\nu}_1\cdots \tilde{\nu}_J\}}(\tilde{p})\cdot \hat{p}_{\tilde{\nu}_{l+1}}\cdots \hat{p}_{\tilde{\nu}_{J}}] \ ,
\end{aligned}
\end{align}
Because $\widehat{\Pi}^{(l,l)}_{\tilde{\mu}_1\cdots\tilde{\mu}_J;\tilde{\nu}_1\cdots\tilde{\nu}_J}(\hat{p})$ is the SO$(d-1)$ symmetric and traceless projector of rank $l$ onto the $(d-1)$-dimensional spacelike subspace, the second line of \eqref{eq:Wightman positivity for 2pt mom} is nothing but the absolute value of some complex function and is positive for arbitrary $f^{\mu_1\cdots\mu_J}(\tilde{x})$.

Therefore, the necessary and sufficient conditions for Wightman positivity come from the first line of \eqref{eq:Wightman positivity for 2pt mom}:
\begin{align}\label{eq:unitarity condition for spinning}
    (-1)^{J-l}\,\mathcal{A}_{l}(\Delta,J)=\frac{2^{d+1-2\Delta}\pi^{d/2+1}}{\Gamma\left(\Delta-\frac{d-2}{2}\right)\Gamma(\Delta+J)}\cdot(\Delta-1)_l\,(\Delta-J-d+2)_{J-l}\geq 0 \ ,
\end{align}
for $0\leq l\leq J$.
The condition sufficient for this inequality \eqref{eq:unitarity condition for spinning} is to require all arguments to be positive:
\begin{align}\label{eq:spinning unitarity bound d>1}
    \Delta\geq J+d-2\qquad\text{for spin-$J$ symmetric and traceless tensors}\ .
\end{align}
When the conformal dimension saturates this bound, we have:
\begin{align}
    (\Delta-J-d+2)_{J-l}\xrightarrow[]{\Delta\to J+d-2} (0)_{J-l}=\delta_{J,l}
\end{align}
Consequently, only $l=J$ component $\widehat{\Pi}^{(J,J)}_{\mu_1\cdots\mu_J;\nu_1\cdots \nu_J}(\hat{p})$ survives. Then, the nilpotency condition \eqref{eq:projector J=l component} implies the conservation law:
\begin{align}
   \tilde{p}^{\tilde{\mu}_1}\,\CO_{\Delta,\tilde{\mu}_1\cdots\tilde{\mu}_J} (\tilde{p})=0  \quad \leftrightarrow \quad   \tilde{\partial}^{\tilde{\mu}_1}\,\CO_{\Delta,\tilde{\mu}_1\cdots\tilde{\mu}_J} (\tilde{x})=0 \qquad\text{for}\quad      \Delta= J+d-2\ .
\end{align}
One still has some parameter regions where the inequality \eqref{eq:unitarity condition for spinning} can possibly be satisfied. However, one hits the singularity in the momentum space integral \eqref{eq:Wightman positivity for 2pt mom} near the light cone $\tilde{p}^2=0$ and fails to have a finite value. Therefore, we conclude that the necessary and sufficient condition for Wightman positivity for symmetric and traceless tensors is \eqref{eq:spinning unitarity bound d>1}.

\chapter{Conformal block expansion of bulk-defect-defect three-point function}\label{app:Conformal block expansion of bulk-defect-defect three-point function}
In this appendix, we derive the conformal block expansion of bulk-defect-defect three-point functions consisting of a bulk scalar primary and two SO$(d-p)$ singlet defect scalars in two ways: One is to sum the DOE of a bulk scalar to all orders. The other is to solve the Casimir differential equation using the leading behavior of the DOE as its boundary condition. 
The authors of \cite{Karch:2018uft} firstly carried out this conformal block expansion for the case of BCFT in appendix A of their paper, but with errors in choosing proper boundary conditions for the Casimir equation.
Their derivation is corrected and generalized to DCFT in \cite[appendix C]{Lauria:2020emq} by summing over the DOE to all orders.

\section{Derivation using Defect Operator Expansion}\label{app:Derivation using Defect Operator Expansion}
In what follows, we derive the conformal block expansion of the bulk-defect-defect three-point function by applying the DOE of the bulk scalar \eqref{eq:DOE of bulk scalar} inside the correlator.
First of all, notice that, of all the possible terms in the DOE of the bulk scalar, only SO$(d-p)$ singlet defect scalar primaries can contribute to the three-point functions, as the defect three-point function of the form $\langle\, \widehat{\CO}_{\widehat{\Delta}_1}\,\widehat{\CO}_{\widehat{\Delta}_2}\,\widehat{\CO}_{\widehat{\Delta},i_1\cdots i_s}\,\rangle$ must be zero identically due to the defect conformal invariance \eqref{eq:defect three-point}. 

We start by performing Fourier transformation for defect local operators $\CO_{\Delta}(x)$ with respect to parallel coordinates to the defect:
\begin{align}
\widehat{\CO}_{\widehat{\Delta}}(\hat{x})=\int\frac{\d^p \hat{q}}{(2\pi)^p}\, e^{\i\,\hat{q}\cdot \hat{x}}\cdot \widehat{\CO}_{\widehat{\Delta}}(\hat{q}) \ .
\end{align}
Then, the scalar channel DOE of $\CO_{\Delta}(x)$ can be rewritten as follows:\footnote{We have used the standard definition of the modified Bessel function of the first kind $I_\nu(z)$:
\begin{align}
    I_\nu(z)=\sum_{n=0}^{\infty}\,\frac{1}{\Gamma(\nu+n+1)\,n!}\,\left(\frac{z}{2}\right)^{\nu+2n}\ .
\end{align}}
\begin{align}\label{eq:momentum space expression btd}
\begin{aligned}
 \CO_{\Delta}(x)&\supset\frac{b( \CO_{\Delta},\widehat{\CO}_{\widehat{\Delta}})/c(\widehat{\CO}_{\widehat{\Delta}},\widehat{\CO}_{\widehat{\Delta}})}{|x_\perp|^{\Delta-\widehat{\Delta}}}\cdot \Gamma(\widehat{\Delta}+1-p/2)\cdot \left(\frac{|x_\perp|}{2}\right)^{p/2-\widehat{\Delta}} \\
 &\qquad\qquad\cdot  \int\,\frac{\d^p \hat{q}}{(2\pi)^p}\, e^{\i\,\hat{q}\cdot \hat{x}}\,|\hat{q}|^{p/2-\widehat{\Delta}}\cdot  I_{\widehat{\Delta}-p/2}(|x_\perp|\,|\hat{q}|) \cdot \widehat{\CO}_{\widehat{\Delta}}(\hat{q})\ .
\end{aligned}
\end{align}
Plugging this into the bulk-defect-defect three-point function using the expression for the partial Fourier transformed three-point functions (see e.g., \cite[equation (3.41)]{Fradkin:1996is}):\footnote{One can readily show this identity by using Schwinger parametrization \eqref{eq:Schwinger parametrization} similarly to the half-space integral \eqref{eq:Integral over half-space} (see e.g., \cite[appendix A.1]{Chen:2019fvi} for a derivation).}
{\small\begin{align}
\begin{aligned}
 \langle\, &\widehat{\CO}_{\widehat{\Delta}}(\hat{q})\,\widehat{\CO}_{\widehat{\Delta}_1}(\hat{y}_1)\,\widehat{\CO}_{\widehat{\Delta}_2}(\hat{y}_2) \,\rangle=\int\d^p \hat{x}\,e^{-\i\,\hat{q}\cdot\hat{x}}\,\langle\, \widehat{\CO}_{\widehat{\Delta}}(\hat{x})\,\widehat{\CO}_{\widehat{\Delta}_1}(\hat{y}_1)\,\widehat{\CO}_{\widehat{\Delta}_2}(\hat{y}_2) \,\rangle\\
&=\frac{c(\widehat{\CO}_{\widehat{\Delta}},\widehat{\CO}_{\widehat{\Delta}_1},\widehat{\CO}_{\widehat{\Delta}_2})}{|\hat{y}_{12}|^{\Delta^+_{12}-\widehat{\Delta}}}\cdot \frac{2\,\pi^{p/2}}{\Gamma\left(\frac{\widehat{\Delta}+\widehat{\Delta}^-_{12}}{2}\right)\Gamma\left(\frac{\widehat{\Delta}+\widehat{\Delta}^-_{21}}{2}\right)}\cdot \left(\frac{|\hat{q}|}{2\,|\hat{y}_{12}|}\right)^{\widehat{\Delta}-p/2}\\
&\quad\cdot \int_0^1\d\xi\,e^{-\i\,\hat{q}\cdot [\xi\,\hat{y}_1+(1-\xi)\,\hat{y}_2]}\cdot \xi^{\frac{\widehat{\Delta}^-_{12}+p/2}{2}-1}\,(1-\xi)^{\frac{\widehat{\Delta}^-_{21}+p/2}{2}-1}\cdot K_{\widehat{\Delta}-p/2}\left(\sqrt{\xi(1-\xi)}\,|\hat{q}|\,|\hat{y}_{12}|\right)\ ,
\end{aligned}
\end{align}}
we arrive at:
{\small\begin{align}\label{eq:three point explicit cal 1}
\begin{aligned}
     & \langle\, \CO_{\Delta}(x)\,\widehat{\CO}_{\widehat{\Delta}_1}(\hat{y}_1)\,\widehat{\CO}_{\widehat{\Delta}_2}(\hat{y}_2) \,\rangle \\
     &= \sum_{\widehat{\CO}} \,\frac{b(\CO_{\Delta},\widehat{\CO}_{\widehat{\Delta}})\,c(\widehat{\CO}_{\widehat{\Delta}},\widehat{\CO}_{\widehat{\Delta}_1},\widehat{\CO}_{\widehat{\Delta}_2})}{c(\widehat{\CO}_{\widehat{\Delta}},\widehat{\CO}_{\widehat{\Delta}})}\\
      &\qquad\cdot \frac{2\,\pi^{p/2}\,\Gamma(\widehat{\Delta}+1-p/2)}{|\hat{y}_{12}|^{\widehat{\Delta}^+_{12}-p/2}\,|x_\perp|^{\Delta-p/2}\,\Gamma\left(\frac{\widehat{\Delta}+\widehat{\Delta}^-_{12}}{2}\right)\Gamma\left(\frac{\widehat{\Delta}+\widehat{\Delta}^-_{21}}{2}\right)}\cdot \int_0^1\d\xi\,\xi^{\frac{\widehat{\Delta}^-_{12}+p/2}{2}-1}\,(1-\xi)^{\frac{\widehat{\Delta}^-_{21}+p/2}{2}-1}\\
      &\qquad\cdot\int\frac{\d^p \hat{q}}{(2\pi)^p}\,e^{\i\,\hat{q}\cdot [\xi(\hat{x}-\hat{y}_1)+(1-\xi)(\hat{x}-\hat{y}_2)]}\cdot  I_{\widehat{\Delta}-p/2}(|x_\perp|\,|\hat{q}|) \cdot K_{\widehat{\Delta}-p/2}\left(\sqrt{\xi(1-\xi)}\cdot |\hat{q}|\,|\hat{y}_{12}|\right)\ ,
\end{aligned}
\end{align}}
with $K_{\mu}(x)$ being the modified Bessel function of the second kind.
The integral over the whole $\hat{q}$-space reduces to a single integral over $|\hat{q}|$:\footnote{The symbol $\mathrm{Vol}\,(\mathbb{S}^n)=2\,\pi^{\frac{n+1}{2}}/\Gamma\left(\frac{n+1}{2}\right)$ stands for the volume of unit $n$-sphere. We also used the following formula for the Bessel function of the first kind:
\begin{align}
\begin{aligned}
      \int_0^\pi\,\d\theta\,(\sin\theta)^a\,e^{-\i\,b\,\cos\theta}&=\int_0^\pi\,\d\theta\,(\sin\theta)^a\,\cos(b\cos\theta)\\
      &=\frac{2^{a/2}\,\sqrt{\pi}\, \Gamma\left(\frac{a+1}{2}\right)}{b^{a/2}}\cdot J_{a/2}(b)\ .
\end{aligned}
\end{align}}
\begin{align}
\begin{aligned}
     \int\frac{\d^p \hat{q}}{(2\pi)^p}\, e^{-\i\,\hat{q}\cdot\hat{x}}\cdots&=\mathrm{Vol}\,(\mathbb{S}^{p-2})\,\int_0^\infty \d |\hat{q}|\, |\hat{q}|^{p-1}\, \int_0^\pi \d\theta \,(\sin\theta)^{p-2}\,e^{-\i\,|\hat{q}| \,|\hat{x}|\,\cos\theta}\cdots\\
     &=\frac{1}{|\hat{x}|^{p/2-1}}\cdot \int_0^\infty\frac{\d |\hat{q}|}{(2\pi)^{p/2}}\, |\hat{q}|^{p/2}\, J_{p/2-1}(|\hat{x}|\,|\hat{q}|)\cdots\ ,
\end{aligned}
\end{align}
Then, the last line of \eqref{eq:three point explicit cal 1} turns out to take the form of the triple Bessel integral  (see e.g., \cite[equation (6.578.11)]{zwillinger2014table}):
\begin{align}
\begin{aligned}
        \int_0^\infty\d x\, x^{\nu+1}\,K_{\mu}(ax)\,I_{\mu}(bx)\,J_{\nu}(cx)&=\frac{\Gamma(\mu+\nu)}{2^{\mu+1}\,\Gamma(\mu+1)}\cdot\frac{c^{\nu}}{(ab)^{\nu+1}\,u^{\mu+\nu+1}}\\
        &\cdot {}_2F_1\left(\frac{\mu+\nu+1}{2},\frac{\mu+\nu+2}{2};\mu+1;\frac{1}{u^2}\right)\ ,
\end{aligned}
\end{align}
with $2 a b u=a^2+b^2+c^2$.
Hence, we have:
\begin{align}
\begin{aligned}
&(\text{The last line of \eqref{eq:three point explicit cal 1}})\\
  &= \frac{\Gamma(\widehat{\Delta})\,[\xi(1-\xi)]^{-\frac{p/2}{2}}}{2^{\widehat{\Delta}+1}\,\pi^{p/2}\,\Gamma(\widehat{\Delta}-p/2+1)\,|x_\perp|^{p/2}\,|\hat{y}_{12}|^{p/2}}\cdot \frac{1}{\tilde{u}^{\widehat{\Delta}}}\cdot {}_2F_1\left(\frac{\widehat{\Delta}}{2},\frac{\widehat{\Delta}+1}{2};\widehat{\Delta}-\frac{p}{2}+1;\frac{1}{\tilde{u}^2}\right) \ ,
\end{aligned}
\end{align}
where $\tilde{u}$ is defined by the relation:
\begin{align}
    \tilde{u}=\frac{|x_\perp|^2+|\hat{x}-\hat{y}_1|^2\,\xi+|\hat{x}-\hat{y}_2|^2\,(1-\xi)}{2\,|x_\perp|\,|\hat{y}_{12}|\,\sqrt{\xi(1-\xi)}} \ .
\end{align}
To address the remaining $\xi$-integral in \eqref{eq:three point explicit cal 1}, we make use of the Mellin–Barnes representation of Gauss's hypergeometric function:
\begin{align}\label{eq:Mellin 2F1}
{}_2F_1(a,b;c;z)=\frac{\Gamma(c)}{\Gamma(a)\Gamma(b)}\cdot\int_{-\i\,\infty}^{\i\,\infty}\frac{\d s}{2\pi\i}\cdot\frac{\Gamma(a+s)\Gamma(b+s)\Gamma(-s)}{\Gamma(c+s)}\,(-z)^s\ .
\end{align}
Combining this with the following formula:
\begin{align}\label{eq:2F1 application}
    \int_0^1\d x\, \frac{x^{\mu-1}(1-x)^{\nu-1}}{[ax+b\,(1-x)+c]^{\mu+\nu}}=\frac{\Gamma(\mu)\,\Gamma(\nu)}{(a+c)^\mu (b+c)^\nu\,\Gamma(\mu+\nu)}\ ,
\end{align}
one can complete the $\xi$-integral to find that:
{\small\begin{align}\label{eq:geodesic integral}
\begin{aligned}
&(\text{The last two lines of \eqref{eq:three point explicit cal 1}})\\
          &=  T^{\widehat{\Delta}_1,\widehat{\Delta}_2}_{\Delta}(x,\hat{y}_1,\hat{y_2})\cdot \upsilon^{\widehat{\Delta}}\\
  &\qquad\cdot \frac{\Gamma(\widehat{\Delta}-p/2+1)}{\Gamma\left(\frac{\widehat{\Delta}+\widehat{\Delta}^-_{12}}{2}\right)\Gamma\left(\frac{\widehat{\Delta}+\widehat{\Delta}^-_{21}}{2}\right)}\, \int_{-\i\,\infty}^{\i\,\infty}\frac{\d s}{2\pi\i}\cdot\frac{\Gamma\left(\frac{\widehat{\Delta}+\widehat{\Delta}^-_{12}}{2}+s\right)\Gamma\left(\frac{\widehat{\Delta}+\widehat{\Delta}^-_{21}}{2}+s\right)\Gamma(-s)}{\Gamma(\widehat{\Delta}-p/2+s+1)}\cdot(-\upsilon)^s\ .
\end{aligned}
\end{align}}
One can identify the integral concerning $s$ with the Mellin–Barnes representation of Gauss's hypergeometric function \eqref{eq:Mellin 2F1}. Putting all together, we end up with the expression for the conformal block expansion of the bulk-defect-defect three-point functions presented in the main text (see the last paragraph of section \ref{sec:Correlation functions in DCFT}).

\section{Derivation from Casimir equation}
We here would like to perform conformal block expansion of the bulk-defect-defect three-point function $ \langle\, \CO_{\Delta}(x)\,\widehat{\CO}_{\widehat{\Delta}_1}(\hat{y}_1)\,\widehat{\CO}_{\widehat{\Delta}_2}(y_2) \,\rangle$ by use of the Casimir differential equation and the limiting form of the DOE:
\begin{align}\label{eq:DOE of bulk scalar lim}
    \CO_{\Delta}(x)\supset \frac{b(\CO_{\Delta},\widehat{\CO}_{\widehat{\Delta},s})/c(\widehat{\CO}_{\widehat{\Delta},s},\widehat{\CO}_{\widehat{\Delta},s})}{|x_\perp|^{\Delta-\widehat{\Delta}+s}}\cdot x^{i_1}_{\perp}\cdots x^{i_s}_{\perp}\cdot \widehat{\CO}_{\widehat{\Delta},i_1\cdots i_s}(\hat{x})+(\text{subleading in }|x_\perp|)\ .
\end{align}
We start by taking radial quantization origin on the defect such that the unit $(d-1)$-sphere includes the bulk primary, isolating the rest (see figure \ref{fig:completeness rel}).\footnote{It is always possible to go to this configuration using the defect conformal transformations.} Thanks to \eqref{eq:defect three spin}, we can ignore the contribution of defect primaries with transverse spin indices. We then expand the correlation function by inserting projectors onto conformal multiplets of defect scalar primaries $\widehat{\CO}_{\widehat{\Delta}}$:
\begin{align}\label{eq:block exp 1}
\begin{aligned}
             \langle\, \CO_{\Delta}(P)\,&\widehat{\CO}_{\widehat{\Delta}_1}(Q_1)\,\widehat{\CO}_{\widehat{\Delta}_2}(Q_2) \,\rangle\\
             &=\sum_{\widehat{\CO}}\, \langle\widehat{\CD}^{(p)}|\,\mathrm{R}\{ \widehat{\CO}_{\widehat{\Delta}_1}(Q_1)\,\widehat{\CO}_{\widehat{\Delta}_2}(Q_2)\}\,|\,\widehat{\CO}_{\widehat{\Delta}}\,|\, \CO_{\Delta}(P)\,|\widehat{\CD}^{(p)}\rangle/\langle\,\CD^{(p)}\,\rangle\ .
\end{aligned}
\end{align}
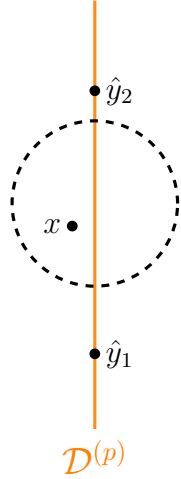
\begin{figure}[ht!]
    \centering
       \begin{tikzpicture}
          \draw[very thick, orange, opacity=0.9] (0,-3)  -- (0,2.7);    
                  \node[below, orange] at (0,-3) {\large $\CD^{(p)}$};
        
        \coordinate[label=0:$\hat{y}_1$] (A) at (0,-2) {};
        \coordinate[label=180:$x$] (B) at (-0.3,-0.3) {};

        \coordinate[label=0:$\hat{y}_2$] (C) at (0,1.5) {};
                \filldraw[black,very thick] (A) circle (0.05);
                                \filldraw[black,very thick] (B) circle (0.05);
                                                \filldraw[black,very thick] (C) circle (0.05);          
        
\draw[dashed,very thick]  (0,0) circle [radius=1.1];
     \end{tikzpicture} 
    \caption{Our choice of the radial quantization origin.}
    \label{fig:completeness rel}
\end{figure}

The Casimir operator associated with parallel conformal group $\frac{1}{2}\,\mathbf{J}^{AB}\mathbf{J}_{AB}$ acts on defect scalar primaries $\widehat{\CO}_{\widehat{\Delta}}$ as follows:\footnote{The generators of the parallel conformal group $\mathbf{J}_{AB}$ act on defect scalar primaries as in \eqref{eq:generator action embed defect local 1}. Recall that the homogeneity condition \eqref{eq:homogeneity condition defect local} implies that $Q\bullet\,\frac{\partial}{\partial Q}\,\widehat{\CO}_{\widehat{\Delta}}(Q)=-\widehat{\Delta}\,\widehat{\CO}_{\widehat{\Delta}}(Q)$.}
\begin{align}
\begin{aligned}
        \frac{1}{2}\,\mathbf{J}^{AB}\mathbf{J}_{AB}\,\widehat{\CO}_{\widehat{\Delta}}(Q)&=-Q\bullet\,\frac{\partial}{\partial Q}\,\left(p+Q\bullet\,\frac{\partial}{\partial Q}\right)\,\widehat{\CO}_{\widehat{\Delta}}(Q)\\
        &=-\widehat{\Delta}\,(\widehat{\Delta}-p)\,\widehat{\CO}_{\widehat{\Delta}}(Q)\ .
\end{aligned}
\end{align}
Combining this with the state/operator correspondence, we find that the Casimir eigenvalue of the projector is given by: 
\begin{align}\label{eq:eigenvector of defect conformal group}
\frac{1}{2}\,\mathbf{J}^{AB}\mathbf{J}_{AB}\,|\,\widehat{\CO}_{\widehat{\Delta}}\,|=-\widehat{\Delta}\,(\widehat{\Delta}-p)\,|\,\widehat{\CO}_{\widehat{\Delta}}\,|\ .
\end{align}
Let us define the conformal block associated with this bulk-defect-defect three-point function through the relation:
\begin{align}\label{eq:block exp 2}
\begin{aligned}
        \langle\widehat{\CD}|\,\CR\{ &\widehat{\CO}_{\widehat{\Delta}_1}(Q_1)\,\widehat{\CO}_{\widehat{\Delta}_2}(Q_2)\}\,|\,\widehat{\CO}_{\widehat{\Delta}}\,|\, \CO_{\Delta}(P)\,|\Omega\rangle\\
        &=\frac{ b(\CO_{\Delta},\widehat{\CO}_{\widehat{\Delta}})\,c(\widehat{\CO}_{\widehat{\Delta}},\widehat{\CO}_{\widehat{\Delta}_1},\widehat{\CO}_{\widehat{\Delta}_2})/c(\widehat{\CO}_{\widehat{\Delta}},\widehat{\CO}_{\widehat{\Delta}}) }{(P\circ P)^{\frac{\Delta}{2}} \,(-2P\bullet Q_1)^{\frac{\widehat{\Delta}^-_{12}}{2}}\,(-2P\bullet Q_2)^{\frac{\widehat{\Delta}^-_{21}}{2}}\,(-2Q_1\bullet Q_2)^{\frac{\widehat{\Delta}^+_{12}}{2}} }\cdot G^{\widehat{\Delta}^-_{12}}_{\widehat{\Delta}}(\upsilon)\ ,
\end{aligned}
\end{align}
so that it is subject to the boundary condition required from the leading DOE of $\CO_{\Delta}$ \eqref{eq:DOE of bulk scalar lim}:\footnote{Notice that $\upsilon\to0$ as $|x_\perp|\to0$.}
\begin{align}\label{eq:conformal block boundary condition}
    G^{\widehat{\Delta}^-_{12}}_{\widehat{\Delta}}(\upsilon)\xrightarrow[]{\upsilon\to0}\,\upsilon^{\widehat{\Delta}/2}\ .
\end{align}

Let us plug the $\mathrm{SO}(1,p+1)$ Casimir $\frac{1}{2}\,\mathbf{J}^{AB}\mathbf{J}_{AB}$ onto the right of $|\,\widehat{\CO}_{\widehat{\Delta}}\,|$ in \eqref{eq:block exp 2}. Then, from \eqref{eq:eigenvector of defect conformal group} and \eqref{eq:generator action embed}, one finds that:
\begin{align}
    \begin{aligned}
 \langle\widehat{\CD}|\,&\CR\{ \widehat{\CO}_{\widehat{\Delta}_1}(Q_1)\,\widehat{\CO}_{\widehat{\Delta}_2}(Q_2)\}\,|\,\widehat{\CO}_{\widehat{\Delta}}\,|\,\left(\frac{1}{2}\,\mathbf{J}^{AB}\mathbf{J}_{AB}\right)\, \CO_{\Delta}(P)\,|\Omega\rangle\\
  &=  -\widehat{\Delta}\,(\widehat{\Delta}-p)\,  \langle\widehat{\CD}|\,\CR\{ \widehat{\CO}_{\widehat{\Delta}_1}(Q_1)\,\widehat{\CO}_{\widehat{\Delta}_2}(Q_2)\}\,|\,\widehat{\CO}_{\widehat{\Delta}}\,|\, \CO_{\Delta}(P)\,|\Omega\rangle\\
  &=    \frac{1}{2}\,\CJ_{AB}(P)\,\CJ^{AB}(P)\,  \langle\widehat{\CD}|\,\CR\{ \widehat{\CO}_{\widehat{\Delta}_1}(Q_1)\,\widehat{\CO}_{\widehat{\Delta}_2}(Q_2)\}\,|\,\widehat{\CO}_{\widehat{\Delta}}\,|\, \CO_{\Delta}(P)\,|\Omega\rangle\ .
    \end{aligned}
\end{align}
This equality simplifies into the following expression:\footnote{Be aware that the differential operator $\CJ_{AB}(P)$ commutes with $P\circ P$ and $(-2Q_1\bullet Q_2)$.}
\begin{align}\label{eq:Casimir eq 1}
  \left[  \frac{1}{2}\,\CJ_{AB}(P)\,\CJ^{AB}(P)+\widehat{\Delta}\,(\widehat{\Delta}-p)\right]\cdot \frac{G^{\widehat{\Delta}^-_{12}}_{\widehat{\Delta}}(\upsilon)}{(-2P\bullet Q_1)^{\frac{\widehat{\Delta}^-_{12}}{2}}(-2P\bullet Q_2)^{\frac{\widehat{\Delta}^-_{21}}{2}} }=0\ .
\end{align}
After doing some calculations, one obtains the following differential equation for the conformal block $G^{\widehat{\Delta}^-_{12}}_{\widehat{\Delta}}(\upsilon)$:\footnote{Recall the explicit form of the defect cross ratio $\upsilon$ expressed in terms of embedding space coordinates \eqref{eq:defect cross ration emb}.}
\begin{align}\label{eq:Conformal Casimir equation}
\begin{aligned}
    & \left\{4\upsilon^2(1-\upsilon)\partial_\upsilon^2+\left[4(1-\upsilon)-2p\right]\,\upsilon\,\partial_\upsilon+(\widehat{\Delta}^-_{12})^2\,\upsilon-\widehat{\Delta}\,(\widehat{\Delta}-p)\right\}\cdot G^{\widehat{\Delta}^-_{12}}_{\widehat{\Delta}}(\upsilon)=0\ .
\end{aligned}
\end{align}
There are two solutions to \eqref{eq:Conformal Casimir equation}, but the one that favors the boundary condition \eqref{eq:conformal block boundary condition} is given by:
\begin{align}
    G^{\widehat{\Delta}^-_{12}}_{\widehat{\Delta}}(\upsilon)=\,\upsilon^{\widehat{\Delta}/2}\cdot {}_2F_1\left(\frac{\widehat{\Delta}+\widehat{\Delta}^-_{12}}{2},\frac{\widehat{\Delta}-\widehat{\Delta}^-_{12}}{2};\widehat{\Delta}+1-\frac{p}{2};\upsilon\right)\ ,
\end{align}
as anticipated.

\bibliographystyle{ytamsalpha}
\baselineskip=.95\baselineskip
\bibliography{ref}

\end{document}